\documentclass{aa}

\usepackage{morefloats}
\usepackage{multirow}
\usepackage{multicol}
\usepackage{amsmath}
\usepackage{amssymb}
\usepackage{afterpage}
\usepackage{float}
\usepackage{placeins}

\usepackage[version=3]{mhchem}   
\usepackage[nottoc, notlof, notlot]{tocbibind}   
\usepackage{enumitem}
\usepackage{graphicx}
\usepackage{color}
\usepackage{colortbl}
\usepackage{txfonts}

\def\as {\ifmmode {\rlap.}$\,$''$\,$\! \else ${\rlap.}$\,$''$\,$\!$\fi}

\bibpunct{(}{)}{;}{a}{}{,}             

\begin{document}
\title{Small-scale physical and chemical structure of diffuse  and \\translucent molecular clouds along the line of sight to Sgr B2}
\author{V. Thiel\inst{1} \and A. Belloche\inst{1} \and K. M. Menten\inst{1} \and A. Giannetti\inst{2} \and H. Wiesemeyer\inst{1} \and B. Winkel\inst{1} \and P. Gratier\inst{3} \and H. S. P. M\"uller\inst{4} \and \\D. Colombo\inst{1} \and R. T. Garrod\inst{5}}
  \institute{Max-Planck-Institut f\"ur Radioastronomie, Auf dem H\"ugel 69, 53121 Bonn, Germany \and INAF-Istituto di Radioastronomia, and Italian ALMA Regional Centre, via P. Gobetti 101, 40129 Bologna, Italy \and Laboratoire d'Astrophysique de Bordeaux, Univ. Bordeaux, CNRS, B18N, All\'ee Geoffroy Saint-Hilaire, 33615 Pessac, France \and I. Physikalisches Institut, Universit\"at zu K\"oln, Z\"ulpicher Str. 77, 50937 K\"oln, Germany \and Departments of Chemistry and Astronomy, University of Virginia, Charlottesville, VA 22904, USA}

\abstract
{The diffuse and translucent molecular clouds traced in absorption along the line of sight to strong background sources have so far been investigated mainly in the spectral domain  because of limited angular resolution or small sizes of the background sources.}
{We aim to resolve and investigate the spatial structure of molecular clouds traced by several molecules detected in absorption along the line of sight to Sgr\,B2(N).}
{We have used spectral line data from the EMoCA survey performed with the Atacama Large Millimeter/submillimeter Array (ALMA), taking advantage of its high sensitivity and angular resolution. The velocity structure across the field of view is investigated by automatically fitting synthetic spectra to the detected absorption features, which allows us to decompose them into individual clouds located in the Galactic centre (GC) region and in spiral arms along the line of sight. We compute opacity maps for all detected molecules. We investigated the spatial and kinematical structure of the individual clouds with statistical methods and perform a principal component analysis to search for correlations between the detected molecules. To investigate the nature of the molecular clouds along the line of sight to Sgr\,B2, we also used archival Mopra data.}
{We identify, on the basis of c-C$_3$H$_2$, 15 main velocity components along the line of sight to Sgr~B2(N) and several components associated with the envelope of Sgr~B2 itself. The c-C$_3$H$_2$ column densities reveal two categories of clouds. Clouds in Category I (3~kpc arm, 4~kpc arm, and some GC clouds) have smaller c-C$_3$H$_2$ column densities, smaller linewidths, and smaller widths of their column density PDFs than clouds in Category II (Scutum arm, Sgr arm, and other GC clouds). We derive opacity maps for the following molecules: c-C$_3$H$_2$, H$^{13}$CO$^+$, $^{13}$CO, HNC and its isotopologue HN$^{13}$C, HC$^{15}$N, CS and its isotopologues C$^{34}$S and $^{13}$CS, SiO, SO, and CH$_3$OH. These maps reveal that most molecules trace relatively homogeneous structures that are more extended than the field of view defined by the background continuum emission (about $15\arcsec$, that is 0.08 to 0.6~pc depending on the distance). SO and SiO show more complex structures with smaller clumps of size $\sim$5--8$\arcsec$. Our analysis suggests that the driving of the turbulence is mainly solenoidal in the investigated clouds.}
{On the basis of HCO$^+$, we conclude that most line-of-sight clouds towards Sgr~B2 are translucent, including all clouds where complex organic molecules were recently detected. We also conclude that CCH and CH are good probes of H$_2$ in both diffuse and translucent clouds, while HCO$^+$ and c-C$_3$H$_2$ in translucent clouds depart from the correlations with H$_2$ found in diffuse clouds.}

\titlerunning{Structure of diffuse and translucent clouds towards Sgr\,B2}
\authorrunning{Thiel et al.}

\maketitle

\section{Introduction}
Molecular clouds can be categorised based on their physical conditions into dense, translucent, and diffuse molecular clouds \citep[][and references therein]{snow2006}. The boundaries between the three different phases are loose. Dense molecular clouds are mostly protected from UV radiation which can destroy molecules, while diffuse molecular clouds are more exposed to this radiation which often results in a lower molecular fraction of hydrogen and lower abundances of molecules. Translucent molecular clouds are the transition regions between dense and diffuse molecular clouds, not completely shielded against UV radiation. Molecular clouds are usually a mixture of all these three types. The kinetic temperature is between about 30 and 100\,K in diffuse molecular clouds, higher than about 15\,K in translucent clouds, and between about 10\,K and 50\,K in dense clouds. The typical hydrogen densities are $100-500$\,cm$^{-3}$, $500-5000$\,cm$^{-3}$ and $>10^4$\,cm$^{-3}$, respectively \citep[][and references therein]{snow2006}. 

Performing absorption studies in the direction of strong background sources offers the opportunity to study the chemical and physical structure of diffuse and translucent molecular clouds along the line of sight. Diffuse molecular clouds make up a large part of the interstellar medium in our galaxy and in other spiral galaxies \citep[e.g.][]{pety2013}. Their extended structures are thought to be the main component of interarm regions in spiral galaxies \citep{sawada2012}. A thick diffuse disk may be present in spiral galaxies, as detected in M51 \citep{pety2013}. In addition, diffuse and translucent molecular clouds form the envelopes of giant molecular clouds (GMCs) in which star formation occurs. Hence, diffuse and translucent molecular clouds play an important role for the interaction between stars and the surrounding gas \citep[e.g.][]{arnett1971}. 

Due to the low densities in diffuse and translucent molecular clouds the excitation temperature of most molecular transitions is close to the temperature of the cosmic microwave background (CMB) radiation, that is 2.73 K \citep{greaves1992}. Rotational lines are thus sub-thermally excited, very weak, and difficult to detect in emission. 

The GMC Sagittarius B2 (Sgr\,B2) emits strong continuum radiation that can be used as an extended background source to investigate the spatial structure of the diffuse and translucent clouds located along the line of sight. Sgr~B2 is located near the Galactic centre (GC) with a projected distance of about 100\,pc. The GC has a distance of $8.34\pm0.16$\,kpc to the Sun \citep{reid2014}. The diameter of Sgr\,B2 is about $40$\,pc and its mass is about $10^7$M$_\odot$ \citep{lis1990}. Here, we focus on the dense molecular core Sgr\,B2(N) that contains several \ion{H}{II} regions \citep{gaume1995} as well as several hot molecular cores \citep{bonfand2017, sanchez2017}. The continuum emission of Sgr~B2(N) in the millimetre wavelength range consists of free-free radiation and thermal dust emission \citep[e.g.][]{liu1999}.

In the past, several molecular absorption studies along the line of sight to Sgr\,B2(N) and Sgr\,B2(M) were made using single-dish telescopes \citep[e.g.][]{greaves1994,neufeld2000,polehampton2005,hieret2005,lis2010,monje2011,corby2018}. The profiles of the detected absorption features were modelled to investigate the molecular content of the material along the line of sight. The angular resolution of these previous studies was not high enough to resolve the continuum structure of Sgr\,B2(N). For instance, \citet{corby2018} investigated simple molecules along the line of sight to Sgr\,B2 using the Green Bank Telescope. Their data covered the frequency range between 1 and 50\,GHz with a resolution between $13^{\prime}$ and $15^{\prime\prime}$. 
\citet{corby2015} performed a spectral survey of Sgr~B2 with the Australia Telescope Compact Array (ATCA) between 30 and 50~GHz with an angular resolution of 5--10$\arcsec$ that starts to resolve the continuum emission of Sgr~B2(N). They reported variations in the column densities of several molecules seen in absorption across the field of view of their observations but they did not have enough resolution elements to perform a detailed study of the spatial structure of the clouds seen in absorption along the line of sight. \citet{mills2018} also used the ATCA between 23 and 37 GHz to observe Sgr~B2(N) with an angular resolution of 3$\arcsec$ in ammonia and methanol, which they detect mainly in emission. They focused their analysis on Sgr~B2(N) itself and its hot cores. 

\begin{figure}[t]
   \resizebox{\hsize}{!}{\includegraphics[width=0.5\textwidth,trim = 2.cm 2.0cm 8.9cm 1.8cm, clip=True]{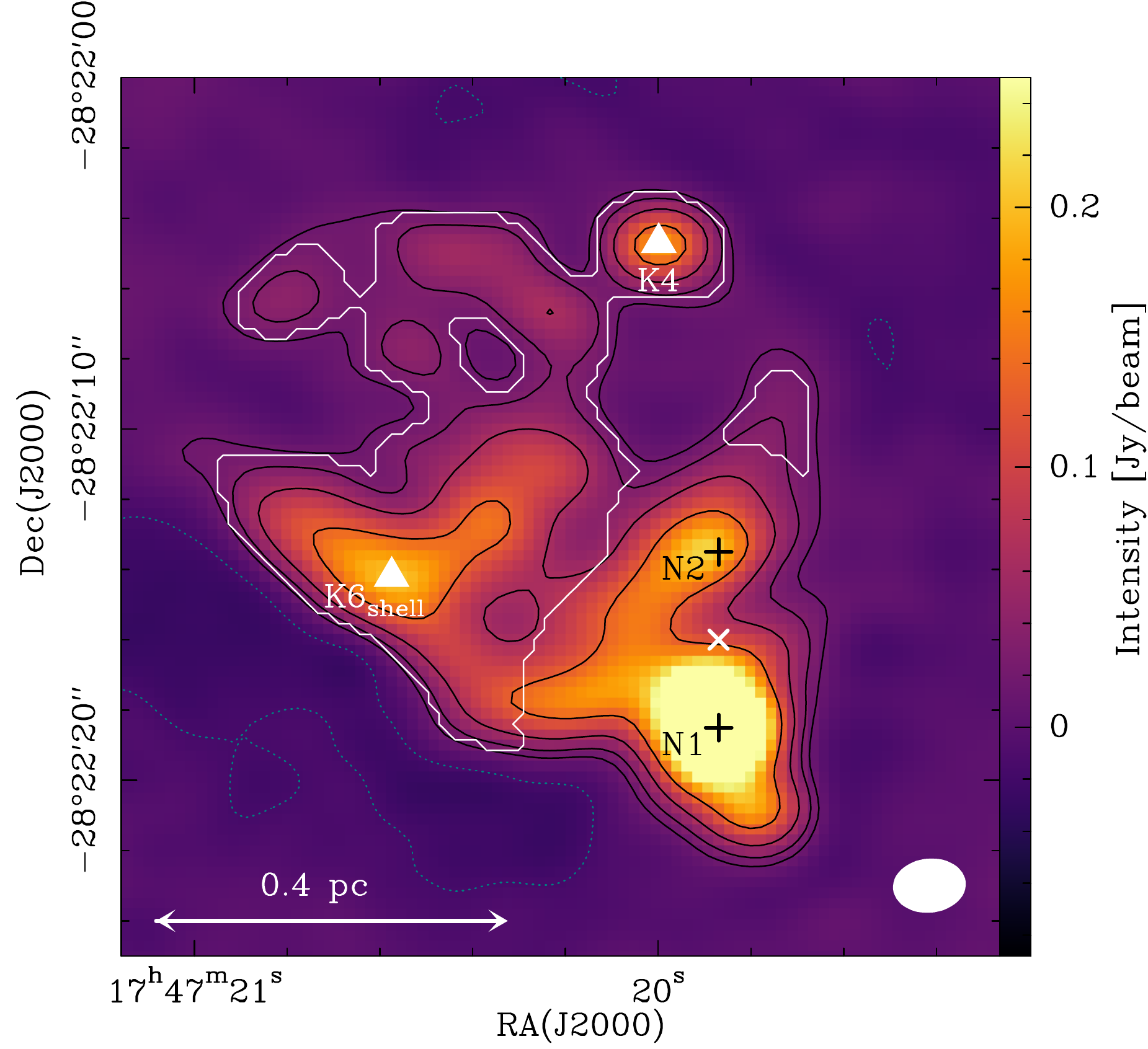}}
 \caption{ALMA continuum map of Sgr\,B2(N) at 85\,GHz. The black contour lines show the flux density levels at 3$\sigma$, 6$\sigma$, 12$\sigma$, and 24$\sigma$ and the dotted ones mark -3$\sigma$, with $\sigma$ the rms noise level of 5.4~mJy~beam$^{-1}$. The black crosses mark the positions of the hot cores Sgr~B2(N1) and Sgr~B2(N2), the white cross the phase centre (EQ\,J2000: $17^\mathrm{h}47^\mathrm{m}19.87^\mathrm{s},-28^\circ$22$^{\prime}$16$^{\prime\prime}$), and the white triangles the ultra compact HII region K4 and the peak in the shell of the HII region K6 \citep{gaume1995}. The white ellipse in the lower right corner is the synthesised beam. The white contour encloses the region selected for the analysis of the absorption features, for the particular case of ortho c-C$_3$H$_2$. The pixel size in this image is $0\as3$.}\label{contmap}
\end{figure}

\begin{figure}[t]
   \resizebox{\hsize}{!}{\includegraphics[width=0.5\textwidth,trim = 0.cm 14.cm 0.cm 3.5cm, clip=True]{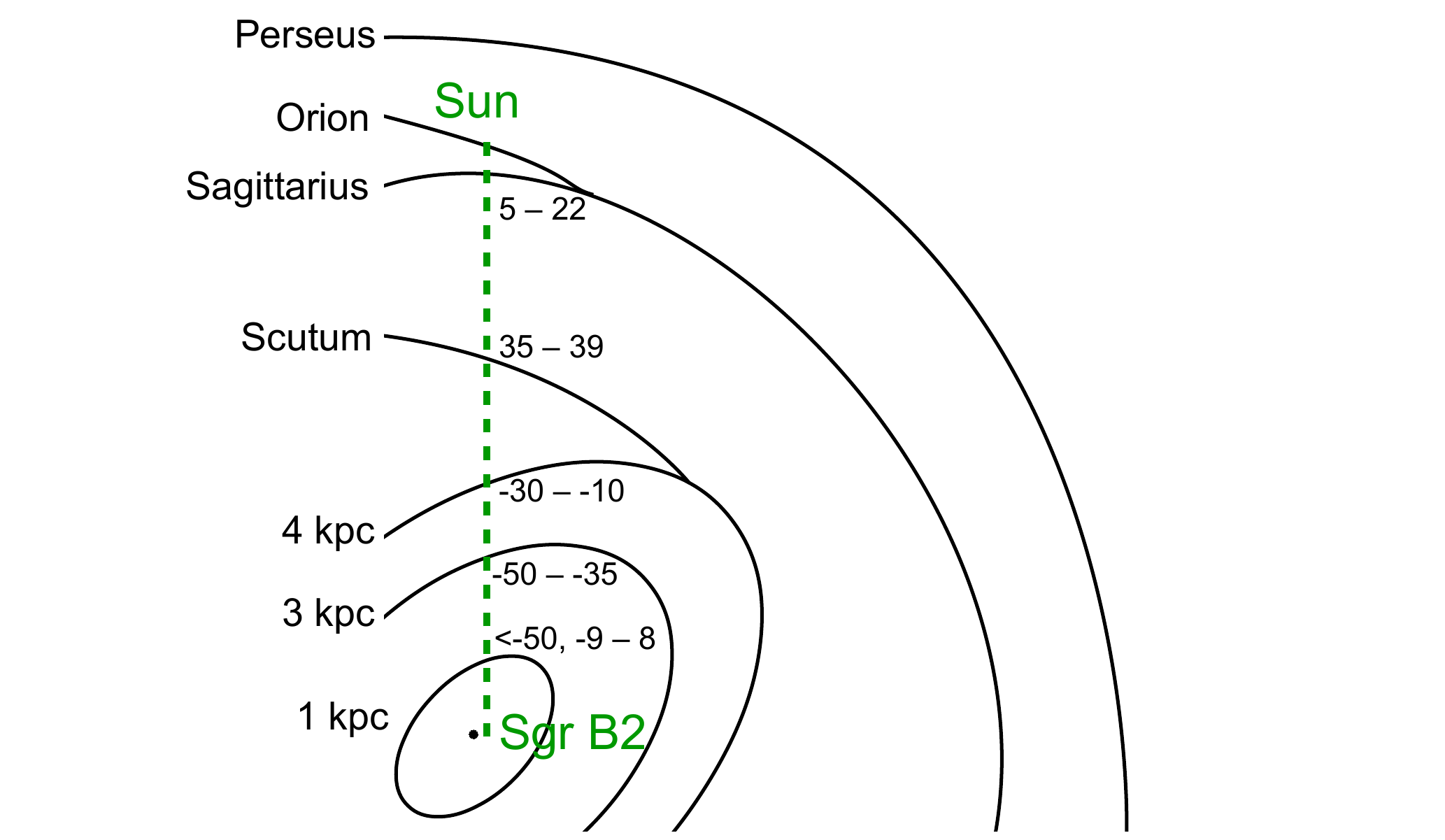}}
 \caption{Sketch of the Milky Way and the spiral arms along the line of sight to Sgr\,B2 \citep[based on][]{greaves1994}. The velocity ranges of the diffuse and translucent clouds are indicated in km~s$^{-1}$.}\label{sketch_mw}
\end{figure}

We use the EMoCA (\textit{Exploring Molecular Complexity with ALMA}) survey for our analysis. The aim of the survey is to explore and expand our knowledge of the chemical complexity of the interstellar medium \citep{belloche2016}. This survey was performed towards Sgr B2(N) with the Atacama Large Millimeter/submillimeter Array (ALMA). The angular resolution of this survey is high enough to resolve the continuum emission of Sgr B2(N) (see Fig.~\ref{contmap}). Hence, we can investigate absorption lines at positions where the continuum is still strong enough but which are sufficiently far away from the hot cores towards which absorption features are blended with numerous emission lines. The survey was carried out in the 3\,mm wavelength range (covering frequencies from 84.1 to 114.4\,GHz). Many important as well as abundant molecular species have transitions in this frequency regime that are suitable for absorption studies. Therefore, this unbiased line survey provides an excellent opportunity to study structures on sub-parsec scales not only in Sgr B2 itself, but also along the whole 8\,kpc long line of sight to the Galactic centre. The line of sight to Sgr\,B2 passes through the Sagittarius, Scutum, 3\,kpc, and 4\,kpc arms as well as the Galactic centre (GC) clouds up to a distance of about 2\,kpc from the GC \citep[e.g.][see Fig.~\ref{sketch_mw}]{greaves1994, menten2011}.

In this paper we investigate the spatial structure of the molecular clouds traced by several molecules detected in absorption along the line of sight to Sgr\,B2 at much better angular resolution, namely $1\as6$. In Sect.~\ref{sect_obs} we briefly describe the dataset we used for this work. The different techniques we adopted to analyse the data are presented in Sect~\ref{sect_method}. We present the results in Sect.~\ref{sect_result} and discuss them in Sect.~\ref{sect_discuss}. We give a summary in Sect.~\ref{sect_conclusion}.
\section{Observations}\label{sect_obs}
We analysed the absorption lines detected in the EMoCA survey \citep{belloche2016}. This spectral line survey was observed with ALMA in Cycles 0 and 1. It was pointed towards Sgr\,B2(N) with the phase centre located half way between the two main hot cores N1 and N2 at EQ\,J2000: $17^\mathrm{h}47^\mathrm{m}19.87^\mathrm{s},-28^\circ$22$^{\prime}$16$^{\prime\prime}$ (see Fig.~\ref{contmap}). The survey covers the frequency range from 84.1 to 114.4\,GHz with a spectral resolution of 488 kHz (1.7 to 1.3 km~s$^{-1}$) at a median angular resolution of $1\as6$. The median largest angular scale is $21\as4$. The average noise level is \textasciitilde 3\,mJy\,beam$^{-1}$ per channel. Details about the calibration and deconvolution of the data are reported in \citet{belloche2016}.

We also analyse the emission lines of the molecules HCO$^+$, HNC, CS, and $^{13}$CO detected in the 3\,mm imaging spectral survey of Sgr\,B2 performed by \citet{jones2008}. The observations were carried out with the 22\,m Mopra Millimetre Telescope in June 2006 in on-the-fly mode, covering an area of 5$\arcmin$ by 5$\arcmin$ centred on the J2000 equatorial position $17^\mathrm{h}47^\mathrm{m}19.8^\mathrm{s},-28^\circ$22$^{\prime}$17$^{\prime\prime}$ which is close to Sgr B2(N). The survey covers the full frequency range between 82 and 114\,GHz with a spectral resolution of 2.2~MHz (6.4\,km\,s$^{-1}$ at 100 GHz). Additional narrow-band spectra with a high resolution of 33 kHz (0.10\,km\,s$^{-1}$ at 100~GHz) were also taken. The angular resolution of the data is about $36^{\prime\prime}$ and the RMS noise level of the broad-band spectra in main-beam temperature scale is 0.12--0.42~K depending on the frequency.

\section{Methods}\label{sect_method}
\subsection{Selected data sample}\label{data_sample}
To have enough sensitivity, we analyse the absorption features towards positions where the continuum emission is brighter than four times the RMS noise level. The noise level is derived from a Gaussian fit to the flux density distribution of all pixels in the continuum map not corrected for primary beam attenuation. For this, we use the command \textit{go noise} in the GILDAS package GREG\footnote{see https://www.iram.fr/IRAMFR/GILDAS/}. Positions close to the hot cores Sgr~B2(N1) and Sgr~B2(N2) have spectra full of emission lines of organic molecules \citep[e.g.][]{bonfand2017}. Therefore, in order to minimise the contamination of the absorption features by emission lines, we perform the analysis towards the positions that are far enough from these hot cores by excluding pixels inside ellipses drawn around them. Because of the slightly different beams and noise levels in the different spectral windows \citep[see][]{belloche2016}, the mask resulting from these two criteria may differ from molecule to molecule. As an example, we show the area selected for c-C$_3$H$_2$ on top of the ALMA continuum map in Fig.~\ref{contmap}. The selected data sample has no emission lines contaminating the absorption features. With a sampling of $0.6\arcsec$, the Nyquist-sampling condition is still fulfilled and we do not lose information. This results in 322 pixels for c-C$_3$H$_2$, which represents an area of about 32 independent beams.

\subsection{Opacity cubes}\label{opac-cubes}
When the excitation temperature of a transition seen in absorption is equal to the temperature of the CMB (see Eq.~\ref{equation_radiative_transfer} below), the opacity, $\tau$, of the absorption line is directly related to the line intensity, $I_\mathrm{l}$, and the continuum level, $I_\mathrm{c}$, through the following equation:
\begin{equation}
\tau(\nu) = -\ln\left(1+\frac{I_\mathrm{l}(\nu)}{I_\mathrm{c}}\right),
\end{equation}
where $I_\mathrm{c}$ is the level of the baseline (representing the continuum emission) in the original spectrum and $I_\mathrm{l}(\nu)$ is the intensity of the absorption line measured in the baseline-subtracted spectrum at a certain frequency. With this definition, $I_\mathrm{l}$ is negative for an absorption line. This formula only yields meaningful values for $\tau$ when the absorption is not too optically thick, otherwise the value in the parentheses gets close to zero and the logarithm diverges. We can then compute the column density of the molecule from the derived opacity (see Eq.~\ref{equation_tau_weeds} below). 

Because the size of our data sample is small, we create many realisations of the opacity cube by injecting noise to $I_l$ and $I_c$ (with $\sigma_{I_l}=\sigma_\mathrm{RMS}\sim 3$\,mJy\,beam$^{-1}$ and $\sigma_{I_c}<<\sigma_{I_l}$) in order to evaluate the impact of the noise on our subsequent analyses. Thereby, we assume the uncertainties on $I_l$ and $I_c$ to have a Gaussian distribution. $I_c$ contributes only little to the uncertainty of $\tau$. Using this assumption we randomly create 1000 opacity cubes from the original line intensity cube. We set the opacity of all pixels with $-I_\mathrm{l}\geq I_\mathrm{c}$ to infinity. Due to the tolerance limit of \texttt{python}\footnote{\url{https://www.python.org}}, the opacity of pixels with $0 < 1+\frac{I_\mathrm{l}}{I_\mathrm{c}} < 10^{-16}$ is also set to infinity. The upper limit corresponds to an opacity of 37.

We keep all pixels with  $I_\mathrm{l}\leq 3.1\sigma_\mathrm{RMS}$. The resulting data still contain noisy pixels. With the assumption of a Gaussian noise distribution, this threshold means that less than $0.1\%$ of pure-noise pixels are excluded (the ones with $I_\mathrm{l}\geq 3.1\sigma_\mathrm{RMS}$). In addition, we use the error propagation law to create a cube containing the uncertainties on the opacity, $\sigma_\tau$.  
\subsection{Modelling}\label{modelling}
In order to identify and characterise the velocity components present in the absorption spectra, we fit synthetic spectra consisting of a collection of Gaussian opacity distributions.
We model the spectra with Weeds \citep{maret2011} which solves the radiative transfer equation under the assumption of local thermodynamic equilibrium and takes into account the finite angular resolution of the observations. 

Because our data sample contains several hundreds of spectra per molecule, the spectra are fitted automatically. For this, we use the fitting routine MCWeeds \citep{giannetti2017}, which combines the python package PyMC2 \citep{patil2010} and Weeds \citep{maret2011}. MCWeeds adjusts the parameters for a given number of velocity components and delivers the best result along with uncertainties. For all fitted parameters a set of initial guesses has to be given, along with their probability distribution and the range over which they should be varied.

The synthetic spectra are computed by Weeds in the following way. For a baseline-subtracted spectrum, the intensity of an absorption line in a medium with constant excitation filling the beam is:  

\begin{equation}\label{equation_radiative_transfer}
T_\mathrm{B}(\nu) = \left[J_\nu(T_\mathrm{ex})-T_\mathrm{c,\nu}-J_\nu(T_\mathrm{CMB})\right]\cdot\left(1-\mathrm{e}^{-\tau(\nu)}\right),
\end{equation}
with $T_\mathrm{B}(\nu)$ the brightness temperature at the frequency $\nu$, $T_\mathrm{CMB}$ the CMB temperature, $T_\mathrm{ex}$ the excitation temperature of the line, $T_\mathrm{c,\nu}$ the baseline level in the spectrum before baseline-subtraction, $\tau(\nu)$ the opacity, and $J_\nu(T) = \frac{h\nu/k}{\mathrm{e}^{h\nu/kT} -1}$, with $h$ the Planck constant and $k$ the Boltzmann constant.
The opacity $\tau(\nu)$ is calculated as: 
\begin{equation}\label{equation_tau_weeds}
\tau(\nu) = \frac{c^2}{8\pi\nu^2}\frac{A g_u}{Q(T_\mathrm{ex})} \sum_i N_\mathrm{tot}^i \mathrm{e}^{-E_u/kT_\mathrm{ex}}\left(\mathrm{e}^{h\nu_0/kT_\mathrm{ex}}-1\right)\phi^i,
\end{equation}
with $c$ the speed of light, $N_\mathrm{tot}^i$ the column density of the molecule, $Q(T_\mathrm{ex})$ the rotational partition function at temperature $T_\mathrm{ex}$ (in LTE, the rotational temperature is equal to the excitation temperature). $A$ the Einstein coefficient for spontaneous emission of the transition, $g_u$ the degeneracy factor of the upper level, $E_u$ the upper level energy, $\nu_0$ the rest frequency, $\phi^i$ the line profile function, and $\sum_i$ the summation over the velocity components contributing to the absorption. The line profile function is assumed to be Gaussian:
\begin{equation}
\phi^i = \frac{1}{\sqrt{2\pi}\sigma_i}\mathrm{e}^{-\frac{(\nu+\frac{-\Delta \nu}{\Delta \varv}\Delta \varv_\mathrm{off}^i-\nu_0^i)^2}{2\sigma_i^2}}, 
\end{equation}
with $\sigma_i$ the standard deviation of the Gaussian, $\Delta \varv_\mathrm{off}^i$ the velocity offset of the velocity component, $\Delta \varv$ the channel width in velocity, and $\Delta \nu$ the channel width in frequency. From $\sigma_i$ the full width at half maximum ($FWHM_i$) in velocity units can be calculated:
\begin{equation}
FWHM_i = \frac{c}{\nu_0^i}\sqrt{8\ln 2}\cdot\sigma_i.
\end{equation}
We assume the excitation temperature to be equal to the temperature of the CMB (2.73\,K) because we focus on the diffuse and translucent clouds along the line of sight to Sgr\,B2 excluding those physically associated with Sgr~B2 itself. For comparison, \citet{godard2010} determined a range of excitation temperatures of 2.7--3\,K  using transitions of HCO$^+$ for four different lines of sight. Previous absorption studies also assumed excitation temperatures in this range \citep[e.g.][]{greaves1994,lucas1999,liszt2012,wiesemeyer2016,ando2016}. 

\begin{figure}[t]
     \resizebox{\hsize}{!}{\includegraphics[width=0.5\textwidth,trim = 1.7cm 0.5cm 3cm 13.7cm, clip=True]{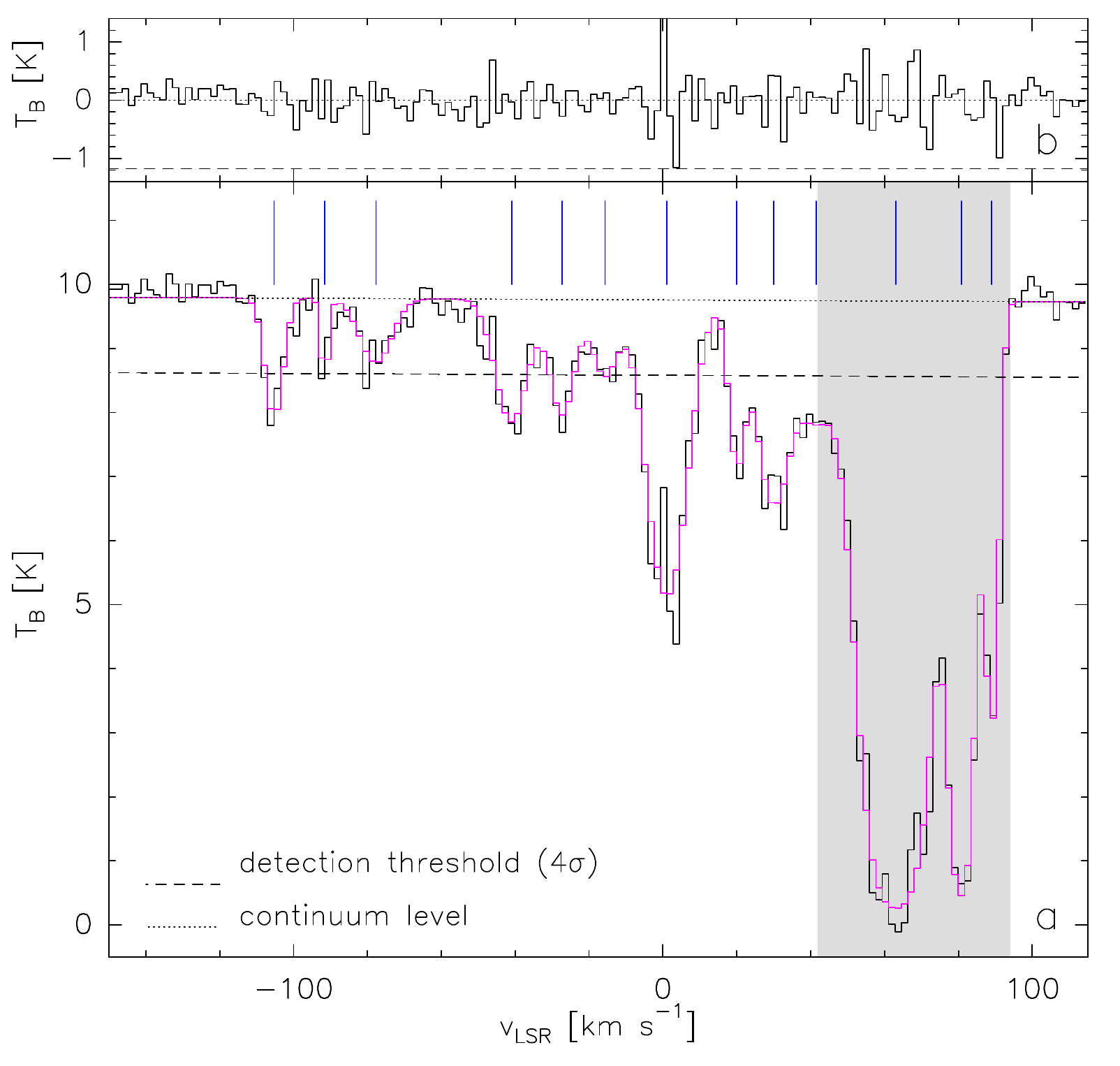}}
 \caption{\textbf{a} ALMA spectrum of ortho c-C$_3$H$_2$ $2_{1,2}-1_{0,1}$ at 85338.896\,MHz towards the UCHII region K4 \citep{gaume1995}. The spectrum in magenta is the synthetic spectrum obtained with MCWeeds. The blue lines show the central velocities of the fitted components. The dotted line represents the continuum level and the dashed line marks the 4$\sigma$ line detection threshold. The grey area marks the velocity range of the envelope of Sgr\,B2 ($\varv_\mathrm{LSR}>42$~km\,s$^{-1}$). \textbf{b} Residuals, that is the difference between the observed and synthetic spectra. The dashed lines indicate the $-4\sigma$ level. The spectrum is corrected for the primary beam attenuation.}\label{examplespec}
\end{figure}

The fitted parameters are the column density $N_\mathrm{tot}$, the width $FWHM$, and the centroid velocity $\varv_0$ of each velocity component. An example of synthetic spectrum of ortho c-C$_3$H$_2$ $2_{1,2}$--$1_{0,1}$ fitted with MCWeeds towards the ultracompact (UCHII) region K4 \citep{gaume1995} is shown in Fig.~\ref{examplespec}. It contains 13 velocity components.

\subsection{Minimisation method}\label{powellmethod}
The absorption features detected in our data consist of many velocity components (see Fig.~\ref{examplespec}). Therefore, many parameters have to be fitted at the same time. A good method to deal with a large number of free parameters is the Powell minimisation method \citep[][see Appendix~\ref{appendix_powell} for more details]{powell1964}. We use a modified version of this method (\textit{fmin\_powell} from \texttt{scipy}\footnote{see https://scipy.org}) with MCWeeds. Because this method finds a local minimum of the problem, good initial guesses have to be given to the fitting routine (see Sect.~\ref{program}). 
\subsection{Automatisation programme}\label{program}

We wrote a \texttt{python} programme that searches for appropriate initial guesses and runs the minimisation with MCWeeds in an automatic way for all selected positions. It also automatically determines the number of velocity components that are required to fit the spectrum of each position. The algorithm is described in detail in Appendix~\ref{appendix_program}. We applied this automatisation programme only to ortho c-C$_3$H$_2$. 
\subsection{Two-point auto-correlation of opacity maps}
To analyse the structure of the cloud probed in absorption, we calculate the two-point auto-correlation function of the opacity maps. As in Sect.~\ref{data_sample}, we use a data sampling of $0.6\arcsec$. The two-point auto-correlation function $C(r)$ is calculated for a sample of pixel separations $r_k$ with $0\leq r_k\leq r_{max}$, where $r_{max}$ is the maximal possible separation of two pixels in the opacity maps (about 17$\arcsec$). The value of the two-point auto-correlation function at pixel separation $r_k$ is the average scalar product of the opacities of the pixel pairs that have a separation $r_k$:
\begin{equation}
C(r_k)=\frac{1}{N}\sum_{i=1}^N \tau(x_{1,i},y_{1,i})\tau(x_{2,i},y_{2,i}),
\end{equation}
with $r_{k-1}^2<(x_{1,i}-x_{2,i})^2+(y_{1,i}-y_{2,i})^2\leq r_k^2$, $x_{1,i}$, $x_{2,i}$ and $y_{1,i}$, $y_{2,i}$ the pixel coordinates, and $N$ the number of pixel pairs fulfilling this condition. To get sufficiently high statistics, $C(r_k)$ is only determined if at least 50 pairs of pixels are available in the range [$r_{k-1},r_k$].

We computed the two-point auto-correlation functions of 1000 realisations of the opacity cubes produced in Sect.~\ref{opac-cubes} to estimate their uncertainties. 

We take the average of the 1000 two-point auto-correlation functions as the best estimate and their dispersion as the uncertainty. We also compute $C(r)$ for channels containing only noise (see Appendix~\ref{sect_noise_auto_corr}).

\subsection{Probability distribution function of the optical depth}
To investigate further cloud properties such as turbulence, we calculate the probability distribution function (PDF) of the opacity maps. We use the following normalisation \citep[see, e.g.][]{schneider2013}:  
\begin{equation}
\eta = \ln\left(\frac{\tau}{\bar{\tau}}\right),
\end{equation}
with $\bar{\tau}$ the mean opacity in the map. Here, we ignore all pixels with a negative opacity resulting from the noise because the normalisation $\eta$ is only defined for positive opacities. We determine the normalised PDF for each of the 1000 realisations of the opacity cubes. For this, we calculate the PDF for bins in $\eta$ of width $0.1$. We calculate the mean value of the PDF and the standard deviation as uncertainty for each bin. The presence of pixels containing only noise results in a broader PDF \citep{ossenkopf2016}. To minimise the effect of the noise, we compute the PDF using only the pixels with opacities above the 3$\sigma$ level implying only positive values for $\eta$ (see Appendix~\ref{sect_noise_pdf}). We fit a normal distribution to the PDF:
\begin{equation}
 p(\eta) = \frac{A}{\sqrt{2\pi}\sigma}\cdot\exp\left(-\frac{(\eta-\mu)^2}{2\sigma^2}\right).
\end{equation}
$A$ is the integral below the curve, $\sigma$ is the dimensionless dispersion, and $\mu$ the mean. For a perfect log-normal distribution, $\mu$ should be equal to 0 and $A$ to 1.

The fit is only performed if the number of counts per log-bin at the peak of the PDF is higher than 10 and if no more than 10 percent of the available pixels have an opacity value set to infinity. 
\subsection{Principal component analysis}\label{pca_theory}
To search for correlations or anti-correlations between the opacity maps of the different molecules, we use the principal component analysis (PCA) \citep[see, e.g.][]{heyer1997,neufeld2015, spezzano2017, gratier2017}. The PCA applies an orthogonal transformation to a dataset to produce a set of components which are linearly uncorrelated, the so-called principal components (PC). The PCs are orthogonal to each other and make up a new coordinate system to which the data are transformed. The first PC goes in the direction of the largest variance in the data. The number of PCs that are considered has to be smaller than or equal to the number of dimensions of the original data set. In our case the number $n$ of molecules used for the PCA is the dimension of the data set and also the number of calculated PCs. We apply the PCA to opacity maps at a given velocity. In our case, the original data set consists of one-dimensional arrays, one for each molecule, which contain the opacities of the selected pixels. Before starting the PCA, the array $\vec{a}_i$ of each molecule $i$ is normalised by subtracting the mean $\bar{a}_i$ and by dividing by the standard deviation $\sigma_i$ \citep[e.g.][]{neufeld2015}:
\begin{equation}
 a_{i,j,\mathrm{normed}} = \frac{a_{i,j}-\bar{a}_i}{\sigma_i},
\end{equation}
with $j$ the pixel position in the array. After this preparation, the PCA is performed with the \texttt{Python} package \texttt{scikit-learn} \citep{scikit-learn}. The procedure computes the principal components as well as the eigenvalues of the decomposition. The powers (eigenvalue divided by sum of eigenvalues) give the contribution of the different components calculated from the eigenvalues. 

To determine the contribution $C_{i,k}$ of each principal component $\vec{PC}_k$ to each molecule array $\vec{a}_{i,\mathrm{normed}}$, the following system of linear equations has to be solved: 
\begin{equation}
 \vec{a}_{i,\mathrm{normed}} =  b_i \sum_k C_{i,k}\cdot \vec{PC}_{k},
\end{equation}
with $b_i$ a constant factor, the normalisation condition $\sum_k C_{i,k}^2=1$, and with the principal components having a length of 1 and a standard deviation of 1.

To estimate the uncertainties of the contributions $C_{i,k}$, we apply the PCA to the 1000 realisations of the opacity cubes. Two conditions have to be fulfilled to exploit the outcome of these 1000 PCAs. First, the pixel lists must be the same. Therefore, we ignore pixels which have an opacity value set to infinity in any of the realisations. The second condition is that the PCs that represent the axes of the new coordinate system have to be aligned to each other. This is not necessarily the case when calculating the PCs for the different realisations of the opacity cubes. To address this, we take the original cubes as reference for the PCA and we align the new coordinate systems (PCs) of the 1000 realisations to this reference by applying with the \texttt{Python} package \texttt{scipy.linalg}\footnote{see https://scipy.org} an orthogonal procrustes rotation as described by \citet{babamoradi2013}. After this, we determine the contributions $C_{i,k}$ as explained above and calculate the mean and the standard deviation.

The noise can have a significant influence on the outcome of the PCA. The normalisation can increase the impact of the noise in cases where a molecule has a relatively homogeneous opacity over the field of view or when the absorption is weak and most of the field of view is dominated by noise. To avoid this problem, we select the molecules depending on the dynamic range of their opacity maps. The peak signal-to-noise ratio must be at least 10 and there must be at least 125 pixels (which corresponds to about five beams of the sample) with a signal-to-noise ratio higher than 5. With these selection criteria we ignore molecules which may have only one compact, strong peak. A meaningful use of the PCA at a given velocity requires at least four molecules.

\section{Results}\label{sect_result}
\subsection{Identification of molecules}
We performed the identification of the molecules on the basis of the spectroscopic information provided in the Cologne Database for Molecular Spectroscopy \citep[CDMS,][]{endres2016,mueller2005,mueller2001} and the Jet Propulsion Laboratory (JPL) molecular spectroscopy catalogue \citep{pickett1998}. In total, we identified 19 molecules seen in absorption in the diffuse and translucent molecular clouds along the line of sight to Sgr~B2(N): C$^{13}$O, CS, CN, SiO, SO, HCO$^+$, HOC$^+$, HCN, HNC, CCH, N$_2$H$^+$, HNCO, H$_2$CS, c-C$_3$H$_2$, HC$_3$N, CH$_3$OH, CH$_3$CN, NH$_2$CHO, and CH$_3$CHO. We also detected the following less abundant isotopologues: C$^{18}$O, C$^{17}$O, C$^{34}$S, $^{13}$CS, C$^{33}$S, $^{13}$CN, $^{29}$SiO, $^{30}$SiO, H$^{13}$CO$^{+}$, HC$^{18}$O$^{+}$, H$^{13}$CN, HC$^{15}$N, HN$^{13}$C, H$^{15}$NC, and $^{13}$CH$_3$OH. A report on the complex organic molecules detected in absorption in this survey is given in \citet{thiel2017}.

\subsection{Identification of velocity components based on c-C$_3$H$_2$}\label{results_cloud_properties}

We selected the molecule c-C$_3$H$_2$ to decompose the absorption features into individual velocity components with our automatisation programme and thereby identify the clouds detected in absorption along the line of sight to Sgr~B2(N). Absorption from the 85.3~GHz ortho c-C$_3$H$_2$ line covers almost the complete velocity range in which the clouds along the line of sight are detected. An advantage compared to other molecules is that the absorption is optically thin, except in parts of the envelope of Sgr\,B2 (highlighted in grey in Fig.~\ref{examplespec}). We note that absorption from c-C$_3$H$_2$ has long been known to trace diffuse interstellar clouds, among others along sight lines to the GC \citep{cox1988}.

To identify the velocity components, we investigate the distribution of linewidths, $FWHM$, centroid velocities, $\varv$, and column densities, $N_\mathrm{tot}$, obtained for ortho c-C$_3$H$_2$ from the fits to all positions where the molecule is detected. In total, 2838 velocity components are detected towards the 322 selected positions. Between two and six velocity components are detected in the envelope of Sgr\,B2 at each position and up to 14 velocity components are found by the programme in the clouds along the line of sight.

\begin{figure*}
   \includegraphics[width=17cm,trim = 0.8cm 1.cm 2.5cm 2.2cm, clip=True]{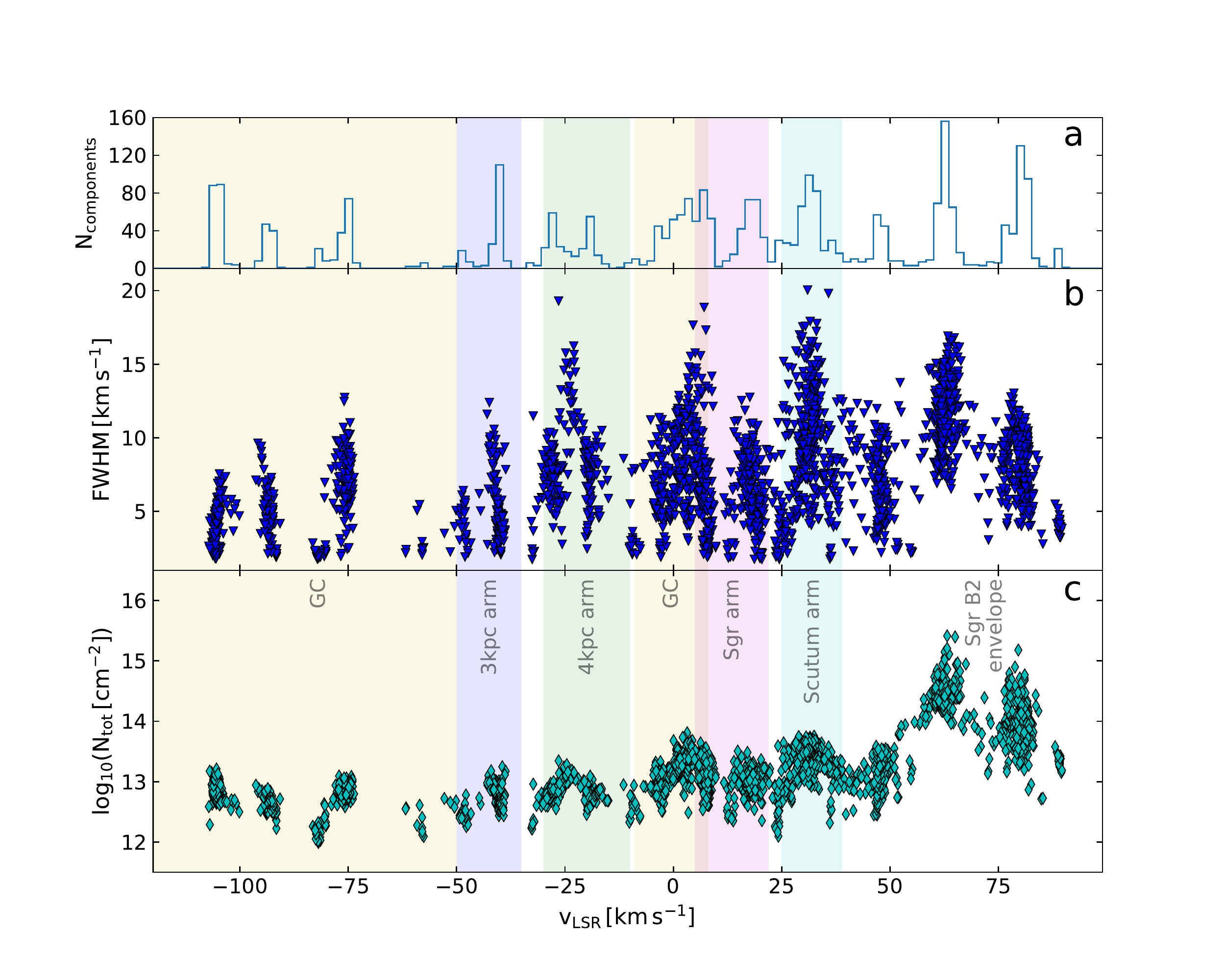}
 \caption{\textbf{a} Number of velocity components found with ortho c-C$_3$H$_2$ in the selected field as a function of centroid velocity. The bin width is one channel, 1.74~km~$s^{-1}$. \textbf{b} Distribution of linewidths. \textbf{c} Distribution of ortho c-C$_3$H$_2$ column densities. The velocity ranges are colour coded in the background of every panel (see Table~\ref{vel_spiralarms} for references).}\label{fwhm}
\end{figure*}

The number of velocity components detected with c-C$_3$H$_2$ in the selected field is shown as a function of centroid velocity in Fig.~\ref{fwhm}a. The distributions of widths and column densities are plotted in panels b and c, respectively. The velocity ranges of the spiral arms, the diffuse Galactic centre clouds and the envelope of Sgr\,B2 are colour coded in the background of Fig.~\ref{fwhm} (see Table~\ref{vel_spiralarms} for references and Fig.~\ref{sketch_mw} for a sketch). There is an ambiguity between the GC and the local spiral arm for the velocities around 0~km\,s$^{-1}$. Due to the compact structure of the absorption component around 0~km~s$^{-1}$ along the line of sight to the Galactic centre, \citet{whiteoak1978} suggested this absorption is not caused by local gas. Later, \citet{gardner1982} determined a low isotopic ratio $\frac{^{12}{\rm C}}{^{13}{\rm C}}$ of 22 for this component, which strongly suggests that it belongs to the Galactic centre region. Hence, we assume that the strong absorption around 0~km\,s$^{-1}$ belongs to the Galactic centre region and in the following the velocity range from $-9$ to 8\,km~s$^{-1}$ will be treated as part of the GC region. c-C$_3$H$_2$ is detected in each group of line-of-sight (l.o.s.) molecular clouds which makes it an excellent molecule for a comparative study of these diffuse and translucent molecular clouds.
\begin{table}
\caption{Velocity ranges and associated locations of the clouds along the line of sight to Sgr\,B2.}
\label{vel_spiralarms}
\centering
\begin{tabular}{l c}       
\hline               
location & $\varv_\mathrm{LSR}$ \\

 & [km\,s$^{-1}$] \\
\hline\hline          
Galactic centre & $<-50$\\
3kpc arm & $-50$ to $-35$\\
4kpc arm & $-30$ to $-10$\\
Galactic centre & $-9$ to $8$\\
Sgr arm & $5$ to $22$\\
Scutum arm & $25$ to $39$\\
\hline                      
\end{tabular}
\tablefoot{References: \citet[][]{greaves1994,neufeld2000,lis2010,menten2011,monje2011}; and references therein. The combination of the velocity ranges mentioned in these papers results in some overlapping ranges. We treat clouds falling in the overlapping velocity range between 5 and 8\,km\,s$^{-1}$ as GC clouds. }
\end{table}

We identify each elongated structure in Fig.~\ref{fwhm}b and each corresponding peak in Fig.~\ref{fwhm}a as a single cloud. In some cases such as the GC clouds in the range $-110$ to $-70$~km\,s$^{-1}$, it is easy to differentiate the clouds, because the velocity components are well separated. For the GC clouds around 0~km\,s$^{-1}$ it is more difficult. In the case of the 3\,kpc arm we see mainly two clouds, but sometimes only one component with a width larger than the two narrow components detected at other positions. After inspecting the spectra we found out that in these cases the programme could not find two different components because they overlap each other in such a way that they cannot be separated along the velocity axis. The same happens for the Scutum and 4~kpc arms. 

\begin{figure*}[t]
   \includegraphics[width=17cm,trim = 0.7cm 1.1cm 1.8cm 2.5cm, clip=True]{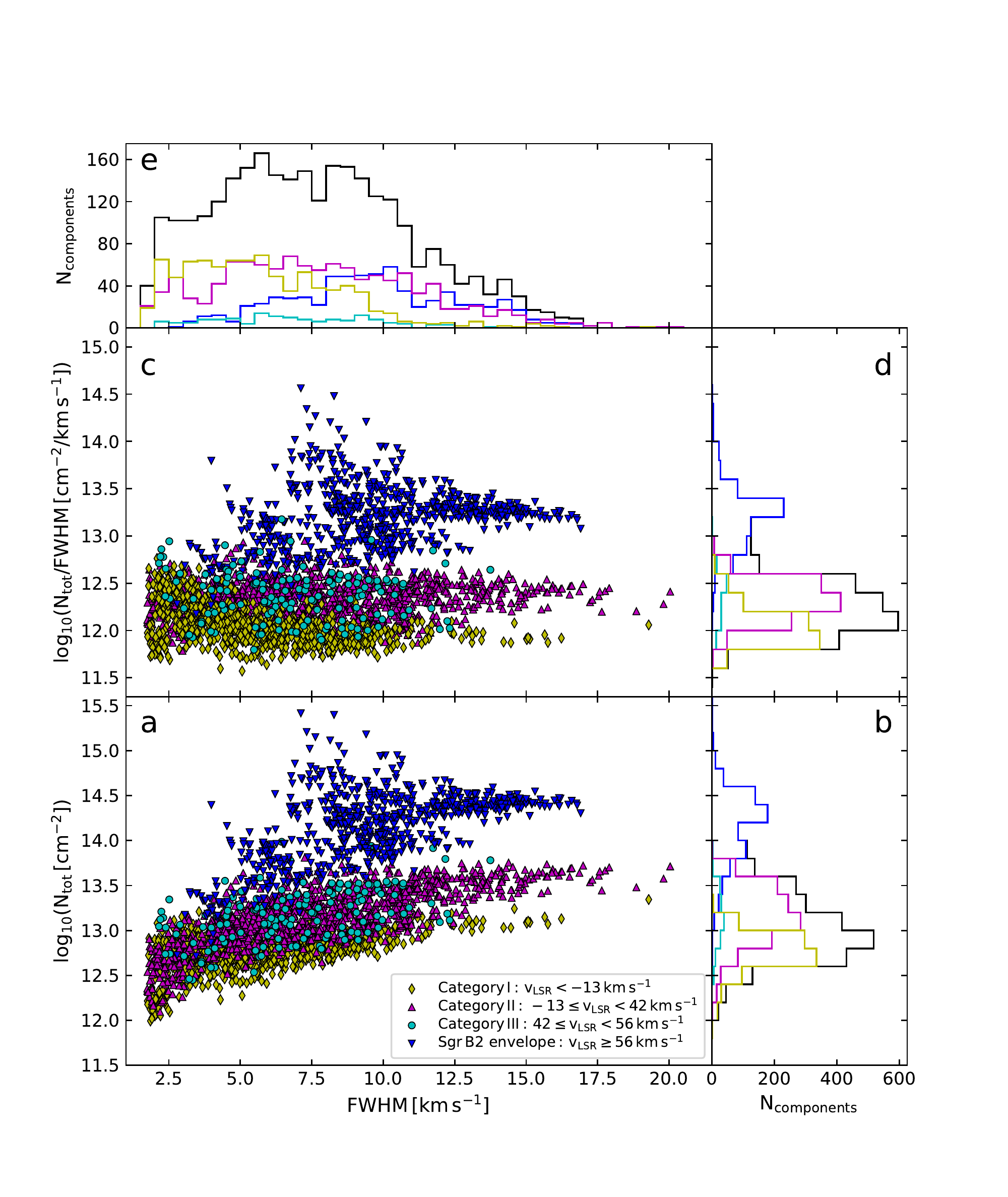}
 \caption{\textbf{a} Distribution of column densities as a function of linewidth. \textbf{b} Number of components as a function of column density. \textbf{c} Distribution of column densities divided by linewidth as a function of linewidth. \textbf{d} Number of components as a function of column density divided by linewidth. \textbf{e} Number of velocity components as a function of linewidth. The bin width is 0.5~km~s$^{-1}$. In all panels the two categories of line-of-sight clouds are coloured in yellow (Category I) and magenta (Category II), the clouds at about 50~km~s$^{-1}$ in cyan (Category III), and the components tracing the diffuse envelope of Sgr\,B2 in blue. In panels b, d, and e, the black histogram represents the full sample of detected components. }\label{ntot_fwhm}
\end{figure*}

The distribution of column densities of ortho c-C$_3$H$_2$ as a function of linewidth is plotted in Fig.~\ref{ntot_fwhm}a. We also show the column density divided by the linewidth in Fig.~\ref{ntot_fwhm}c because, in this representation, the detection limit is roughly horizontal. Another advantage of the latter representation is that it reduces the bias due to the clouds that partially overlap in velocity and could not be fitted separately. 

We divide the diffuse and translucent clouds along the line of sight to Sgr\,B2  with $\varv_\mathrm{LSR}$<42~km\,s$^{-1}$ into two main categories based on their ortho c-C$_3$H$_2$ column densities (see Fig.~\ref{ntot_fwhm}b and d). We call Category I the l.o.s. clouds with velocities up to $-13$~km\,s$^{-1}$ and Category II the ones with velocities between $-$13~km\,s$^{-1}$ and 42~km\,s$^{-1}$.
The absorption at velocities between 50 and 90~km~s$^{-1}$ is usually considered to be caused by the envelope of Sgr~B2 \citep{neill2014}. The envelope of Sgr\,B2 ($\varv_\mathrm{LSR} \geq 56$~km\,s$^{-1}$) contains two main velocity components at about $64$~km\,s$^{-1}$ and $80$~km\,s$^{-1}$ \citep[e.g.][]{huettemeister1995, lang2010}. The velocity component at about 48~km\,s$^{-1}$ is usually also associated with the envelope of the Sgr~B2 complex \citep[e.g.][and references therein]{garwood1989}. We plot it in cyan in Fig.~\ref{ntot_fwhm} because it stands out with lower column densities compared to the two main components of the Sgr~B2 envelope. We call the velocity range between 42~km\,s$^{-1}$ and 56~km\,s$^{-1}$ Category III.

The distribution of linewidths is shown in black in Fig.~\ref{ntot_fwhm}e. The lower limit is set by the channel width of $1.74$~km\,s$^{-1}$. The linewidths cover the range between this lower limit and 20~km\,s$^{-1}$. The ortho c-C$_3$H$_2$ column densities cover a range of three orders of magnitude from $10^{12}$~cm$^{-2}$ to $10^{15}$~cm$^{-2}$ (Figs.~\ref{ntot_fwhm}a and b). Each histogram of Fig.~\ref{ntot_fwhm} is also split into the four categories of components introduced above. The median linewidths and column densities of these four categories are listed in Table~\ref{median_fwhm_ntot}.

\begin{table*}
\caption{Median linewidths and column densities of the four categories shown in Fig.~\ref{ntot_fwhm}. }
\label{median_fwhm_ntot}
\centering
\begin{tabular}{l c c c c }       
\hline               
Category & $\varv_\mathrm{LSR}$ & $FWHM$ & $\log_{10}{N_\mathrm{tot}}$ & $\log_{10}{\frac{N_\mathrm{tot}}{FWHM}}$ \\

& [km\,s$^{-1}$] & [km\,s$^{-1}$] & [cm$^{-2}$] & [cm$^{-2}$~km$^{-1}$~s]\\
\hline\hline 
Category I & $\varv_\mathrm{LSR}<-13$ & 5.4 & 12.8 & 12.0 \\
Category II & $-13\leq \varv_\mathrm{LSR}<42$ & 7.5 & 13.2 & 12.3 \\
Category III & $42\leq \varv_\mathrm{LSR}<56$ & 6.7 & 13.2 & 12.4 \\         
Sgr~B2 envelope & $\varv_\mathrm{LSR}\geq 56$ & 9.6 & 14.2 & 13.2 \\
\hline                      
\end{tabular}
\end{table*}

The majority of l.o.s. clouds have a linewidth smaller than 10~km\,s$^{-1}$, but there is a tail up to 20~km\,s$^{-1}$. We believe that most of these broader components represent two or more overlapping components with narrower widths that could not be fitted individually. Hence, these components could contain several cloud entities. The l.o.s. clouds can be divided into two categories (yellow and magenta in Fig.~\ref{ntot_fwhm}). Category I has a median linewidth of 5.4~km\,s$^{-1}$. It contains the GC clouds with a velocity lower than $-50$~km\,s$^{-1}$ and the clouds of the 3~kpc and 4~kpc arms. The GC clouds around 0~km\,s$^{-1}$ and the clouds in the Scutum and the Sagittarius arms (Category II) have a somewhat larger median linewidth of 7.5~km\,s$^{-1}$. The components in the envelope of Sgr\,B2 have an even larger median linewidth of 9.6~km\,s$^{-1}$. This larger value may partly be due to the optical thickness. The high opacities affecting these components make it indeed sometimes difficult to fit individual velocity components. The components in Category III have a median linewidth of 6.7~km~s$^{-1}$, in between the ones of Categories I and II.

The median column densities of ortho c-C$_3$H$_2$, both before and after normalisation by the linewidth, of Categories I, II, and III are similar, in the order of $10^{13}$~cm$^{-2}$ and $10^{12}$ cm$^{-2}$~km$^{-1}$~s, respectively, with Category I lying slightly below Categories II and III. While Categories II and III are more affected than Category I by overlapping components that cannot be fitted separately, their higher median column densities do not result from this because they still lie above Category I by a factor of $\sim$2 after normalisation by the linewidth. The components in the Sgr~B2 envelope are characterised by much higher column densities, about one order of magnitude compared to Categories II and III, both before and after normalisation by the linewidth. 

Overall, the components around 50 km~s$^{-1}$ (Category III) have similar properties (linewidths and ortho c-C$_3$H$_2$ column densities) as the ones in the Scutum and Sagittarius arms (Category II).

In the following, we ignore the components belonging to the envelope of Sgr~B2 because of their high optical depths. In addition, because the velocity component of Category III is blended with the one of the envelope of Sgr~B2 at 64~km\,s$^{-1}$ (see grey shaded area in Fig.~\ref{examplespec}), we focus our subsequent analyses on the clouds belonging to Categories I and II. They can be described with 15 components whose centroid velocities are derived from the peaks in Fig.~\ref{fwhm}a. These 15 components are listed in Table~\ref{velocities}.

\begin{table}
\caption{Velocities and localisation of the diffuse and translucent molecular clouds detected with c-C$_3$H$_2$ along the line of sight to Sgr~B2, excluding Category III and the envelope of Sgr~B2. }
\label{velocities}
\centering
\begin{tabular}{r l c c }       
\hline               
$\varv_\mathrm{LSR}$ & Location\tablefootmark{a} & Category\tablefootmark{b} & $d$\tablefootmark{c} \\

\multicolumn{1}{c}{[km\,s$^{-1}$]} & & & \multicolumn{1}{c}{[kpc]} \\
\hline\hline          
$-105.9$ & Galactic centre & I & 7.0 \\
$-93.7$ & Galactic centre & I & 7.0 \\
$-81.5$ & Galactic centre & I & 7.0 \\
$-74.6$ & Galactic centre & I & 7.0 \\
$-48.4$ & 3\,kpc arm & I & 5.5 \\
$-39.7$ & 3\,kpc arm & I & 5.5 \\
$-27.6$ & 4\,kpc arm & I & 4.3 \\
$-18.9$ & 4\,kpc arm & I & 4.3 \\
$-3.2$ & Galactic centre & II & 7.0 \\
2.0 & Galactic centre & II & 7.0 \\
7.3 & Galactic centre & II & 7.0 \\
17.7 & Sagittarius arm & II & 1.0 \\
24.7 & Scutum arm & II & 2.8 \\
31.6 & Scutum arm & II & 2.8 \\
36.9 & Scutum arm & II & 2.8 \\
\hline                      
\end{tabular}
\tablefoot{\tablefoottext{a}{Location of the clouds.}\tablefoottext{b}{See Table~\ref{median_fwhm_ntot}.}\tablefoottext{c}{Approximate distance to the Sun.}}
\end{table}
\subsection{Opacity maps}

\begin{table*}
\caption{Rotational transitions used in this work.}
\label{transitions}
\centering
\begin{tabular}{l c r r c c}       
\hline               
Molecule & Transition & \multicolumn{1}{c}{$\nu_0$\tablefootmark{a}} & \multicolumn{1}{c}{$E_\mathrm{up}/k$\tablefootmark{b}} & $A_\mathrm{u,l}$\tablefootmark{c} & References \\
 & & \multicolumn{1}{c}{[MHz]} & \multicolumn{1}{c}{[K]} & [s$^{-1}$] & \\
\hline\hline 
ortho c-C$_3$H$_2$ & $2_{1,2}$--$1_{0,1}$ & $85338.894$ & 4.1 & $2.32\times 10^{-5}$ & 1 \\
HC$^{15}$N & 1--0 & $86054.966$ & $4.1$ & $2.20\times 10^{-5}$ & 10\\
H$^{13}$CO$^+$ & 1--0 & $86754.288$ & 4.2 & $3.85\times 10^{-5}$ & 2 \\  
SiO & 2--1 & $86846.985$ & $6.3$ & $2.93\times 10^{-5}$ & 7\\
HN$^{13}$C & 1--0 & $87090.825$ & $4.2$ & $2.38\times 10^{-5}$ & 9 \\
HNC & 1--0 & $90663.568$ & $4.4$ & $2.69\times 10^{-5}$ & 8\\
$^{13}$CS & 2--1 & $92494.308$ & $6.7$ & $1.41\times 10^{-5}$ & 5 \\          
C$^{34}$S & 2--1 & $96412.950$ & 6.9 & $1.60\times 10^{-5}$ & 4,5 \\
CH$_3$OH A$^*$ & $2_0$--$1_0$ & $96741.371$ & 7.0 & $3.41\times 10^{-6}$ & 11\\
CS & 2--1 & $97980.953$ & 7.1 & $1.68\times 10^{-5}$ & 4,5 \\
SO & $2_3$--$1_2$ & $99299.870$ & $9.2$ & $1.13\times 10^{-5}$ & 6\\
$^{13}$CO & 1--0 & $110201.354$ & $5.3$ & $6.33\times 10^{-8}$ & 2,3 \\
\hline                      
\end{tabular}
\tablefoot{The spectroscopic information for the molecule marked with a star is taken from JPL, otherwise from CDMS. \tablefoottext{a}{Rest frequency.}\tablefoottext{b}{Upper level energy.} \tablefoottext{c}{Einstein coefficient for spontaneous emission from upper level $u$ to lower level $l$.}}
\tablebib{(1)~\citet{spezzano2012}; (2)~\citet{schmid2004}: (3)~\citet{cazzoli2004}; (4)~\citet{gottlieb2003}; (5)~\citet{bogey1982}; (6)~\citet{tiemann1974}; (7)~\citet{mueller2013}; (8)~\citet{saykally1976}; (9)~\citet{vandertak2009}; (10)~\citet{cazzoli2005}; (11)~\citet{mueller2004}.}
\end{table*}

\begin{figure*}
\centering
\includegraphics[width=17cm, trim = 1.8cm 2.4cm 0.4cm 3.1cm, clip=True]{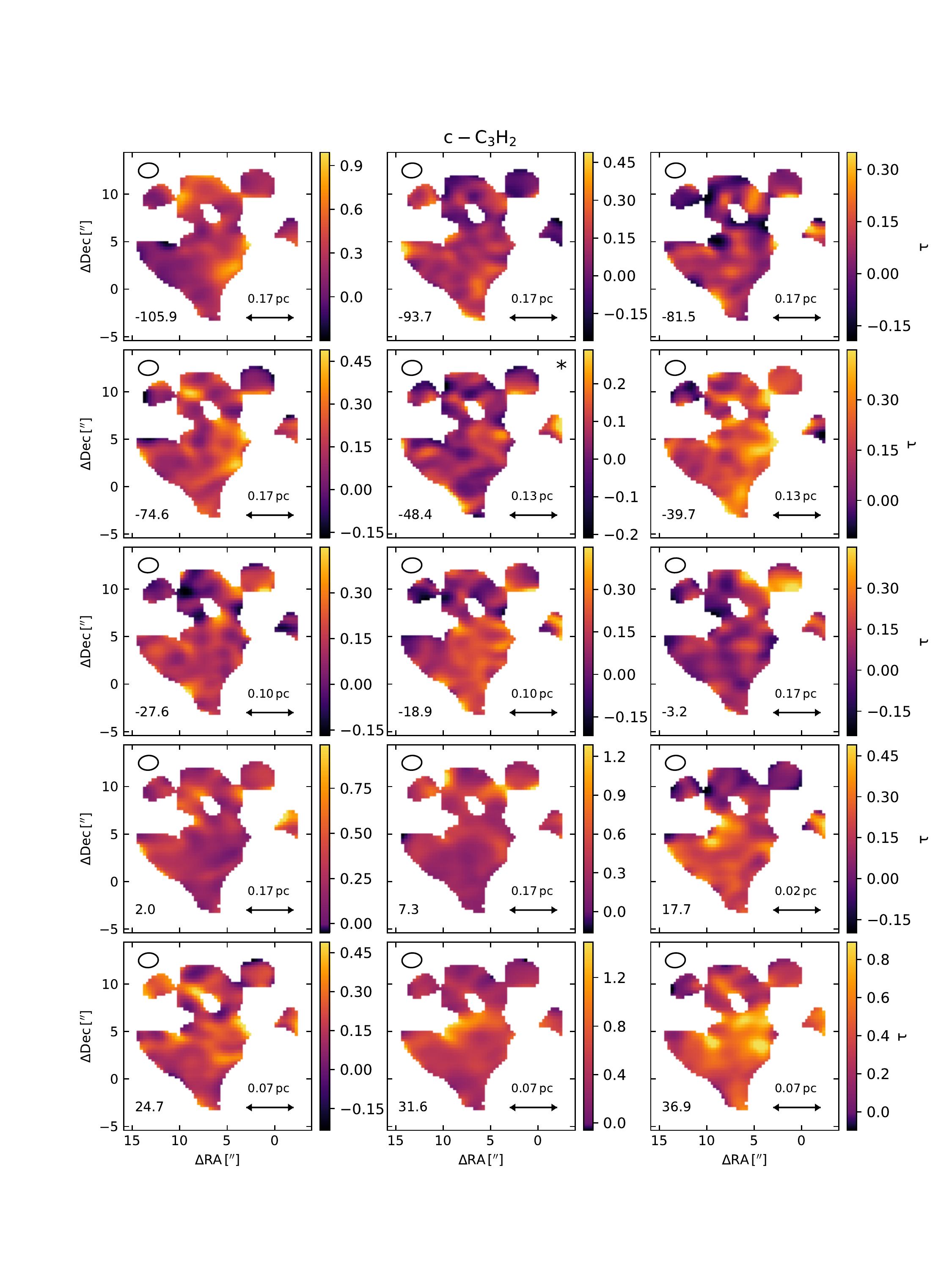}
\caption{Opacity maps of c-C$_3$H$_2$ $2_{1,2}$--$1_{0,1}$ for all 15 investigated velocity components. In each panel, the velocity of the channel is given in km\,s$^{-1}$ in the bottom left corner, the beam ($HPBW$) is shown as an ellipse in the upper left corner, and an approximate physical scale for the assumed distance of the cloud (see Table~\ref{velocities}) is indicated in the bottom right corner. A star in the upper right corner marks the components with a maximum signal-to-noise ratio $\tau$/$\sigma_\tau$ smaller than 5 (see Fig.~\ref{snr_opacity_c-c3h2}). The equatorial offsets are relative to the phase centre.} 
\label{opacity_c-c3h2}
\end{figure*}

\begin{figure*}
\centering
\includegraphics[width=17cm, trim = 1.8cm 2.4cm 0.8cm 3.1cm, clip=True]{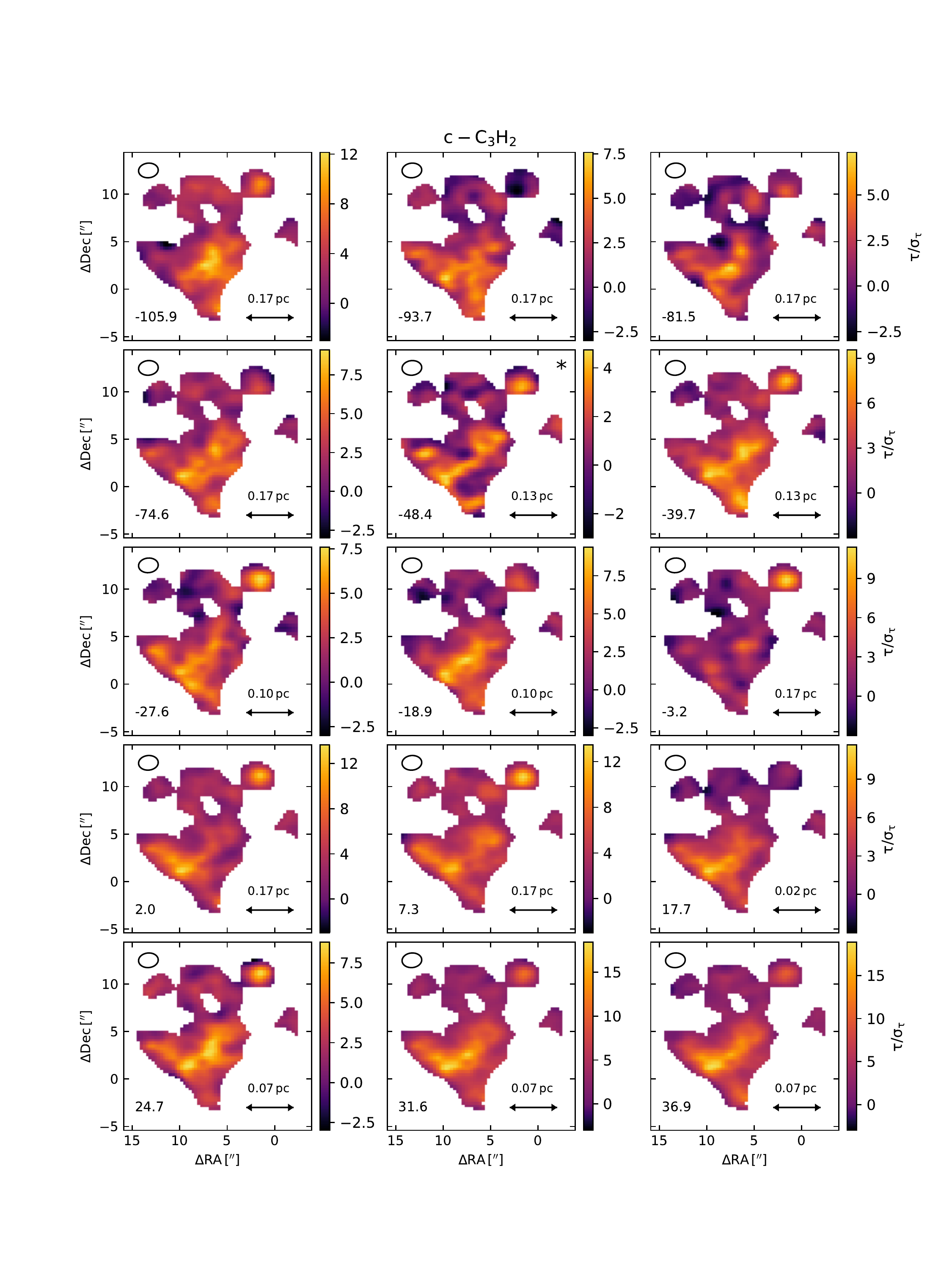}
\caption{Same as Fig.~\ref{opacity_c-c3h2} but for the signal-to-noise ratio $\tau/\sigma_\tau$ of c-C$_3$H$_2$ $2_{1,2}$--$1_{0,1}$.}
\label{snr_opacity_c-c3h2}
\end{figure*}

To investigate the spatial structure of the clouds we look for molecules that reveal absorption over an extended area of the field of view. We do not consider molecules with a resolved hyperfine structure that makes velocity assignments more complicated without fitting. Out of all molecules detected along the line of sight to Sgr~B2, eight molecules fulfil these criteria: H$^{13}$CO$^+$, $^{13}$CO, HNC and its isotopologue HN$^{13}$C, HC$^{15}$N, CS and its isotopologues C$^{34}$S and $^{13}$CS, SiO, SO, and CH$_3$OH. For some components the less abundant isotopologues are useful when the main isotopologue is optically thick. The spectroscopic parameters of the transitions of these selected molecules are listed in Table~\ref{transitions} and the example spectra towards the two positions K4 and K6$_\mathrm{shell}$ (see Fig.~\ref{contmap}) are shown in Fig.~\ref{spectra}.  

We show in Fig.~\ref{opacity_c-c3h2} the opacity maps of c-C$_3$H$_2$ at the 15 velocities listed in Table~\ref{velocities}, and in Fig.~\ref{snr_opacity_c-c3h2} the maps of signal-to-noise ratio (SNR). It is important to consider the SNR maps when interpreting the opacity maps because the noise level is not uniform due to the variations of the background continuum emission. The SNR maps are strongly correlated to the continuum map (see Fig.~\ref{contmap}): the stronger the continuum the lower the opacity noise level. 

At first sight, large-scale structures are detected in the opacity maps of nearly all velocity components (Fig.~\ref{opacity_c-c3h2}). The components at $-48.4$~km~s$^{-1}$ and $-3.2$~km~s$^{-1}$ do not show such extended structures but this may simply result from a lack of sensitivity: their SNR maps indicate that the peak SNR is low (less than about 5 if we exclude K4) and only few positions have a SNR above 3. 

The component at $\varv_\mathrm{LSR}=36.9$\,km\,s$^{-1}$ looks more clumpy in Fig.~\ref{opacity_c-c3h2}, with three seemingly prominent, unresolved structures. However, all three opacity peaks have low SNR ($\sim 5$) in Fig.~\ref{snr_opacity_c-c3h2}. They may be noise artefacts and may not trace real compact structures. 

The opacity and SNR maps of the other molecules are shown in Figs.~\ref{opacity_h13cop}--\ref{snr_opacity_ch3oh}. The velocities of the channels differ slightly from the ones of c-C$_3$H$_2$ because of the discrete sampling of the frequency axis. The channels selected for these figures are the ones nearest to the velocities listed in Table~\ref{velocities}. The pixels that have an opacity set to infinity (see Sect.~\ref{opac-cubes}) are masked (in cyan).

The type of structures seen in the opacity maps is similar for all molecules. In many cases, extended structures are present. Compact clumps that are present in some maps often have a low SNR and may simply be noise artefacts. The SNR of low-abundance molecules such as $^{13}$CS is too low to characterise the structural properties of the clouds. A better sensitivity would be needed for these tracers. 

Because of the high number of opacity maps, we use in the following sections statistical tools to analyse and quantify the structure of the clouds traced in absorption towards Sgr~B2(N).
%
%
%
%
%
%
%
%
%

\subsection{Cloud sub-structure}\label{autocorrel}

\begin{figure*}
\centering
\includegraphics[width=17cm, trim = 3.cm 1.2cm 4.cm 2.0cm, clip=True]{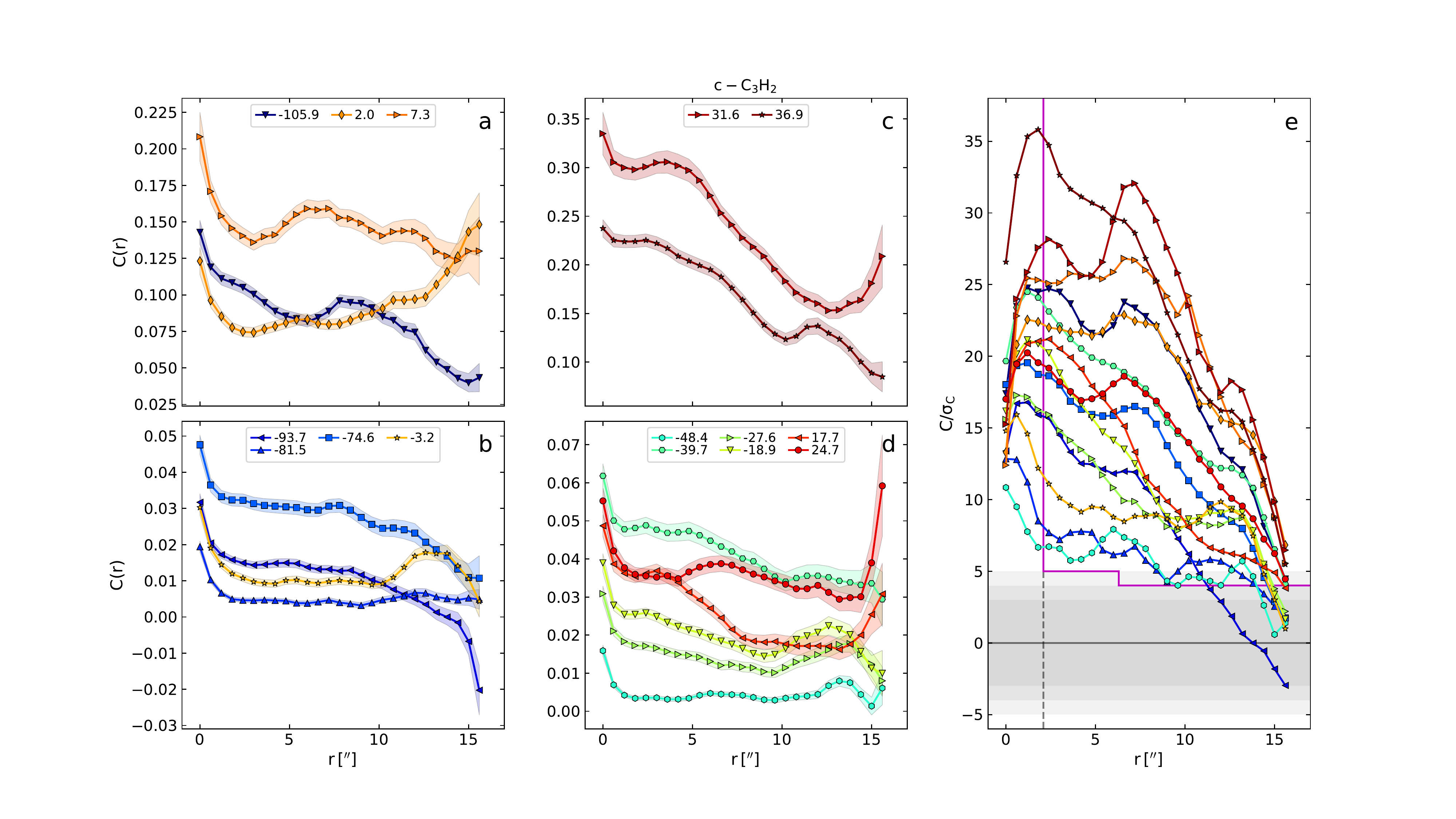}
\caption{\textbf{a--d} Two-point auto-correlation functions $C(r)$ as a function of pixel separation $r$ for the velocity components traced with c-C$_3$H$_2$. The points give the mean values of the 1000 realisations and the colour-shaded regions represent the standard deviations ($1\sigma$). The centroid LSR velocities of the clouds are indicated in km~s$^{-1}$ at the top of each panel. \textbf{e} Signal-to-noise ratio (SNR) of the two-point auto-correlation functions shown in panels a--d. The SNR levels of $\pm$3, $\pm$4 and $\pm$5 are highlighted in shades of grey. The colours and symbols are the same as in panels a--d. The vertical dashed line shows the size of the beam (HPBW). The area below and left of the magenta line represents the area where the two-point auto-correlation functions are not significant.} 
\label{autocorr_c-c3h2}
\end{figure*}

\begin{figure*}
\centering
\includegraphics[width=17cm, trim = 3.cm 1.2cm 4.cm 2.0cm, clip=True]{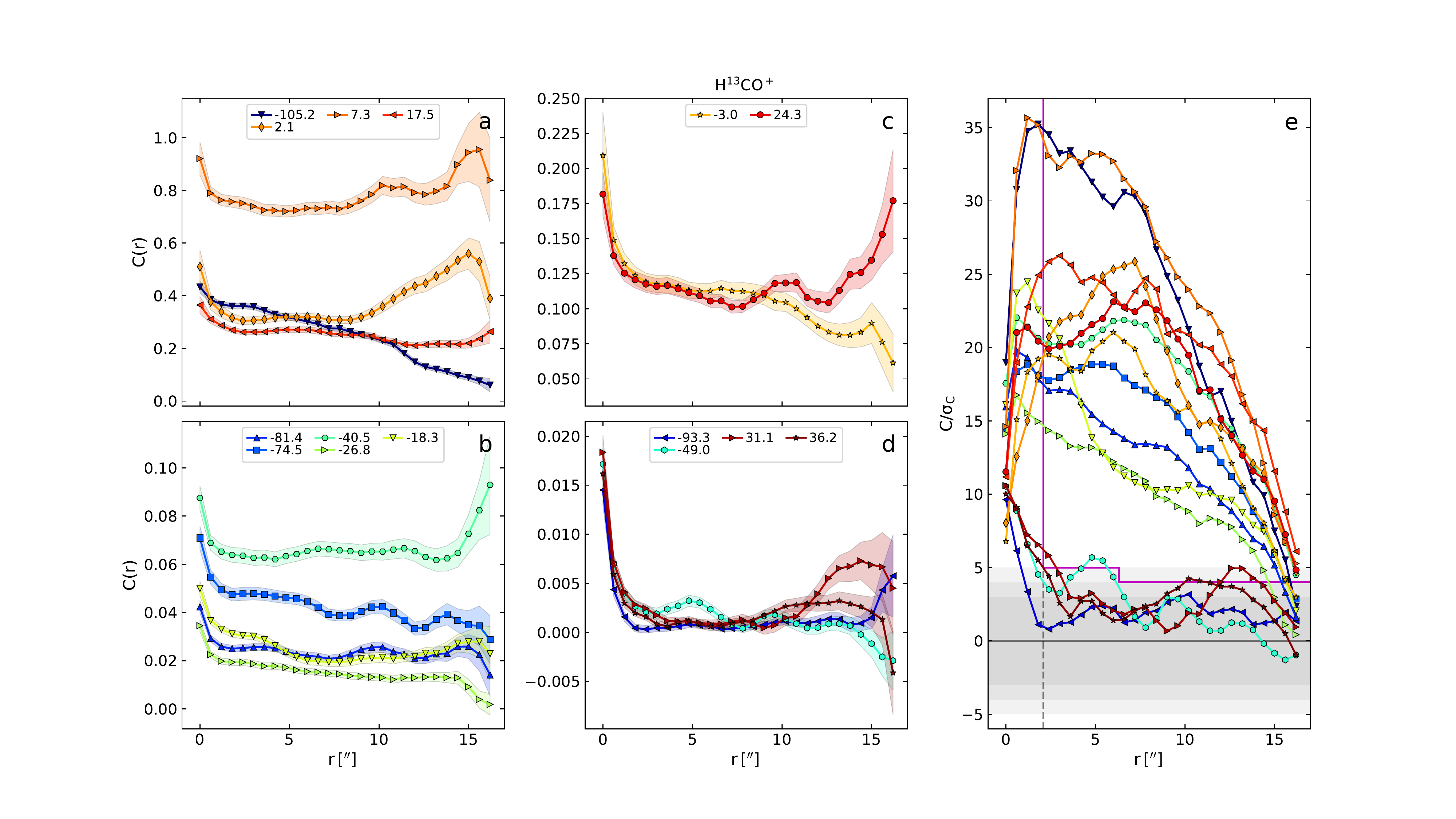}
\caption{Same as Fig.~\ref{autocorr_c-c3h2}, but for H$^{13}$CO$^+$.} 
\label{autocorr_h13cop}
\end{figure*}

The two-point auto-correlation functions of c-C$_3$H$_2$ and H$^{13}$CO$^+$ are shown in Figs.~\ref{autocorr_c-c3h2} and \ref{autocorr_h13cop}, respectively. The two-point auto-correlation functions of the other molecules are displayed in Figs.~\ref{autocorr_13co}--\ref{autocorr_ch3oh}. Panels a--d of each of these figures show the two-point auto-correlation functions $C(r)$ of the velocity components and panel (e) their SNR ($C/\sigma_C$). The analysis of the two-point auto-correlation function of noise channels performed in Appendix~\ref{sect_noise_auto_corr} indicates that only SNR values higher than 5 and 4 for pixel separations below and above 6$\arcsec$, respectively, are significant. In addition, the true two-point auto-correlation function cannot be evaluated below a separation corresponding to the size of the beam ($HPBW$). As a result, the values of the two-point auto-correlation functions are significant only in the upper-right part of their SNR curves, above and right of the magenta demarcation in panels (e).

The two-point auto-correlation functions show various shapes: flat, decreasing towards larger pixel separations, or stronger correlation at small and large separations with a dip in between. Flat curves indicate structures that are more extended than the region sampled with our data. Decreasing curves characterise clouds with structures that are somewhat more compact than the extent of the sampled region. The third type of shapes could result from the presence of several compact structures. 

The maximum angular separation, $\Delta r_{\rm max}$, at which $C(r)$ drops below the significance threshold (magenta line in panels e) is given for each velocity component and each molecule in Table~\ref{structuresizes}. When the SNR is too low, the opacity map is dominated by noise and no statement can be made about the sizes of the detected structures. The components with a peak SNR $\tau/\sigma_\tau$ smaller than five are therefore marked with a star in Table~\ref{structuresizes}. Most components with a $\Delta r_{\rm max}$ smaller than the beam ($<2^{\prime\prime}$) are in this situation. When $\Delta r_{\rm max}$ is equal to the largest available pixel separation, only a lower limit for the size of the structures can be determined. We convert these angular sizes to physical sizes in Table~\ref{structuresizes_physical}, using the approximate distances listed in Table~\ref{velocities}.

\begin{table*}
\caption{Angular sizes of cloud structures derived from two-point auto-correlation functions.}
\label{structuresizes}
\centering
\begin{tabular}{rrrrrrrrrrrrr}       
\hline               
velocity\tablefootmark{a} & c-C$_3$H$_2$ & H$^{13}$CO$^+$ & $^{13}$CO & CS & C$^{34}$S & $^{13}$CS & SO & SiO & HNC & HN$^{13}$C & HC$^{15}$N & CH$_3$OH\\

[km\,s$^{-1}$] & & & & & & & & & & & & \\
\hline\hline
\multicolumn{13}{c}{Galactic centre}\\
$-105.9$ & $>16.2$ & $16.2$ & $>15.6$ & $>16.2$ & $13.2$ & $4.8$ & $n^*$ & $8.4$ & $>15.6$ & $13.8$ & $n^*$ & $12.6$ \\
$-93.7$ & $11.4$ & $n^*$ & $15.0^*$ & $15.6$ & $n^*$ & $n^*$ & $n^*$ & $n^*$ & $14.4$ & $n^*$ & $n^*$ & $n^*$ \\
$-81.5$ & $14.4$ & $15.6$ & $15.0$ & $>16.2$ & $13.8$ & $n^*$ & $n^*$ & $3.6^*$ & $>15.6$ & $n^*$ & $>16.8$ & $n^*$ \\
$-74.6$ & $15.0$ & $16.2$ & $>15.6$ & $>16.2$ & $9.0^*$ & $2.4^*$ & $n^*$ & $2.4$ & $>15.6$ & $n^*$ & $15.0$ & $n^*$ \\
$-3.2$ & $15.0$ & $16.2$ & $6.0$ & $>16.2$ & $n^*$ & $n^*$ & $7.8$ & $15.6$ & $>15.6$ & $n$ & $10.8$ & $15.0$ \\
$2.0$ & $>16.2$ & $>16.8$ & $12.6$ & $15.0$ & $15.0$ & $16.2$ & $>16.2$ & $>16.2$ & $>15.6$ & $6.6$ & $15.6$ & $15.6$ \\
$7.3$ & $>16.2$ & $>16.8$ & $15.0$ & $>16.2$ & $15.0$ & $16.2$ & $>16.2$ & $>16.2$ & $>15.6$ & $16.2$ & $15.0$ & $15.6$ \\
\hline
\multicolumn{13}{c}{3\,kpc arm}\\
$-48.4$ & $14.4^*$ & $n^*$ & $>15.6$ & $4.8$ & $n^*$ & $n^*$ & $n^*$ & $n^*$ & $>15.6$ & $n^*$ & $n^*$ & $n^*$ \\
$-39.7$ & $>16.2$ & $>16.8$ & $>15.6$ & $>16.2$ & $15.0$ & $n^*$ & $>16.2$ & $n^*$ & $>15.6$ & $n^*$ & $15.6^*$ & $n^*$ \\
\hline
\multicolumn{13}{c}{4\,kpc arm}\\
$-27.6$ & $15.0$ & $15.0$ & $>15.6$ & $>16.2$ & $n^*$ & $n^*$ & $n^*$ & $n$ & $>15.6$ & $n^*$ & $n^*$ & $n^*$ \\
$-18.9$ & $15.0$ & $16.2$ & $>15.6$ & $>16.2$ & $4.8^*$ & $n^*$ & $4.8$ & $n^*$ & $>15.6$ & $n^*$ & $n^*$ & $n^*$ \\
\hline
\multicolumn{13}{c}{Scutum arm}\\
$24.7$ & $>16.2$ & $>16.8$ & $13.2$ & $>16.2$ & $n^*$ & $n^*$ & $4.2$ & $15.6$ & $>15.6$ & $n^*$ & $n^*$ & $15.6$ \\
$31.6$ & $>16.2$ & $3.0$ & $14.4$ & $15.6$ & $n^*$ & $n^*$ & $4.8$ & $2.4$ & $>15.6$ & $n^*$ & $n^*$ & $>16.2$ \\
$36.9$ & $>16.2$ & $2.4$ & $12.6$ & $15.6$ & $4.8^*$ & $n^*$ & $n$ & $5.4$ & $>15.6$ & $n^*$ & $n^*$ & $15.0$ \\
\hline
\multicolumn{13}{c}{Sagittarius arm}\\
$17.7$ & $15.6$ & $>16.8$ & $14.4$ & $>16.2$ & $9.0$ & $n^*$ & $12.0$ & $>16.2$ & $>15.6$ & $7.8$ & $10.8^*$ & $13.8$ \\
\hline                      
\end{tabular}
\tablefoot{The sizes are given in arcseconds. Channels with unresolved structures are marked with $n$ and channels with a SNR $\tau$/$\sigma_\tau$ smaller than 5 are marked with a star.\\ \tablefoottext{a}{Cloud centroid velocities derived from c-C$_3$H$_2$.}}
\end{table*}

\begin{table*}
\caption{Physical sizes of cloud structures derived from two-point auto-correlation functions.}
\label{structuresizes_physical}
\centering
\begin{tabular}{rrrrrrrrrrrrr}       
\hline               
velocity\tablefootmark{a} & c-C$_3$H$_2$ & H$^{13}$CO$^+$ & $^{13}$CO & CS & C$^{34}$S & $^{13}$CS & SO & SiO & HNC & HN$^{13}$C & HC$^{15}$N & CH$_3$OH\\

[km\,s$^{-1}$] & & & & & & & & & & & & \\
\hline\hline
  \multicolumn{13}{c}{Galactic centre}\\
$-105.9$ & $>0.55$ & $0.55$ & $>0.53$ & $>0.55$ & $0.45$ & $0.16$ & $n^*$ & $0.29$ & $>0.53$ & $0.47$ & $n^*$ & $0.43$\\
$-93.7$ & $0.39$ & $n^*$ & $0.51^*$ & $0.53$ & $n^*$ & $n^*$ & $n^*$ & $n^*$ & $0.49$ & $n^*$ & $n^*$ & $n^*$\\
$-81.5$ & $0.49$ & $0.53$ & $0.51$ & $>0.55$ & $0.47$ & $n^*$ & $n^*$ & $0.12^*$ & $>0.53$ & $n^*$ & $>0.57$ & $n^*$\\
$-74.6$ & $0.51$ & $0.55$ & $>0.53$ & $>0.55$ & $0.31^*$ & $0.08^*$ & $n^*$ & $0.08$ & $>0.53$ & $n^*$ & $0.51$ & $n^*$\\
$-3.2$ & $0.51$ & $0.55$ & $0.2$ & $>0.55$ & $n^*$ & $n^*$ & $0.26$ & $0.53$ & $>0.53$ & $n$ & $0.37$ & $0.51$\\
$2.0$ & $>0.55$ & $>0.57$ & $0.43$ & $0.51$ & $0.51$ & $0.55$ & $>0.55$ & $>0.55$ & $>0.53$ & $0.22$ & $0.53$ & $0.53$\\
$7.3$ & $>0.55$ & $>0.57$ & $0.51$ & $>0.55$ & $0.51$ & $0.55$ & $>0.55$ & $>0.55$ & $>0.53$ & $0.55$ & $0.51$ & $0.53$\\
 \hline
  \multicolumn{13}{c}{3\,kpc arm}\\
$-48.4$ & $0.38^*$ & $n^*$ & $>0.42$ & $0.13$ & $n^*$ & $n^*$ & $n^*$ & $n^*$ & $>0.42$ & $n^*$ & $n^*$ & $n^*$\\
$-39.7$ & $>0.43$ & $>0.45$ & $>0.42$ & $>0.43$ & $0.4$ & $n^*$ & $>0.43$ & $n^*$ & $>0.42$ & $n^*$ & $0.42^*$ & $n^*$\\
 \hline
  \multicolumn{13}{c}{4\,kpc arm}\\
$-27.6$ & $0.31$ & $0.31$ & $>0.33$ & $>0.34$ & $n^*$ & $n^*$ & $n^*$ & $n$ & $>0.33$ & $n^*$ & $n^*$ & $n^*$\\
$-18.9$ & $0.31$ & $0.34$ & $>0.33$ & $>0.34$ & $0.1^*$ & $n^*$ & $0.1$ & $n^*$ & $>0.33$ & $n^*$ & $n^*$ & $n^*$\\
 \hline
  \multicolumn{13}{c}{Scutum arm}\\
$24.7$ & $>0.22$ & $>0.23$ & $0.18$ & $>0.22$ & $n^*$ & $n^*$ & $0.06$ & $0.21$ & $>0.21$ & $n^*$ & $n^*$ & $0.21$\\
$31.6$ & $>0.22$ & $0.04$ & $0.2$ & $0.21$ & $n^*$ & $n^*$ & $0.07$ & $0.03$ & $>0.21$ & $n^*$ & $n^*$ & $>0.22$\\
$36.9$ & $>0.22$ & $0.03$ & $0.17$ & $0.21$ & $0.07$ & $n^*$ & $n^*$ & $0.07$ & $>0.21$ & $n^*$ & $n^*$ & $0.2$\\
 \hline
  \multicolumn{13}{c}{Sagittarius arm}\\
$17.7$ & $0.08$ & $>0.08$ & $0.07$ & $>0.08$ & $0.04$ & $n^*$ & $0.06$ & $>0.08$ & $>0.08$ & $0.04$ & $0.05^*$ & $0.07$\\
\hline                      
\end{tabular}
\tablefoot{The sizes are given in pc. Channels with unresolved structures are marked with $n$ and channels with a SNR of smaller than 5 in the opacity maps are marked with a star. \tablefoottext{a}{Cloud centroid velocities derived from c-C$_3$H$_2$.}}
\end{table*}

The two-point auto-correlation functions of the molecules c-C$_3$H$_2$, H$^{13}$CO$^+$, $^{13}$CO, HNC and its isotopologue HN$^{13}$C, HC$^{15}$N, CS and its isotopologues C$^{34}$S and $^{13}$CS, and CH$_3$OH are discussed in detail in Appendix~\ref{auto_appendix_mol}. The opacity maps suggest that most detected structures are extended on the scale of our field of view, $\sim$15$\arcsec$, or beyond. In a few cases, the two-point auto-correlation functions indicate the presence of smaller structures of sizes $\sim$4--6$\arcsec$. These structures are mostly seen for less abundant species for which most of the opacity map is dominated by noise. For example, the two GC clouds at $2.0$\,km\,s$^{-1}$ and $7.3$\,km\,s$^{-1}$ are detected with high sensitivity for all investigated molecules and display structures that are more extended than $15^{\prime\prime}$ ($\sim$0.5~pc). The only exception is HN$^{13}$C, but the shorter correlation length revealed by this tracer results from its lower abundance, hence a lower sensitivity, compared to the main isotopologue, HNC. 

The other two molecules, SO and SiO, present a more complex picture (Figs.~\ref{opacity_so} and \ref{opacity_sio}). Among the components with peak SNR higher than 5 in their opacity map (Figs.~\ref{snr_opacity_so} and \ref{snr_opacity_sio}), the following ones reveal large-scale structures of the size of the field of view or larger: 2.4, 6.9, and -39.1~km~s$^{-1}$ in SO, and $-$3.2, 1.9, 7.0, 24.0, and 17.2~km~s$^{-1}$ in SiO. The structures traced with SO for three velocity components with a peak SNR of 6--8 in their opacity maps, at $-3.5$, 32.1, and 36.6~km~s$^{-1}$, are more compact, with correlation lengths of $\sim$8$\arcsec$, 5$\arcsec$, and unresolved, respectively. A similar type of compact structures with correlation lengths of $\sim$8$\arcsec$, unresolved, and 5$\arcsec$ is revealed in SiO for the velocity components at $-105.5$, $-27.1$, and 37.7~km~s$^{-1}$ with peak SNR in their opacity maps of $\sim$11, 6, and 6, respectively.

The two-point auto-correlation functions of SO and SiO at $7.3$\,km\,s$^{-1}$ decrease first and increase again at pixel separations larger than about $10^{\prime\prime}$ (see Figs.~\ref{autocorr_so}-\ref{autocorr_sio}). In the opacity maps two smaller structures of sizes of $5^{\prime\prime}$ appear at offsets of about ($1\as 5,11\as0$) and ($12\as0 ,10\as0$), with an angular separation of about $10\as5$ (see Figs.~\ref{opacity_so} and \ref{opacity_sio}). Because they are at the edges of the available field of view, these structures may be more extended.

The two-point auto-correlation functions plotted depending on the physical distance are shown for the eight strongest molecules (c-C$_3$H$_2$, H$^{13}$CO$^+$, $^{13}$CO, CS, SO, SiO, HNC, and CH$_3$OH) in Fig.~\ref{autocorr_c-c3h2_small}--\ref{autocorr_ch3oh_small}. For the seven molecules c-C$_3$H$_2$, $^{13}$CO, CS, SO, SiO, HNC, and CH$_3$OH the auto-correlation functions decrease strongly for the cloud at a velocity of about 18~km\,s$^{-1}$. The physical sizes derived for this cloud which is located in the Sagittarius arm are between 0.04 and 0.08\,pc. The auto-correlation functions for this cloud have sometimes the same shape as the first part of the two-point auto-correlation functions seen for other clouds, for example for CS for the velocities of -2.7 and 18.4\,km\,s$^{-1}$ (see Fig.~\ref{autocorr_cs_small}). Hence, we may only see a smaller part of the cloud located closer to us, but with the same properties as of those clouds more distant from us. On the other hand, structures with sizes smaller than 0.04--0.08\,pc (structure size in the Sagittarius arm) cannot be resolved in the more distant GC clouds. A better resolution is needed to investigate if there are smaller structures present.

\subsection{Turbulence in diffuse and translucent clouds}\label{results_pdf}

\begin{figure*}
\centering
\includegraphics[width=17cm, trim = 1.2cm 2.2cm 0.7cm 3.cm, clip=True]{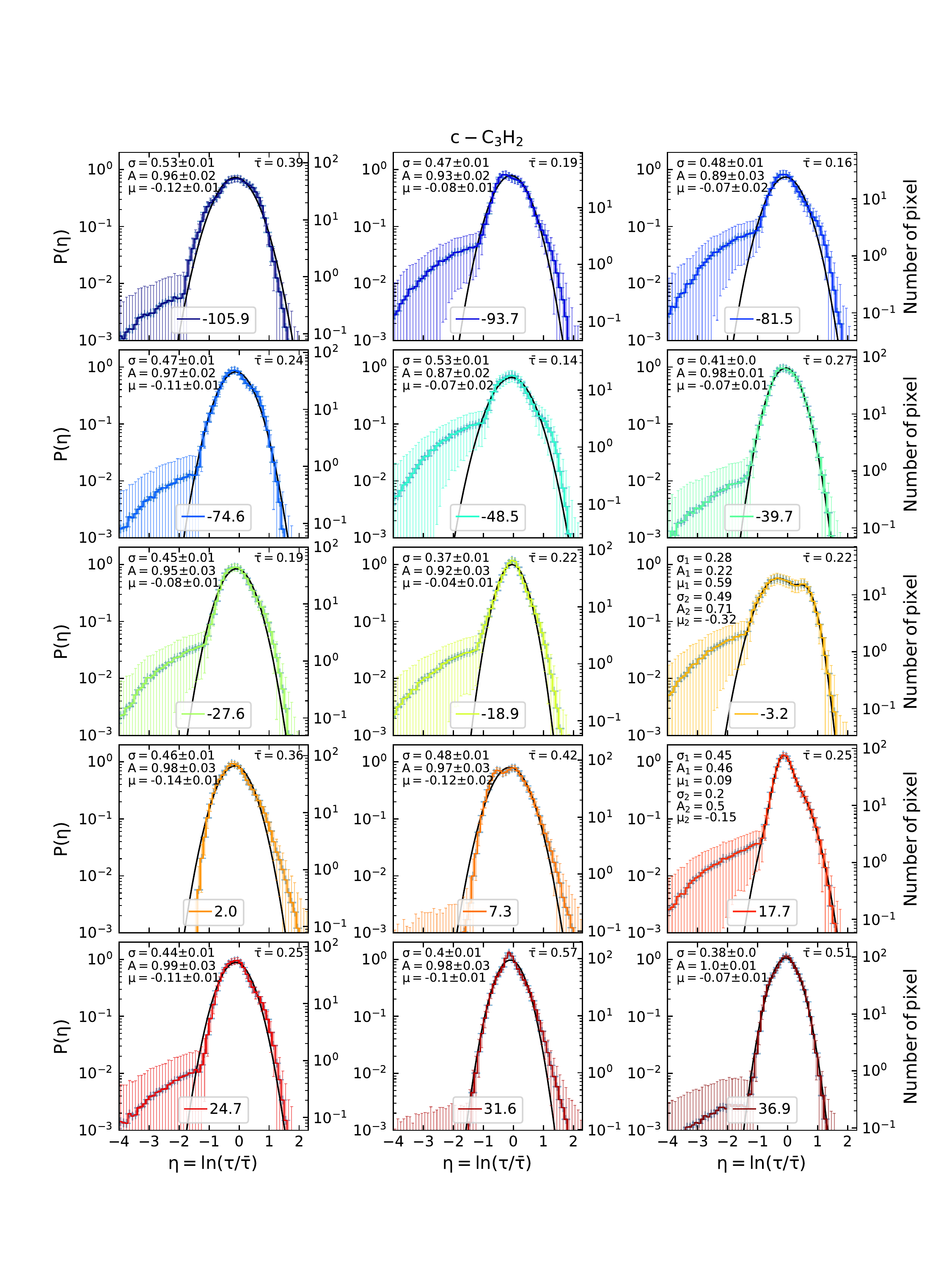}
\caption{Probability distribution functions $P(\eta)$ of the velocity components probed with c-C$_3$H$_2$. The velocity of the component is indicated at the bottom of each panel in km\,s$^{-1}$. The right y-axis indicates the number of pixels counted in each bin. The mean opacity $\bar{\tau}$ is given in the upper right corner and the parameters of the fitted Gaussian(s) in the upper left corner: dispersion $\sigma$, integral $A$, and centre $\mu$.} 
\label{pdf_c-c3h2}
\end{figure*}

In order to investigate the turbulence properties of the clouds detected in absorption towards Sgr~B2, we analyse the PDFs of their opacity maps. To reduce the influence of the noise on the Gaussian fitted to the PDFs (see Section~\ref{sect_noise_pdf}), we use a threshold of $3\sigma_\mathrm{noise}$ to analyse the profile of the PDFs. The PDFs $P(\eta)$ of all 15 velocity components probed with c-C$_3$H$_2$ are shown in Fig.~\ref{pdf_c-c3h2}. The PDFs of the other molecules are plotted in Figs.~\ref{pdf_c34s}--\ref{pdf_13cs} in the Appendix. The number of Gaussians fitted to each PDF is indicated in Table~\ref{number_gaussians}. The results of the Gaussian fits to the PDFs are displayed in Figs.~\ref{sigma_pdf_vel_gauss} and \ref{sigma_pdf_mol_gauss}, and the mean and median widths for each velocity component and for each molecule are listed in Tables~\ref{sigma_pdf_spiralarms_gauss} and \ref{sigma_pdf_molecules_gauss}, respectively. The velocity components that are optically thick and the ones dominated by the noise are marked in Table~\ref{number_gaussians}. The number of fitted Gaussians seems to depend neither on the molecule nor on the velocity component. However, the molecules for which the PDFs are most often well fitted with a single Gaussian are HNC and c-C$_3$H$_2$, with only one and two velocity component(s) fitted with two Gaussians, respectively.

\citet{tremblin2014} investigated the structure of the dense gas in several molecular clouds and explained the presence of two log-normal profiles or an enlarged shape of the PDF of a cloud as two different zones existing in the cloud. In the case they studied the turbulent molecular gas creates the low density part in the PDF and the second peak describes a compression zone created by the expansion of ionised gas into the molecular cloud. Another possibility is that the PDFs containing two log-normal profiles result from two different clouds overlapping along the line of sight at different distances from the observer but with the same velocity. We used opacity maps of only one channel and no integrated maps for the calculation of the PDFs to reduce the possibility of two clouds contributing to the same opacity map but such an overlap may still occur. Furthermore, our limited field of view that is set by the strength of the background continuum emission may have an effect on the shape of the PDFs. Because no velocity component shows a PDF with a two-Gaussian shape for all molecules, we believe that this particular shape does probably not characterise the true physical structure of the component. Therefore, to avoid being biased by the decomposition of the PDFs into two Gaussians, in the following we measure the width of each PDF by directly calculating its standard deviation, excluding the noise tail towards lower values of $\eta$. This cut may result in a slightly underestimated width of the PDF.

\begin{figure}
   \resizebox{\hsize}{!}{\includegraphics[width=0.5\textwidth,trim = 0.cm 0.cm 0cm 0cm, clip=True]{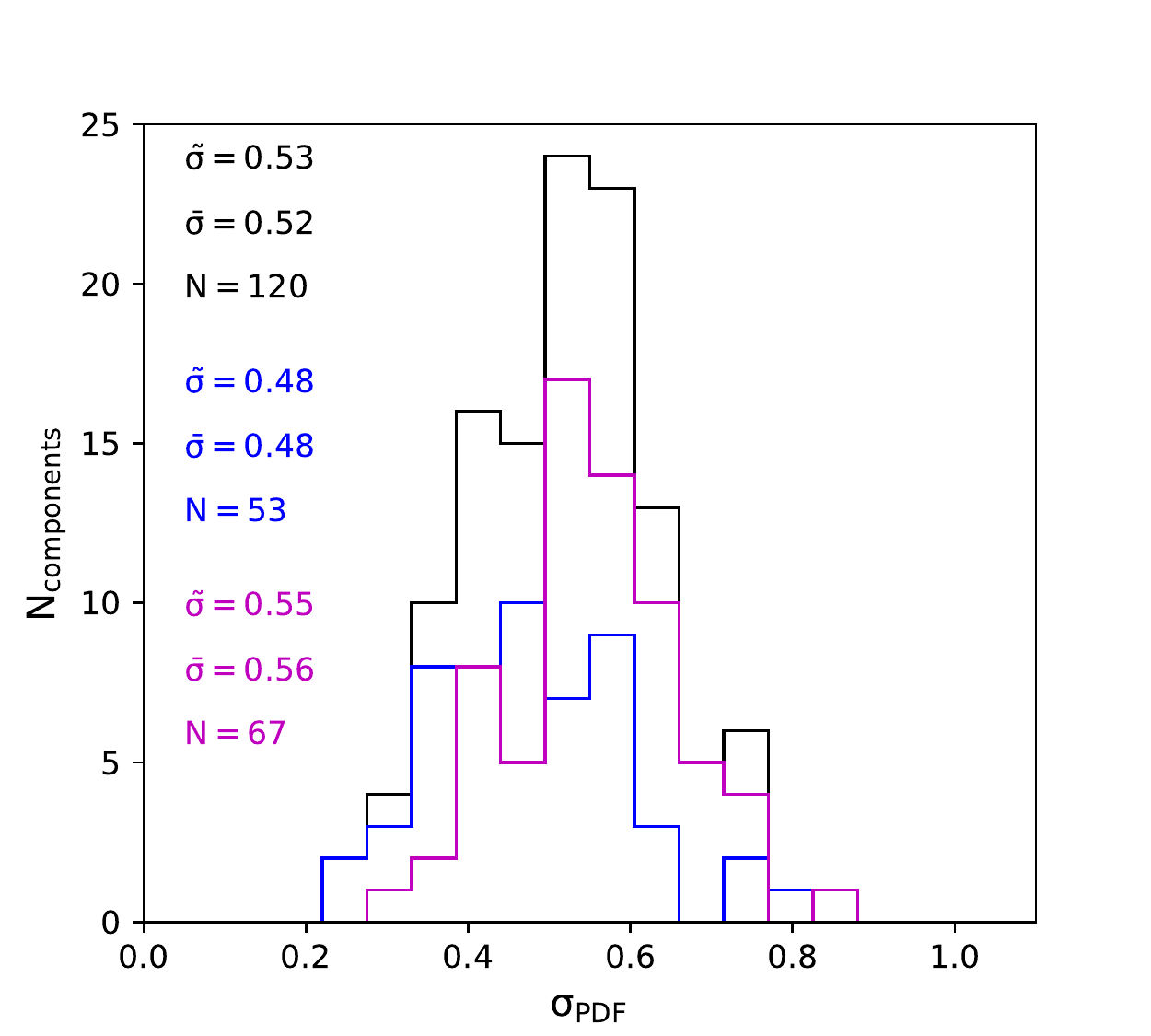}}
 \caption{Distribution of standard deviations (width) of the PDFs of all molecules for all 15 velocity components. $N$ gives the total number of PDFs used for each histogram, $\bar{\sigma}$ the mean value, and $\tilde{\sigma}$ the median value. The distribution of Categories I and II are shown in blue and magenta, respectively.}\label{pdf_total}
\end{figure}

The distribution of PDF widths derived from the direct calculation is plotted in black in Fig.~\ref{pdf_total}. The median and mean values are similar, with values of 0.52 and 0.53 for the total distribution, respectively. The distributions corresponding to Categories I and II defined in Sect.~\ref{results_cloud_properties} are plotted in blue and magenta. Category I has a mean width of 0.48, somewhat smaller than Category II (0.56).

\begin{figure}
   \resizebox{\hsize}{!}{\includegraphics[width=0.5\textwidth,trim = 0.cm 0.cm 0cm 0cm, clip=True]{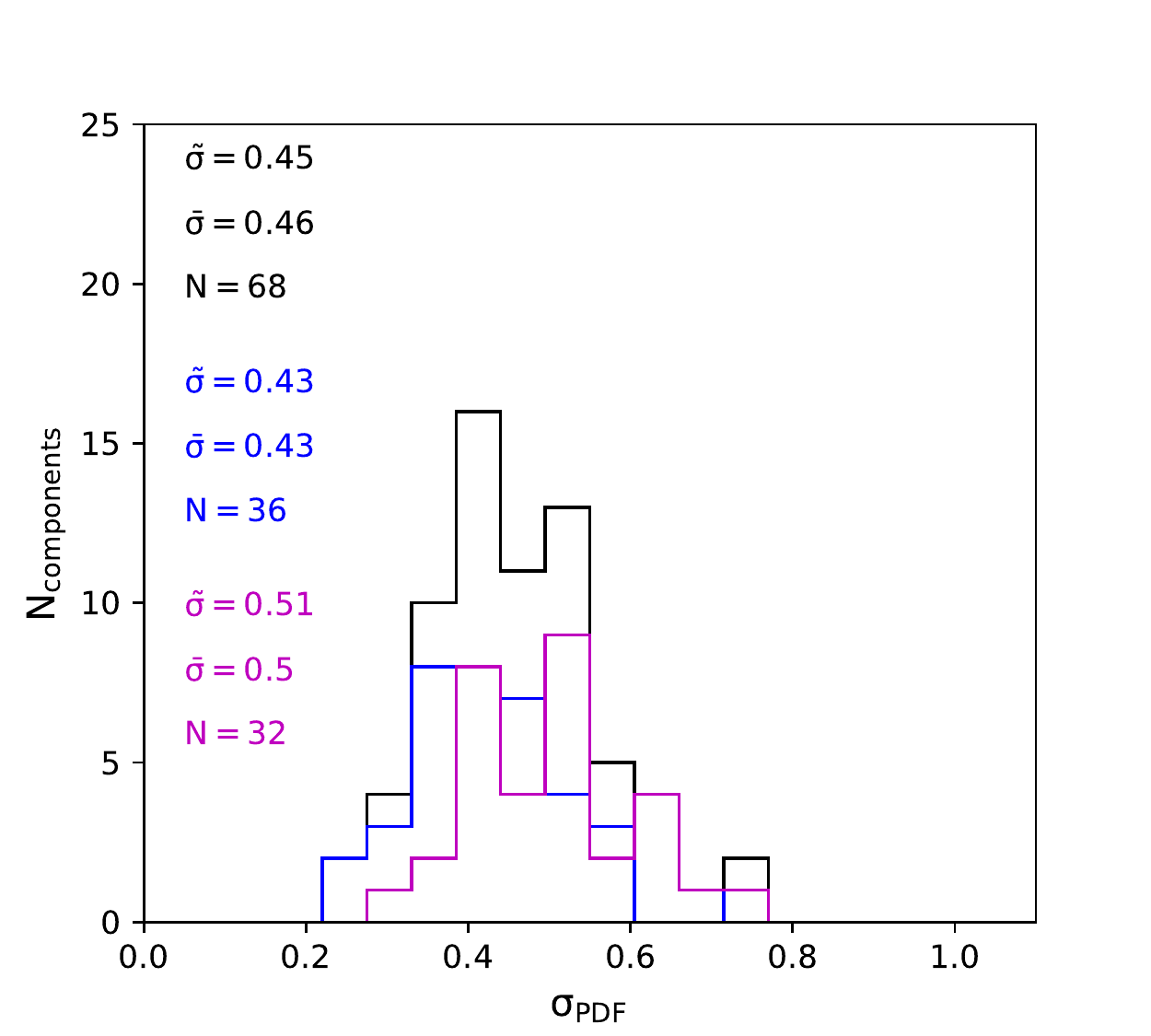}}
 \caption{Same as Fig.~\ref{pdf_total} but for the sub-sample containing only c-C$_3$H$_2$, H$^{13}$CO$^+$, $^{13}$CO, CS, and HNC.}
\label{pdf_total_subsample}
\end{figure}

To investigate whether the shift between the two groups results from the different samples of molecules used for the different velocity components, we determine the distribution of PDF widths for the following five molecules only: c-C$_3$H$_2$, H$^{13}$CO$^+$, $^{13}$CO, CS, and HNC (see Fig.~\ref{pdf_total_subsample}). These molecules are well detected over the field of view for almost all velocity components. The other molecules are not detected for some of the components. With this reduced sample of molecules, the two categories of velocity components still have mean PDF widths that differ, with values of 0.43 (Category I) and 0.50 (Category II). The widths are smaller than for the sample including all molecules. This is most likely due to the noise affecting the molecules that show weak absorption because the noise tends to broaden the PDF \citep{ossenkopf2016}. 

\begin{figure}
   \resizebox{\hsize}{!}{\includegraphics[width=0.5\textwidth,trim = 1.cm 1.0cm 2cm 2.95cm, clip=True]{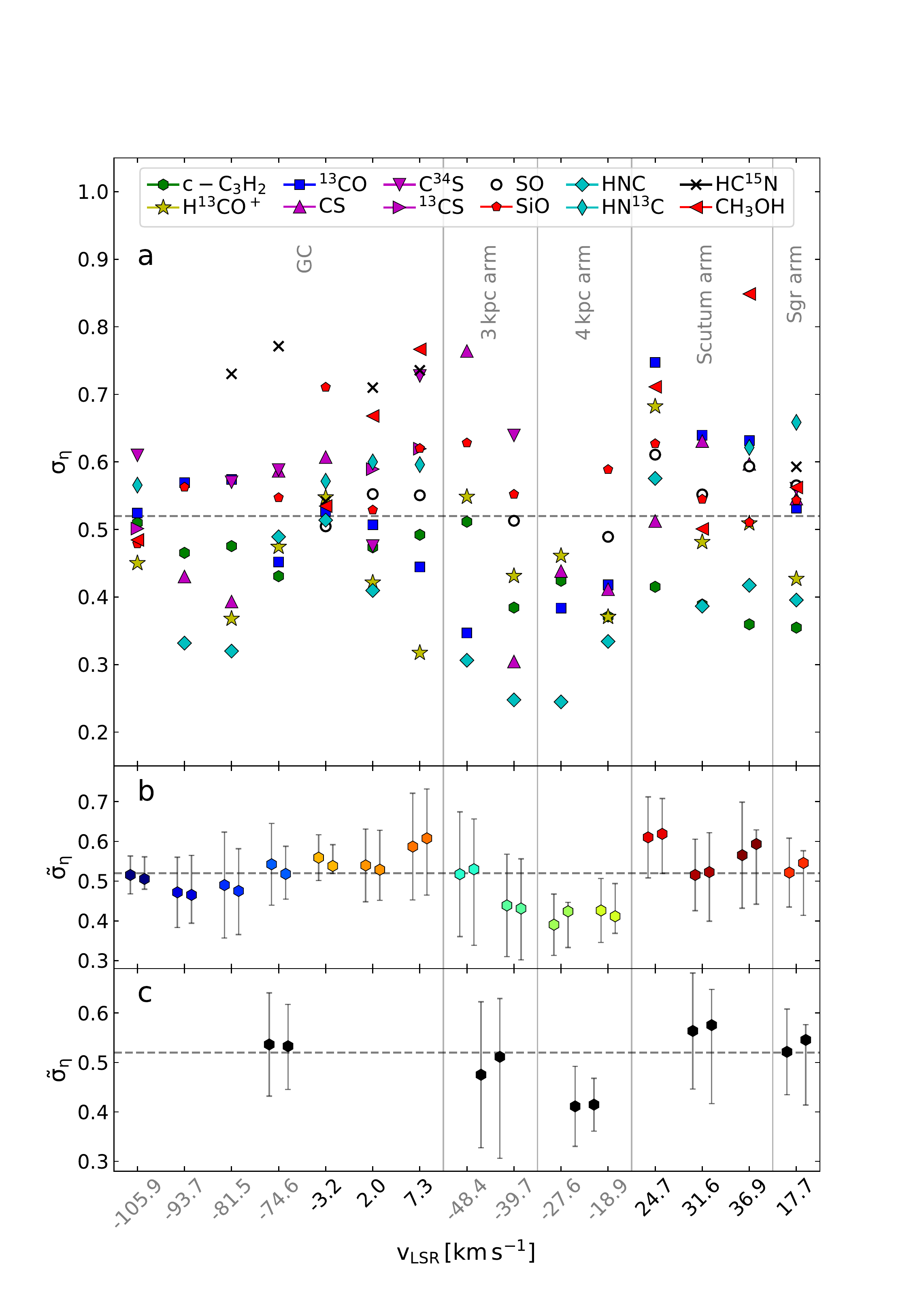}}
 \caption{\textbf{a} Widths of the PDFs of all molecules for the 15 velocity components, roughly sorted by their distance to the Galactic centre. \textbf{b} Mean (left) and median (right) values for each velocity component. \textbf{c} Mean (left) and median (right) values for each sub-sample of clouds, from left to right: Galactic centre, 3~kpc arm, 4~kpc arm, Scutum arm, Sagittarius arm. The uncertainties represent the standard deviation for the mean and the corresponding percentiles for the median. The dashed line in each panel marks the mean value of all data points shown in panel a. Velocity components belonging to Category I and II are coloured in grey and black, respectively.}\label{sigma_pdf_vel}
\end{figure}

The widths of the PDFs of all molecules are plotted in Fig.~\ref{sigma_pdf_vel} for all velocity components, sorted by their rough distance to the GC, and are listed in Table~\ref{sigma_pdf_spiralarms_allmol}. To investigate whether there are systematic differences between the velocity components, we plot the mean and median values of each velocity component in panel b. We also show the mean and median values of each spiral arm and the GC in panel c. These values are also listed in Table~\ref{sigma_pdf_spiralarms_allmol}. The median and mean values match each other within the uncertainties.

\begin{table}
\caption{Mean ($\bar{\sigma}$) and median ($\tilde{\sigma}$) widths directly computed from the PDFs for each velocity component.}
\label{sigma_pdf_spiralarms_allmol}
\centering
\begin{tabular}{r c c c c}       
\hline               
\multicolumn{1}{c}{$\varv_\mathrm{LSR}$\tablefootmark{a}} & $\bar{\sigma}$ & $\tilde{\sigma}$ & $\bar{\sigma}$\tablefootmark{b} & $\tilde{\sigma}$\tablefootmark{b}\\

 [km\,s$^{-1}$] & & & & \\
\hline\hline   
\multicolumn{5}{c}{Galactic centre}\\
$-105.9$ & $0.52\pm 0.05$ & $0.48\substack{+0.06 \\ -0.03}$ & \multirow{7}{*}{$0.54\pm0.10$} & \multirow{7}{*}{$0.53\substack{+0.08 \\ -0.09}$}\\
 $-93.7$ & $0.47\pm 0.09$ & $0.47\substack{+0.10 \\ -0.07}$ & &\\
 $-81.5$ & $0.49\pm 0.13$ & $0.48\substack{+0.11 \\ -0.11}$ & &\\
 $-74.6$ & $0.54\pm 0.10$ & $0.52\substack{+0.07 \\ -0.06}$ & &\\
 $-3.2$ & $0.56\pm 0.06$ & $0.54\substack{+0.05 \\ -0.02}$ & &\\
 $2.0$ & $0.54\pm 0.09$ & $0.53\substack{+0.10 \\ -0.08}$ & &\\
 $7.3$ & $0.59\pm 0.13$ & $0.61\substack{+0.12 \\ -0.14}$ & &\\
 \hline
 \multicolumn{5}{c}{3\,kpc arm}\\
$-48.4$ & $0.52\pm 0.16$ & $0.53\substack{+0.13 \\ -0.19}$ & \multirow{2}{*}{$0.48\pm0.15$} & \multirow{2}{*}{$0.51\substack{+0.12 \\ -0.21}$}\\
$-39.7$ & $0.44\pm 0.13$ & $0.43\substack{+0.13 \\ -0.13}$ & &\\
 \hline
 \multicolumn{5}{c}{4\,kpc arm}\\
 $-27.6$ & $0.39\pm 0.08$ & $0.42\substack{+0.02 \\ -0.09}$ & \multirow{2}{*}{$0.41\pm0.08$} & \multirow{2}{*}{$0.41\substack{+0.05 \\ -0.05}$}\\
$-18.9$ & $0.43\pm 0.08$ & $0.41\substack{+0.08 \\ -0.04}$ & &\\
 \hline
  \multicolumn{5}{c}{Scutum arm}\\
$24.7$ & $0.61\pm 0.10$ & $0.62\substack{+0.09 \\ -0.10}$ & \multirow{3}{*}{$0.56\pm0.12$} & \multirow{3}{*}{$0.58\substack{+0.07 \\ -0.16}$}\\
 $31.6$ & $0.52\pm 0.09$ & $0.52\substack{+0.10 \\ -0.12}$ & &\\
 $36.9$ & $0.57\pm 0.13$ & $0.59\substack{+0.04 \\ -0.15}$ & &\\
 \hline
  \multicolumn{5}{c}{Sagittarius arm}\\
$17.7$ & $0.52\pm 0.09$ & $0.55\substack{+0.03 \\ -0.13}$ & \multirow{1}{*}{$0.52\pm0.09$} & \multirow{1}{*}{$0.55\substack{+0.03 \\ -0.13}$}\\
\hline                      
\end{tabular}
\tablefoot{\tablefoottext{a}{Channel velocities of c-C$_3$H$_2$.}\tablefoottext{b}{Mean and median values for each sub-sample of clouds for all molecules.}}
\end{table}

As seen in Fig.~\ref{pdf_total} there is a difference between Categories I and II. Category I contains the clouds in the 3~kpc and 4~kpc arms and the GC in the velocity range between $-106$ and $-75$~km~s$^{-1}$. They have narrower PDF widths than the clouds in Category II. The GC clouds belonging to Category I have widths between 0.47 and 0.52, the GC clouds of Category II have widths between 0.54 and 0.61. Especially the clouds in the 4~kpc arm have a narrower mean width, 0.41$\pm0.08$, which is somewhat smaller than the overall mean value (0.52).  

\begin{figure}
   \resizebox{\hsize}{!}{\includegraphics[width=0.5\textwidth,trim = 1.cm 0.3cm 2cm 2.2cm, clip=True]{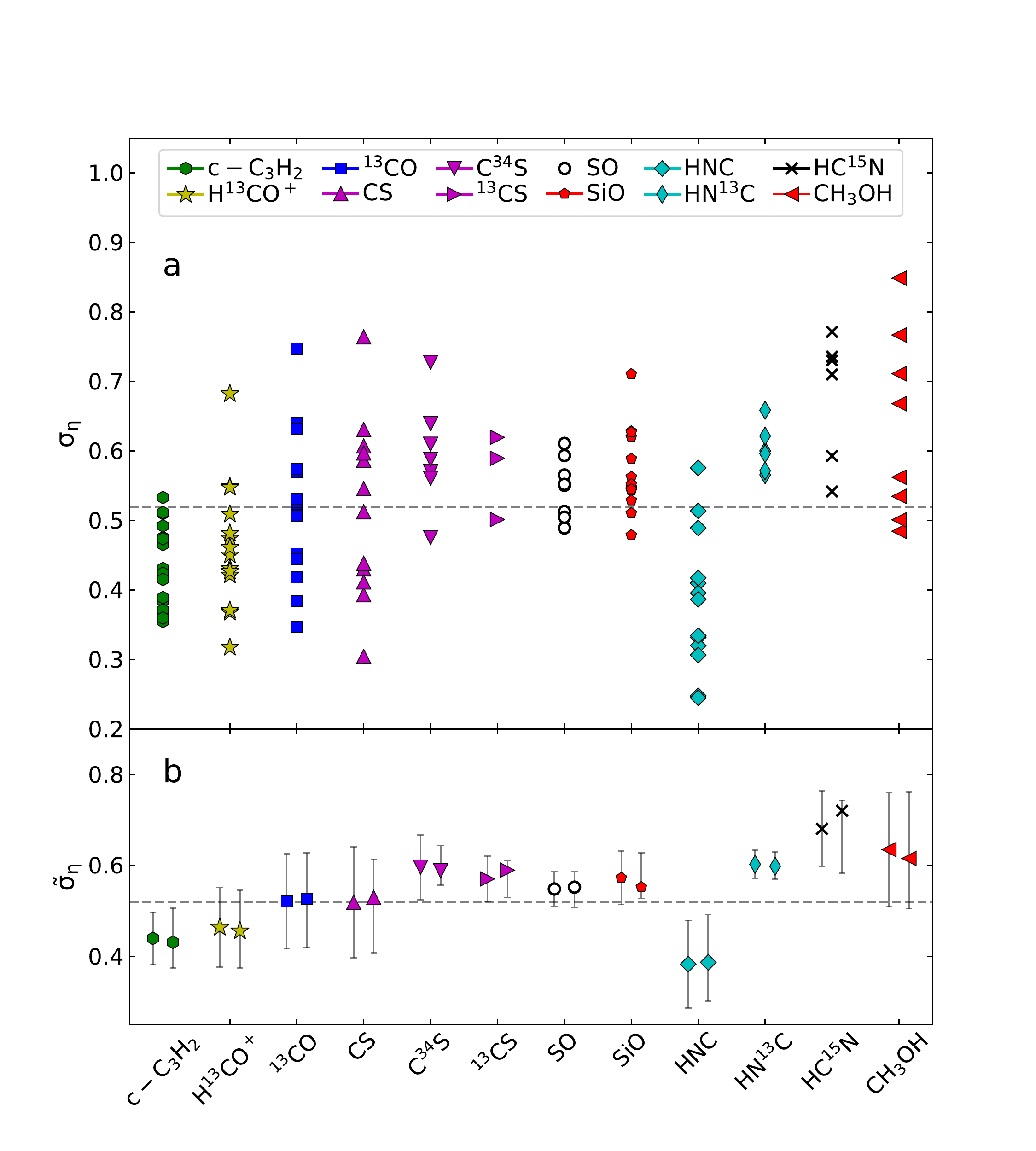}}
 \caption{\textbf{a} Widths of the PDFs of 15 different velocity components along the line of sight sorted by molecule. \textbf{b} Mean (left) and median (right) values for each molecule. The uncertainties represent the standard deviation for the mean and the corresponding percentiles for the median. The dashed line in both panels marks the mean and the median value for all components (see Fig.~\ref{pdf_total}).}\label{sigma_pdf_mol}
\end{figure}

We also investigate whether the width of the PDFs depends on the molecule. Figure~\ref{sigma_pdf_mol}a shows the distribution of PDF widths as a function of molecule. The mean and median values are plotted in panel b and listed in Table~\ref{sigma_pdf_molecules}. The molecules c-C$_3$H$_2$, H$^{13}$CO$^+$, and HNC have the narrowest widths. The less abundant isotopologues C$^{34}$S, $^{13}$CS, HN$^{13}$C, and HC$^{15}$N and the less abundant molecules SO, SiO, and CH$_3$OH have systematically broader PDF widths than the previous, more abundant molecules. This explains the difference seen between Figs.~\ref{pdf_total} and \ref{pdf_total_subsample}. 

\begin{table*}
\caption{PDF widths and parameters describing the turbulence for all investigated molecules.}
\label{sigma_pdf_molecules}
\centering
\begin{tabular}{l c c c c c c c c}       
\hline               
molecule & $\bar{\sigma}$\tablefootmark{a} & $\tilde{\sigma}$\tablefootmark{b} & \multicolumn{3}{c}{$\zeta$\tablefootmark{c}} & \multicolumn{3}{c}{$b$\tablefootmark{d}}\\

& & & 20~K & 40~K & 80~K & 20~K & 40~K & 80~K \\
\hline\hline   
c-C$_3$H$_2$ & $0.44\pm 0.06$ & $0.43\substack{+0.08 \\ -0.06}$ & x & x & x & $0.14\substack{+0.09 \\ -0.07}$ &$0.20\substack{+0.12 \\ -0.10}$ & $0.28\substack{+0.07 \\ -0.14}$ \\
H$^{13}$CO$^+$ & $0.46\pm 0.09$ & $0.46\substack{+0.09 \\ -0.08}$ & x & x & x & $0.13\substack{+0.12 \\ -0.06}$ & $0.18\substack{+0.14 \\ -0.08}$ & $0.26\substack{+0.12 \\ -0.11}$ \\         
$^{13}$CO & $0.52\pm 0.10$ & $0.53\substack{+0.10 \\ -0.11}$ & x & x & 0.93 & $0.17\substack{+0.12 \\ -0.07}$ & $0.25\substack{+0.08 \\ -0.09}$ & $0.33\substack{+0.05 \\ -0.11}$ \\
CS & $0.52\pm 0.12$ & $0.53\substack{+0.08 \\ -0.12}$ & x & x & 0.78 & $0.20\substack{+0.07 \\ -0.10}$ & $0.28\substack{+0.05 \\ -0.15}$ & $0.33\substack{+0.04 \\ -0.14}$ \\
C$^{34}$S & $0.60\pm 0.07$ & $0.59\substack{+0.6 \\ -0.03}$ & 0.79 & 0.52 & 0.38 & $0.33\substack{+0.01 \\ -0.13}$ & $0.36\substack{+0.04 \\ -0.08}$ & $0.45\substack{+0.06 \\ -0.13}$ \\
$^{13}$CS & $0.57\pm 0.05$ & $0.59\substack{+0.02 \\ -0.06}$ & x & 0.99 & 0.58 & $0.23\substack{+0.04 \\ -0.05}$ & $0.33\substack{+0.01 \\ -0.07}$ & $0.34\substack{+0.01 \\ -0.01}$ \\
SO & $0.55\pm 0.04$ & $0.55\substack{+0.03 \\ -0.05}$ & x & x & 0.74 & $0.21\substack{+0.04 \\ -0.08}$ & $0.29\substack{+0.03 \\ -0.11}$ & $0.33\substack{+0.02 \\ -0.08}$ \\
SiO & $0.57\pm 0.06$ & $0.55\substack{+0.08 \\ -0.03}$ & x & x & 0.66 & $0.21\substack{+0.12 \\ -0.07}$ & $0.30\substack{+0.07 \\ -0.10}$ & $0.33\substack{+0.12 \\ -0.06}$ \\
HNC & $0.38\pm 0.10$ & $0.39\substack{+0.11 \\ -0.09}$ & x & x & x & $0.09\substack{+ 0.07\\ -0.04}$ & $0.13\substack{+0.11 \\ -0.05}$ & $0.19\substack{+0.14 \\ -0.07}$ \\
HN$^{13}$C & $0.60\pm 0.03$ & $0.60\substack{+0.03 \\ -0.03}$ & x & 0.92 & 0.60 & $0.25\substack{+0.07 \\ -0.04}$ & $0.33\substack{+0.01 \\ -0.04}$ & $0.34\substack{+0.04 \\ -0.01}$ \\
HC$^{15}$N & $0.68\pm 0.08$ & $0.72\substack{+0.02 \\ -0.14}$ & 0.80 & 0.60 & 0.44 & $0.33\substack{+0.07 \\ -0.10}$ & $0.36\substack{+0.12 \\ -0.04}$ & $0.43\substack{+0.17 \\ -0.09}$ \\
CH$_3$OH & $0.63\pm 0.13$ & $0.62\substack{+0.15 \\ -0.11}$ & x & 0.97 & 0.64  & $0.23\substack{+0.14 \\ -0.05}$ & $0.32\substack{+0.16 \\ -0.06}$ & $0.33\substack{+0.27 \\ -0.01}$ \\
\hline                      
\end{tabular}
\tablefoot{\tablefoottext{a}{Mean width of the PDF.}\tablefoottext{b}{Median width of the PDF.}\tablefoottext{c}{Forcing parameter $\zeta$. x means that there is no intersection of the two functions $b(\zeta)$. In these cases, we assume $f=2.9$ to derive $b$.}\tablefoottext{d}{Forcing parameter $b$.}}
\end{table*}

\begin{table*}
\caption{PDF widths, linewidths, Mach numbers, and parameters describing the turbulence of the velocity components.}
\label{sigma_pdf_spiralarms}
\centering
\begin{tabular}{r c c c c c c c c c c c c c }       
\hline               
\multicolumn{1}{c}{$\varv_\mathrm{LSR}$\tablefootmark{a}} & $\bar{\sigma}$\tablefootmark{b} & $\tilde{\sigma}$\tablefootmark{c} & $FWHM$\tablefootmark{e} & \multicolumn{3}{c}{$M$\tablefootmark{f}} & \multicolumn{3}{c}{$\zeta$\tablefootmark{g}}& \multicolumn{3}{c}{$b$\tablefootmark{h}} \\

 [km\,s$^{-1}$] & & &  [km\,s$^{-1}$] & 20~K & 40~K & 80~K & 20~K & 40~K & 80~K  & 20~K & 40~K & 80~K\\
\hline\hline   
\multicolumn{13}{c}{Galactic centre}\\
$-105.9$ & $0.49\pm 0.03$ & $0.48\substack{+0.03 \\ -0.02}$ & 4.2 & 11.7 & 8.1 & 5.8 & x & x & 0.63 & $0.22\substack{+0.03 \\ -0.02}$ & $0.31\substack{+0.02 \\ -0.02}$ &  $0.34\substack{+0.02 \\ -0.01}$ \\
 $-93.7$ & $0.47\pm 0.09$ & $0.47\substack{+0.10 \\ -0.07}$ & 4.6 & 12.8 & 8.9 & 6.4 & x & x & 0.85 & $0.18\substack{+0.11 \\ -0.05}$ & $0.26\substack{+0.07 \\ -0.07}$ & $0.33\substack{+0.04 \\ -0.06}$  \\
 $-81.5$ & $0.43\pm 0.09$ & $0.39\substack{+0.12 \\ -0.04}$ & 2.1 & 5.8 & 4.1 & 2.9 & x & 0.66 & 0.43 & $0.28\substack{+0.07 \\ -0.05}$ & $0.33\substack{+0.12 \\ -0.02}$ & $0.41\substack{+0.18 \\ -0.05}$ \\
 $-74.6$ & $0.50\pm 0.05$ & $0.48\substack{+0.07 \\ -0.03}$ & 7.1 & 19.7 & 13.7 & 9.9 & x & x & x & $0.13\substack{+0.06 \\ -0.02}$ & $0.18\substack{+0.08 \\ -0.03}$ & $0.25\substack{+0.08 \\ -0.04}$ \\
 $-3.2$ & $0.56\pm 0.06$ & $0.53\substack{+0.07 \\ -0.02}$ & 6.4 & 17.8 & 12.4 & 8.9 & x & x & 0.89 & $0.18\substack{+0.07 \\ -0.02}$ & $0.26\substack{+ 0.07\\ -0.02}$ & $0.33\substack{+0.01 \\ -0.01}$ \\
 $2.0$ & $0.51\pm 0.08$ & $0.51\substack{+0.05 \\ -0.09}$ & 9.4 & 26.1 & 18.2 & 13.1 & x & x & x & $0.11\substack{+0.03 \\ -0.04}$ & $0.15\substack{+0.04 \\ -0.05}$ & $0.22\substack{+0.06 \\ -0.07}$ \\
 $7.3$ & $0.53\pm 0.14$ & $0.52\substack{+0.13 \\ -0.10}$ & 6.1 & 16.9 & 11.8 & 8.5 &  & x & 0.89 & $0.18\substack{+0.13 \\ -0.07}$ & $0.25\substack{+0.10 \\ -0.10}$ & $0.32\substack{+0.08 \\ -0.09}$ \\
 all\tablefootmark{d} & $0.54\pm0.10$ & $0.53\substack{+0.08 \\ -0.09}$ &  &  &  &  &  & x &  &  &  &  \\
 \hline
 \multicolumn{13}{c}{3\,kpc arm}\\
$-48.4$ & $0.52\pm 0.16$ & $0.53\substack{+0.13 \\ -0.19}$ & 4.2 & 11.7 & 8.1 & 5.8 & x & 0.82 & 0.50 & $0.27\substack{+0.08 \\ -0.16}$ & $0.33\substack{+0.08 \\ -0.17}$ & $0.37\substack{+ 0.16\\ -0.14}$ \\
$-39.7$ & $0.41\pm 0.11$ & $0.41\substack{+0.11 \\ -0.11}$ & 4.1 & 11.4 & 7.9 & 5.7 & x & x & x & $0.16\substack{+0.11 \\ -0.06}$ & $0.22\substack{+0.11 \\ -0.09}$ & $0.30\substack{+0.06 \\ -0.12}$ \\
all\tablefootmark{d} & $0.48\pm0.15$ & $0.51\substack{+0.12 \\ -0.21}$ &  &  &  &  &  &  &  &  &  &  \\
 \hline
 \multicolumn{13}{c}{4\,kpc arm}\\
 $-27.6$ & $0.39\pm 0.08$ & $0.42\substack{+0.02 \\ -0.09}$ & 7.6 & 21.1 & 14.7 & 10.6 & x & x & x & $0.09\substack{+0.01 \\ -0.03}$ & $0.13\substack{+0.01 \\ -0.04}$ &  $0.18\substack{+0.02 \\ -0.06}$ \\
$-18.9$ & $0.43\pm 0.08$ & $0.41\substack{+0.08 \\ -0.04}$ & 8.2 & 22.8 & 15.9 & 11.4 & x & x & x & $0.08\substack{+0.04 \\ -0.01}$ & $0.11\substack{+0.05 \\ -0.02}$ & $0.16\substack{+0.07 \\ -0.03}$ \\
all\tablefootmark{d} & $0.41\pm0.08$ & $0.41\substack{+0.05 \\ -0.05}$ &  &  &  &  &  &  &  &  &  &  \\
 \hline
  \multicolumn{13}{c}{Scutum arm}\\
$24.7$ & $0.61\pm 0.10$ & $0.62\substack{+0.09 \\ -0.10}$ & 4.1 & 11.4 & 7.9 & 5.7 & 0.72 & 0.48 & 0.38 & $0.33\substack{+0.05 \\ -0.07}$ & $0.37\substack{+0.10 \\ -0.04}$ & $0.47\substack{+0.14 \\ -0.11}$ \\
 $31.6$ & $0.52\pm 0.09$ & $0.52\substack{+0.10 \\ -0.12}$ & 10.2 & 28.3 & 19.7 & 14.2 & x & x & x & $0.11\substack{+0.07 \\ -0.05}$ & $0.15\substack{+0.10 \\ -0.07}$ & $0.21\substack{+0.11 \\ -0.09}$ \\
 $36.9$ & $0.56\pm 0.14$ & $0.55\substack{+0.08 \\ -0.12}$ & 7.4 & 20.5 & 14.3 & 10.3 & x & x & 0.94 & $0.18\substack{+0.08 \\ -0.08}$ &  $0.25\substack{+0.08 \\ -0.11}$& $0.30\substack{+0.04 \\ -0.11}$ \\
 all\tablefootmark{d} & $0.56\pm0.12$ & $0.58\substack{+0.07 \\ -0.16}$ &  &  &  &  &  &  &  &  &  &  \\
 \hline
  \multicolumn{13}{c}{Sagittarius arm}\\
$17.7$ & $0.49\pm 0.08$ & $0.54\substack{+0.02 \\ -0.14}$ & 6.5 & 18.0 & 12.6 & 9.0 & x & x & 0.89 & $0.18\substack{+0.02 \\ -0.09}$ & $0.26\substack{+0.03 \\ -0.12}$ & $0.33\substack{+0.01 \\ -0.14}$ \\
\hline                      
\end{tabular}
\tablefoot{The molecules used for this analysis are c-C$_3$H$_2$, H$^{13}$CO$^+$, $^{13}$CO, CS, SO, SiO, HNC, and CH$_3$OH. \tablefoottext{a}{Channel velocities of c-C$_3$H$_2$.}\tablefoottext{b}{Mean width of the PDF.}\tablefoottext{b}{Median width of the PDF.}\tablefoottext{d}{Mean and median values of the PDF width for each sub-sample of clouds.}\tablefoottext{e}{Median $FWHM$ for each velocity component determined with c-C$_3$H$_2$.}\tablefoottext{f}{Mach number for three assumed temperatures.}\tablefoottext{g}{Forcing parameter $\zeta$. x means that there is no intersection of the two functions $b(\zeta)$. In these cases, we assume $f=2.9$ to derive $b$.}\tablefoottext{h}{Forcing parameter $b$.}}
\end{table*}

We use the widths of the PDFs to investigate the turbulent properties of the clouds probed in absorption by calculating the forcing parameter $b$ \citep[e.g.][]{federrath2010}. For this calculation, we need the Mach number $M$ which is defined as:
\begin{equation}
 M = (\sqrt{3} FWHM)/(c_\mathrm{s}\sqrt{8\,\ln(2)})
\end{equation}  
with $FWHM$ the linewidth of the molecule\footnote{We neglect the thermal contribution to the linewidth of the molecules, which is justified given the large measured linewidths, even for a kinetic temperature of 100~K.} and $c_\mathrm{s}$ the sound speed
\begin{equation}
 c_\mathrm{s} = \sqrt{\frac{k_\mathrm{B}T_\mathrm{kin}}{\bar{\mu}m_\mathrm{H}}}
\end{equation} 
with $T_\mathrm{kin}$ the kinetic temperature, $k_\mathrm{B}$ the Boltzmann constant, $m_\mathrm{H}$ the mass of the hydrogen atom, and $\bar{\mu}$ the mean molecular weight \citep[2.37, see, e.g.][]{kauffmann2008}. \citet{snow2006} quote kinetic temperatures between 30 and 100\,K for diffuse molecular clouds and  between 15 and 50\,K for translucent molecular clouds. Here, we assume temperatures of 20, 40, and 80\,K. We determine the median $FWHM$ for each velocity component and calculate the Mach number for the assumed temperatures (Table~\ref{sigma_pdf_spiralarms}). We obtain Mach number values between 5.8 and 28.3 for $T_\mathrm{kin}=20$~K and between 2.9 and 14.2 for 80~K. 

The forcing parameter $b$ relates the velocity and density fields in a cloud \citep{padoan1997,federrath2008}: 
\begin{equation}\label{equ_b}
 \sigma_\mathrm{s}^2 = f^2\sigma_\mathrm{\eta}^2 = \ln(1+b^2 M^2) 
\end{equation}
with $\sigma_\mathrm{s}$ the standard deviation of the volume density fluctuations and $\sigma_\mathrm{\eta}$ the dispersion of the two-dimensional column density or opacity fluctuations. The relation is derived from numerical simulations of magnetohydrodynamics (MHD) and hydrodynamics \citep{padoan1997,passot1998}.
In \citet{federrath2010}, the value of $f$ is investigated in the extreme cases of purely solenoidal forcing (divergence free) and purely compressive forcing (curl-free): they obtain $2.9$ for solenoidal forcing ($f=\sigma_\mathrm{s}/\sigma_\mathrm{\eta}=1.32/0.46$) and $2.0$ for compressive forcing ($3.04/1.51$). They also define a parameter $\zeta$ that sets the power of compressive forcing with respect to the total power of the turbulence forcing. $\zeta$ takes values between 0 (purely compressive) and 1 (purely solenoidal). They show that $b$ is a function of $\zeta$ (see their Fig. 8).

\begin{figure*}
  \centering
\includegraphics[width=17cm,trim = 1.5cm .2cm 1.0cm 1.cm, clip=True]{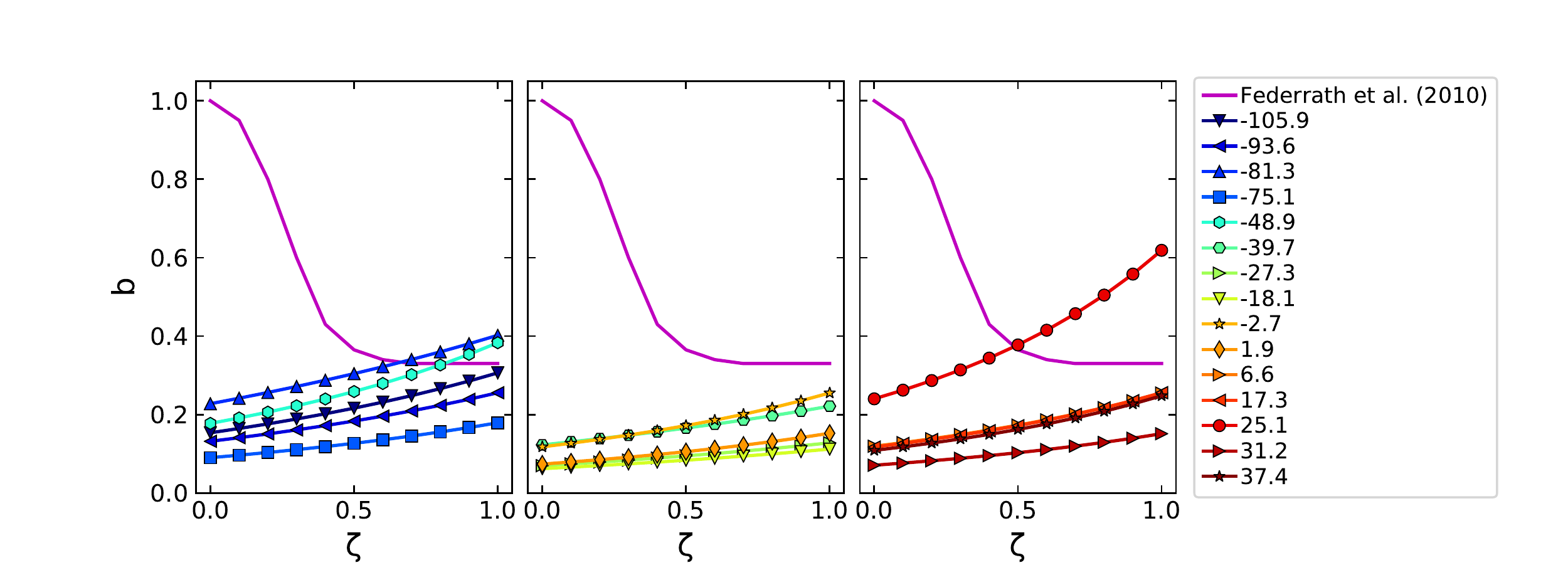}
 \caption{Forcing parameter $b$ plotted against $\zeta$ for $T_\mathrm{kin}=40$~K for all 15 velocity components. The different velocities are colour coded and given in km~s~$^{-1}$. The magenta line represents the function derived from hydrodynamic simulations by \citet{federrath2010}.}\label{zeta}
\end{figure*}

To calculate $b$ from Eq.~\ref{equ_b}, we need to know $f$. Given that $f$ does not vary much between the two extreme forcing cases investigated by \citet{federrath2010}, we assume that it is a simple linear function of $\zeta$ and parametrise it as $f=2.9\times\zeta+(1-\zeta)\times2.0$ (linear interpolation between the values of $f$ obtained for the extreme cases $\zeta$=1 and $\zeta$=0). Equation~\ref{equ_b} then gives us $b$ as a function of $\zeta$. The intersection of this function with the relation found by \citet{federrath2010} gives the solution ($b$,$\zeta$), when it exists. As an example, these functions are plotted for the different velocity components for a kinetic temperature of $40$~K in Fig.~\ref{zeta}. In many cases the two curves do not intersect for $\zeta$ between 0 and 1, but they come the closest to each other for $\zeta = 1$. In these cases, we assume a value of 2.9 for $f$ to derive $b$. In the other cases, the intersection gives us $b$ and $\zeta$.

We consider only the eight molecules with highest SNR to derive $b$ for each velocity component: c-C$_3$H$_2$, H$^{13}$CO$^+$, $^{13}$CO, CS, SO, SiO, HNC, and CH$_3$OH. The median values are listed in Table~\ref{sigma_pdf_spiralarms}. We also compute for each molecule the median value of $b$ over all velocity components (see Table~\ref{sigma_pdf_molecules}).

\begin{figure}
   \resizebox{\hsize}{!}{\includegraphics[width=0.5\textwidth,trim = 0.7cm 1.0cm 2cm 3.65cm, clip=True]{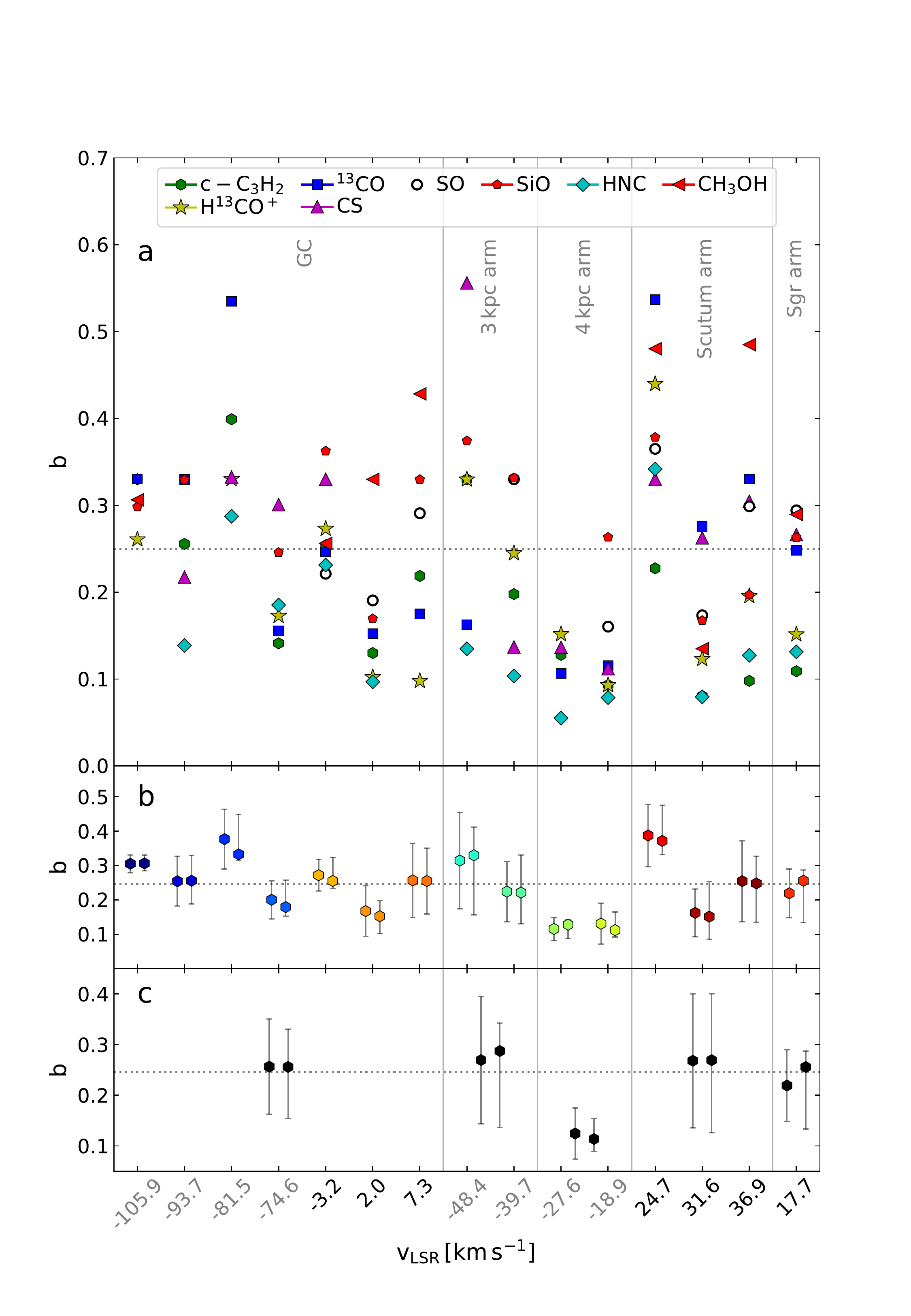}}
 \caption{\textbf{a} Forcing parameter $b$ of eight molecules, assuming $T_\mathrm{kin}=40$~K for the 15 velocity components roughly sorted by their distance to the Galactic centre. \textbf{b} Mean (left) and median (right) values for each velocity component. \textbf{c} Mean (left) and median (right) values for each sub-sample of clouds, from left to right: Galactic centre, 3~kpc arm, 4~kpc arm, Scutum arm, Sagittarius arm. The uncertainties represent the standard deviation for the mean and the corresponding percentiles for the median. The dashed line in each panel marks the median value of all data points shown in panel a. The velocities of the components belonging to Categories I and II are coloured in grey and black, respectively.}\label{b_pdf_vel}
\end{figure}

The distribution of forcing parameter $b$ is shown as a function of velocity component in Fig.~\ref{b_pdf_vel} as an example for $T_\mathrm{kin}=40$~K. The median value of the forcing parameter $b$ is 0.26, indicated by the dashed line. $b$ is higher for the velocity components that have low SNR for most molecules, which may be a bias due to the lack of sensitivity ($\varv_\mathrm{LSR}=-81.5$~km~s~$^{-1}$, $-48.4$~km~s~$^{-1}$, and 24.7~km~s~$^{-1}$). The values for the 4~kpc arm are significantly lower than the averaged value. The uncertainties for these values are relatively low. For a kinetic temperature of 40~K most values of $b$ fall in the range 0.11 to 0.37. The forcing parameters $b$ are smaller if we assume $T_\mathrm{kin}=20$~K (0.08--0.33) and larger for 80~K (0.16--0.47). 

\begin{figure}
   \resizebox{\hsize}{!}{\includegraphics[width=0.5\textwidth,trim = 0.7cm 0.3cm 2cm 2.75cm, clip=True]{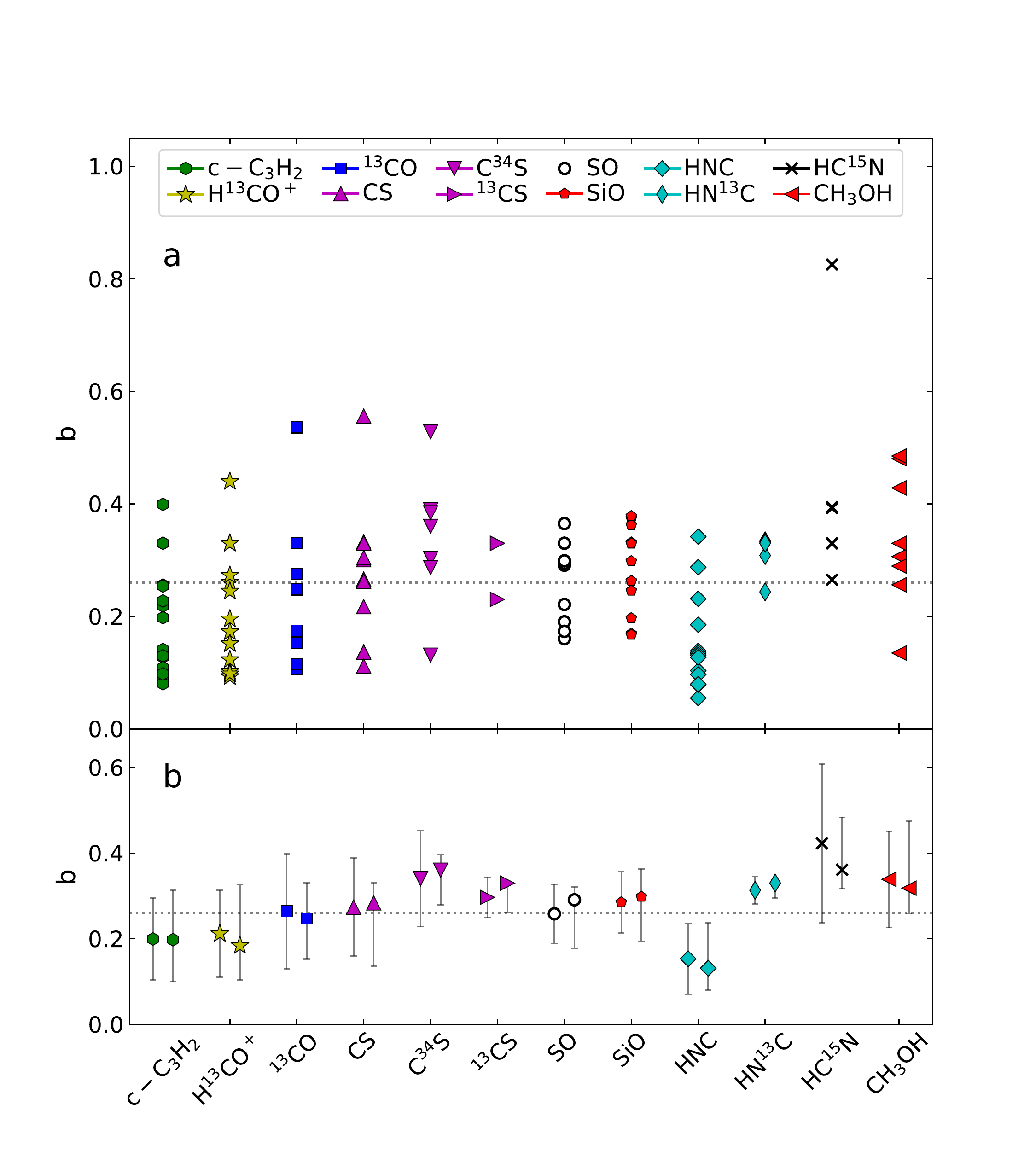}}
 \caption{\textbf{a} Forcing parameter $b$ of up to 15 velocity components sorted by molecule and assuming $T_\mathrm{kin}=40$~K. \textbf{b} Mean (left) and median (right) values for each molecule. The uncertainties represent the standard deviation for the mean and the corresponding percentiles for the median. The dashed line in both panels marks the median value of the sub-sample of molecules used in Fig.~\ref{b_pdf_vel}.}\label{b_pdf_mol}
\end{figure}

The distribution of forcing parameter $b$ as a function of molecule is displayed in Fig.~\ref{b_pdf_mol}. Most molecules show similar values of $b$. Exceptions are C$^{34}$S and HC$^{15}$N, which lie above the other ones, probably due to their low SNR, and HNC, which lies below the average.
\subsection{Principal component analysis}
Six of the 15 velocity components fulfil the selection criteria defined in Sect.~\ref{pca_theory} to perform a principal component analysis. The velocities of these components are: $-105.9$\,km\,s$^{-1}$, 2.0\,km\,s$^{-1}$, and 7.3\,km\,s$^{-1}$ in the GC, 24.7\,km\,s$^{-1}$ and 31.6\,km\,s$^{-1}$ in the Scutum arm, and 17.7\,km\,s$^{-1}$ in the Sagittarius arm. The molecules used for each component are listed in Table~\ref{pca_table}. The PCA is performed on the opacity maps after removing the average signal and scaling the standard deviation to 1. Hence, the PCA is sensitive only to the variance on scales smaller than the field of view.

To investigate the influence of the noise on the results of the PCA we performed a PCA on channels that contain only noise (see Appendix~\ref{sect_noise_pca}). We used six molecules. The powers, that is the contributions of the principal components (PCs) to the total variance, are similar for the first three components and on the level of 20--30\%. We conclude from this test that powers of the first PCs much higher than 20--30\% are required to be considered as significant. 

Because our field of view is limited by the extent of the background continuum emission, we performed several tests to examine the robustness of the PCA applied to our data (see Appendix~\ref{robustness_pca}). For these tests we changed the grid size, the number of selected pixels, and the number of selected molecules. The PCA seems to be robust to these changes. However, when no clear structure is dominant for all molecules, decreasing the number of pixels results in more changes in the values of the PC coefficients. 

\begin{table}
\caption{Molecules used for the PCA for six velocity components.}
\label{pca_table}
\centering
\begin{tabular}{l c c c c c c c }       
\hline    
& \multicolumn{6}{c}{Velocity component (km~s$^{-1}$)}\\            
molecule & $-105.9$ & $2.0$ & $7.3$ & $24.7$ & $31.6$ & $17.7$\\

& \multicolumn{6}{c}{[km\,s$^{-1}$]}\\
\hline\hline   
c-C$_3$H$_2$ & x & x & x & - & x & x \\
H$^{13}$CO$^+$ & x & x & x & x & - & x \\         
CS             & - & - & - & x & x & x \\
C$^{34}$S      & - & x & - & - & - & - \\
SiO            & x & x & x & - & - & x \\
HNC            & - & - & - & x & x & x \\
HN$^{13}$C     & - & x & x & - & - & - \\
CH$_3$OH       & x & x & x & x & x & x \\
\hline                      
\end{tabular}
\tablefoot{The molecules used for the PCA are marked with x.}
\end{table}

\begin{figure*}
\centering
\includegraphics[width=17cm, trim = 4.cm 21.5cm 4.cm 3.2cm, clip=True]{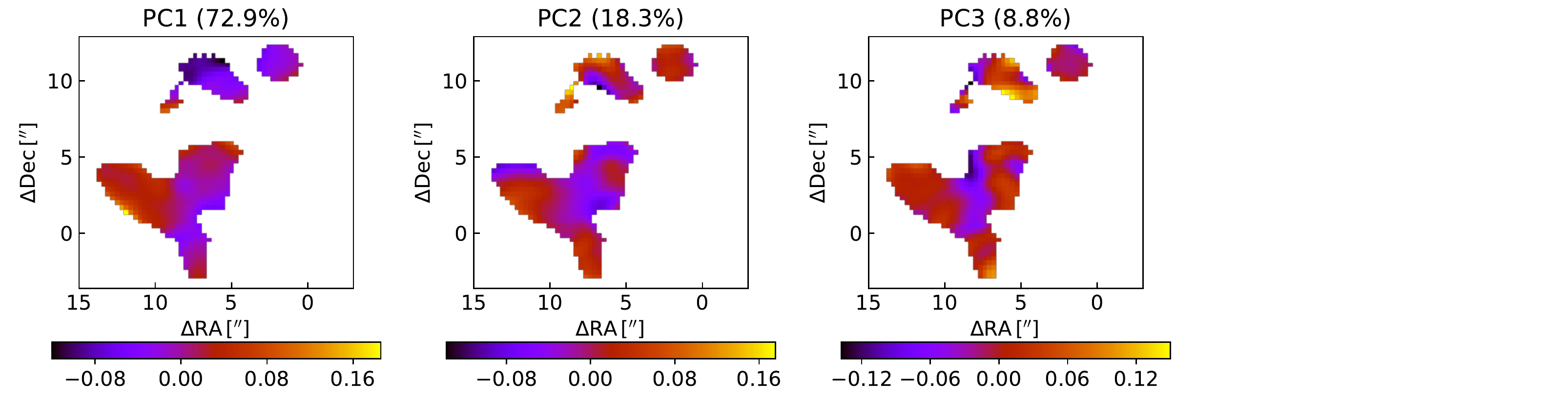}
\caption{Principal components determined at $\varv_\mathrm{LSR}= 24.7$\,km\,s$^{-1}$. The colours give the intensity. The absolute value of each PC is normalised to 1 as explained in Sect.~\ref{pca_theory}. The percentages in parentheses give the contributions of the PCs to the total variance.} 
\label{pca_components_channel12}
\end{figure*}

\begin{figure*}
\centering
\includegraphics[width=17cm, trim = 4.cm 18.9cm 4.cm 3.cm, clip=True]{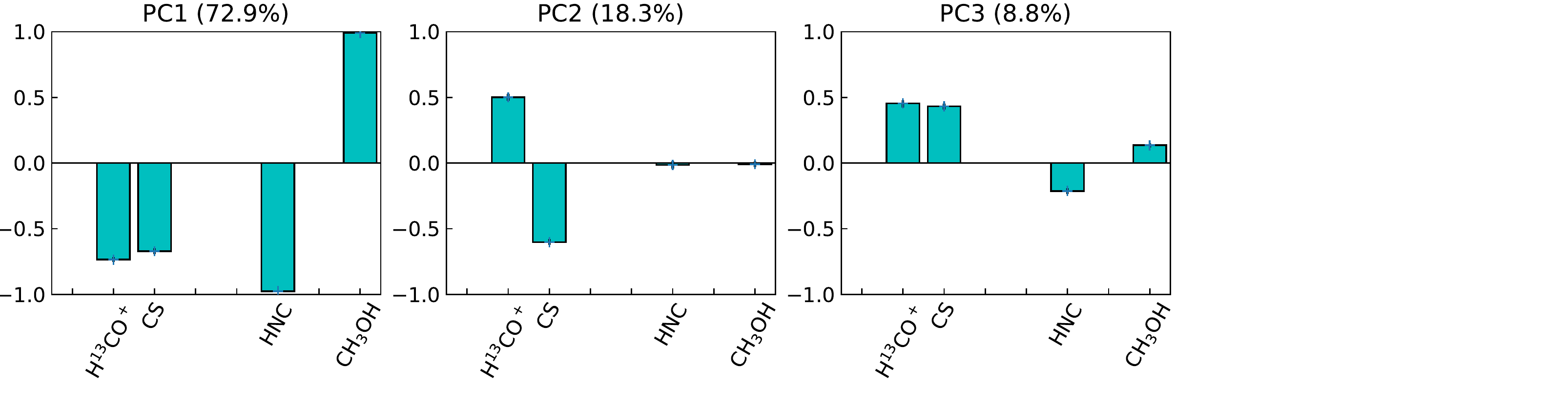}
\caption{Principal component coefficients determined at $\varv_\mathrm{LSR}= 24.7$\,km\,s$^{-1}$. The contribution factors of each molecule are normalised such that the sum of the squares is equal to 1. The percentages in parentheses give the contributions of the PCs to the total variance.}
\label{pca_contr_channel12}
\end{figure*}

\begin{figure*}
\centering
\begin{minipage}{0.49\textwidth}
\textbf{a}\\
  \includegraphics[width=8.5cm,trim = 3.cm 19.cm 21.cm 4.cm, clip=True]{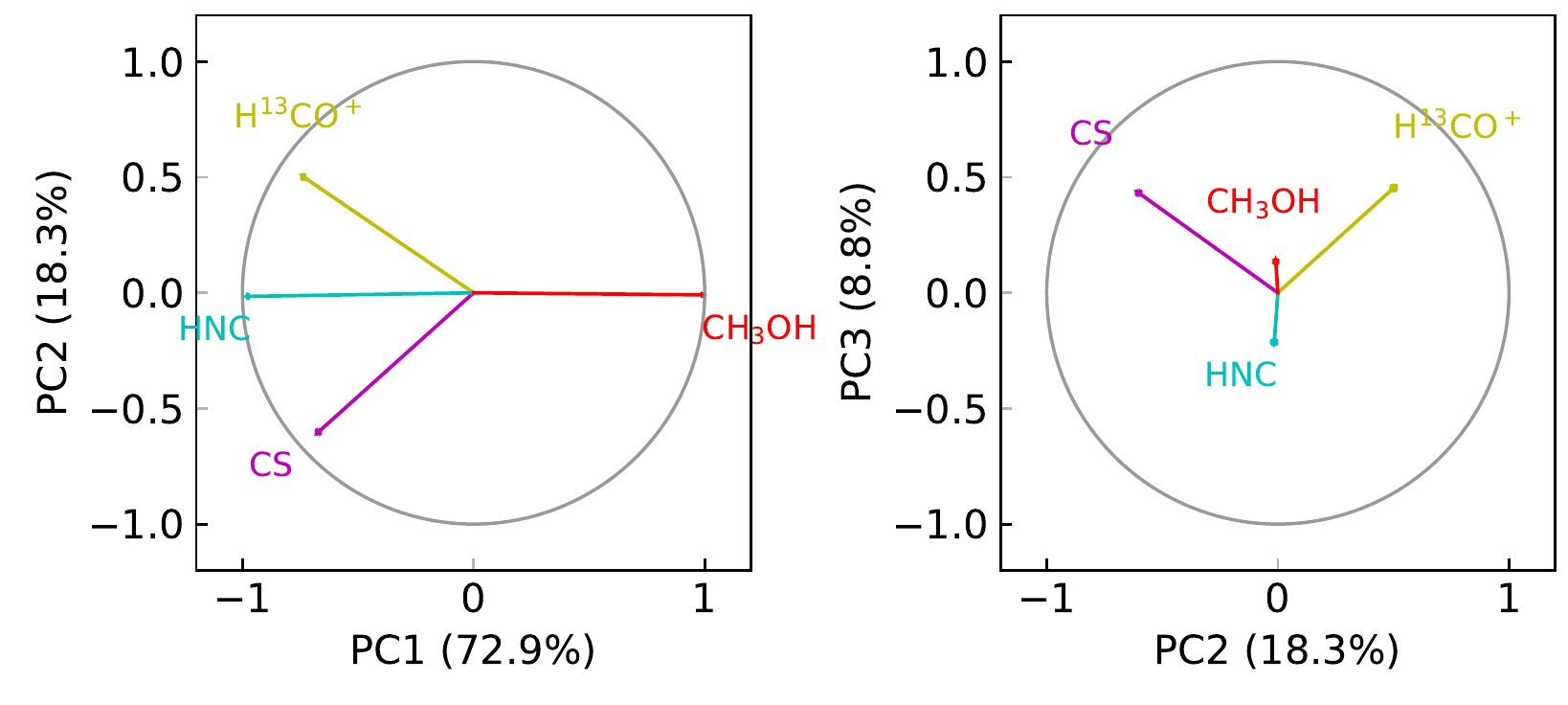}\\
  \textbf{c}\\
  \includegraphics[width=8.5cm,trim = 3.cm 19.cm 21cm 4cm, clip=True]{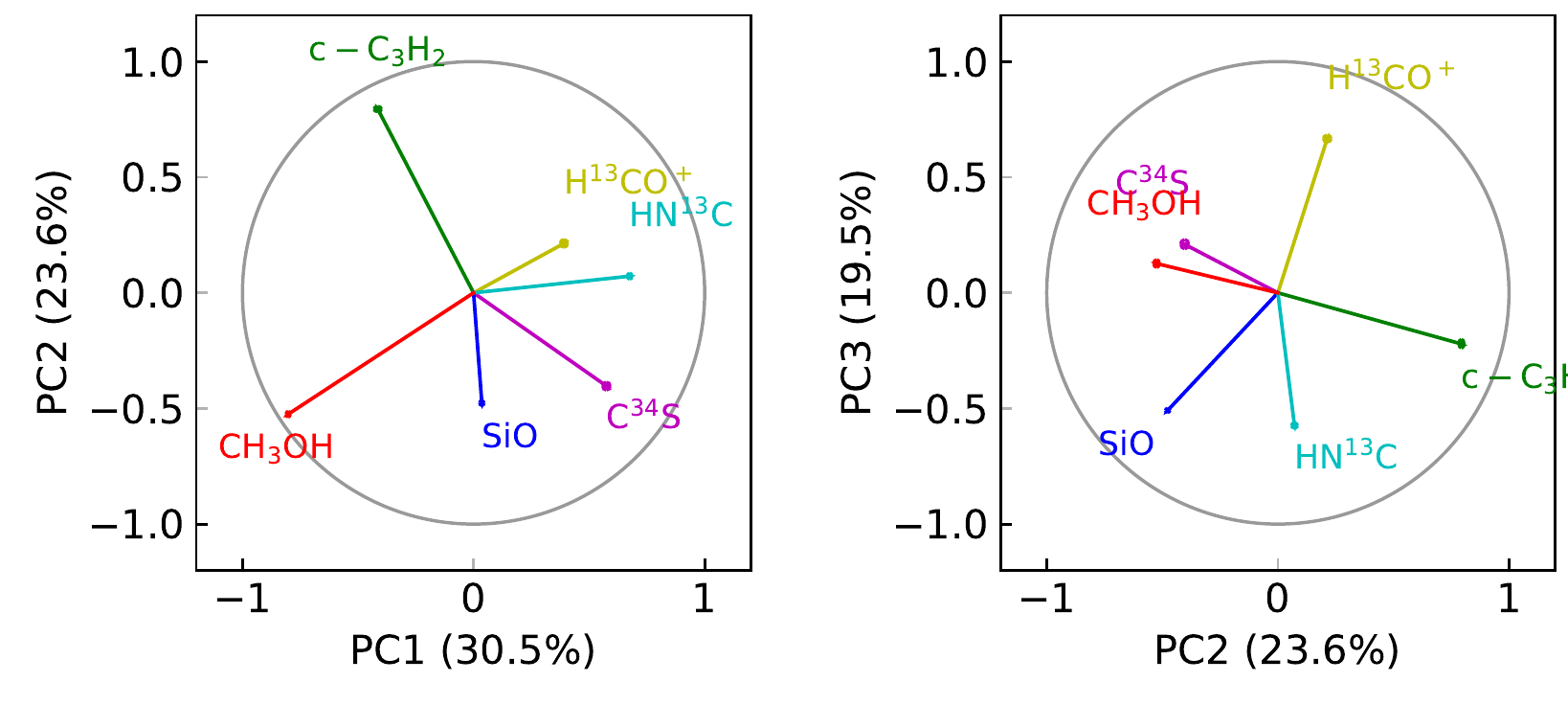}\\
  \textbf{e}\\
  \includegraphics[width=8.5cm,trim = 3.cm 19.cm 21cm 4cm, clip=True]{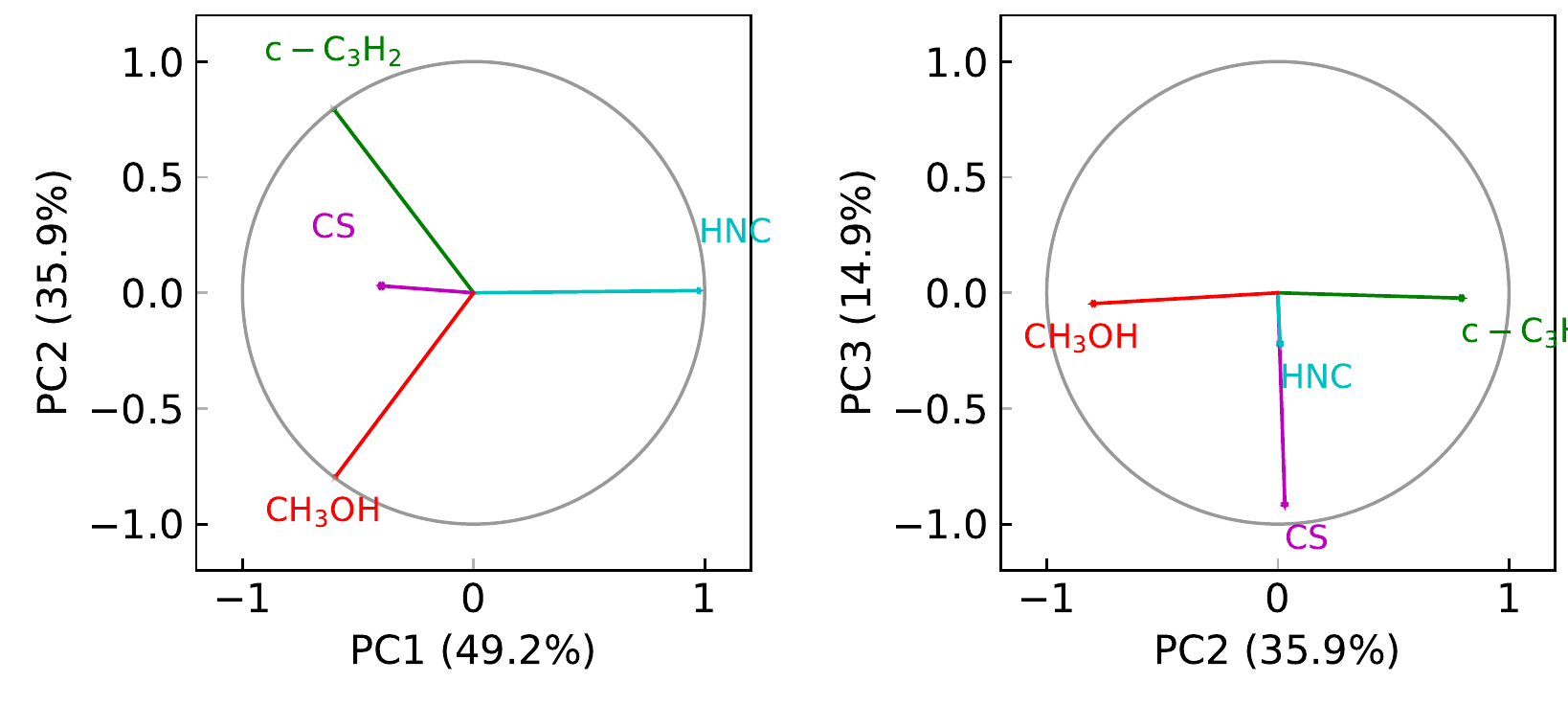}
\end{minipage}
\hfill
\begin{minipage}{0.49\textwidth}
  \textbf{b}\\
  \includegraphics[width=8.5cm,,trim = 3.cm 19.cm 21cm 4cm, clip=True]{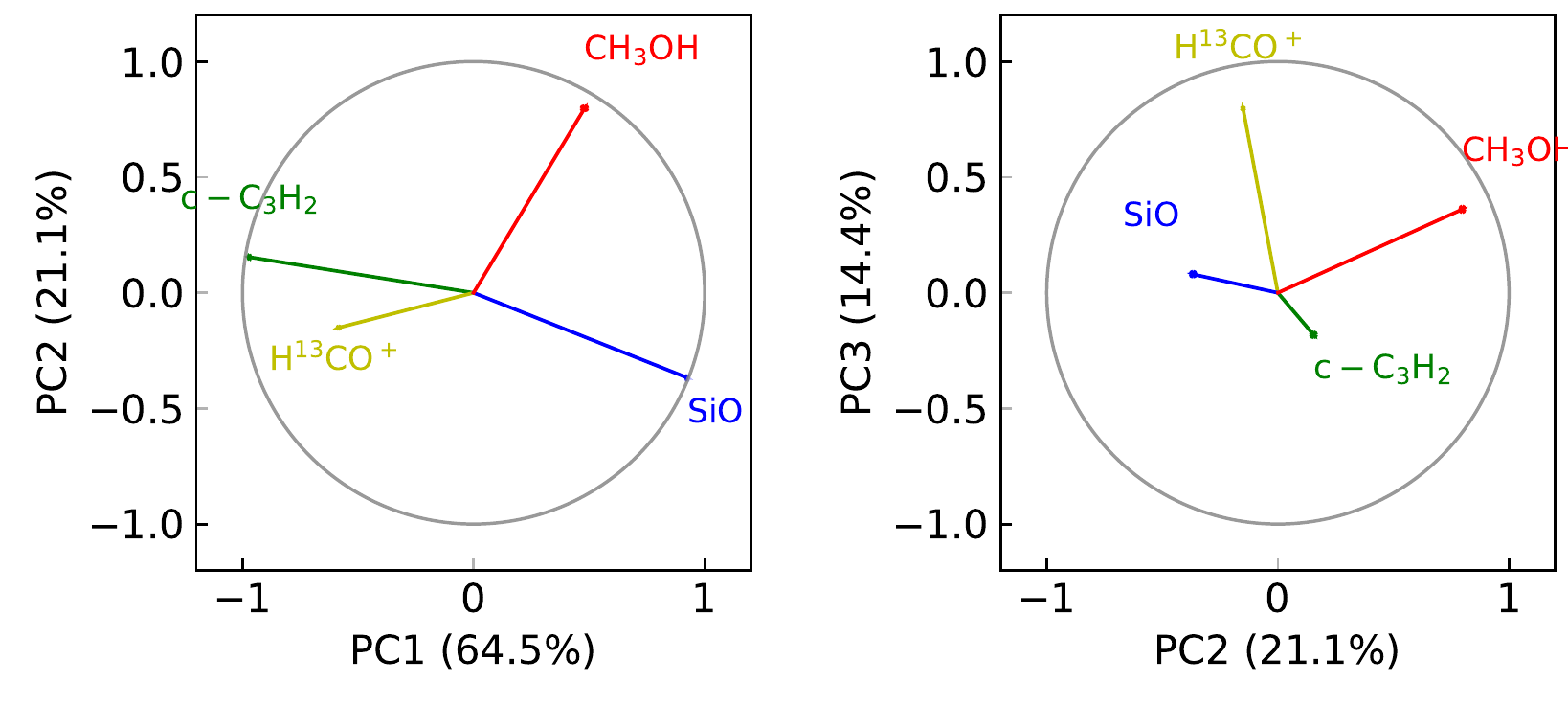}\\
    \textbf{d}\\
  \includegraphics[width=8.5cm,trim = 3.cm 19.cm 21cm 4cm, clip=True]{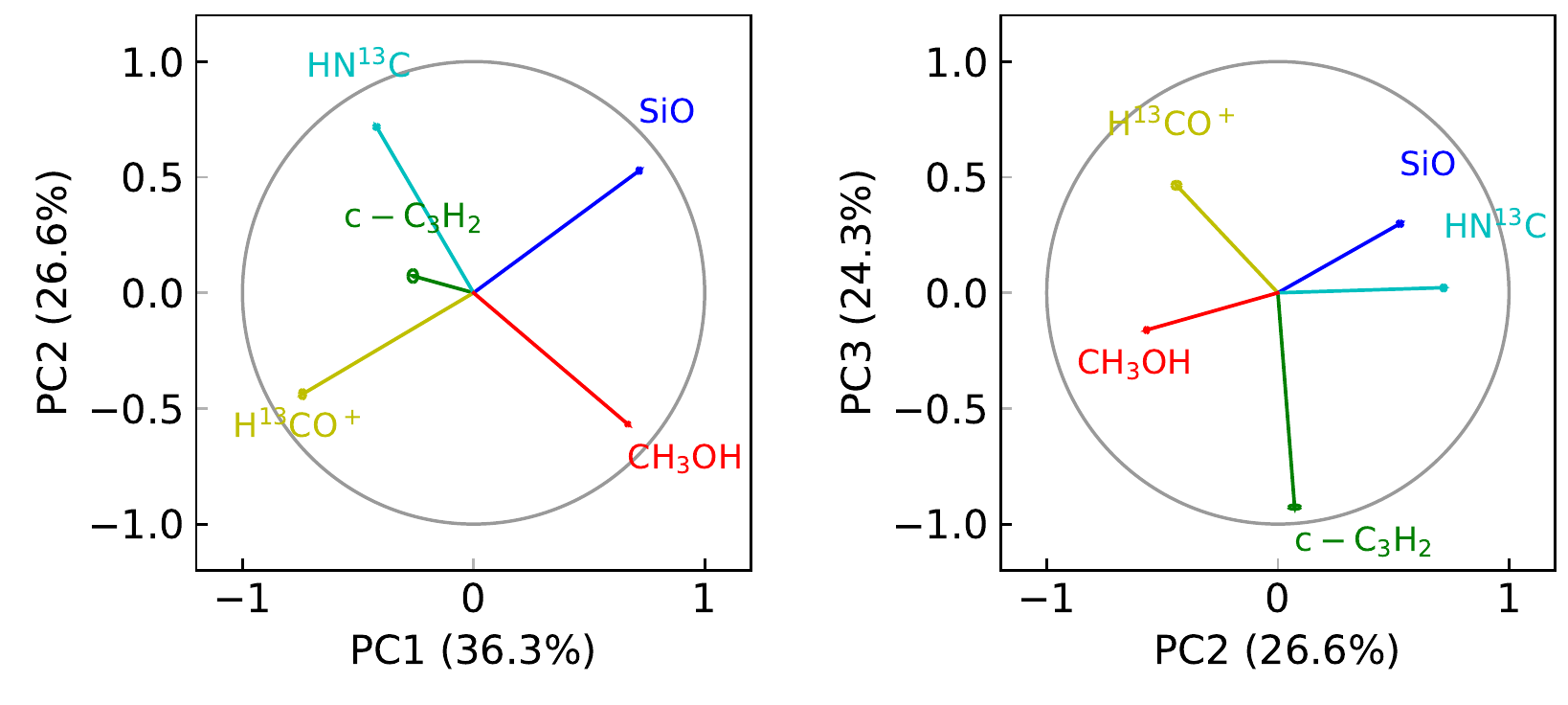}\\
    \textbf{f}\\
  \includegraphics[width=8.5cm,trim = 3.cm 19.cm 21cm 4cm, clip=True]{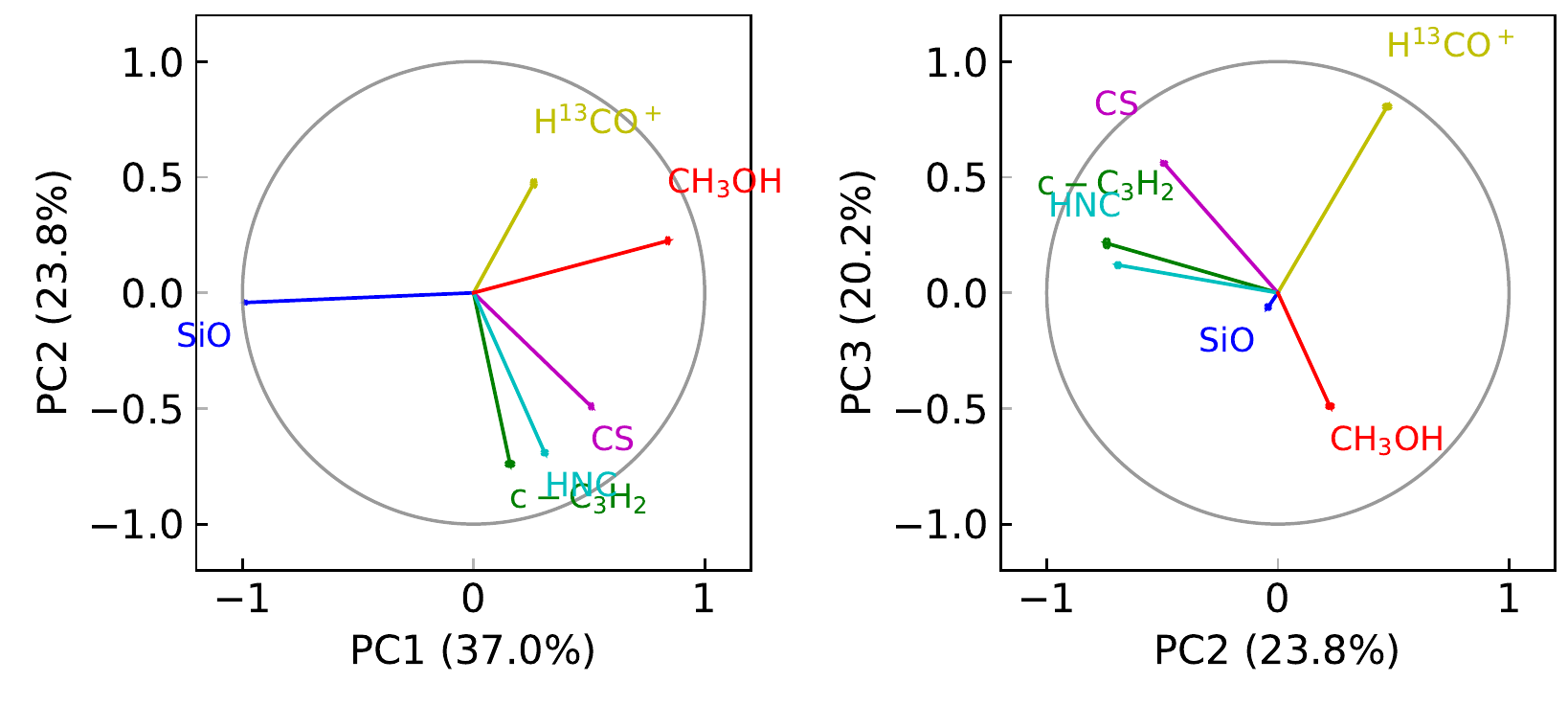}
\end{minipage}
 \caption{Correlation wheels for PCs at $\varv_\mathrm{LSR}= 24.7$\,km\,s$^{-1}$ (panel \textbf{a}), $-105.9$\,km\,s$^{-1}$ (\textbf{b}), 2.0\,km\,s$^{-1}$ (\textbf{c}), 7.3\,km\,s$^{-1}$ (\textbf{d}), 31.6\,km\,s$^{-1}$ (\textbf{e}), and 17.7\,km\,s$^{-1}$ (\textbf{f}). The percentages in parentheses give the contributions of the PCs to the total variance. The ellipses around the arrow heads show the uncertainties estimated from 1000 realisations of the opacity cubes.}
 \label{pca_corr_wheels}
\end{figure*}

The PCs calculated for $\varv_\mathrm{LSR}= 24.7$\,km\,s$^{-1}$ are shown in Fig.~\ref{pca_components_channel12}. The fourth component has a very small power of $1\times10^{-29}$. Hence, it can be neglected and is not displayed. The contribution factors of each PC to the selected molecules are shown in Fig.~\ref{pca_contr_channel12}. The error bars represent the standard deviation calculated from the 1000 realisations of the opacity cubes. They are relatively small and barely visible. For this velocity component, $73\%$ of the total variance in the data is described by the first principal component. This means a prominent structure is present for most molecules. The second and third PCs describe only small parts of the total variance 18\% and 9\%, respectively. The first two correlation wheels are plotted in Fig.~\ref{pca_corr_wheels}a. H$^{13}$CO$^+$, HNC, and CS are anti-correlated to CH$_3$OH for PC1. H$^{13}$CO$^+$ and CS are correlated for PC1 and PC3, but anti-correlated for PC2.

The correlation wheels for the other velocity components are displayed in Figs.~\ref{pca_corr_wheels}b--f, the corresponding PCs and coefficients in Figs.~\ref{pca_components_channel0}--\ref{pca_contr_channel11}. The power of the fourth PC is always very low, in the order of $10^{-29}$. Hence, the fourth PC is not displayed in these figures. At $\varv_\mathrm{LSR}= 31.6$\,km\,s$^{-1}$ (Fig.~\ref{pca_corr_wheels}e), HNC is anti-correlated to CH$_3$OH,  c-C$_3$H$_2$, and CS for PC1. The first component has only a contribution of $49\%$ to the total variance. In PC2 ($36\%$) c-C$_3$H$_2$ is strongly anti-correlated to CH$_3$OH. CS is mostly described by PC3. At $\varv_\mathrm{LSR}=-105.9$\,km\,s$^{-1}$ (Fig.~\ref{pca_corr_wheels}b), c-C$_3$H$_2$ and H$^{13}$CO$^+$ are anti-correlated with CH$_3$OH and SiO with respect to the first PC. H$^{13}$CO$^+$ is mostly described by the third PC and CH$_3$OH mostly by the second one. Here, the first PC has a high contribution of $65\%$. 

The other velocity components have powers of their PCs in the order of 20--30\%, similar to those obtained for pure noise channels. The correlation wheels of these components are therefore most likely not significant.

\subsection{Nature of the detected line-of-sight clouds}\label{radex_models}
\begin{figure}[t]
   \resizebox{\hsize}{!}{\includegraphics[width=0.5\textwidth,trim = 1.0cm 2.7cm 5.cm 2.3cm, clip=True]{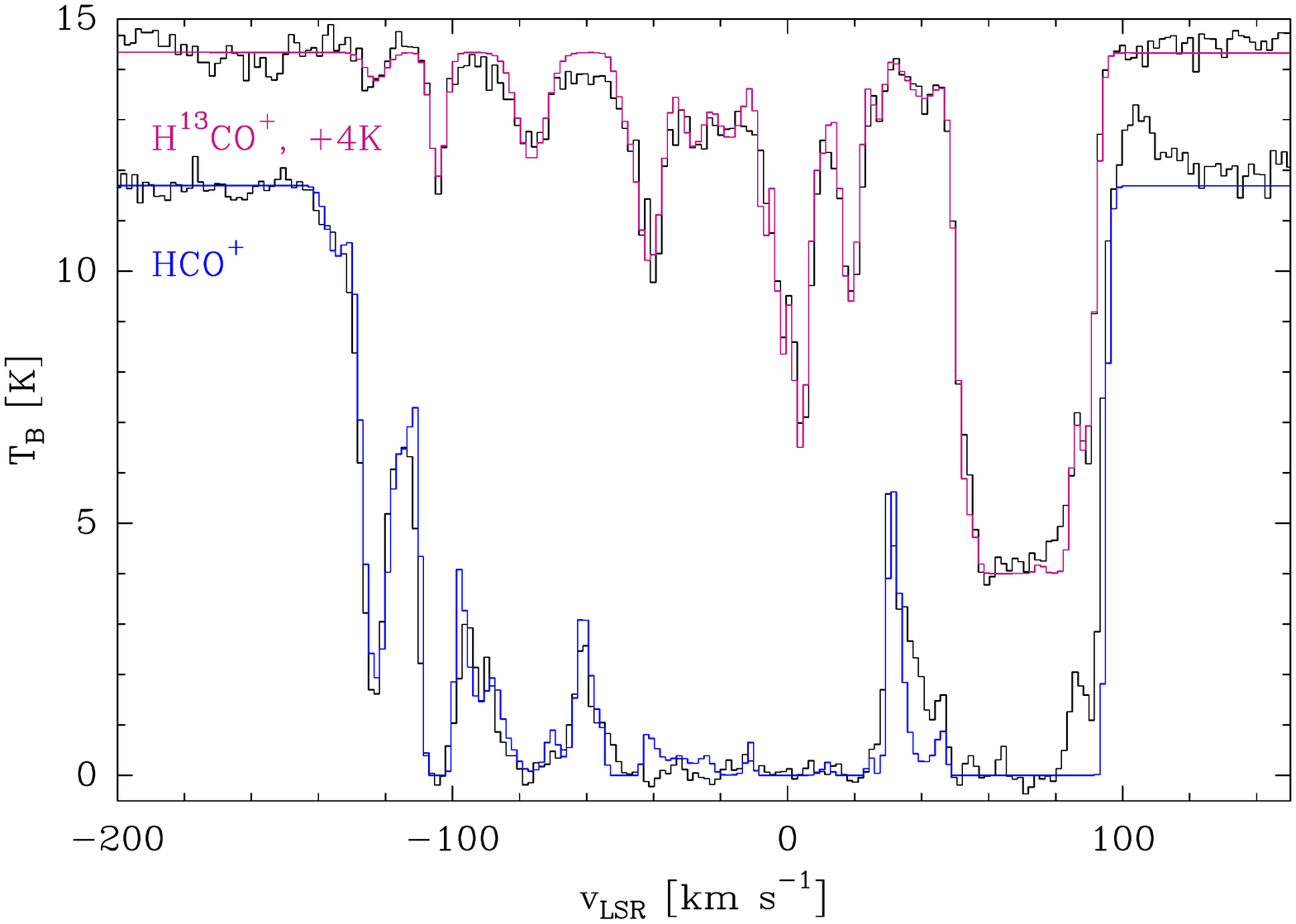}}
 \caption{ALMA spectra of HCO$^+$ 1--0 and H$^{13}$CO$^+$ 1--0 (shifted by +4\,K) in the direction of K4 in black. The synthetic spectra of HCO$^+$ and H$^{13}$CO$^+$ computed with Weeds are overplotted in blue and magenta, respectively.}
\label{spectrum_alma_hcop_h13cop}
\end{figure}

In order to understand the nature of the line-of-sight clouds detected towards Sgr\,B2(N), that is whether they are diffuse or translucent, we want to estimate their H$_2$ column densities and visual extinctions, $A_{\rm v}$. HCO$^+$ has been shown to be a good tracer of H$_2$ in diffuse clouds, with $N$(HCO$^+$)/$N$(H$_2$)$= 3 \times 10^{-9}$ \citep{liszt2010}. Here we use the EMoCA spectrum towards K4 to derive the HCO$^+$ column densities of the clouds detected in absorption. For the velocity components for which HCO$^+$ 1--0 is optically thick, we model H$^{13}$CO$^+$ 1--0 and assume the same $^{12}$C/$^{13}$C ratios as \citet{belloche2013} to derive the HCO$^+$ column densities (see their Table~2). We use Weeds \citep{maret2011} to model the velocity components detected towards K4 in absorption. The resulting parameters are listed in Table~\ref{h2_densities_alma} and the synthetic spectra are shown in Fig.~\ref{spectrum_alma_hcop_h13cop}. We obtain H$_2$ column densities ranging from $3.3\times10^{20}$ to $9.0\times10^{22}$\,cm$^{-2}$, which corresponds to $A_{\rm v}$ between 0.4 and 96~mag (columns 4 and 5 in Table~\ref{h2_densities_alma}).

\begin{table*}[t]
\caption{H$_2$ column densities densities and visual extinctions derived from the HCO$^+$ column densities determined in the ALMA spectrum in the direction of K4.}
\label{h2_densities_alma}
\centering
\begin{tabular}{r c c c r c r}       
\hline               
\multicolumn{1}{c}{$\varv_\mathrm{LSR}$\tablefootmark{a}} & \multicolumn{1}{c}{N(H$^{13}$CO$^+$)\tablefootmark{b}} & \multicolumn{1}{c}{N(HCO$^+$)\tablefootmark{c}} &  \multicolumn{1}{c}{N(H$_2$)\tablefootmark{d}} & \multicolumn{1}{c}{$A_\mathrm{v}$\tablefootmark{e}} & \multicolumn{1}{c}{N(H$_2$)\tablefootmark{f}} & \multicolumn{1}{c}{$A_\mathrm{v}$\tablefootmark{e}}\\

 \multicolumn{1}{c}{[km\,s$^{-1}$]} & \multicolumn{1}{c}{[cm$^{-2}$]} & \multicolumn{1}{c}{[cm$^{-2}$]} & \multicolumn{1}{c}{[cm$^{-2}$]} & \multicolumn{1}{c}{[mag]} & \multicolumn{1}{c}{[cm$^{-2}$]} & \multicolumn{1}{c}{[mag]} \\
\hline\hline 
40.8 & 1.3(12) & 5.2(13)$^*$ & 1.7(22) & 18.4 & 3.6(21) & 3.8 \\
27.0 & 6.2(11) & 2.4(13)$^*$ & 8.0(21) & 8.5 & 2.5(21) & 2.7 \\
18.6 & 4.5(12) & 2.7(14)$^*$ & 9.0(22) & 95.7 & 1.4(22) & 14.4 \\
9.8 & 1.5(12) & 3.0(13)$^*$ & 1.0(22) & 10.6 & 2.8(21) & 3.0 \\
3.8 & 8.0(12) & 1.6(14)$^*$ & 5.3(22) & 56.7 & 8.0(21) & 8.5 \\
-1.7 & 3.8(12) & 7.6(13)$^*$ & 2.5(22) & 27.0 & 4.3(21) & 4.5 \\
-6.9 & 2.0(12) & 8.0(13)$^*$ & 2.7(22) & 28.4 & 4.4(21) & 4.6 \\
-17.2 & 2.1(12) & 8.4(13)$^*$ & 2.8(22) & 29.8 & 4.5(21) & 4.7 \\
-28.1 & 1.8(12) & 3.2(13)$^*$ & 1.1(22) & 11.3 & 2.9(21) & 3.0 \\
-35.2 & -- & 2.4(13) & 8.0(21) & 8.5 & 2.5(21) & 2.7 \\
-41.2 & 5.1(12) & 2.1(14)$^*$ & 7.0(22) & 74.5 & 1.1(22) & 11.1 \\
-48.8 & 5.7(11) & 2.3(13)$^*$ & 7.7(21) & 8.2 & 2.5(21) & 2.6 \\
-56.3 & -- & 2.1(13) & 7.0(21) & 7.4 & 2.4(21) & 2.5 \\
-66.3 & -- & 2.3(13) & 7.7(21) & 8.2 & 2.5(21) & 2.6 \\
-76.4 & 2.7(12) & 5.6(13)$^*$ & 1.9(22) & 19.9 & 3.7(21) & 3.9 \\
-83.9 & -- & 8.6(12) & 2.9(21) & 3.0 & 1.6(21) & 1.7 \\
-91.9 & -- & 2.6(13) & 8.7(21) & 9.2 & 2.8(21) & 2.8 \\
-104.3 & 1.7(12) & 9.2(13) & 3.1(22) & 32.6 & 4.7(21) & 5.0 \\
-114.1 & -- & 5.0(12) & 1.7(21) & 1.8 & 1.2(21) & 1.3 \\
-122.8 & 4.8(11) & 1.5(13) & 5.0(21) & 5.3 & 2.0(21) & 2.1 \\
-134.6 & -- & 1.0(12) & 3.3(20) & 0.4 & 3.3(20) & 0.4 \\
\hline                      
\end{tabular}
\tablefoot{$X(Y)$ corresponds to $X\times 10^Y$. The excitation temperature is assumed to be 2.73\,K. \tablefoottext{a}{Centroid velocity of the Gaussian component.}\tablefoottext{b}{H$^{13}$CO$^+$ column density. A dash indicates the cases where H$^{13}$CO$^+$ is too weak.}\tablefoottext{c}{HCO$^+$ column density. A star indicates the cases when it is derived from H$^{13}$CO$^+$.}\tablefoottext{d}{H$_2$ column density derived from HCO$^+$ assuming an HCO$^+$ abundance of $3 \times 10^{-9}$ relative to H$_2$ typical for diffuse clouds \citep{liszt2010}.}\tablefoottext{e}{Visual extinction computed from the previous column with the formula $A_\mathrm{v}$(mag) = $N({\rm H}_2$)/($9.4\times 10^{20}$\,cm$^{-2}$).}\tablefoottext{f}{H$_2$ column density derived from HCO$^+$ assuming a HCO$^+$ abundance of $3 \times 10^{-9}$ for $N$(HCO$^+$) $< 2.8 \times 10^{12}$~cm$^{-2}$, $2 \times 10^{-8}$ for $N$(HCO$^+$) $> 9.4 \times 10^{13}$~cm$^{-2}$, and an interpolated value (in log-log space) in between.}}

\end{table*}

\begin{figure*}
\centering
\includegraphics[width=17cm, trim = 1.5cm 11.cm 4.cm 3.cm, clip=True]{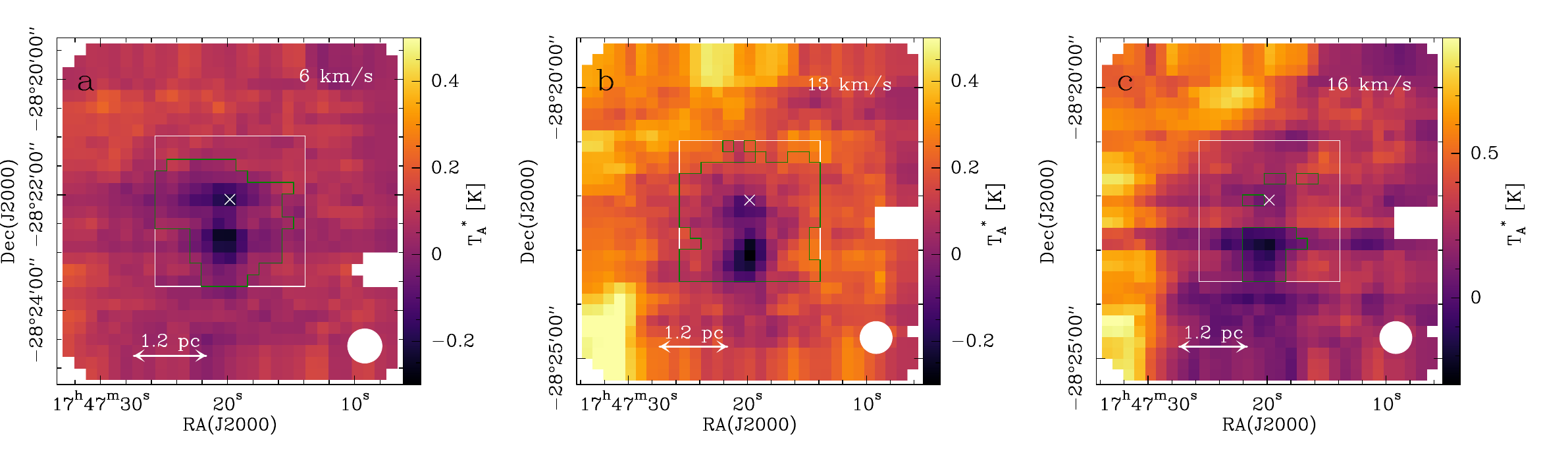}
\caption{Mopra channel maps in the direction of Sgr\,B2 at about 14\,km\,s$^{-1}$ for HCO$^+$ 1--0 (panel \textbf{a}), HNC 1--0 (\textbf{b}), and CS 2--1 (\textbf{c}) as observed by \citet{jones2008}. In each panel, the white box shows the region selected to calculate the averaged spectrum and the green contour encloses the pixels that were masked to avoid absorption. The white circle in the lower right corner of each panel represents the beam and the cross indicates the position of K4. The pixel size is $12\as$.} 
\label{hcop_hnc_cs_mopra_map}
\end{figure*}

\begin{figure}[t]
   \resizebox{\hsize}{!}{\includegraphics[width=0.5\textwidth,trim = 1.0cm 3.cm 5.cm 2.5cm, clip=True]{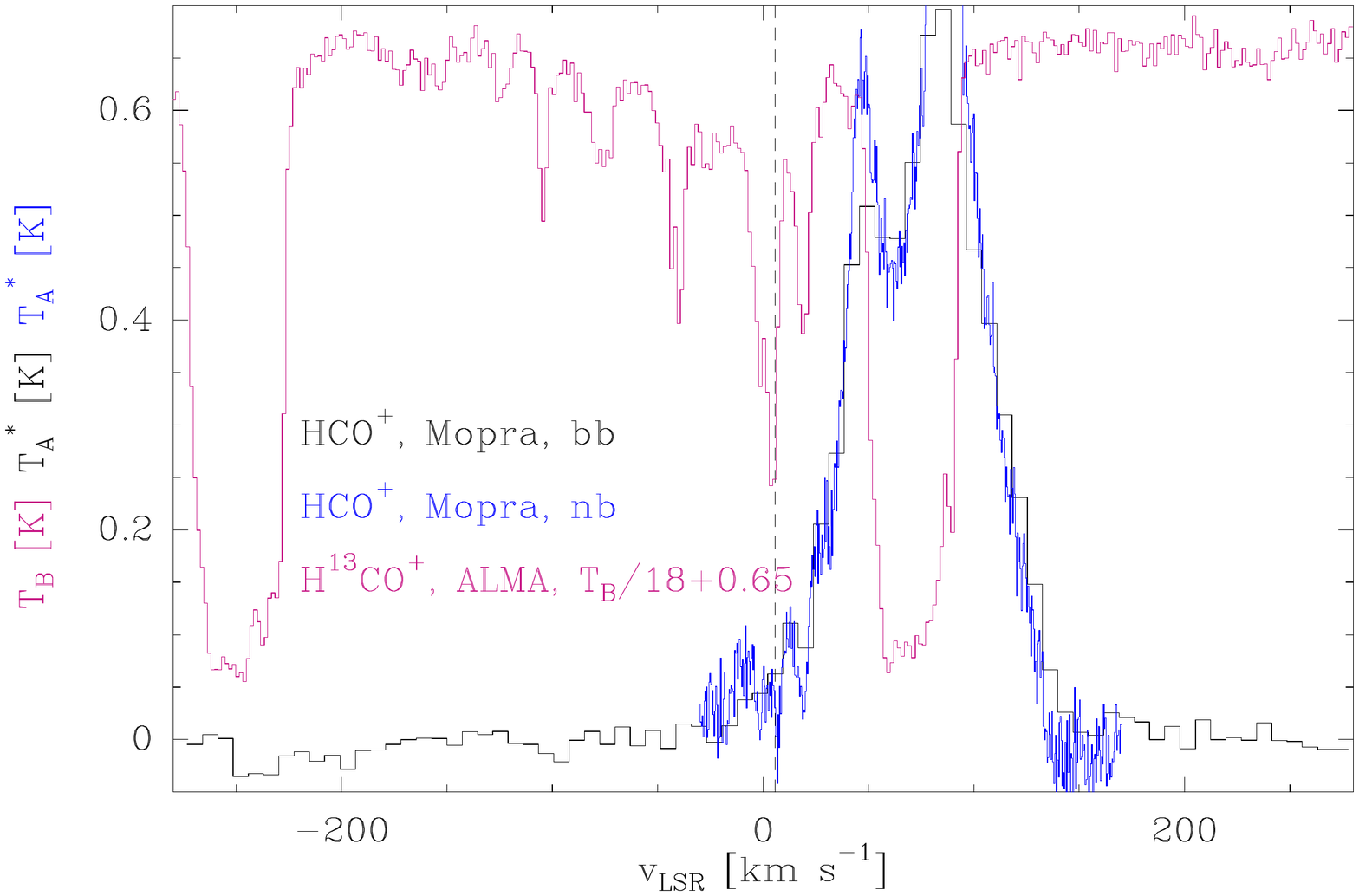}}
 \caption{Mopra average spectrum of HCO$^+$ 1--0 towards Sgr~B2(N), extracted from the imaging survey of \citet{jones2008} after excluding pixels with absorption (see Fig.~\ref{hcop_hnc_cs_mopra_map}). The black and blue spectra were obtained with the broad-band and narrow-band backends, respectively. They are displayed in $T_{\rm A}^*$ scale. The purple spectrum is the EMoCA spectrum of H$^{13}$CO$^+$ 1--0 towards K4 in brightness temperature scale. The feature between -280 and -210~km~s$^{-1}$ is caused by SiO in the envelope of Sgr\,B2. The dashed line marks the channel that corresponds to a $^{13}$CO emission peak in Fig.~\ref{spectrum_mopra_13co}.}
\label{spectrum_mopra_hcop}
\end{figure}

\begin{figure}[t]
   \resizebox{\hsize}{!}{\includegraphics[width=0.5\textwidth,trim = 1.0cm 3.cm 5.cm 2.5cm, clip=True]{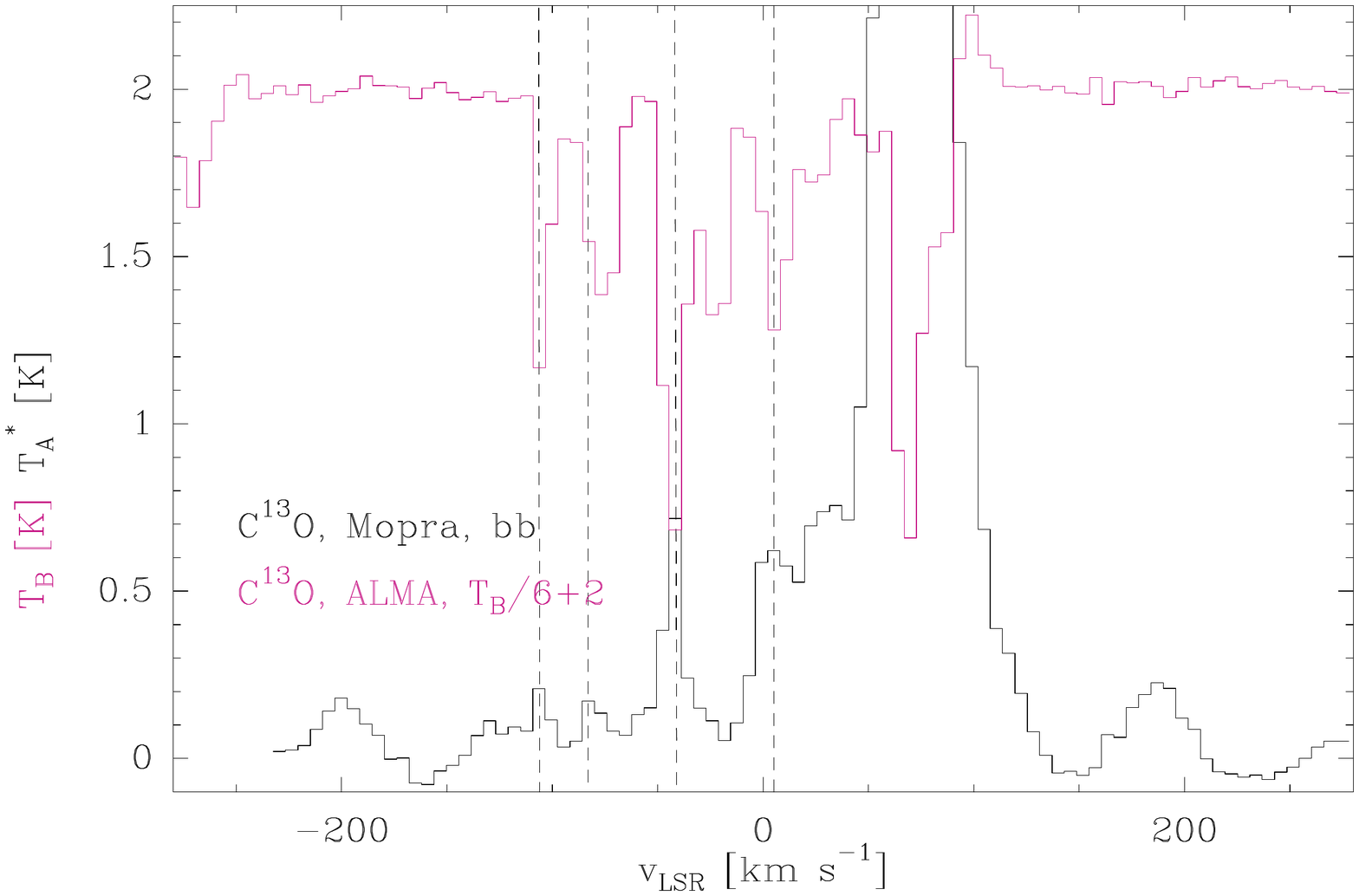}}
 \caption{Same as Fig.~\ref{spectrum_mopra_hcop}, but for $^{13}$CO 1--0. The ALMA spectrum is resampled to the spectral resolution of the Mopra spectrum. The dashed lines mark emission peaks that match well absorption features of the EMoCA spectrum outside the velocity range of the Sgr~B2 envelope.}
 \label{spectrum_mopra_13co}
\end{figure}

With the HCO$^+$ abundance relative to H$_2$ assumed above, all but three components (at $\varv_\mathrm{LSR}=-83.9$~km~s$^{-1}$, $-114.1$~km~s$^{-1}$, and $-134.6$~km~s$^{-1}$) would have visual extinctions higher than 5~mag, which would imply that they are dense molecular clouds. If this were indeed the case, then we would expect to see these clouds in emission towards positions without strong continuum background. To test this, we cannot use the EMoCA survey because of the spatial filtering of the interferometer. Instead, we check the imaging survey of Sgr~B2 performed by \citet{jones2008} with Mopra at 3~mm. We select the following transitions: HCO$^+$ 1--0, HNC 1--0, CS 2--1, and $^{13}$CO 1--0. The HCO$^+$, HNC, and CS lines are partly seen in absorption in the Mopra data. We mask the pixels where absorption is detected in the velocity range from $-100$~km\,s$^{-1}$ to 0~km\,s$^{-1}$. For each of these three species, we compute the average Mopra spectrum within a square box of size 156$\arcsec$ centred on the J2000 equatorial position $17^\mathrm{h}47^\mathrm{m}19.8^\mathrm{s},-28^\circ$22$^{\prime}$17$^{\prime\prime}$, excluding the masked pixels (see Fig.~\ref{hcop_hnc_cs_mopra_map}). $^{13}$CO 1--0 does not show any absorption in the Mopra spectra and we take the average spectrum over a square box of size 84$\arcsec$ centred on the same position (see Fig.~\ref{13co_mopra_map}). The resulting Mopra spectra are compared to the EMoCA spectra in Figs.~\ref{spectrum_mopra_hcop}, \ref{spectrum_mopra_13co}, \ref{spectrum_mopra_hnc}, and \ref{spectrum_mopra_cs}.

The $^{13}$CO Mopra average spectrum shows two strong velocity components in emission that match well the position of components seen in absorption in the EMoCA spectrum (at $-42$~km~s$^{-1}$ and 5~km~s$^{-1}$, see Fig.~\ref{spectrum_mopra_13co}). Two weaker emission peaks at $-83$~km~s$^{-1}$ and $-107$~km~s$^{-1}$ also match absorption components seen with ALMA. None of the components at velocities below $-30$~km~s$^{-1}$ are detected in emission in the HCO$^+$, HNC, and CS Mopra spectra, but these species show emission at velocities above $\sim -30$~km~s$^{-1}$, which may be at least in part associated with the absorption components seen in the ALMA spectra.

\begin{figure*}
\centering
\includegraphics[width=17cm, trim = 2.9cm 0.7cm 1.5cm 1.2cm, clip=True]{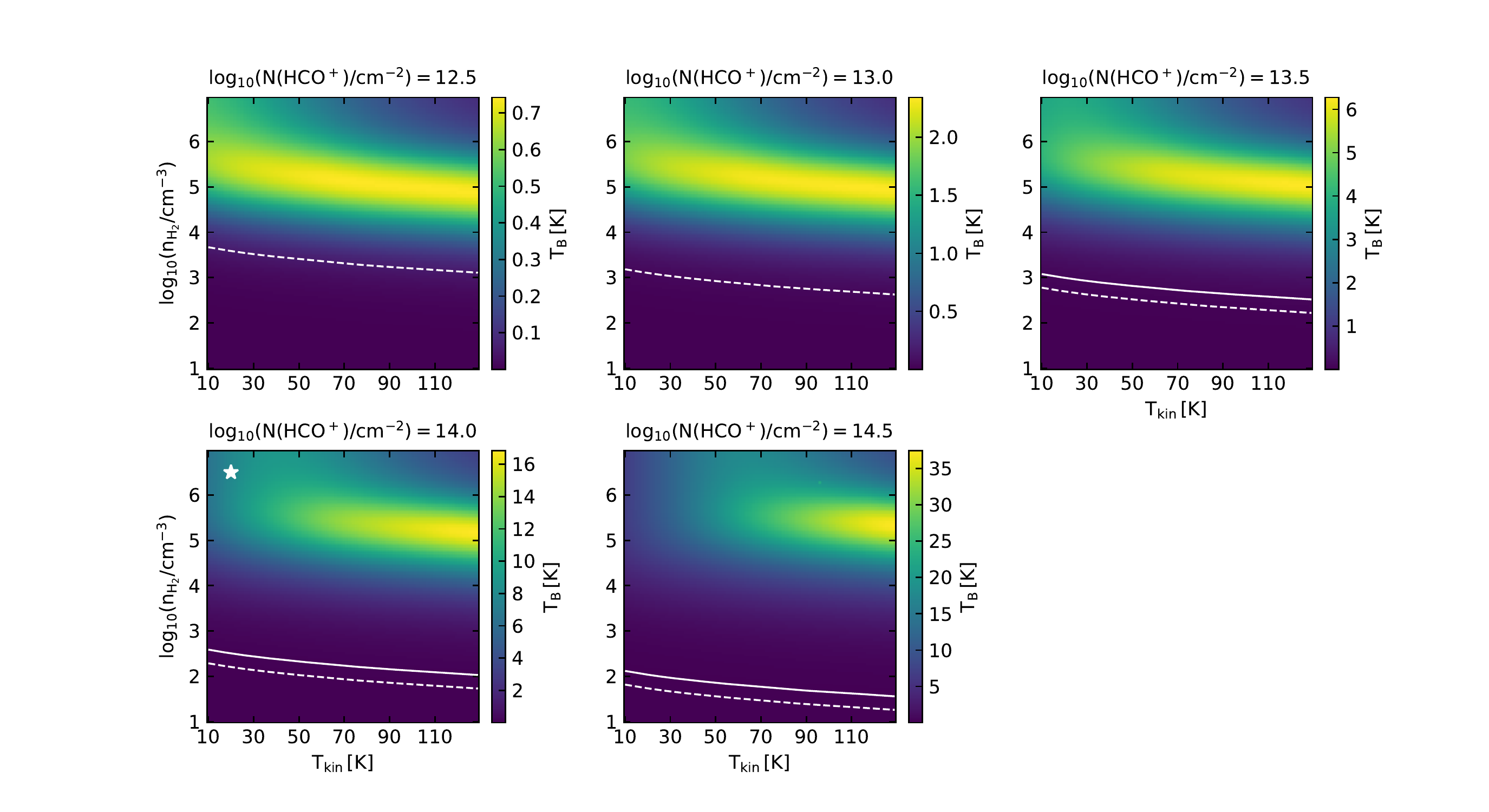}
\caption{HCO$^+$ 1--0 brightness temperature computed with RADEX as a function of kinetic temperature (x-axis) and H$_2$ density (y-axis), assuming different column densities. In each panel, the solid line corresponds to the level of emission measured in the Mopra spectrum around 14\,km\,s$^{-1}$ and the dashed line to the upper limit ($3\sigma$) over the velocity range from $-110$~km~s$^{-1}$ to $-10$~km~s$^{-1}$. The solid line is displayed only for the column densities that fall in the range determined from the ALMA absorption map at this velocity and the white star marks the median column density obtained from the ALMA absorption map at this velocity.}
\label{radex_hcop}
\end{figure*}

We perform non-LTE radiative transfer calculations with RADEX \citep{vandertak2007} to estimate the densities and kinetic temperatures that are consistent with the emission seen with Mopra, or its upper limits. We take the spectroscopic parameters and collisional rates with H$_2$ from the Leiden Atomic and Molecular Database \citep[LAMDA,][]{schoir2005} for HCO$^+$ \citep{botschwina1993, flower1999, schoir2005}, CS \citep[CDMS, and][]{lique2006}, HNC \citep[CDMS, and][]{dumouchel2010}, and $^{13}$CO \citep[CDMS, JPL, and][]{goorvitch1994,cazzoli2004,yang2010}. We perform the calculations for a wide range of parameters: 10--130~K for the kinetic temperature and 10--10$^7$~cm$^{-3}$ for the H$_2$ density. These ranges cover the values expected for diffuse, translucent, and dense molecular clouds. We explore the following ranges of column densities: 10$^{12.5\text{--}14.5}$~cm$^{-2}$ for HCO$^+$, 10$^{12.0\text{--}14.5}$~cm$^{-2}$ for HNC, 10$^{12.5\text{--}14.5}$~cm$^{-2}$ for CS, and 10$^{15.0\text{--}16.5}$~cm$^{-2}$ for $^{13}$CO. They correspond to the ranges derived from the absorption features detected with ALMA towards six strong continuum positions covered by EMoCA. The brightness temperatures computed with RADEX for the selected transitions are displayed in Figs.~\ref{radex_hcop}, \ref{radex_hnc}, \ref{radex_cs}, and \ref{radex_13co}, respectively. The solid lines plotted in these figures indicate the level of emission detected with Mopra for the component around 14~km~s$^{-1}$. For $^{13}$CO, the dashed lines correspond to the emission component detected around 83~km~s$^{-1}$, while for the other species, they correspond to the emission upper limits (three times the RMS noise level) derived from the Mopra spectra between $-110$~km~s$^{-1}$ and $-10$~km~s$^{-1}$. We converted the Mopra antenna temperatures into brightness temperatures by multiplying them with a factor 1.7, which roughly corresponds to the extended beam efficiency of $\sim 0.6$ measured by \citet{ladd2005} for sources larger than $\sim 80\arcsec$, consistent with the extended emission seen in the Mopra channel maps shown in Figs.~\ref{hcop_hnc_cs_mopra_map} and \ref{13co_mopra_map}.  

\begin{table}[t]
\caption{H$_2$ densities derived from the Mopra data.}
\label{h2_densities}
\centering
\begin{tabular}{l r r r r}       
\hline               
molecule & \multicolumn{1}{c}{$\varv_\mathrm{LSR}$\tablefootmark{a}} & \multicolumn{1}{c}{$T_\mathrm{b}$\tablefootmark{b}} & \multicolumn{2}{c}{$n_\mathrm{H_2}$\tablefootmark{c}} \\

 & \multicolumn{1}{c}{[km\,s$^{-1}$]} & \multicolumn{1}{c}{[K]} & \multicolumn{2}{c}{[cm$^{-3}$]} \\
\hline\hline 
HCO$^+$ & 6 & 0.11 & 4(1)--1(3)& \\ 
& x & $<$0.06 & $<$2(1)--5(3)& \\
HNC & 6 & 0.37 & 7(2)--4(4)& \\ 
& x & $<$0.15 & $<$3(2)--5(5)& [>5(5)] \\
CS & 10 & 0.43 & 4(2)--2(4)&[>1(5)] \\ 
& x & $<$0.06 & $<$5(1)--3(4)&[>1(5)] \\
$^{13}$CO & 5 & 1.06 & 3(1)--5(3)&[>1(3)] \\ 
& $-83$ & 0.35 & $<$5(2)&[$>$5(3)]\\
\hline                      
\end{tabular}
\tablefoot{$X(Y)$ corresponds to $X\times 10^Y$. \tablefoottext{a}{x corresponds to the velocity range from $-110$ to $-10$ km~s$^{-1}$.}\tablefoottext{b}{Measured brightness temperature in the Mopra spectrum, or upper limit of 3$\sigma$.}\tablefoottext{c}{Range of densities consistent with the detected Mopra emission, or upper limit to the densities. The values in brackets give the high-density solution, when it exists.}}
\end{table}

Given the limits assumed for the kinetic temperature, our RADEX analysis provides constraints on the molecular hydrogen densities (see Table~\ref{h2_densities}). For the component around 6~km~s$^{-1}$, the ranges of densities derived from HCO$^+$ and $^{13}$CO are similar, in the order of 30--5000 cm$^{-3}$, while the ranges derived for HNC and CS are shifted by nearly one order of magnitude towards higher densities. For the component at $-83$\,km\,s$^{-1}$detected in emission in $^{13}$CO 1--0, the upper end of the density range is 500~cm$^{-2}$. The upper limits derived for HCO$^+$, HNC, and CS over the range $-110$ to $-10$~km~s$^{-1}$ imply densities below $5 \times 10^3$~cm$^{-3}$, $5\times10^5$~cm$^{-3}$, and $3\times10^4$~cm$^{-3}$, respectively. Some of the RADEX plots also show higher density solutions for CS and $^{13}$CO but these solutions would not be consistent with the constraints set by HCO$^+$. Our RADEX analysis does not bring any constraint on the kinetic temperature.

Our calculations with RADEX do not take the collisional excitation by electrons into account. Electrons can have a significant impact on the excitation of molecules at low densities when the electron fraction is high and the CO fraction is small \citep{liszt2016a}. Taking the collisional excitation by electrons into account in our radiative transfer calculations would lower the densities or upper limits derived in this section.

\section{Discussion}\label{sect_discuss}
\subsection{Types of line-of-sight clouds}\label{subsect_discuss_types}
Assuming that HCO$^+$ has the same abundance relative to H$_2$ as the one established for diffuse clouds ($3 \times 10^{-9}$) leads to the conclusion that most clouds seen in absorption in the EMoCA survey would be dense clouds, with visual extinctions higher than 5~mag (Table~\ref{h2_densities_alma}). However, the maps obtained towards Sgr~B2 with Mopra by \citet{jones2008} reveal only few velocity components in emission in the tracers HCO$^+$ 1--0, HNC 1--0, CS 2--1, and $^{13}$CO 2--1. Our radiative-transfer analysis indicates that the emission component at $\sim 6$~km~s$^{-1}$ detected in $^{13}$CO 2--1 and HCO$^+$ 1--0 must have an H$_2$ density lower than a few times $10^3$~cm$^{-3}$. In addition, our analysis of the ALMA HCO$^+$ 1--0 absorption spectrum towards K4 indicates that, except for the component at $-83.9$ km~s$^{-1}$, the velocity components between $-110$ and $-10$~km~s$^{-1}$, which are not detected in emission in the HCO$^+$ 1--0 Mopra data, have HCO$^+$ column densities higher than 10$^{13}$~cm$^{-2}$ (Table~\ref{h2_densities_alma}). The radiative-transfer calculations then imply that these components have H$_2$ volume densities below $10^3$~cm$^{-3}$ (Fig.~\ref{radex_hcop}). Finally, the component at $\sim -83$~km~s$^{-1}$ detected in $^{13}$CO 2--1 emission with Mopra also has a low density, less than 500~cm$^{-3}$, according to our radiative-transfer analysis (Sect.~\ref{radex_models}). Taking the collisional excitation of molecules by electrons into account would imply even lower densities. All components between $-110$ and $-10$~km~s$^{-1}$ and the one at 6~km~s$^{-1}$ thus have densities that are too low for them to be dense clouds, which are characterised by densities higher than 10$^4$~cm$^{-3}$ \citep{snow2006}. We note that \citet{greaves1995} derived densities in the order of 10$^4$~cm$^-3$ for the clouds at velocities $-102$, $-41$, $-27$, and 3~km~s$^{-1}$ on the basis of HCN 3--2 probed in absorption with the JCMT, while their CS 2--1 and 3--2 observations of the same clouds in absorption indicate densities lower than 600~cm$^{-3}$, in rough agreement with our conclusion above. These authors concluded that a range of densities from $\sim 200$~cm$^{-3}$ up to 10$^4$~cm$^{-3}$ must be present in these clouds.

The components discussed in the previous paragraph are not dense clouds, and they cannot represent diffuse clouds either, otherwise we would obtain visual extinctions lower than 1\,mag when computing their H$_2$ column densities with the standard diffuse-cloud abundance of HCO$^+$. We conclude that the clouds with velocities between $-110$ and $-10$~km~s$^{-1}$ and at 6~km~s$^{-1}$ are translucent clouds, and that the HCO$^+$ abundance relative to H$_2$ must be higher than $3 \times 10^{-9}$ in these clouds, by at least a factor of two, and maybe even a factor of six in order to reconcile the visual extinction of the component at $-104$~km~s$^{-1}$ with its translucent nature ($A_{\rm v}$ should be between 1 and 5~mag). All these conclusions hold only if our assumption that the HCO$^+$ column densities derived from our ALMA absorption spectra are representative of the HCO$^+$ column densities at the scales probed with Mopra.

Our conclusion that HCO$^+$ likely has a higher abundance in translucent clouds compared to diffuse clouds is consistent with the results of the GEMS survey performed by \citet{fuente2018} with the IRAM 30\,m telescope towards the dark cloud TMC1. They report that HCO$^+$ reaches its maximum abundance relative to H$_2$ at a visual extinction of 5~mag, with a value of $\sim 2 \times 10^{-8}$. A visual extinction of 5~mag corresponds to a H$_2$ column density of $4.7 \times 10^{21}$~cm$^{-2}$, which implies a HCO$^+$ column density of $9.4 \times 10^{13}$~cm$^{-2}$ assuming the HCO$^+$ abundance above. The transition between diffuse and translucent clouds at a visual extinction of 1~mag corresponds to an H$_2$ column density of $9.4 \times 10^{20}$~cm$^{-2}$, that is a HCO$^+$ column density of $2.8 \times 10^{12}$~cm$^{-2}$ assuming the $\approx 7$~times lower standard HCO$^+$ abundance for diffuse clouds ($3 \times 10^{-9}$). Most components in Table~\ref{h2_densities_alma} have HCO$^+$ column densities between these two values, implying that they are translucent clouds. The exceptions are the component at $-134.6$~km~s$^{-1}$, which corresponds to a diffuse cloud, and the components at $-41.2$~km~s$^{-1}$, 18.6~km~s$^{-1}$, and 3.8~km~s$^{-1}$, which must be dense clouds. To have a better estimate of the H$_2$ column density and visual extinction of each component, we recompute these quantities assuming a HCO$^+$ abundance of $3 \times 10^{-9}$ for $N$(HCO$^+$) $< 2.8 \times 10^{12}$~cm$^{-2}$, $2 \times 10^{-8}$ for $N$(HCO$^+$) $> 9.4 \times 10^{13}$~cm$^{-2}$, and an interpolated value (in log-log space) in between (see Table~\ref{h2_densities_alma}). 

In \citet{thiel2017}, some of us reported on the basis of the EMoCA survey the detection of complex organic molecules in four velocity components (one in the Scutum arm and three in the GC region) that we ascribed to what we termed diffuse clouds. \citet{liszt2018} questioned the diffuse nature of these clouds on the basis of their high HCO$^+$ column densities. The detailed analysis performed here confirms that these components with COM detections are not diffuse. Instead, we find that the component in the Scutum arm (at 27~km~s$^{-1}$) and two of the GC components (at 9.8~km~s$^{-1}$ and $-1.7$~km~s$^{-1}$) are translucent clouds. The third GC component has a visual extinction of 8.5~mag suggesting it is somewhat dense, but still close to the border between translucent and dense clouds. Therefore, we conclude that complex organic molecules are present in some translucent clouds along the line of sight to Sgr~B2.
\subsection{Suitable tracers of H$_2$ in translucent clouds}
The fact that HCO$^+$ has a higher abundance in translucent clouds compared to diffuse clouds has an impact on the analysis of the CH abundance performed by \citet{qin2010} for the clouds probed in absorption with \textit{Herschel} towards Sgr~B2(M). They took HCO$^+$ as a proxy for H$_2$ and assumed a uniform abundance of $5 \times 10^{-9}$. With this assumption, the distribution of CH and H$_2$ column densities does not follow the correlation found by previous studies for diffuse clouds. However, if we take into account the variation of the HCO$^+$ abundance as described above, we find a correlation between CH and H$_2$ much closer to the diffuse-cloud correlation. The kink at an H$_2$ column density of $10^{21}$~cm$^{-2}$ reported by \citet{qin2010} vanishes (see Fig.~\ref{ch_h2}). The correlation of CH to H$_2$ determined by \citet{sheffer2008} for diffuse molecular clouds is now valid for H$_2$ column densities up to $10^{21.7}$~cm$^{-2}$, which means that CH is a good tracer of H$_2$ up to $A_{\rm v} \sim 5.3$~mag. At $N_\mathrm{H_2}\approx10^{23}$~cm$^{-2}$, we see a deviation from the correlation. Hence, the CH abundance relative to H$_2$ must decrease somewhere between $10^{21.7}$~cm$^{-2}$ and $10^{23}$~cm$^{-2}$. This is consistent with the drop of CH abundance above a H$_2$ column density of $5 \times 10^{21}$~cm$^{-2}$ reported for TMC-1 by \citet{suutarinen2011}.

\begin{figure*}
\centering
\includegraphics[width=17cm, trim = 1.6cm 0.6cm 1.7cm 1.7cm, clip=True]{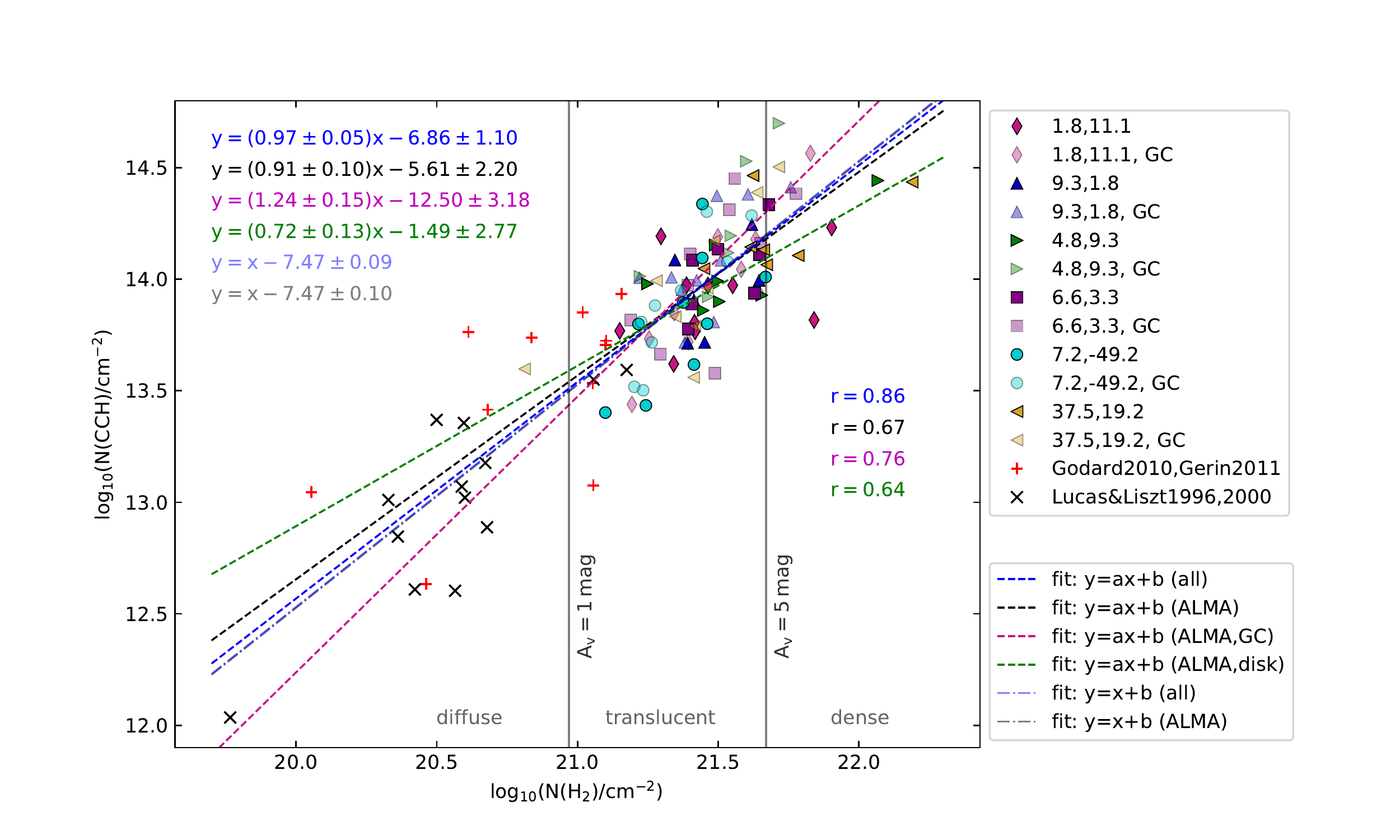}
\caption{Distribution of CCH column densities versus H$_2$ column densities, calculated from HCO$^+$ as described in Sect.~\ref{subsect_discuss_types}. The filled symbols represent the values obtained for the velocity components detected toward six positions with strong continuum background in the ALMA data. Their equatorial offsets are indicated to the right. The fainter filling colours are used for the GC components. The red crosses are measurements from \citet{godard2010} and \citet{gerin2011} and the black ones from \citet{lucas1996,lucas2000}.  The dashed lines are linear fits to the following samples: all data (blue), ALMA data only (black), ALMA GC components only (magenta), and ALMA disk components only (green). The dotted-dashed lines are fits with a slope of unity. The fits results are given in the upper left corner with their uncertainties. The Pearson correlation coefficients $r$ are given on the right side. The vertical lines highlight the limits between diffuse, translucent, and dense clouds.}
\label{cch_hcop}
\end{figure*}

CCH is strongly correlated with CH in diffuse molecular clouds \citep{gerin2010a}. There is also a tight correlation between CCH and c-C$_3$H$_2$ in translucent clouds \citep[for $N($c-C$_3$H$_2) < 10^{12.5}$~cm$^{-2}$,][]{lucas2000, gerin2011}. Here, we want to investigate whether both CCH and c-C$_3$H$_2$ are good tracers of H$_2$ in translucent clouds. Figure~\ref{cch_hcop} shows the distribution of CCH column densities as a function of H$_2$ column densities for the velocity components detected towards six positions with strong continuum background in our survey, along with measurements reported in the literature for other diffuse and translucent clouds. In this plot, the H$_2$ column densities are derived from the HCO$^+$ column densities assuming the same non-uniform HCO$^+$ abundance profile as in Sect.~\ref{subsect_discuss_types}. Overall, we see that CCH is well correlated with H$_2$ for both diffuse and translucent clouds. The slope of unity indicates that it can be used as a good tracer of H$_2$ up to $A_{\rm v} = 5$~mag at least. The correlation is a bit tighter for the GC translucent clouds than for the ones located in the disk of our Galaxy, with a slope slightly higher and lower than unity for the former and latter, respectively.

Figure~\ref{c3h2_hcop} shows the same plot for c-C$_3$H$_2$. If we take all data together, there is an overall correlation between c-C$_3$H$_2$ and H$_2$ with a slope close to unity, but a fit limited to the ALMA data only yields a much flatter correlation, with a large dispersion. c-C$_3$H$_2$ does not seem to correlate as tightly with H$_2$ as CCH in the translucent regime. This larger dispersion is dominated by the clouds located in the Galactic disk for which there is no correlation between c-C$_3$H$_2$ and H$_2$. The correlation is tighter for the GC clouds with, however, a slope higher than unity. In both cases, c-C$_3$H$_2$ thus does not appear as a good tracer of H$_2$.

The column densities of CCH and c-C$_3$H$_2$ are plotted against each other in Fig.~\ref{c3h2_cch}. Here also, while there is an overall correlation with a slope close to unity for the full sample of diffuse and translucent clouds, the fit limited to the ALMA data departs significantly from a slope of unity. Therefore, our data \textbf{do} not reveal a tight correlation between CCH and c-C$_3$H$_2$ beyond the range of C$_3$H$_2$ column densities explored by \citet{lucas2000} and \citet{gerin2011}. This conclusion holds also for the GC or galactic disk clouds taken separately (see magenta and green fits in Fig.~\ref{c3h2_cch}).

\subsection{Velocity components and velocity dispersions}\label{disc_vel_disp}
Previous absorption studies revealed velocity components similar to those we found in our data. \citet{corby2018} used GBT data to investigate ortho c-C$_3$H$_2$ ($1_{1,0}-1_{0,1}$) at $18.343$\,GHz. They detected many narrow features with $FWHM\lesssim3$\,km\,s$^{-1}$ which are superimposed on components with widths between $3$ and $16$\,km\,s$^{-1}$. They suggested at least ten distinguishable line-of-sight absorption components. \citet{garwood1989} used VLA data with a spectral resolution of $2.6$\,km\,s$^{-1}$ to investigate HI in absorption along the line of sight to Sgr\,B2(M). Because the velocities in the direction of Sgr\,B2(N) are similar to those in the direction of Sgr\,B2(M) \citep{greaves1994} we chose these velocities for comparison. The components detected by \citet{corby2018} and \citet{garwood1989} are compared to ours in Table~\ref{vel_comparison}.

All velocity components found by \citet{garwood1989} are detected in this work and in the data examined by \citet{corby2018}. \citet{corby2018} modelled the spectrum using several broad Gaussian components overlaid with narrower Gaussian features resulting in a higher number of velocity components than we investigate here. \citet{corby2018} detected two velocity components, at $-120$\,km\,s$^{-1}$ and $-58$\,km\,s$^{-1}$, for which we see no clear counterparts in our c-C$_3$H$_2$ spectra. The component at $-58$\,km\,s$^{-1}$ is detected towards a few positions but because of the poor statistics, we decided not to investigate this component. The component at $-120$~km~s$^{-1}$ is detected in other molecules such as HCN or HCO$^+$ (see Table~\ref{h2_densities_alma}), so our non-detection of this component in c-C$_3$H$_2$ is due to a lack of sensitivity. \citet{belloche2013} investigated the velocity components of c-C$_3$H$_2$ along the line of sight to Sgr\,B2 using the IRAM 30\,m telescope. They were also able to find a few additional velocity components compared to the components reported here. In our analysis, we only consider the velocity components associated with the strongest peaks (minima) of the absorption of c-C$_3$H$_2$. The programme we wrote to fit synthetic spectra to the 322 investigated positions may underestimate the number of velocity components in some cases because we do not allow peak velocities to be closer to each other than two channels (see Sect.~\ref{program}). This approach still yields good fits to the observed spectra (see Fig.~\ref{examplespec}). In addition, the number of identified velocity components depends on the investigated position. Because our main goal is to investigate the spatial structure of the diffuse and translucent clouds seen in absorption, we selected the velocity components for which a cloud is detected towards a sufficient number of positions. The detection of components depends also on the strength of the continuum. The IRAM 30~m and GBT single-dish data have angular resolutions of $\sim$$30\arcsec$ and $\sim$$40\arcsec$, respectively, which means that they are much more sensitive to the extended continuum emission and they include the two main hot cores Sgr~B2(N1) and Sgr~B2(N2). The stronger continuum of Sgr~B2(N1) (see Fig.~\ref{contmap}) reveals more velocity components in our data but they are too weak to be detected at lower continuum values. These hot-core positions are however affected by contamination from numerous emission lines in the 3~mm atmospheric window, which is the reason why we excluded them from our analysis.   

\begin{table}[t]
\caption{Velocities of diffuse and translucent clouds along the line of sight to Sgr~B2.}
\label{vel_comparison}
\centering
\begin{tabular}{r r r}       
\hline               
$v_\mathrm{this work}$\tablefootmark{a} &$v_\mathrm{Corby}$\tablefootmark{b} &$v_\mathrm{Garwood}$\tablefootmark{c}\\

[km\,s$^{-1}$] & [km\,s$^{-1}$] & [km\,s$^{-1}$]\\
\hline\hline 
$-$ & $-120$ & $-$\\         
$-105.9$ & $-106$ & $-107.6$\\
$-93.7$ & $-92$ & $-$\\
$-81.5$ & $-80$ & $-81.7$\\
$-74.6$ & $-73$ & $-$\\
$-$ & $-58$ & $-$\\
$-48.4$ & $-47$ & $-51.9$/$-44.0$\\
$-39.7$ & $-40$ & $-$\\
$-27.6$ & \multirow{2}{*}{$-23$}  & \multirow{2}{*}{$-24.4$}\\
$-18.9$ &  & \\
$-3.2$ & \multirow{3}{*}{0} & \multirow{3}{*}{1.1} \\
$2.0$ & & \\
$7.3$ & & \\
$17.7$ & \multirow{2}{*}{$20$} & $15.7$\\
$24.7$ &  & \multirow{3}{*}{31.4}\\
$31.6$ &  &\\
$36.9$ &  &\\
\hline                      
\end{tabular}
\tablefoot{\tablefoottext{a}{LSR cloud velocities of ortho c-C$_3$H$_2$ determined in this work.}\tablefoottext{b}{LSR cloud velocities determined by \citet{corby2018} from c-C$_3$H$_2$ at $18.343$\,GHz in absorption.}\tablefoottext{c}{LSR cloud velocities determined by \citet{garwood1989} from HI in absorption.}}
\end{table}

The smallest linewidth we derive for a velocity components traced by c-C$_3$H$_2$ is 1.74\,km\,s$^{-1}$, which is equal to the channel width. The largest width is about 20\,km\,s$^{-1}$ (see Fig.~\ref{fwhm}). \citet{corby2018} found a range of $FWHM$ between $\sim$1 and 10\,km\,s$^{-1}$, plus two broad components with $FWHM \sim 16$\,km\,s$^{-1}$ that probably result from the overlap of several narrower components. The majority (84\%) of the widths derived here is smaller than 11\,km\,s$^{-1}$ (see Fig.~\ref{fwhm}), thus falling in the same interval. \citet{belloche2013} found typical widths of the line-of-sight clouds of 3--5\,km\,s$^{-1}$, and \citet{menten2011} values between 3 and 8\,km\,s$^{-1}$. \citet{gerin2010} studied the diffuse and translucent clouds along the line of sight to G10.6-0.4 and found values between 3 and 6\,km\,s$^{-1}$ for the $FWHM$. The median value of Category I, 5.4\,km\,s$^{-1}$ (Table~\ref{median_fwhm_ntot}), lies at the upper edge of these ranges for the $FWHM$. If we look at the width distribution of Category I (see Fig.~\ref{fwhm}a) the majority (75\%) of the velocity components have $FWHM$ smaller than about $7.5$\,km\,s$^{-1}$. For larger widths the number of velocity components is decreasing. This tail could be caused by overlapping velocity components that could not be fitted separately.

The largest $FWHM$ derived from our fits is 20\,km\,s$^{-1}$. \citet{wiesemeyer2016} investigated OH and OH$^+$ along the lines of sight to other continuum sources than Sgr\,B2 and found velocity components with widths up to 18\,km\,s$^{-1}$ for OH and up to 23\,km\,s$^{-1}$ for OH$^+$ covering the same velocity interval. The distribution of their $FWHM$ shows that most velocity components have smaller widths and only a few have $FWHM$ larger than 10\,km\,s$^{-1}$. Their explanation for the wide variance of determined $FWHM$ is that the absorption of single clouds may cause the narrow features, whereas the larger widths of the velocity components may be caused by the blending of the contribution from several clouds along the lines of sight. Furthermore, the large beams of single dish telescopes may cause spatial blending effectively resulting in larger line widths. As explained in Section~\ref{results_cloud_properties} the programme we used for the modelling of c-C$_3$H$_2$ is not able to distinguish between a single broad Gaussian and multiple overlapping components producing the same shape. In those cases a single Gaussian is assumed, which likely explains the tail we obtain to higher $FWHM$ and a smaller number of velocity components. \citet{corby2018} investigated the OH absorption spectrum towards Sgr B2 at 1612~MHz and 1667~MHz. The former line shows linewidths larger than 10\,km\,s$^{-1}$ (up to 50\,km\,s$^{-1}$) for most velocity components while for the latter half of the components have widths smaller than 10\,km\,s$^{-1}$, but still larger than c-C$_3$H$_2$. Here again, the velocity overlap between clouds probably leads to an overestimate of the OH linewidths in some cases. Summarising, it seems that c-C$_3$H$_2$ traces less diffuse regions with smaller velocity dispersions than OH.%

\subsection{Cloud sizes}
The two-point auto-correlation functions of the opacity maps suggest that most detected structures are extended on the scale of our field of view, $\sim$15$\arcsec$, or beyond (Sect.~\ref{autocorrel}). This means that our assumption of a beam filling factor of 1 to fit the spectra is reasonable. In a few cases, the two-point auto-correlation functions indicate the presence of smaller structures of sizes $\sim$4$\arcsec$--6$\arcsec$. These structures are mostly seen for less abundant species for which most of the opacity map is dominated by noise. For example HN$^{13}$C shows small structures for the GC cloud at 2.0\,km\,s$^{-1}$, but the main isotopologue HNC shows an extended structure at this velocity. Therefore, we believe that most of the compact structures suggested by the two-point auto-correlation functions simply result from a lack of sensitivity. More sensitive observations would likely detect the same underlying extended structures traced by the more abundant molecules.  

The auto-correlation functions of SO and SiO suggest a clumpy structure for some of the velocity components, which cannot be explained by a lack of sensitivity for these molecules. The signal-to-noise ratio is relatively good for SiO at velocities between $-5$~km~s$^{-1}$ and 25~km~s$^{-1}$ and for SO between 0~km~s$^{-1}$ and 10~km~s$^{-1}$. Hence, the clumps seen at these velocities are significant. Especially at the velocity of 7.0~km~s$^{-1}$, which corresponds to a translucent cloud, these two molecules have a high SNR and reveal a smaller structure in the opacity maps of size of $5^{\prime\prime}$ at an offset of ($1\as5 ,11\as0$) (see Figs.~\ref{opacity_so} and \ref{opacity_sio}). The physical size of these clumps is in the order of 0.2~pc. SiO also shows a more complex structure at 17.2~km~s$^{-1}$ in the Sagittarius arm. A denser clump peaked at ($6\as5$,$5\as0$) with a size of about $8^{\prime\prime}$ (0.04~pc) is overlying a more extended structure. Except the clump at ($6\as5$,$5\as0$), the clumps detected in SiO and SO are located close to the boundaries of the available field of view. Hence, we may not see the complete structure of the clump and the determined size is only a lower limit. 

The structures of sizes 0.04--0.08\,pc seen in some molecules at the velocity of the Sagittarius arm can not be spatially resolved for the GC clouds. Observations with higher spatial resolution are needed to investigate if such smaller physical structures are present in the GC clouds. 

The absorption components detected towards Sgr\,B2(N) and Sgr\,B2(M), which are separated by $\sim$50$\arcsec$, have similar velocities \citep{greaves1994}. This suggests that the structures of the foreground clouds detected in absorption are more extended than 50$\arcsec$, which can be seen in $^{13}$CO large-scale emission maps.

\subsection{Turbulence}
The PDFs analysed in Sect.~\ref{results_pdf} can be described by one or two log-normal distributions. A log-normal profile characterises the low density part of the PDFs of star-forming regions \citep[e.g][]{schneider2013}. Isothermal hydrodynamic simulations including turbulence and gravity have shown that a power tail at higher densities arises due to gravitational collapse/contraction \citep[e.g.][]{klessen2000}. When gravity is not included in the simulations, only the log-normal part remains in the PDF \citep{padoan1997, kritsuk2007, federrath2008}. 

The log-normal part of the PDF is not influenced by resolution effects and should be unchanged by using larger beam sizes \citep{schneider2015a, federrath2013}. But in cases where the pixel distribution used for determining the PDF is not covering the complete structure, a fall off at lower densities can appear \citep{schneider2015b}. If the structure is not completely covered by the field of view of the observations, the width can be underestimated and the peak position overestimated \citep{ossenkopf2016}. Due to the limitations set by the background continuum emission on our actual field of view, we most likely do not trace the complete structure of the line-of-sight clouds. Hence, the width of their PDFs is likely underestimated. 

The volume-weighted density PDF in numerical simulations has a wider dispersion than the column density PDF, but shows similar properties \citep{federrath2010}. Both PDFs are nearly perfect log-normal distributions in simulations without gravity, only small effects seen as non-Gaussian features are present \citep[see][]{federrath2010}. The effects can be high- and low-density wings \citep[e.g.][]{passot1998,kritsuk2007, federrath2010}. We sometimes see such weak wings in our data (see, e.g. Fig.~\ref{pdf_cs} at $\varv_\mathrm{LSR}=-81.3$~km~s$^{-1}$), but they do not influence the parameters of the fitted log-normal profiles and also have only weak influence on the calculated dispersion of the profiles.

\citet{federrath2008} investigated the gas density in turbulent supersonic flows by performing numerical simulations. They examined the differences in the column density fields for solenoidal forcing (divergence free) and compressive forcing (curl-free). They fixed the Mach number to 5.5. A result of the analysis is that the compressive forcing leads to a much higher density contrast than the solenoidal forcing despite the same Mach number. Solenoidal forcing is mostly present in more quiescent regions with low star formation rates, which is also seen in observations \citep{federrath2010}. For both the volume-weighted density PDF and the column density PDF, the width of the lognormal part of the PDF is three times larger for compressive forcing than for solenoidal forcing. The compressive forcing of the turbulence as well as a higher Mach number broaden the PDFs \citep{federrath2013}.

The width of the column density PDF for purely solenoidal forcing is determined in simulations to be $\sigma_\eta = 0.46 \pm 0.06 $ \citep{federrath2010}, which is similar to the average value of our sample of 15 diffuse and translucent clouds (0.52). The larger widths we obtained for low-abundance species (C$^{34}$S, $^{13}$CS, HN$^{13}$C, and HC$^{15}$N) is likely due to the lack of sensitivity and the stronger influence of the noise \citep{ossenkopf2016}. The width of column density PDFs for purely compressive forcing would be $\sigma_\eta = 1.51 \pm 0.28 $ \citep{federrath2010}. We therefore conclude that on average, the turbulence of the diffuse and translucent clouds traced in absorption in our data is mainly solenoidally driven, provided the bias resulting from our limited field of view is not too strong.

We derived in Sect.~\ref{results_pdf} the turbulence parameters $\zeta$ and $b$ from the width of the PDFs of all 15 velocity components using c-C$_3$H$_2$, H$^{13}$CO$^+$, $^{13}$CO, CS, SO, SiO, HNC, and CH$_3$OH. Because these parameters depend on the Mach number, the kinetic temperature is needed and we assumed a range of 20~K to 80~K to cover roughly the range of temperature possible in diffuse and translucent molecular clouds. With this range of temperature, the range of Mach number of the 
velocity components we detected is 2.9--28.3.

The question is which kinetic temperature is more likely: 20\,K, 40\,K, or 80\,K. The investigation of H$_2$ using far-UV observations shows a temperature for diffuse clouds of 50--150\,K depending on the investigated line of sight \citep[][and references therein]{snow2006}. The mean value is about $~80$\,K. The temperatures in the direction of $\zeta$\,Persei derived using C$_2$ is $80\pm15$\,K, and using H$_2$ $58\pm8$\,K \citep[][and references therein]{snow2006}. The temperature in the direction of $o$\,Persei is a bit lower, $60\pm20$\,K using C$_2$ and $48\pm5$\,K using H$_2$ \citep[][and references therein]{snow2006}. The temperature seems to show a high variance. However, as we concluded in Sect.~\ref{subsect_discuss_types}, most clouds along the line of sight to Sgr~B2 are translucent. Hence, the kinetic temperature is likely lower than the temperature of diffuse clouds mentioned above. \citep{fuente2018} derived a temperature of 13--14~K for visual extinctions below 7.5~mag in TMC-1. Therefore, we believe that the temperature of the translucent clouds investigated here is between $\sim$15~K and $\sim$50~K. Our radiative-transfer calculations (Sect.~\ref{radex_models}) do unfortunately not deliver any constraint on the kinetic temperature. In summary, for most of the line-of-sight clouds of our sample we favour temperatures of 20~K or 40\,K, but we cannot exclude a higher temperature for the diffuse cloud in our sample.
 
For any of the three temperatures assumed above, the forcing parameter $b$ is always smaller than 0.5 for all components. This confirms the conclusion drawn above that the turbulence is dominated by solenoidal forcing in the translucent clouds investigated here.

Many values of the forcing parameter $b$ are either smaller than the value expected for purely solenoidal forcing or a bit larger, but still relatively close to 1/3. Values of $b<$1/3 are in principle not possible in the framework of the turbulence investigated in the simulations \citep[][]{federrath2010}. The underestimation of $b$ could result from an underestimation of the width of the PDF because of the limited field of view, as mentioned above. A second reason for a low value of $b$ could be an overestimation of the linewidth, for instance due to the overlap of several components along the line of sight. For example, $b$ is relatively small for the following velocity components: $-74.6$, $2.0$, $-27.6$, $-18.9$, and $31.6$~km~s$^{-1}$. The median $FWHM$ for those components is on average larger in comparison with the other velocity components (7.1--10.2~km~s$^{-1}$). 

In summary, our investigation of the PDFs of the clouds along the line of sight to Sgr\,B2 shows that the driving of the turbulence is purely solenoidal, or at least dominated by solenoidal forcing. Solenoidal forcing causes smaller density contrasts \citep{federrath2008}, which results in more homogeneous cloud structures. This is consistent with the lack of sub-structures that we noticed on the basis of our analysis of the two-point autocorrelation functions of the opacity maps.

Studies of three translucent high-latitude clouds, MBM\,16, MBM\,40, and MBM\,3, show that the motions inside the clouds are highly correlated in clumps which have sizes of about 0.5\,pc \citep{magnani1993, shore2003,shore2006, shore2007}. This scale corresponds to the extent of the field of view we have in our ALMA data for the most distant clouds. The clumps in the translucent high-latitude clouds are located in a more diffuse structure. They are not self-gravitating and do not show star-forming activity. Some significant changes in emission at the scale of 0.03\,pc are visible resulting in strong density gradients of a factor up to 10. In our data towards Sgr~B2, we see, for a few molecules, structures at scales of about 0.04--0.08\,pc only for the line-of-sight clouds located closer to us where such scales are resolved. The dynamics in the translucent high-latitude clouds was shown to result from the combination of shear flows and thermal instabilities, with the large-scale shear flow powering the turbulence and the density field maybe caused by thermal instability. Because we see also homogeneous structures on scales up to 0.55\,pc roughly the size of the clumps detected in the high-latitude clouds, maybe the same processes are at work in the line-of-sight clouds investigated here.

\subsection{Meaning of the cloud Categories}
We have divided the clouds detected in absorption along the line of sight to Sgr~B2 into two categories on the basis of their c-C$_3$H$_2$ column densities. Category I, with lower c-C$_3$H$_2$ column densities (10$^{12.8}$~cm$^{-2}$ on average), contains the GC clouds with LSR velocities below $-50$~km~s$^{-1}$ and the clouds belonging to the 3~kpc and 4~kpc arms. The GC clouds with velocities around 0~km~s$^{-1}$, the clouds in the Scutum arm and the ones in the Sagittarius arm belong to Category II, with higher c-C$_3$H$_2$ column densities (10$^{13.2}$~cm$^{-2}$ on average). Clouds belonging to Category I have on average smaller velocity dispersions ($FWHM = 5.4$~km~s$^{-1}$) and a smaller PDF width ($\sigma_\eta=0.48$) than clouds in Category II ($FWHM$=7.5~km~s$^{-1}$ and $\sigma_\eta=0.56$). The statistically higher column densities of Category II do not result from a lack of sensitivity, which could occur if the linewidths are broader, because the column densities per velocity unit show the same trend. 

The spiral arms closer to the GC (3~kpc and 4~kpc arms) belong to Category I whereas the more distant Scutum and Sagittarius arms belong to Category II. 
It is thus a priori surprising that the GC clouds with velocities around 0~km~s$^{-1}$ have similar properties as the clouds in the Scutum and Sagittarius arms. Given that absorption at velocities close to 0~km~s$^{-1}$ could arise from local clouds, we could be tempted to argue that these clouds were misassigned and are in fact local clouds. The assignment was in part based on the $\frac{^{12}\mathrm{C}}{^{13}\mathrm{C}}$ isotopic ratio of 22 determined by \citet{gardner1982} which is consistent with the Galactic centre value and not with the higher ratio that characterises local clouds. Indeed, the $\frac{^{12}\mathrm{C}}{^{13}\mathrm{C}}$ isotopic ratio shows a gradient with galactocentric distance, increasing from about 20 in the GC to about 60--70 for local gas \citep[][]{milam2005,halfen2017}. As a further verification, we used the EMoCA data to measure the $\frac{^{12}\mathrm{C}}{^{13}\mathrm{C}}$ isotopic ratio of the velocity components at $-2.6$, 3.7, and 8~km~s$^{-1}$ (covering the velocity range from $-5$ to 10~km~s$^{-1}$). We used the absorption spectra of CN and $^{13}$CN and derived isotopic ratios of $\frac{^{12}\mathrm{C}}{^{13}\mathrm{C}}=20\pm 3$ for these clouds, confirming once more their location in the GC. This value is robust against optical depth effects because we used the hyperfine components of CN to derive the ratio. \citet{liszt2018a} determined a higher isotopic ratio ($\frac{^{12}\mathrm{C}}{^{13}\mathrm{C}}=64\pm4$) towards J1774 using HCO$^+$ isotopologues in the velocity range of $-33$ to 13~km\,s\,$^{-1}$, which is dominated by an absorption feature between -10\,km\,s$^{-1}$ and 10\,km\,s$^{-1}$. The continuum source J1774 is located at small longitudes near the Galactic centre, with a separation of about 3$^\circ$ from Sgr\,B2. This high isotopic ratio would indicate the clouds producing these absorption features to originate from the disk. The velocity range covered by the velocity components of CN used to determine the ratio $\frac{^{12}\mathrm{C}}{^{13}\mathrm{C}}$ in this work is smaller than their velocity range. Hence, we cannot exclude some gas seen in absorption at velocities between -9 and -5~km~s$^{-1}$ may be located in the disk and not in the GC (see also the discussions in Sect. 3 of \citet{wirstrom2010} and Sect.~5.2 of \citet{corby2015}). But the components traced with CN seem to be located in the GC. Finally, the clouds in Category III have similar c-C$_3$H$_2$ column densities as those in Category II, and a median linewidth (6.7~km~s$^{-1}$) that is closer to the one of Category II than to the one of Category I.

In summary, the GC region seems to contain two distinct populations of line-of-sight clouds (Category I and Categories II+III), but all with turbulence being driven (mainly) solenoidally. In the Galactic disk, the dichotomy correlates with the galactocentric distance: the clouds in Category~II are located further away from the GC than the clouds in Category~I. It is unclear to us whether this correlation with galactocentric distance is fortuitous or not.

In the Mopra data, HCO$^+$~1--0, HNC~1--0, and CS~2--1 are detected in emission over a wide range of velocities, from about $\varv_\mathrm{LSR}=-15$\,km\,s$^{-1}$ up to $\approx 130$\,km\,s$^{-1}$. The lower limit corresponds roughly to the border between Categories I and II (see Figs.~\ref{spectrum_mopra_hcop}, \ref{spectrum_mopra_hnc}, \ref{spectrum_mopra_cs}, respectively). These tracers have higher critical densities than $^{13}$CO 2--1, which is detected in emission in both Categories of clouds (Fig.~\ref{spectrum_mopra_13co}). This suggests that the clouds in Category I have lower volume densities than the clouds in Category II.

\subsection{Principal component analysis}
We performed the PCA for six velocity components. The first principal component contributes strongly to the total variance of the opacity maps for only two of them: $-105.9$\,km\,s$^{-1}$ and $24.7$\,km\,s$^{-1}$. This means that a large part of the structures seen in the opacity maps of the different molecules are strongly correlated or anti-correlated for all molecules. Hence, there are spatial structures in these two clouds with smaller extent than the field of view.

The normalisation (mean$=$0 and standard deviation$=$1) used for the PCA implies that maps with a relatively uniform emission cannot be distinguished by the PCA from the maps of pure noise channels. Hence, if the maps of several molecules are relatively uniform but with a good signal-to-noise ratio, the PCA is not able to give any indications about the large-scale correlation. At $2.0$\,km\,s$^{-1}$ most molecules show indeed extended structures in the opacity maps, especially H$^{13}$CO$^+$ which has a relatively homogeneous opacity over the field of view (see Fig.~\ref{opacity_h13cop}). On the contrary, SiO shows a stronger variation of the opacity over the field of view at this velocity (see Fig.~\ref{opacity_sio}). This molecule dominates the structure in the first PC (see Figs.~\ref{pca_components_channel9} and \ref{pca_contr_channel9}). 

Overall, the PCA indicates that most clouds probed in absorption towards Sgr\,B2(N) have relatively homogeneous structures over the field of view, which is consistent with the conclusions we draw from the analysis of the PDFs and auto-correlation functions of their opacity maps. 
\section{Conclusions}\label{sect_conclusion} 
We used the EMoCA survey performed with ALMA to resolve and investigate the spatial structure of diffuse and translucent clouds traced by several molecules detected in absorption along the line of sight to Sgr B2(N), taking advantage of the high sensitivity and angular resolution of this survey. We investigated the velocity structure over the field of view by fitting the synthetic spectra to the absorption features of ortho c-C$_3$H$_2$. In addition, we investigated the spatial and kinematic structure of the individual clouds using the molecules c-C$_3$H$_2$, H$^{13}$CO$^+$, $^{13}$CO, HNC and its isotopologue HN$^{13}$C, HC$^{15}$N, CS and its isotopologues C$^{34}$S and $^{13}$CS, SiO, SO, and CH$_3$OH. Our main results are summarised as follows:
\begin{enumerate}
\item We found 15 main velocity components along the line of sight to Sgr~B2 on the basis of c-C$_3$H$_2$. The strong velocity components match the ones found by previous studies \citep[e.g.][]{garwood1989,corby2018}. The envelope of Sgr~B2 shows two main components at $\sim 63$\,km\,s$^{-1}$ and $\sim 80$\,km\,s$^{-1}$. In addition, we report the detection of a cloud at $48$\,km\,s$^{-1}$.  

  \item The absorption features along the line of sight can be divided into two categories on the basis of their c-C$_3$H$_2$ column densities: Category I ($\varv_\mathrm{LSR}<-13$~km~s$^{-1}$) contains some GC clouds and clouds belonging to the 3~kpc and 4~kpc arms. Category II ($-13$~km~s$^{-1}\leq \varv_\mathrm{LSR}<42$~km~s$^{-1}$) contains other GC clouds and clouds belonging to the Scutum and Sagittarius arms. The clouds of Category II have larger $FWHM$ and broader PDFs and they seem to have higher volume densities.
  
 \item Most clouds detected along the line of sight to Sgr~B2 are translucent on the basis of their HCO$^+$ column densities, including most clouds where we previously reported the detection of complex organic molecules. This is in agreement with the densities derived in previous single-dish studies of these clouds.

 \item Our analysis of the HCO$^+$ abundance in translucent clouds indicates that HCO$^+$ is not a good tracer of H$_2$ in translucent clouds. We further find that CCH and CH are good probes of H$_2$ in translucent clouds, but that c-C$_3$H$_2$ is not.
 
  \item The two-point auto-correlation functions reveal that the molecules investigated here trace relatively homogeneous structures as extended as 15$\arcsec$ at least. Smaller structures suggested by the two-point auto-correlation functions are in many cases dubious due to sensitivity limitations. 
  
  \item The average width of the column density PDFs is $\sigma_\eta=0.52$, which is close to the value of 0.46 expected for purely solenoidal forcing. The turbulence in these clouds is thus dominated by solenoidal forcing, which may explain the relatively homogeneous structures traced by the molecules investigated here.
 
  \item A principal component analysis indicates in most cases a homogeneous distribution of the molecules. Only two of the six investigated velocity components show a structure smaller than the field of view.

\end{enumerate}

Our analysis of the structure of the translucent clouds detected in absorption towards Sgr~B2 is limited by two factors: the limited size of the continuum background emission, which results from the spatial filtering of the interferometer, and the finite angular resolution. Combining these ALMA main array data with measurements performed with the ACA would allow us to extend the analysis to scales larger than 15$\arcsec$.

\begin{acknowledgements} 
We thank Volker Ossenkopf-Okada and Nicola Schneider for fruitful discussions about PDFs. We thank the referee, Harvey Liszt, for his insightful comments.
This paper makes use of the following ALMA data: ADS/JAO.ALMA\#2011.0.00017.S, ADS/JAO.ALMA\#2012.1.00012.S. ALMA is a partnership of ESO (representing its member states), NSF (USA) and NINS (Japan), together with NRC (Canada), NSC and ASIAA (Taiwan), and KASI (Republic of Korea), in cooperation with the Republic of Chile. The Joint ALMA Observatory is operated by ESO, AUI/NRAO and NAOJ. The interferometric data are available in the ALMA archive at https://almascience.eso.org/aq/. DC acknowledges support by the Deutsche Forschungsgemeinschaft, DFG, through project number SFB956C.
\end{acknowledgements}

\bibliographystyle{aa} 
\bibliography{references} 


\begin{appendix}

\section{Powell's method}\label{appendix_powell}
Powell's conjugate direction method works in an $n$-dimensional space, where $n$ is the number of free parameters. The starting point in this $n$-dimensional space is defined by the initial guesses $x_i$ of the $n$ parameters to fit, with $i$ between 1 and $n$. The initial guesses define a position in the $n$-dimensional space: $\vec{p}_0=(x_1,x_2, ... x_n)$. To converge towards the best-fit parameters, a set of search vectors is needed along which the minimisation is done. The set of $n$ starting vectors $\vec{\xi}_i$ is built by $n$ linearly independent directions. The initial set of search vectors is simply taken parallel to the parameter axes in the $n$-dimensional space.

The Powell minimisation procedure comprises an inner loop and an outer loop. During the inner loop the minimum in each current direction, that is along each search vector $\vec{\xi}_i$, is determined. The resulting minima along all directions give a new position $\vec{p}_1$ in the $n$-dimensional space. The vector difference $\Delta\vec{p}=\vec{p}_1-\vec{p}_0$ between the optimised point $\vec{p}_1$ and the starting point $\vec{p}_0$ is added to the set of search vectors and the vector $\vec{\xi}_i$ which is contributing the most to the new position is removed. The most contributing vector is the one with the smallest angle to the new search vector $\Delta\vec{p}$. The outer loop repeats the process until a given tolerance criterion is met, which can be a maximum number of iterations or a given noise level.

\section{Automatisation programme}\label{appendix_program}
This appendix describes the algorithm of our python programme that fits the absorption spectra automatically (Sect.~\ref{program}). The process is done in several iterations. The parameters of the synthetic spectrum resulting from the previous iteration are taken as initial guesses for the next iteration. The following parameters are fitted: the decimal logarithm of the column density $\log_{10}(N_\mathrm{tot})$, the width $FWHM$, and the centroid velocity $\varv_0$ of each velocity component. Different probability distributions around the initial guesses are chosen for these three parameters. The probability distribution for the velocity is uniform over the chosen interval ($\pm3$~km~s$^{-1}$); the probability distribution for the $FWHM$ is a truncated normal distribution with a width of 3~km~s$^{-1}$ peaked at the initial guess over the range defined by the channel width as a minimum value and a maximum value of 30~km~s$^{-1}$; the probability distribution for the logarithm of the column density is a normal distribution peaked around the initial guess with a width of two (that is two orders of magnitude for the density). The starting points are not the exact initial guesses but randomly selected points around the values following the probability distributions. The procedure follows this sequence:%
\begin{enumerate}
  \item Baseline subtraction for each position using the command base in the GILDAS package CLASS. This yields the continuum level and the spectral line noise level of each spectrum.
  \item Search in the continuum-subtracted spectrum for absorption dips with a peak signal-to-noise ratio higher than 4 ($-I_\mathrm{l}\geq 4\sigma$). The programme searches for all channels fulfilling this condition and keeps only those that have $|I_\mathrm{l}|$ higher than their surrounding channels (two channels in each direction). The velocities $v_i$ of these selected channels are then used to fit the spectrum and obtain initial guesses for the column densities $N_\mathrm{tot,i}$ and linewidths $FWHM_i$.

  \item Fit of the spectrum performed step by step in four partially overlapping velocity ranges ([-121, -58]~km\,s$^{-1}$, [-66, -8]~km\,s$^{-1}$, [-16, 48]~km\,s$^{-1}$, [40, 100]~km\,s$^{-1}$). Only the parameters of the components whose initial guess for the centroid velocity is in the currently fitted interval are fitted. The parameters of the other components are kept fixed to their initial guesses for the components in the next velocity range or to their fitted values for the components in the previous velocity range.
      
The step-by-step fitting in four velocity ranges is a useful procedure to reduce the computing time. If the maximum number of iterations is executed without reaching the tolerance limit of $2\sigma$, the fit is repeated up to five times (with randomly chosen initial guesses), but in the fourth iteration the noise-tolerance limit is increased to a higher value of $10\sigma$. This step is needed in the cases where the absorption caused by the envelope of Sgr\,B2 contains channels with $|I_\mathrm{l}|>I_\mathrm{c}+2\sigma$.

 \item Validation of the fit. The aim is to look for missing components which were not identified as peaks at step 2. The programme looks for peaks above a threshold of $5\sigma$ in the residuals. The peak velocities of these new components are then used and kept fixed to fit the residuals and estimate their column densities and linewidths. These estimates are used as initial guesses for these new components at step 5.
 \item Repeat step 3 once more and then go to step 6.
 \item Final check for missing components. The programme repeats step 4 (and step 3), but with a lower threshold of $4\sigma$, then goes to step 7. 
 \item To get the final synthetic spectrum, the $n$ velocity components are fitted step by step in $n-2$ iterations. Only three components are fitted at once at each iteration, starting with the first three velocity components and then shifting by one component at each subsequent iteration. The best-fit values of the first velocity component of each processed set of three components are kept fixed at the next iterations and the best-fit values of the second and third velocity components are used as initial guesses for the next iteration. This step-by-step fitting procedure decreases the computing time.
 
In the case where the fit was not successful for one of the $n-2$ iterations, 11 more repetitions of this fitting procedure are possible. During the repetitions several changes are made in the fitting conditions. The changes are necessary in some cases in order to get a good result. These changes were included after analysing the problems resulting in an unsuccessful fit using a randomly chosen small sub-sample of the data. The first problem is caused by the optical thickness of absorption lines from the envelope of Sgr~B2. The estimated column densities sometimes exhibit very high values up to $10^{17}$~cm$^{-2}$ which  sometimes results in computing problems. Hence, after the third repetition high initial guesses for the column densities in the envelope of Sgr~B2 are reduced to $10^{13}$~cm$^{-2}$ to avoid starting values higher than $10^{15}$~cm$^{-2}$ for these components. If this still does not result in a successful fit, the reason may be the limited velocity interval in which each peak velocity can be varied to find the best solution. After the sixth repetition the range for the velocity is increased from 6\,km\,s$^{-1}$ to 10\,km\,s$^{-1}$. If this still does not help the tolerance limit is increased after the ninth iteration.  
\end{enumerate}
We check the centroid velocities of the components before each fitting because the presence of two components lying too close to each other leads to calculation problems due to the ambiguity of these components. Therefore, when the velocity difference between two components is smaller than two channels ($\sim$3.5\,km\,s$^{-1}$) we remove the second component. In the case of three components following each other over a velocity range smaller than three channels ($\sim$5.3\,km\,s$^{-1}$), we remove the middle component. Due to this selection of the centroid velocities, the programme seldom fits only one velocity component to the peak at a velocity of about 3~km~s$^{-1}$ (see Fig.~\ref{examplespec}).

\section{Influence of noise} 
\subsection{Two-point auto-correlation functions}\label{sect_noise_auto_corr}

\begin{figure}
   \resizebox{\hsize}{!}{\includegraphics[width=0.5\textwidth,trim = 2.3cm 0.2cm 1.5cm 1.5cm, clip=True]{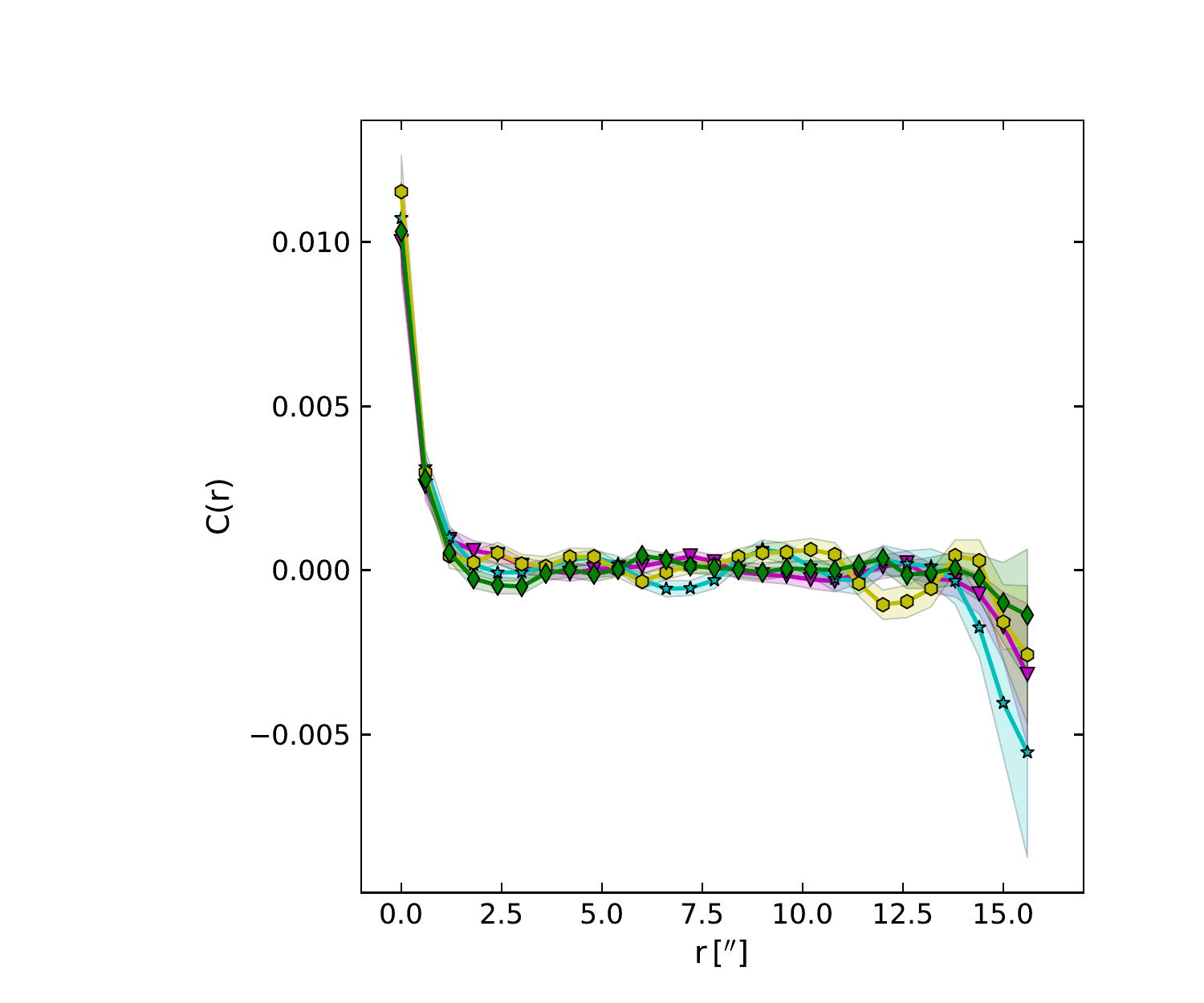}}
 \caption{Two-point auto-correlation function $C(r)$ of four different noise channels next to the transition of c-C$_3$H$_2$ as a function of pixel separation $r$. The coloured regions around the curves represent the standard deviation of the 1000 realisations of the opacity cube.}\label{autocorr_c-c3h2_noise}
\end{figure}

\begin{figure}
   \resizebox{\hsize}{!}{\includegraphics[width=0.5\textwidth,trim = 0.0cm 0.5cm 1.3cm 1.5cm, clip=True]{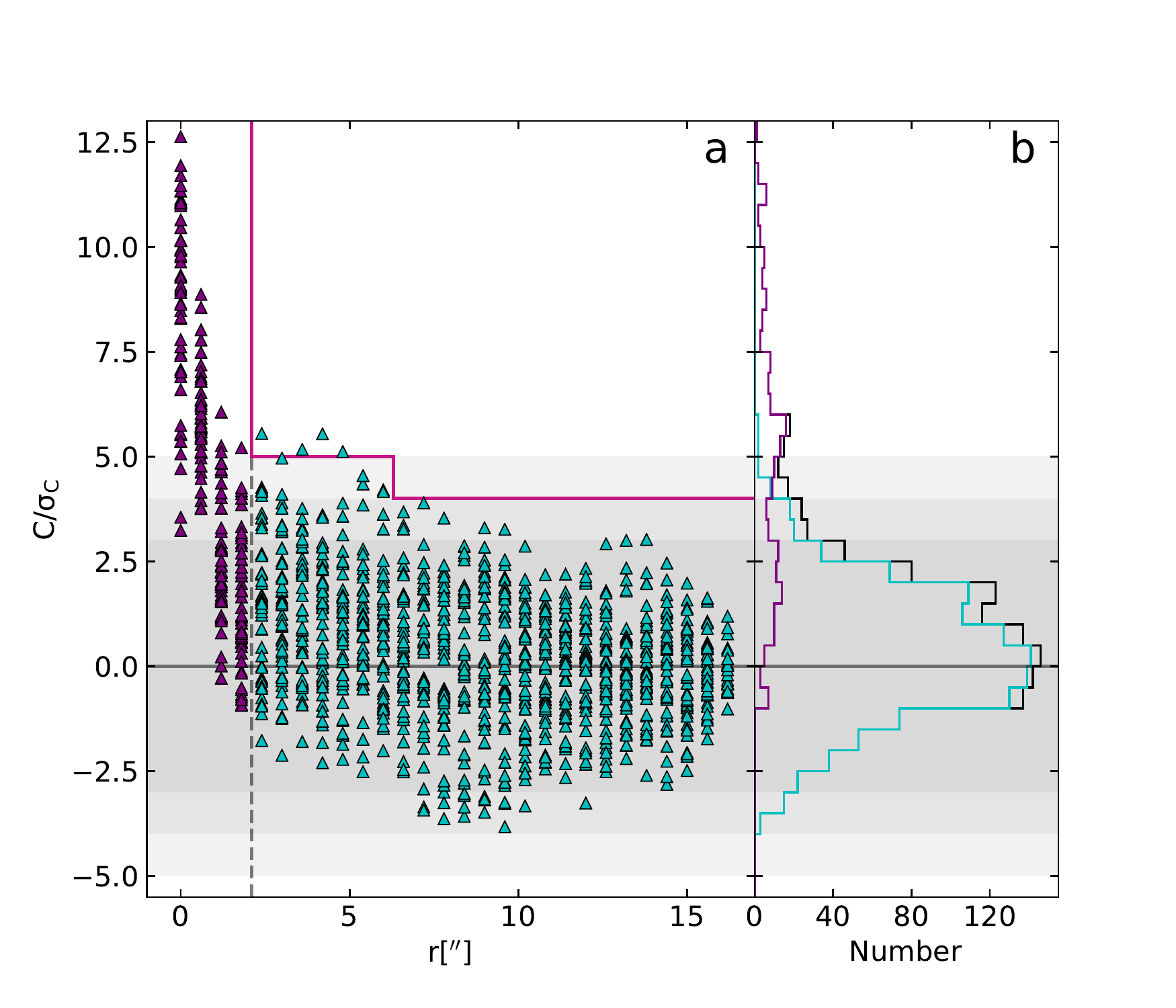}}
 \caption{\textbf{a} Signal-to-noise ratio (SNR) of the two-point auto-correlation functions ($C(r)/\sigma_C$) of 48 noise channels, four per molecule, as a function of pixel separation $r$. All points at pixel separations smaller than the beam are coloured in purple, the other ones in cyan. The vertical dashed line at 2$\arcsec$ corresponds to the major axis of the beam. The area above and to the right of the magenta line is the area where the correlation function of a channel containing signal indicates a significant correlation. \textbf{b} Distribution of $C/\sigma_C$. The regions of SNR smaller than 3, 4, and 5 are highlighted in shades of grey.} \label{autocorr_noise_statistic}
\end{figure}

To evaluate the influence of noise on the two-point auto-correlation functions $C(r)$, we calculated $C$ for several noise channels. In Fig.~\ref{autocorr_c-c3h2_noise} the two-point auto-correlation functions of four noise channels next to the transition of c-C$_3$H$_2$ are depicted. $C(r)$ shows a correlation at pixel separations smaller than about $2^{\prime\prime}$ which corresponds to the size of the major axis of the beam. At larger pixel separations, $C(r)$ is overall consistent with zero within $3\sigma$.

The signal-to-noise ratios ($C(r)/\sigma_C$) of 48 noise channels (four per molecule) are shown in Fig.~\ref{autocorr_noise_statistic}. Beyond the separation of 2$\arcsec$ mentioned above, most SNRs are below 3. For separations between 2$\arcsec$ and 6$\arcsec$, some outliers have SNRs between 3 and 5, and only a handful have SNRs between 5 and 6. Beyond 6$\arcsec$, all outliers have SNRs below 4. Therefore, we conclude that the correlation function of a channel containing signal will show a significant correlation beyond 2$\arcsec$ only when $C(r)/\sigma_C$ is higher than 5 for $r<6\arcsec$ and higher than 4 beyond $6\arcsec$. This condition corresponds to the area located to the right and above the magenta line shown in Fig.~\ref{autocorr_noise_statistic}.

\subsection{Probability distribution functions}\label{sect_noise_pdf}

\begin{figure}
   \resizebox{\hsize}{!}{\includegraphics[width=0.5\textwidth,trim = 0.3cm 0.7cm 0.1cm 1.7cm, clip=True]{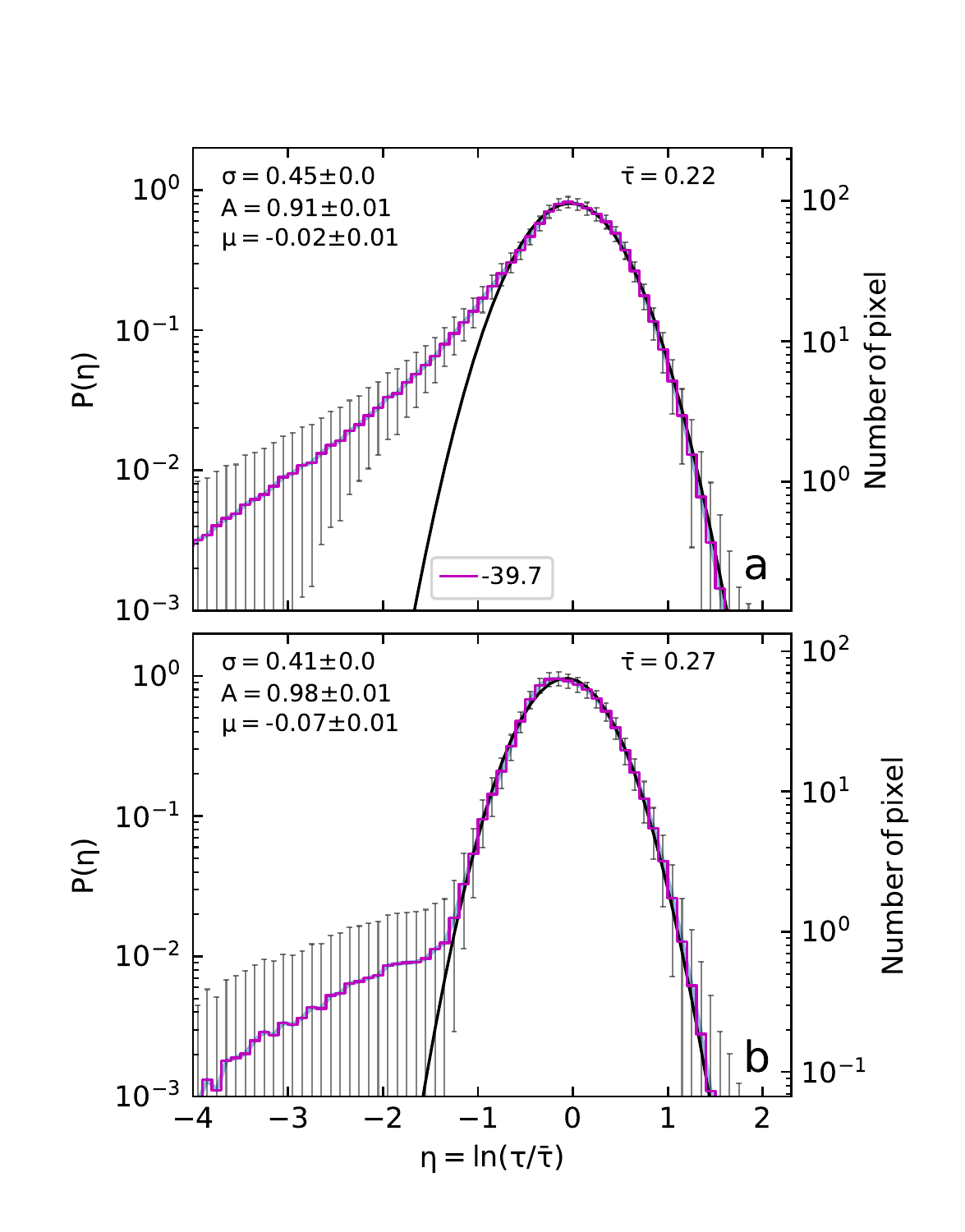}}
 \caption{\textbf{a} PDF of c-C$_3$H$_2$ using all pixels at a velocity of $-39.7$\,km\,s$^{-1}$. \textbf{b} PDF of the same velocity component but using only the pixels $\tau/\sigma_\tau > 3$. The results of the Gaussian fit, integral $A$, dispersion $\sigma$, and centre $\mu$, are indicated in the top left corner of each panel. The mean opacity $\bar{\tau}$ is indicated in the top right corner.}\label{pdf_noise}
\end{figure}

We investigate here the influence of the noise on the PDFs of the opacity maps. The PDF of c-C$_3$H$_2$ at a velocity of $-39.7$\,km\,s$^{-1}$ is plotted in Fig.~\ref{pdf_noise}a using all pixels and in panel b after masking the pixels below $3\sigma_\mathrm{noise}$. The tail that is visible on the left side of the PDF in panel a is much attenuated in panel b. This tail is therefore dominated by the noise. The noise affects the main properties of the PDF as well. The parameters derived from the Gaussian fit change slightly after masking the noisy pixels, in particular the width is reduced by 10\%.

\subsection{PCA}\label{sect_noise_pca}
\begin{figure}
   \resizebox{\hsize}{!}{\includegraphics[width=0.5\textwidth,trim = 3.3cm 19.0cm 20.3cm 3.8cm, clip=True]{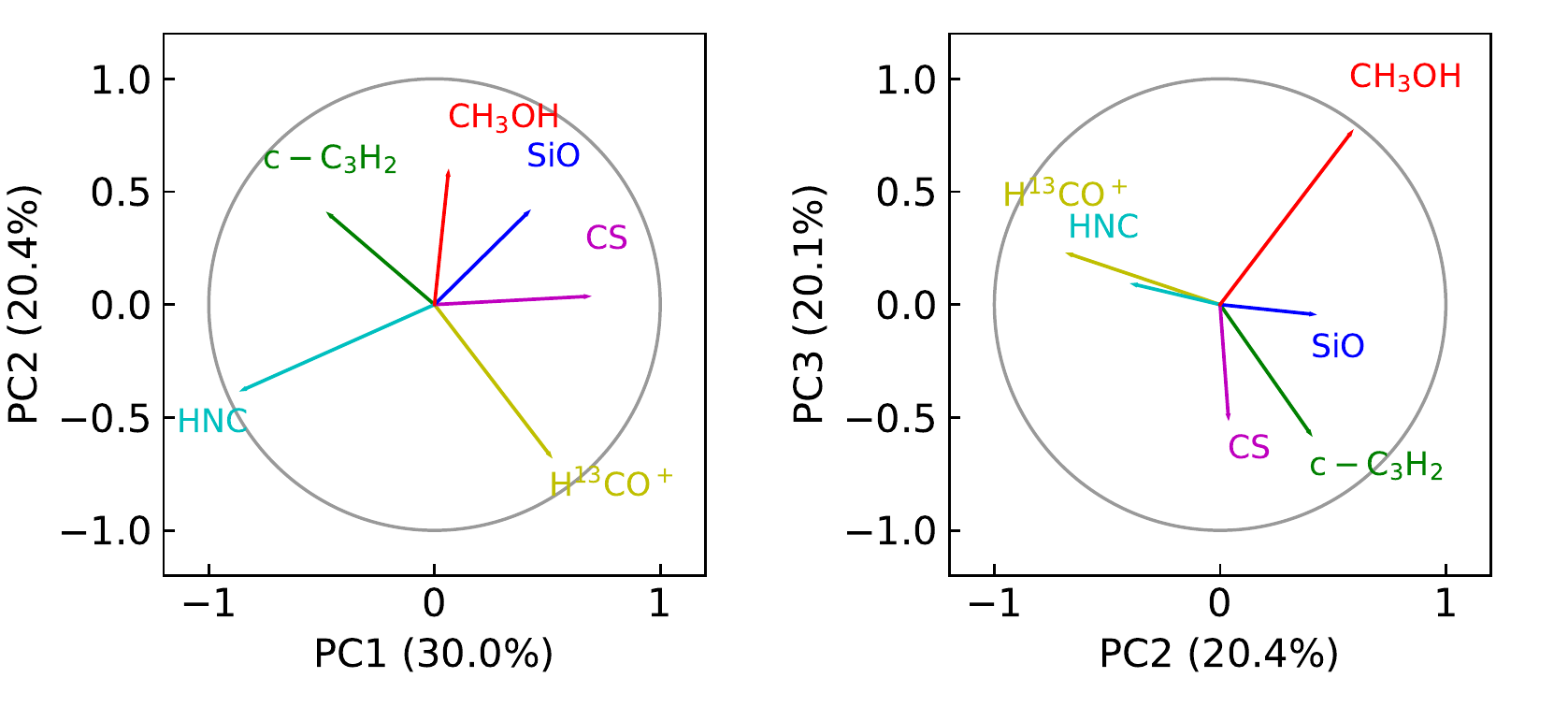}}
 \caption{Correlation wheels for a PCA performed for six molecules using channels that contain only noise. The percentages in parentheses give the contributions of the PCs to the total variance of the data.}
 \label{pca_corr_wheel_channelnoise}
\end{figure}

We investigate here the influence of the noise on the PCA. We perform the PCA for six molecules, using channels that contain only noise. The first two correlation wheels are displayed in Fig.~\ref{pca_corr_wheel_channelnoise}. The power of the first PC is only $30\%$ and the second and third have similar powers of $20.4\%$ and $20.1\%$. No clear correlation is visible in the wheels. The molecules are randomly distributed in the wheels as expected for noise. 

\section{Robustness of PCA}\label{robustness_pca}
\begin{figure}
   \resizebox{\hsize}{!}{\includegraphics[width=0.5\textwidth,trim = 3.cm 19.cm 21cm 3.6cm, clip=True]{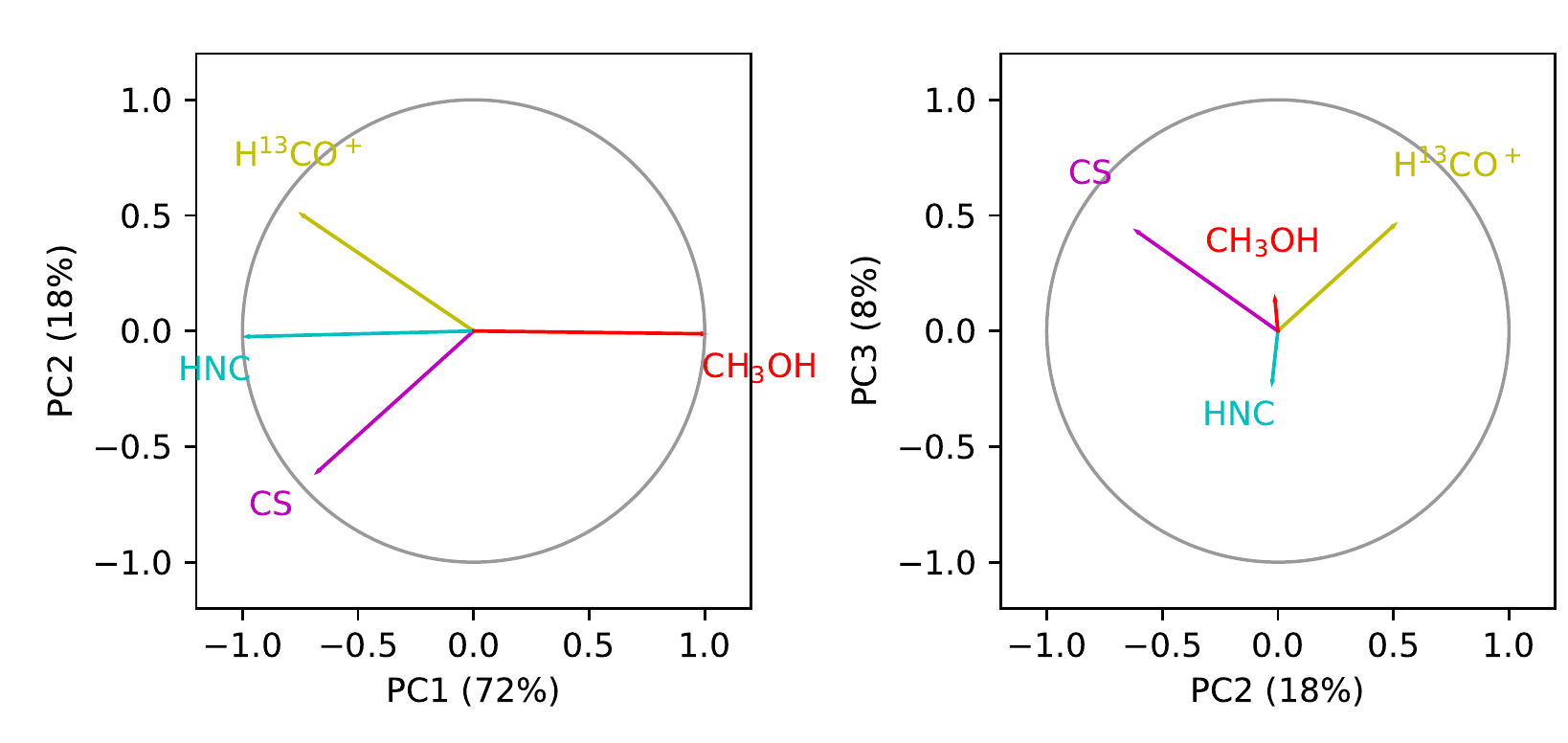}}
    \resizebox{\hsize}{!}{\includegraphics[width=0.5\textwidth,trim = 3.cm 19.cm 21cm 3.6cm, clip=True]{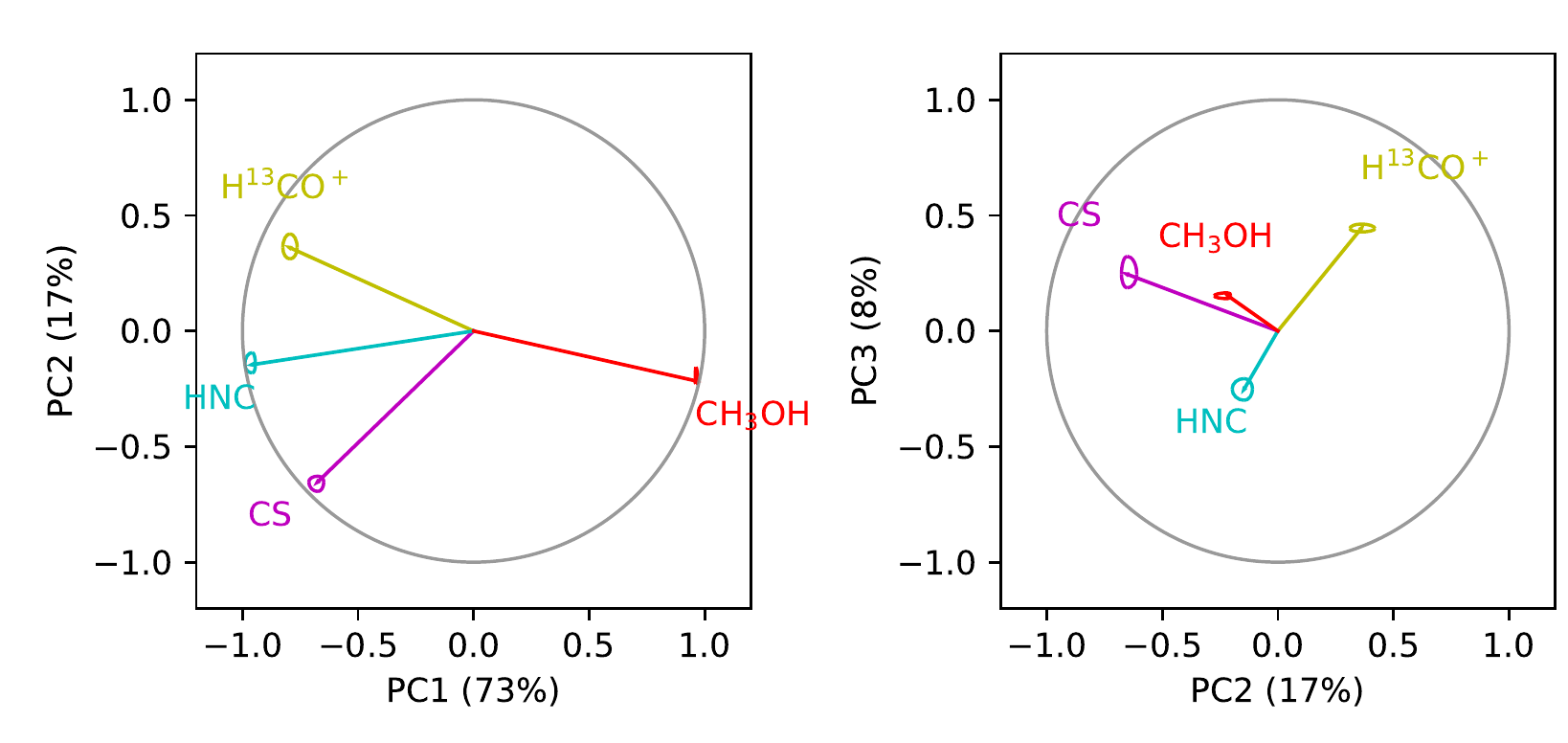}}
     \resizebox{\hsize}{!}{\includegraphics[width=0.5\textwidth,trim = 3.cm 19.cm 21cm 3.6cm, clip=True]{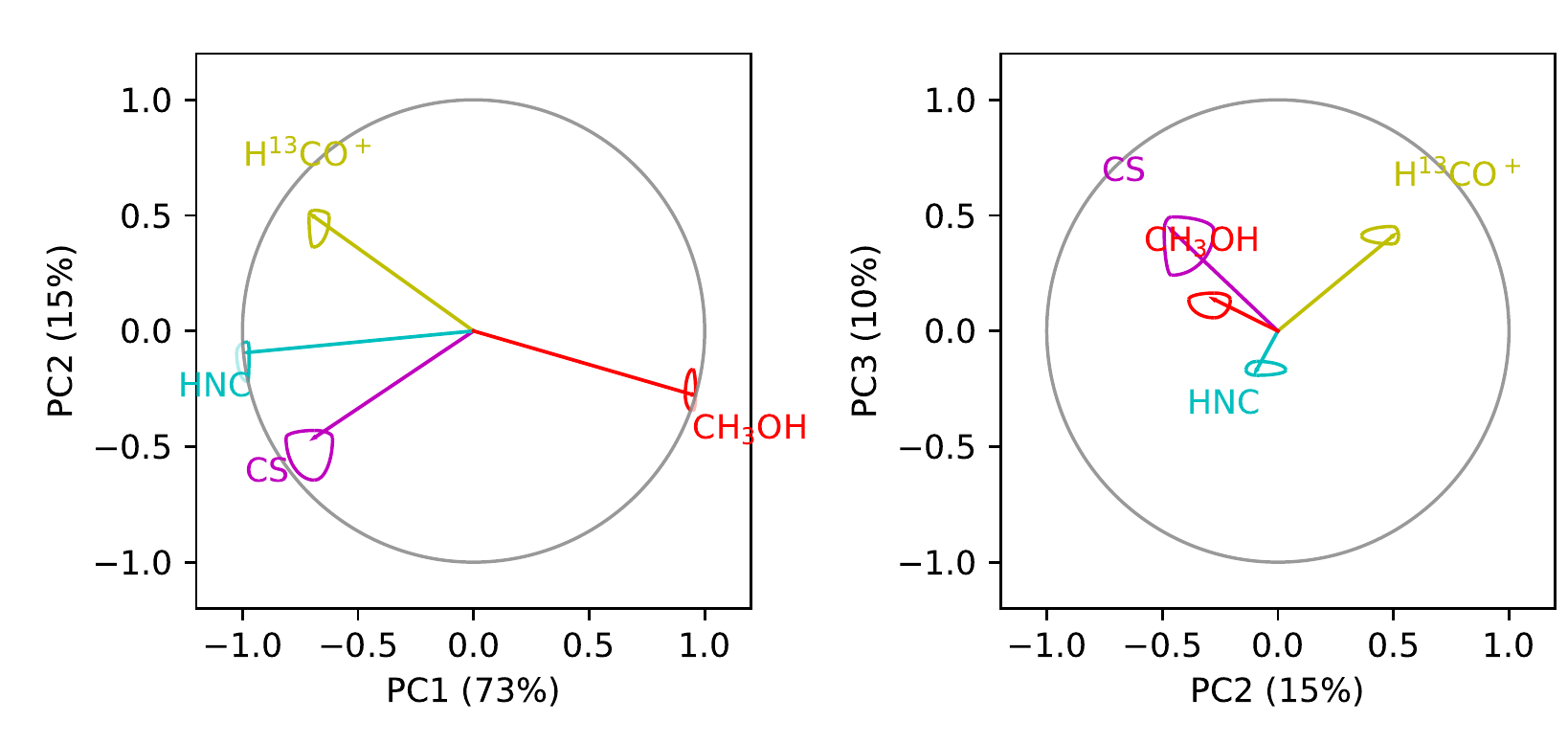}}
 \caption{Correlation wheels for the velocity component at $\varv_\mathrm{LSR}= 24.7$\,km\,s$^{-1}$. The percentages in parentheses give the contributions of the PCs to the total variance of the data. The top, middle, and bottom panels represent a grid size of 1, 2, and 3, respectively. The ellipses around the arrow heads in the middle and bottom panels show the uncertainties estimated by shifting the grids.}
 \label{pca_grids}
\end{figure}

We perform several tests to check the robustness of the principal component analysis applied to our data. First, we test different grid sizes. In the following, a grid size of 1, 2, or 3 means that we use every pixel of the original map, every second pixel, or every third pixel, respectively. With a grid size of 3, some information is lost, because the Nyquist-sampling theorem is not fulfilled. The correlation wheels of the velocity component at $\varv_\mathrm{LSR}= 24.7$\,km\,s$^{-1}$ are shown in Fig.~\ref{pca_grids}. The top, middle, and bottom panels correspond to a grid size of 1, 2, and 3, respectively. The percentages in parentheses give the powers of the PCs (their contributions to the total variance of the data). The ellipses around the arrow heads in the middle and bottom panels represent the uncertainties. They were estimated by shifting the grids by one or several pixels. To allow comparison of the PCs resulting from the different starting points of the grid, we realigned them. We chose the first PCA made in the grid sample to fix the signs of the PCs. We changed the signs of the PCs of the other grids to match these signs. We adopted this procedure following \citet{babamoradi2013}. The correlation wheels shown in the top and middle panels are very similar. We conclude that the PCA results do not depend much on the size of the gris and its starting position, provided the Nyquist-sampling condition is met. However, with a grid size of 3 (bottom panels), the uncertainties become larger and the PCA results start to depend significantly on the starting position of the grid.

\begin{figure}
 \resizebox{\hsize}{!}{\includegraphics[width=0.25\textwidth,trim = 3.cm 19.cm 28.7cm 3.6cm, clip=True]{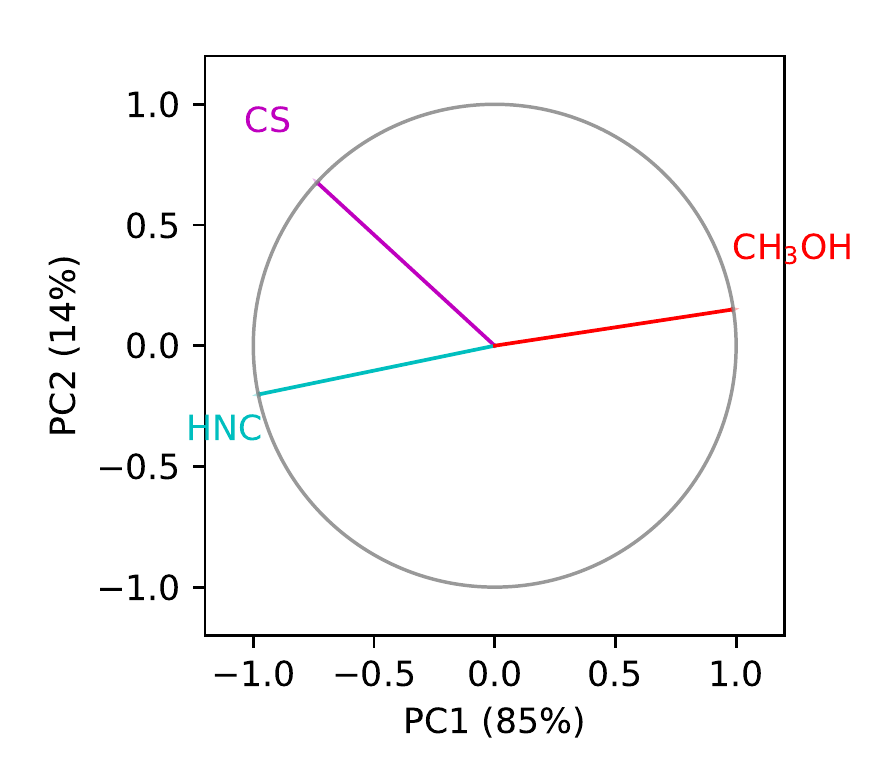}\includegraphics[width=0.25\textwidth,trim = 3.cm 19.cm 28.7cm 3.6cm, clip=True]{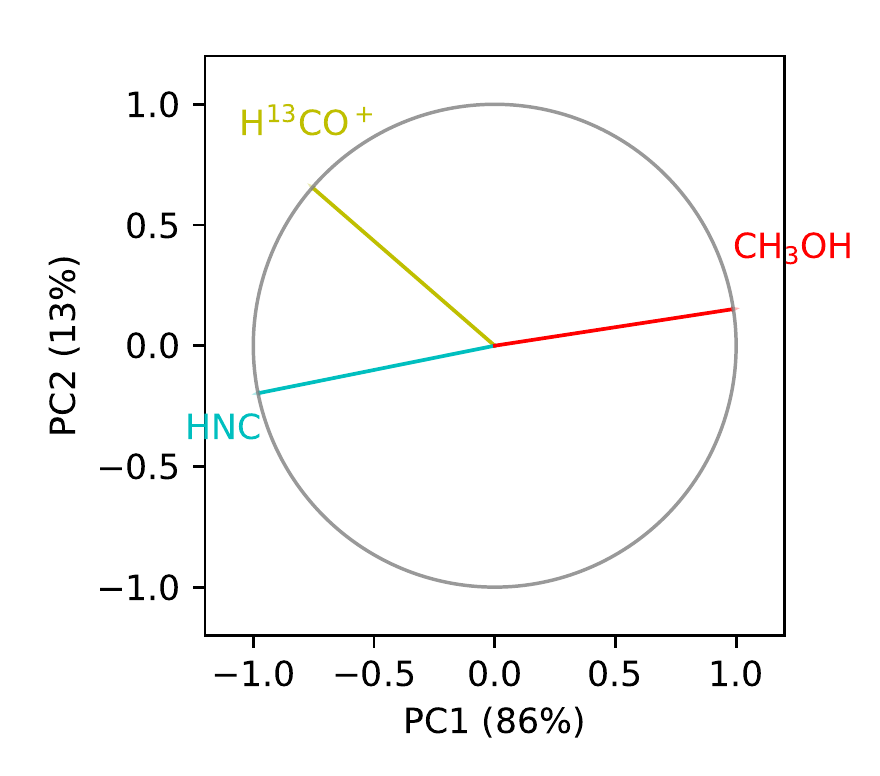}}
   \resizebox{\hsize}{!}{\includegraphics[width=0.25\textwidth,trim = 3.cm 19.cm 28.7cm 3.6cm, clip=True]{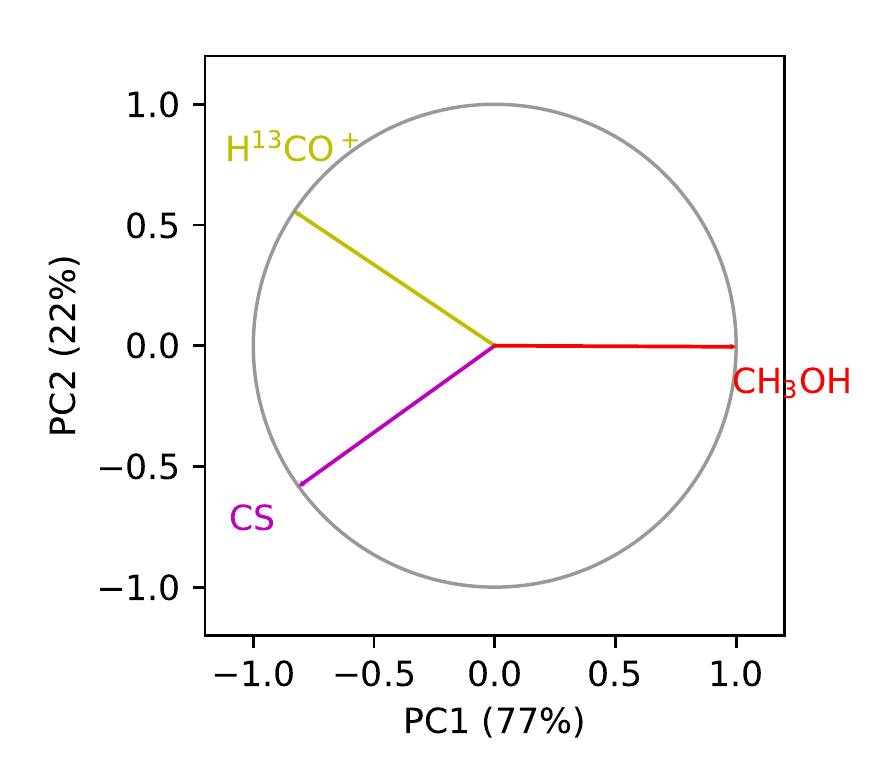}\includegraphics[width=0.25\textwidth,trim = 3.cm 19.cm 28.7cm 3.6cm, clip=True]{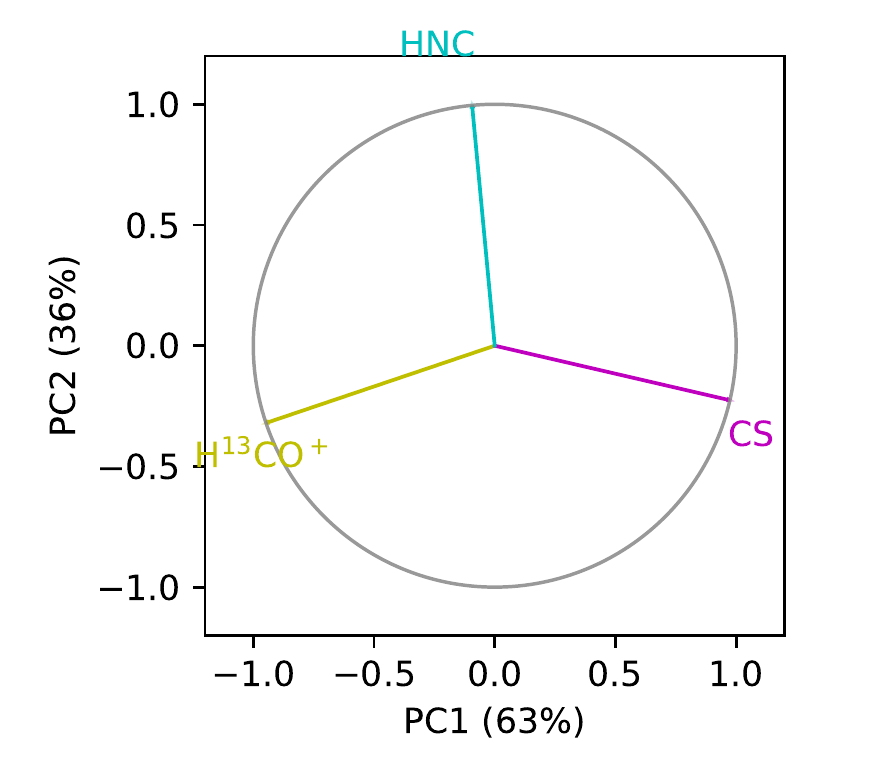}}
 \caption{First correlation wheels for the velocity component at $\varv_\mathrm{LSR}=24.7$\,km\,s$^{-1}$ for grid size 2. Each panel shows the results of the PCA after removing one of the four molecules used in the middle left panel of Fig.~\ref{pca_grids}.}
 \label{pca_mol}
\end{figure}

\begin{figure}
   \resizebox{\hsize}{!}{\includegraphics[width=0.5\textwidth,trim = 3.cm 19.cm 21cm 3.6cm, clip=True]{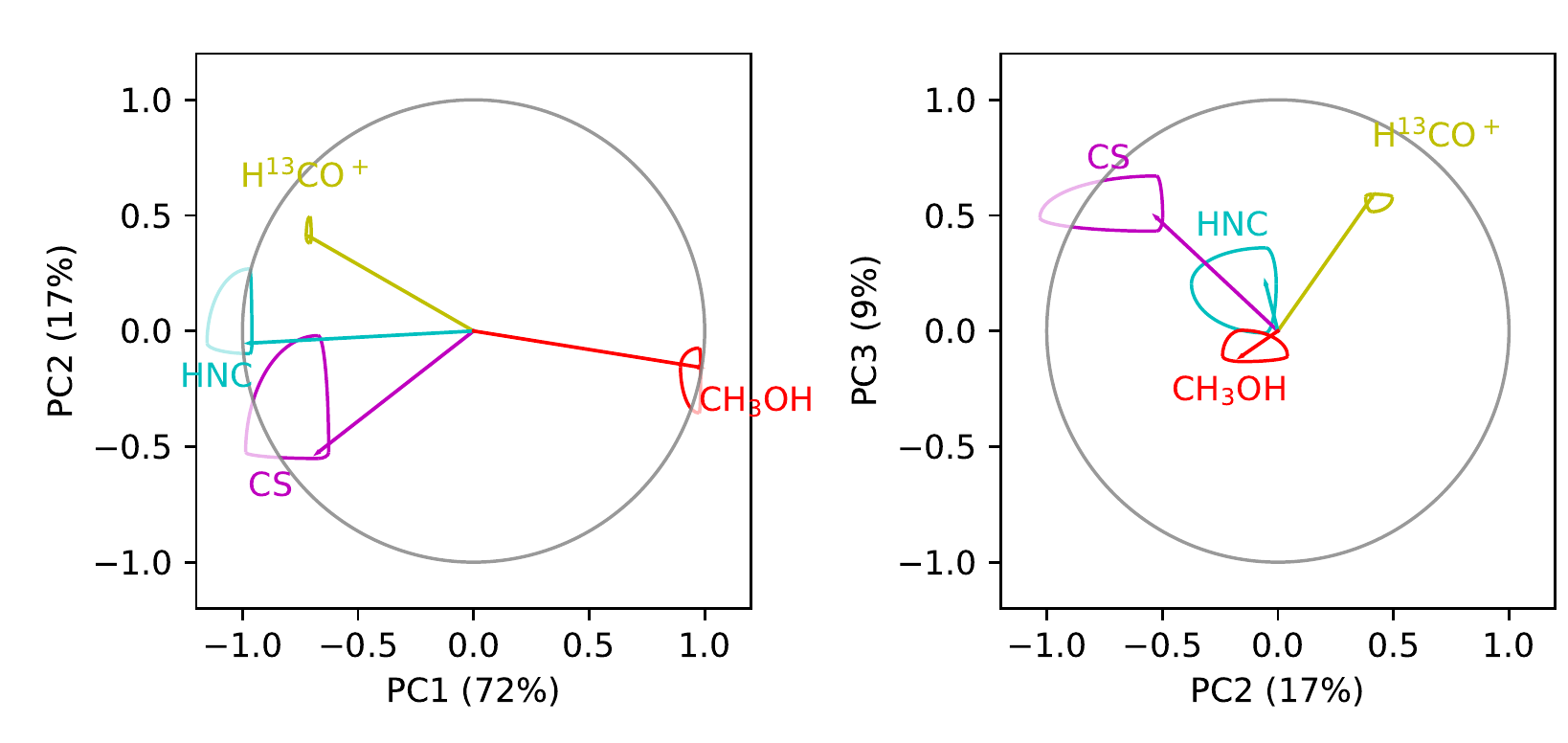}}
      \resizebox{\hsize}{!}{\includegraphics[width=0.5\textwidth,trim = 3.cm 19.cm 21cm 3.6cm, clip=True]{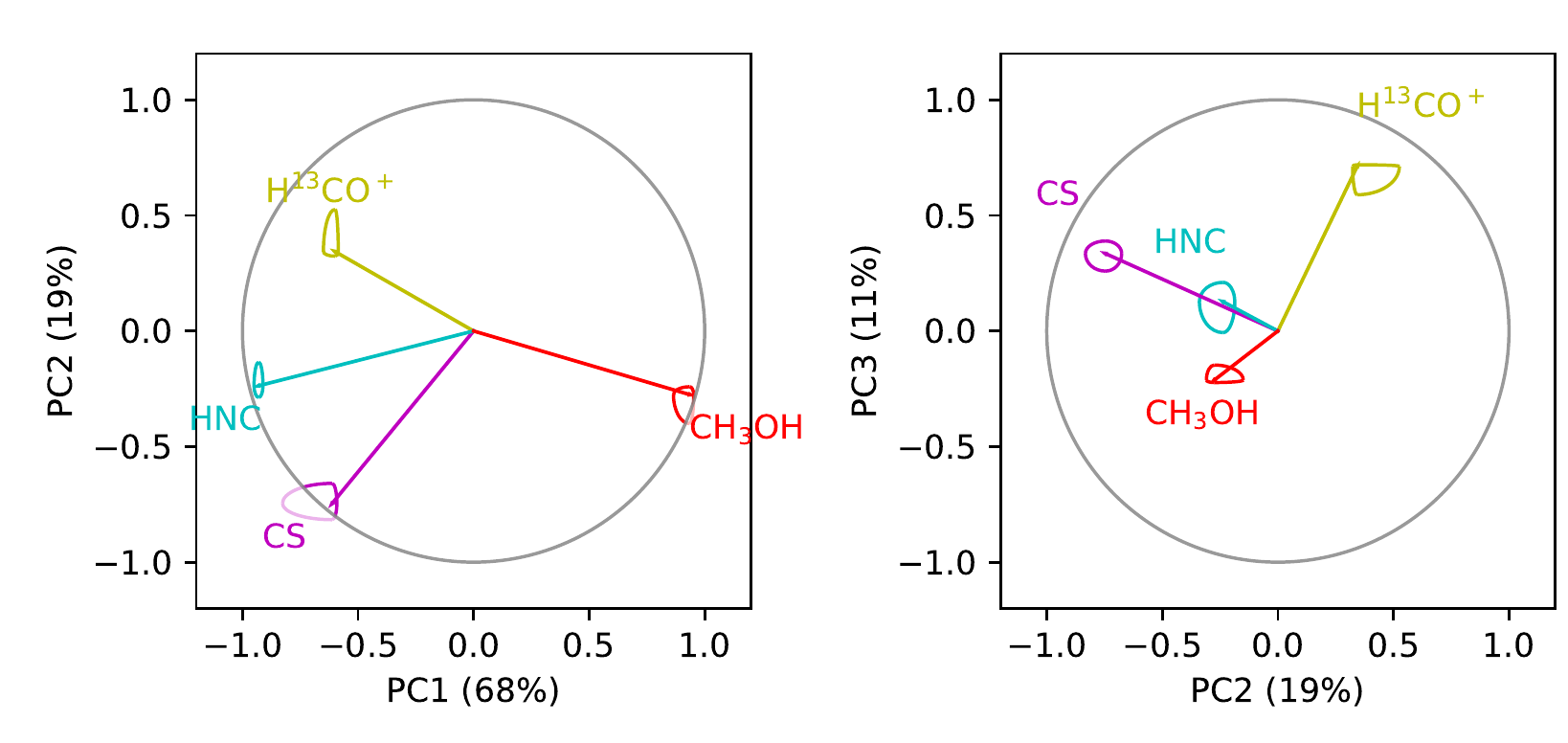}}
 \caption{Correlation wheels for PCs at $\varv_\mathrm{LSR}=$24.7\,km\,s$^{-1}$ for grid size 1 as in Fig.~\ref{pca_grids}. The first two wheels displayed in the upper panels show the results from the PCA excluding 40 pixels. The wheels in the lower panels the results from excluding 80 pixels. The ellipses around the arrow heads show the uncertainties.}
 \label{pca_cut}
\end{figure}

We investigate in Fig.~\ref{pca_mol} the impact of removing one molecule on the PCA results. We show the first correlation wheels at $\varv_\mathrm{LSR}=$24.7\,km\,s$^{-1}$ for grid size 2 in four cases, each one corresponding to one of the four molecules being removed. As long as methanol is used, the PC coefficients are similar to those in Fig.~\ref{pca_grids}. When methanol is excluded, the PC coefficients change significantly and the power of the first PC is reduced. This means that methanol has a dominant structure for this velocity component. 

We investigate in Fig.~\ref{pca_cut} the impact of excluding some pixels on the PCA results. We show the first and second correlation wheels for the velocity component at $\varv_\mathrm{LSR}=$24.7\,km\,s$^{-1}$ after excluding 40 pixels (top panels) and 80 pixels (bottom panels). These pixels were chosen randomly and we computed the PCA for 1000 realisations. The ellipses in Fig.~\ref{pca_cut} correspond to the dispersion of these 1000 realisations. The results are similar to those in Fig.~\ref{pca_grids}, except for the molecules with small ($<0.5$) PC coefficients.

Altogether the PCA seems to be robust to all these tests for the velocity component at 24.7 km~s$^{-1}$, for which the first principal component contains most of the variance of the data. 

%
%
%
%
%
\section{Spectra}
\begin{figure*}
\centering
\includegraphics[width=17cm, trim = 2.cm 1.9cm 1.7cm 3.9cm, clip=True]{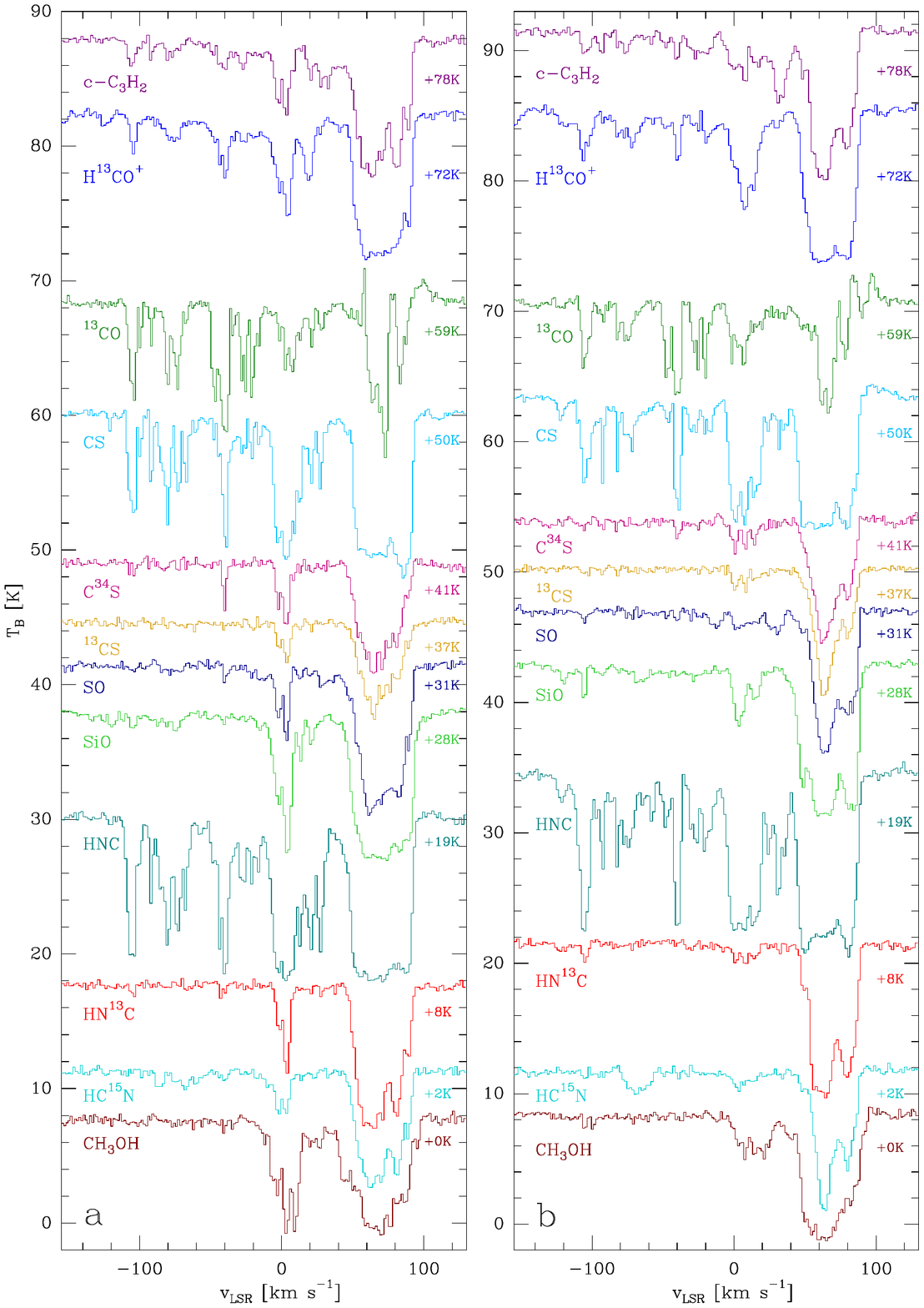}
\caption{Spectra of the 12 investigated molecules towards K4 (a) and the peak of the shell of K6 (b). The spectra are shifted vertically by the amount indicated on the right.}
\label{spectra}
\end{figure*}

The EMoCA spectra of the 12 investigated molecules towards K4 and the peak of the shell of K6 are shown in Fig.~\ref{spectra}.

\section{Opacity maps}
\begin{figure*}
\centering
\includegraphics[width=17cm, trim = 1.8cm 2.4cm 0.4cm 3.cm, clip=True]{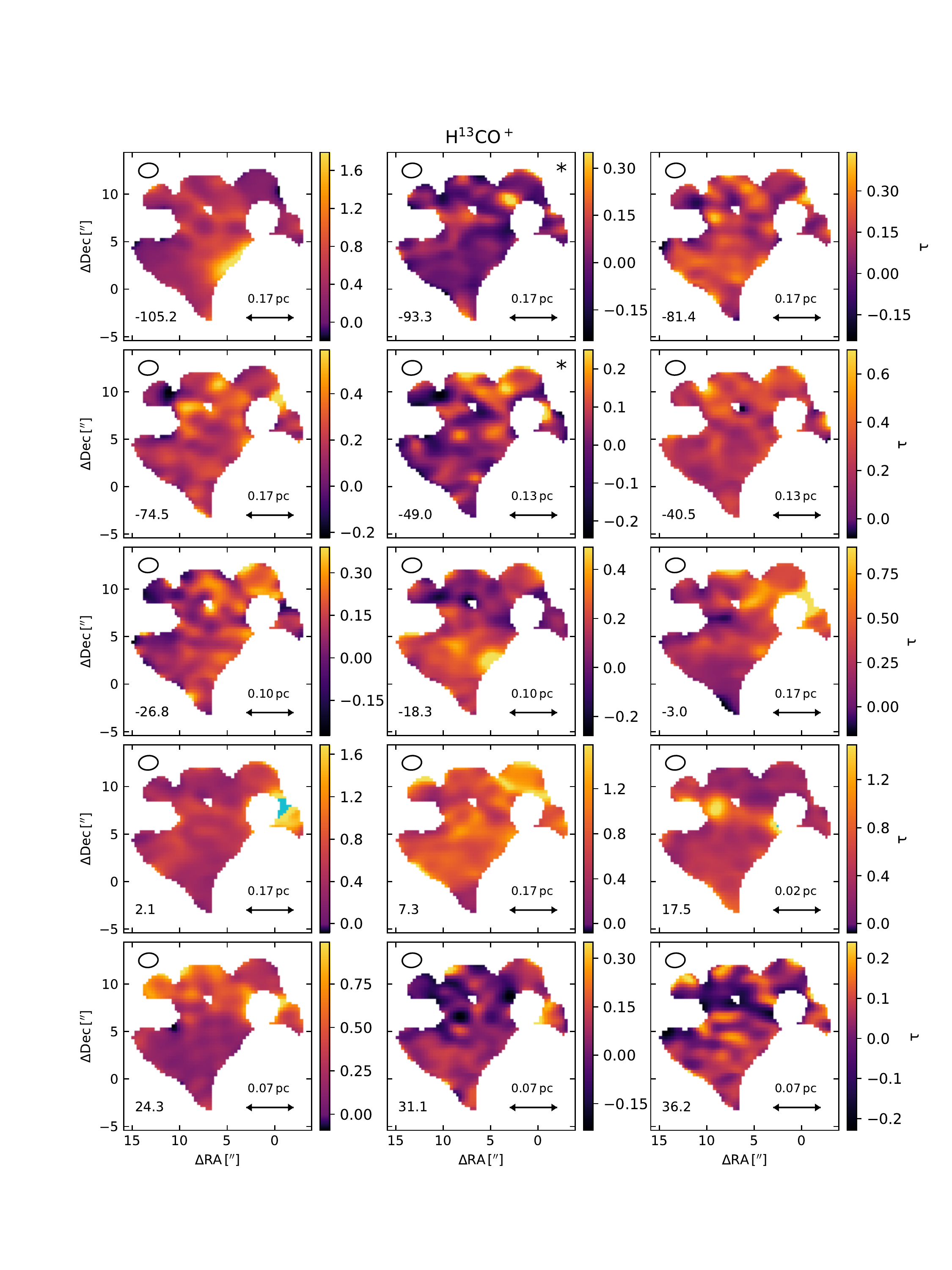}
\caption{Same as Fig.~\ref{opacity_c-c3h2}, but for H$^{13}$CO$^+$. In this figure and the following ones, blue pixels represent pixels that were masked because the absorption is too optically thick.}
\label{opacity_h13cop}
\end{figure*}

\begin{figure*}
\centering
\includegraphics[width=17cm, trim = 1.8cm 2.4cm 0.4cm 3.cm, clip=True]{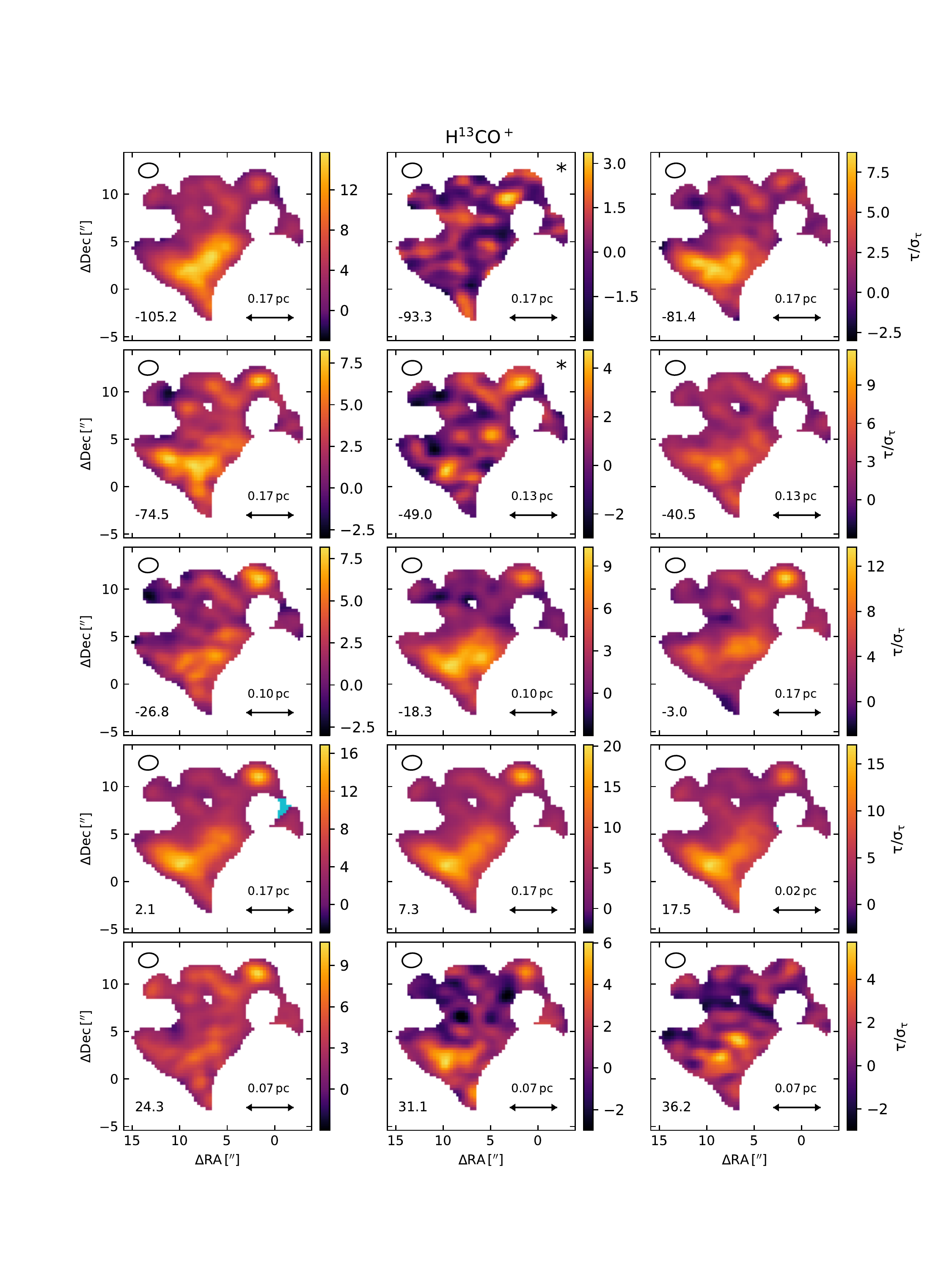}
\caption{Same as Fig.~\ref{snr_opacity_c-c3h2}, but for H$^{13}$CO$^+$.}
\label{snr_opacity_h13cop}
\end{figure*}

\begin{figure*}
\centering
\includegraphics[width=17cm, trim = 1.8cm 2.4cm 0.4cm 3.cm, clip=True]{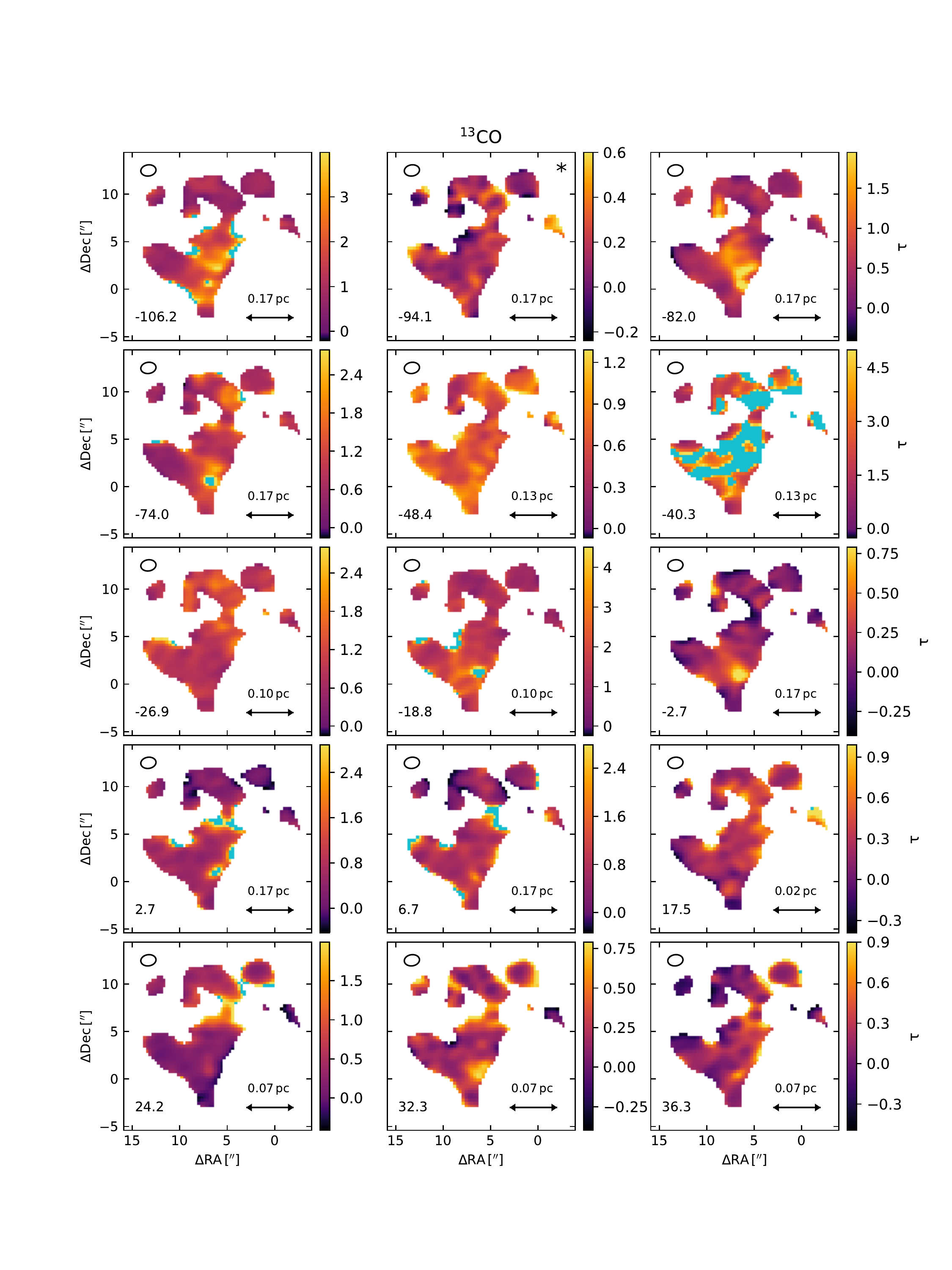}
\caption{Same as Fig.~\ref{opacity_c-c3h2}, but for $^{13}$CO.}
\label{opacity_13co}
\end{figure*}

\begin{figure*}
\centering
\includegraphics[width=17cm, trim = 1.8cm 2.4cm 0.4cm 3.cm, clip=True]{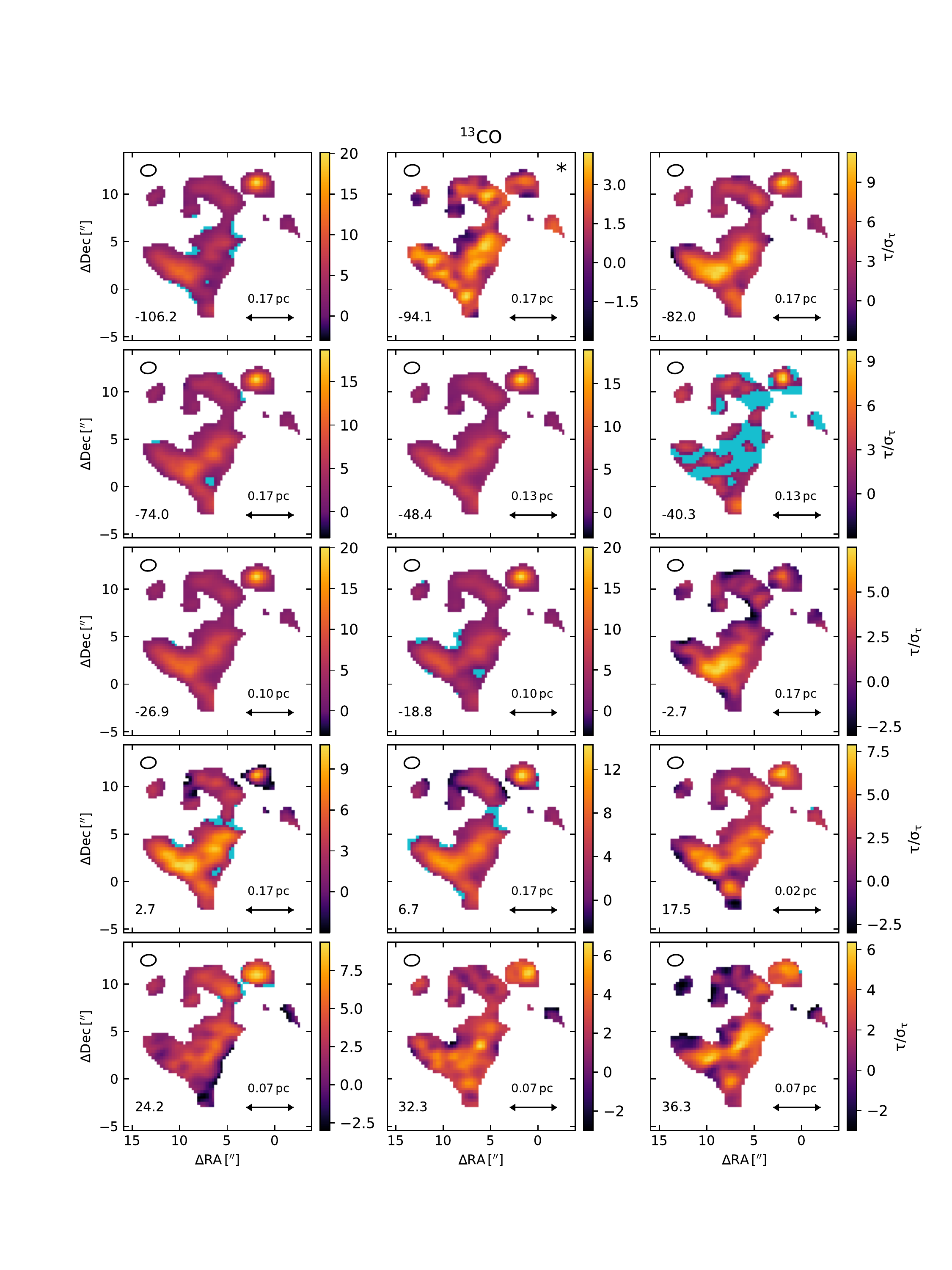}
\caption{Same as Fig.~\ref{snr_opacity_c-c3h2}, but for $^{13}$CO.}
\label{snr_opacity_13co}
\end{figure*}

\begin{figure*}
\centering
\includegraphics[width=17cm, trim = 1.8cm 2.4cm 0.4cm 3.1cm, clip=True]{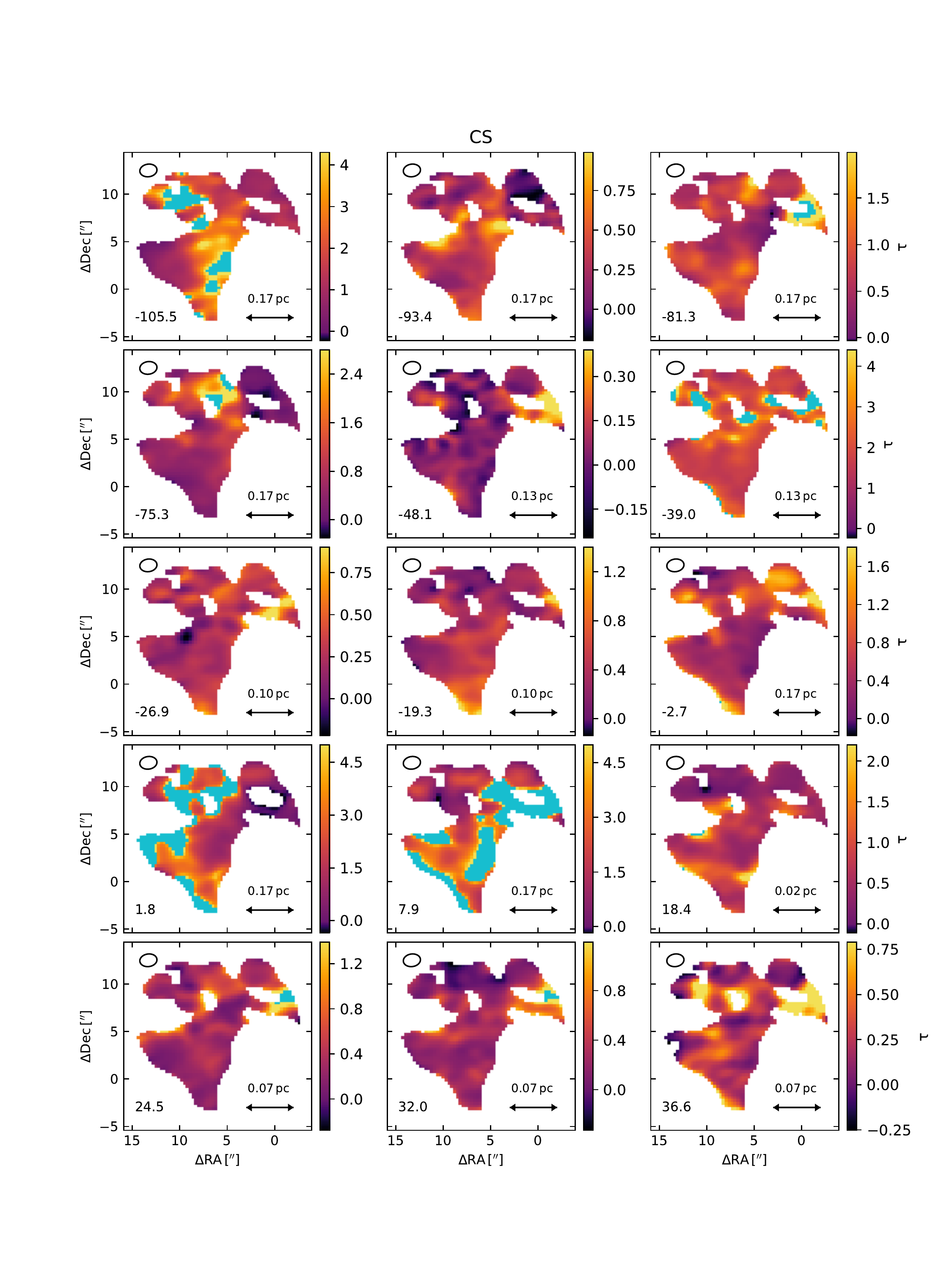}
\caption{Same as Fig.~\ref{opacity_c-c3h2}, but for CS.}
\label{opacity_CS}
\end{figure*}

\begin{figure*}
\centering
\includegraphics[width=17cm, trim = 1.8cm 2.4cm 0.4cm 3.1cm, clip=True]{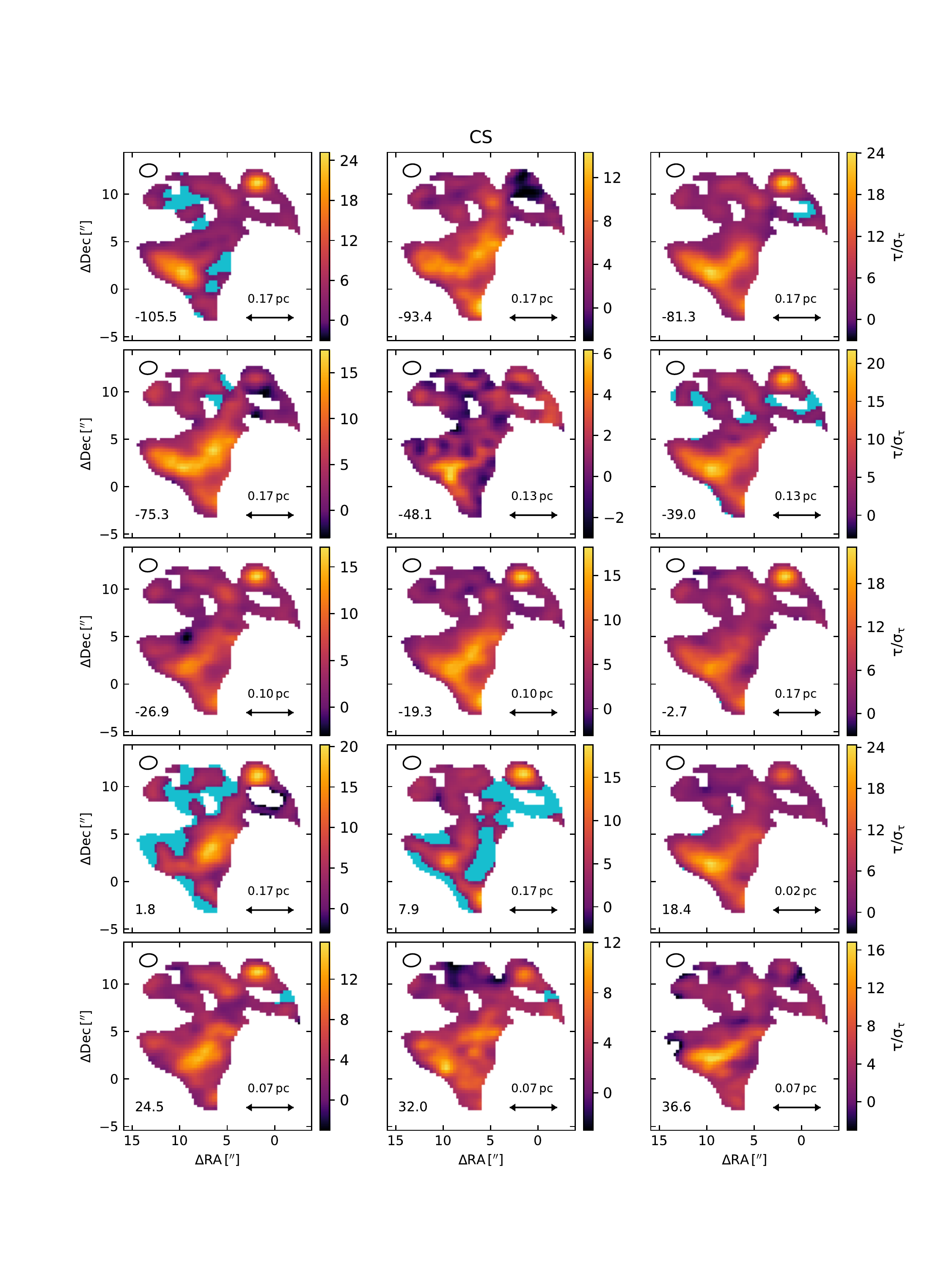}
\caption{Same as Fig.~\ref{snr_opacity_c-c3h2}, but for CS.}
\label{snr_opacity_CS}
\end{figure*}

\begin{figure*}
\centering
\includegraphics[width=17cm, trim = 1.8cm 2.4cm 0.4cm 3.cm, clip=True]{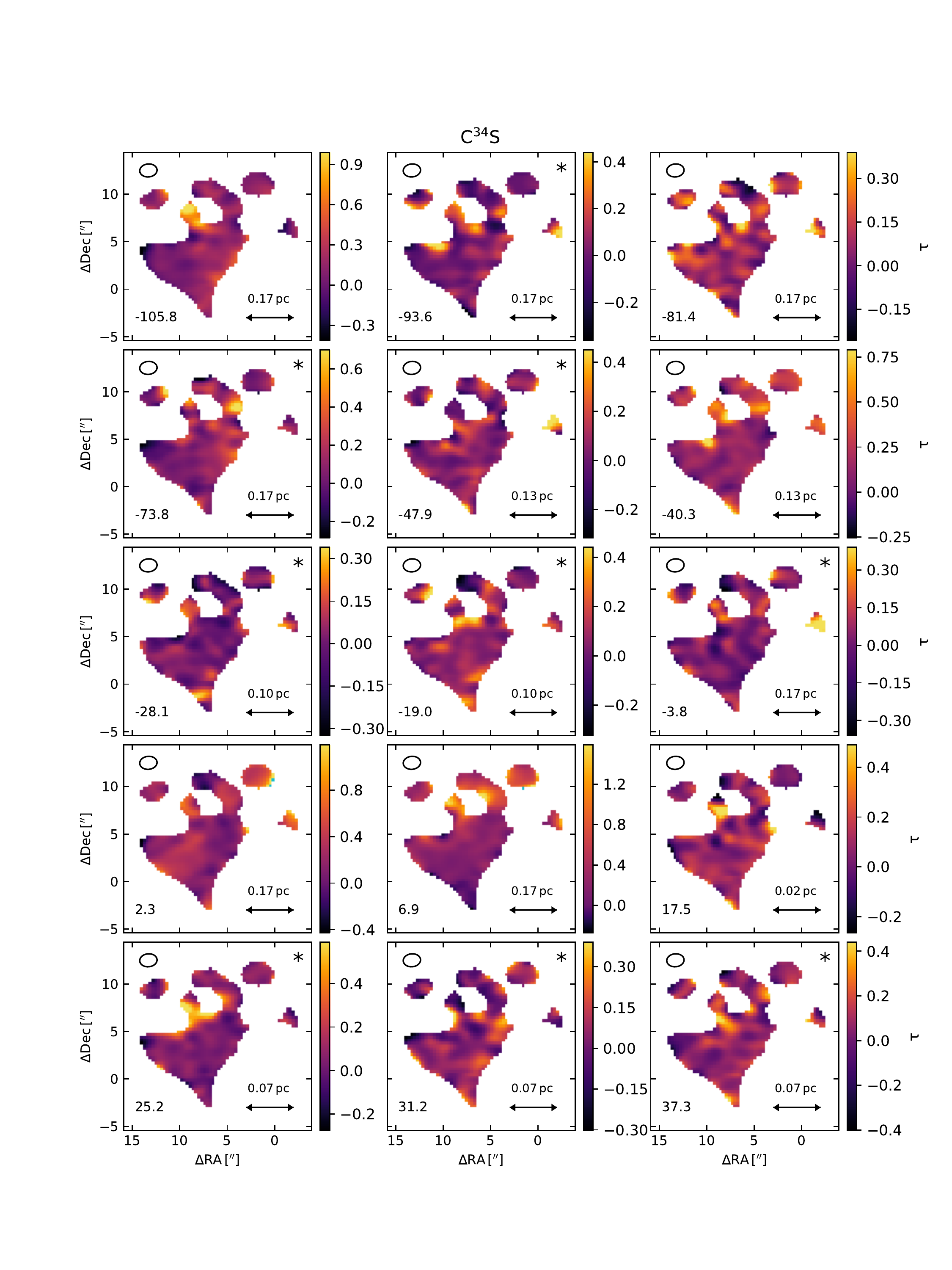}
\caption{Same as Fig.~\ref{opacity_c-c3h2}, but for C$^{34}$S.}
\label{opacity_c34s}
\end{figure*}

\begin{figure*}
\centering
\includegraphics[width=17cm, trim = 1.8cm 2.4cm 0.4cm 3.cm, clip=True]{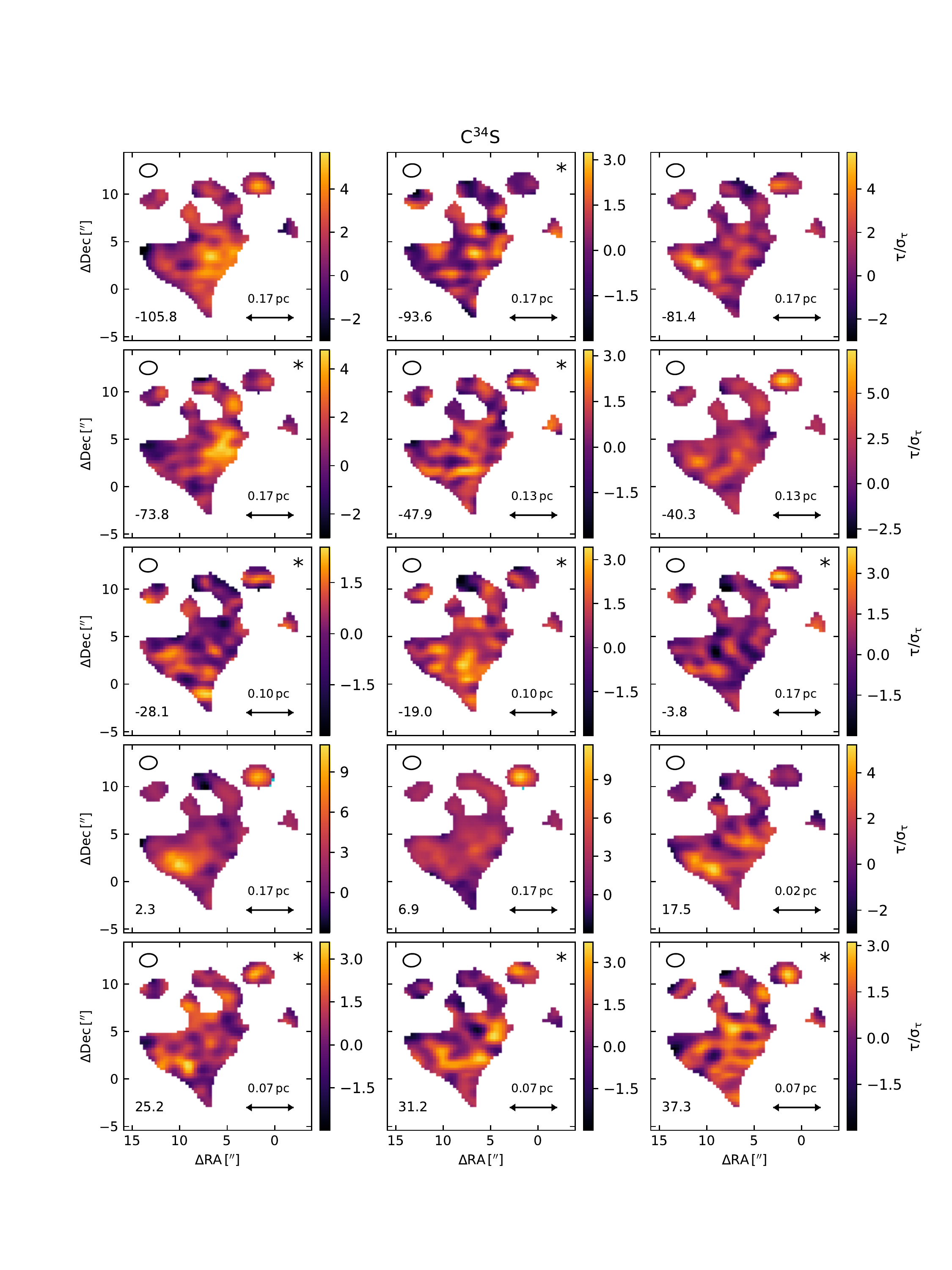}
\caption{Same as Fig.~\ref{snr_opacity_c-c3h2}, but for C$^{34}$S.}
\label{snr_opacity_c34s}
\end{figure*}

\begin{figure*}
\centering
\includegraphics[width=17cm, trim = 1.8cm 2.4cm 0.4cm 3.cm, clip=True]{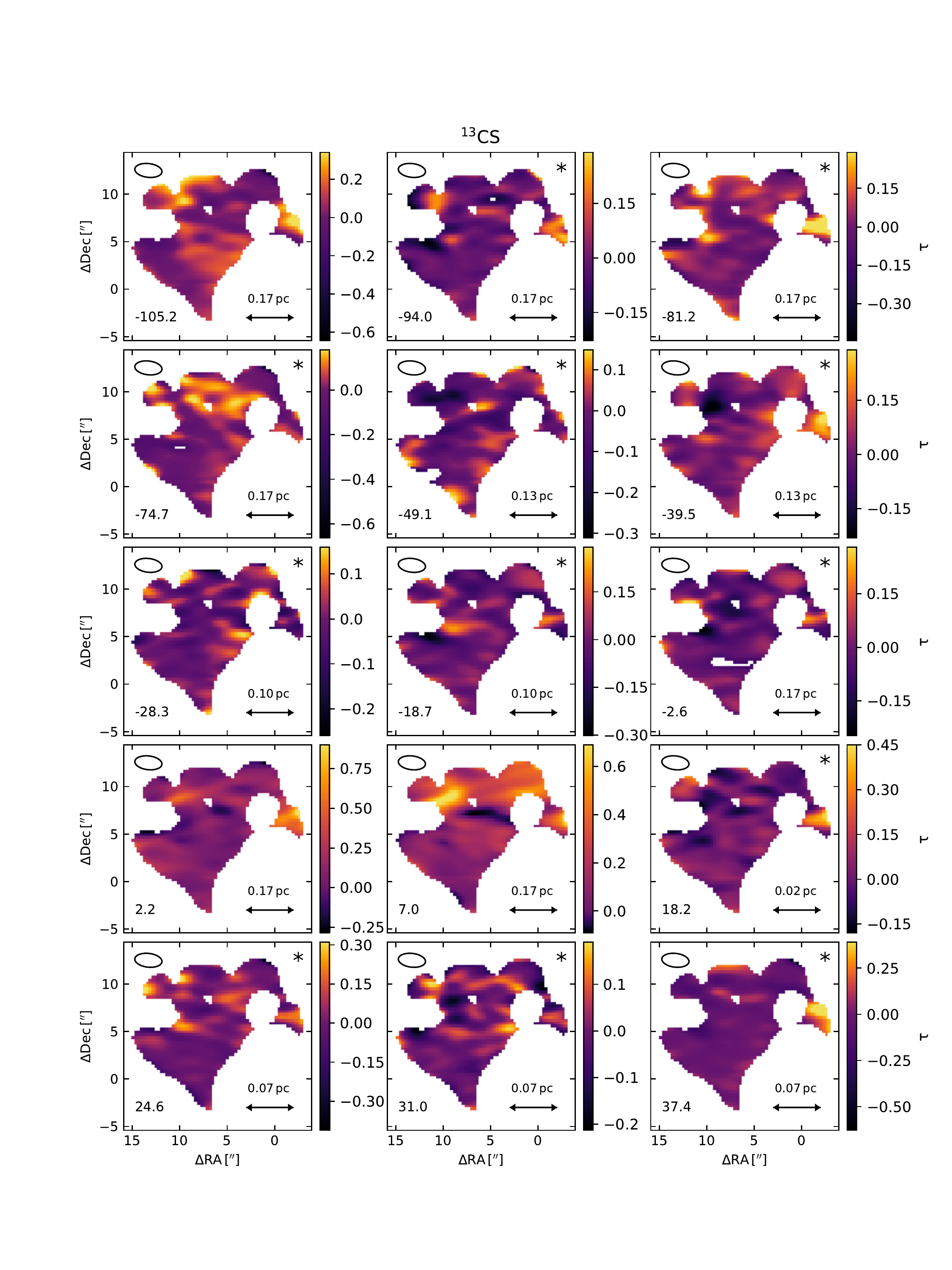}
\caption{Same as Fig.~\ref{opacity_c-c3h2}, but for $^{13}$CS.}
\label{opacity_13cs}
\end{figure*}

\begin{figure*}
\centering
\includegraphics[width=17cm, trim = 1.8cm 2.4cm 0.4cm 3.cm, clip=True]{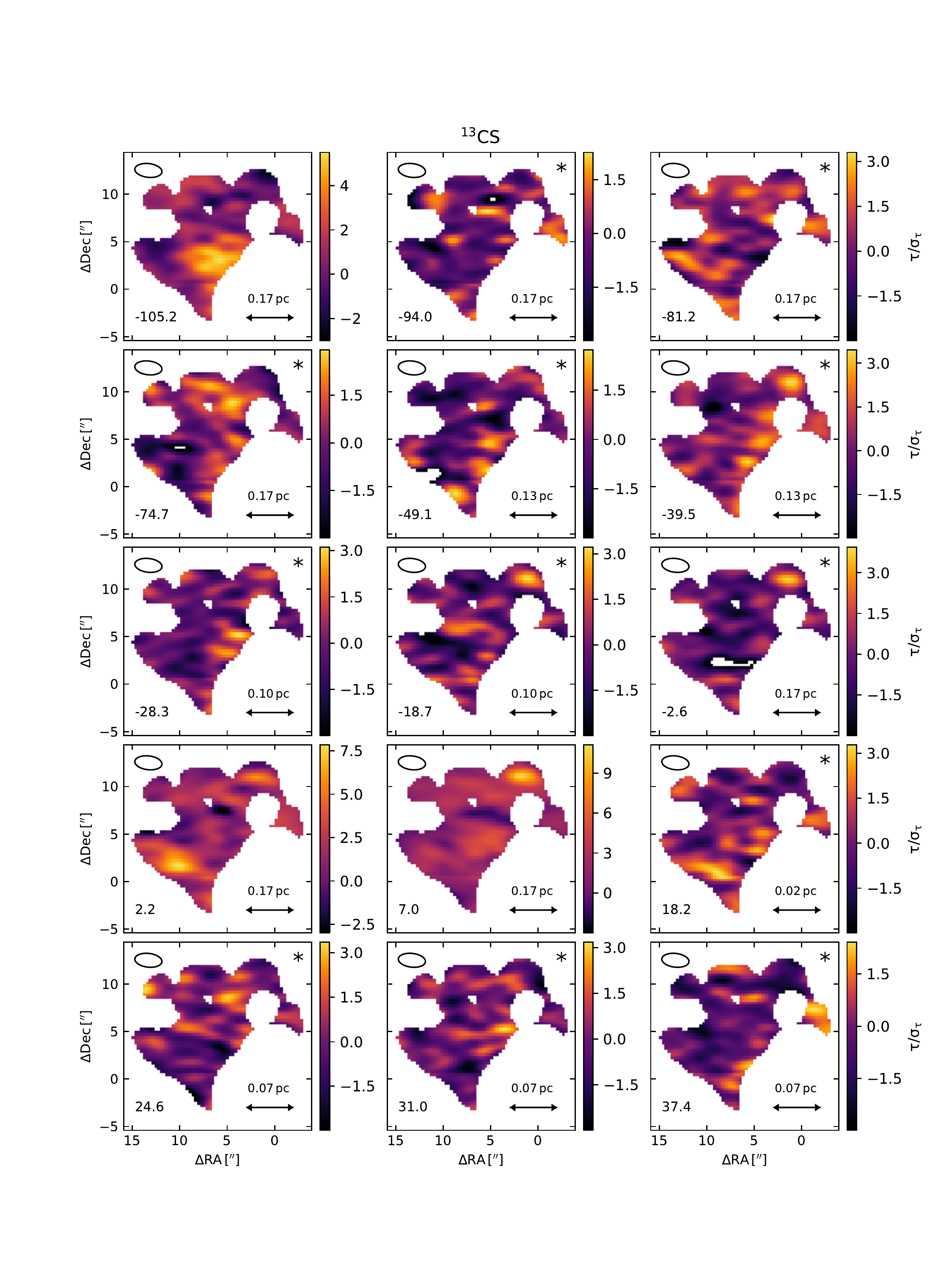}
\caption{Same as Fig.~\ref{snr_opacity_c-c3h2}, but for $^{13}$CS.}
\label{snr_opacity_13cs}
\end{figure*}

\begin{figure*}
\centering
\includegraphics[width=17cm, trim = 1.8cm 2.4cm 0.4cm 3.1cm, clip=True]{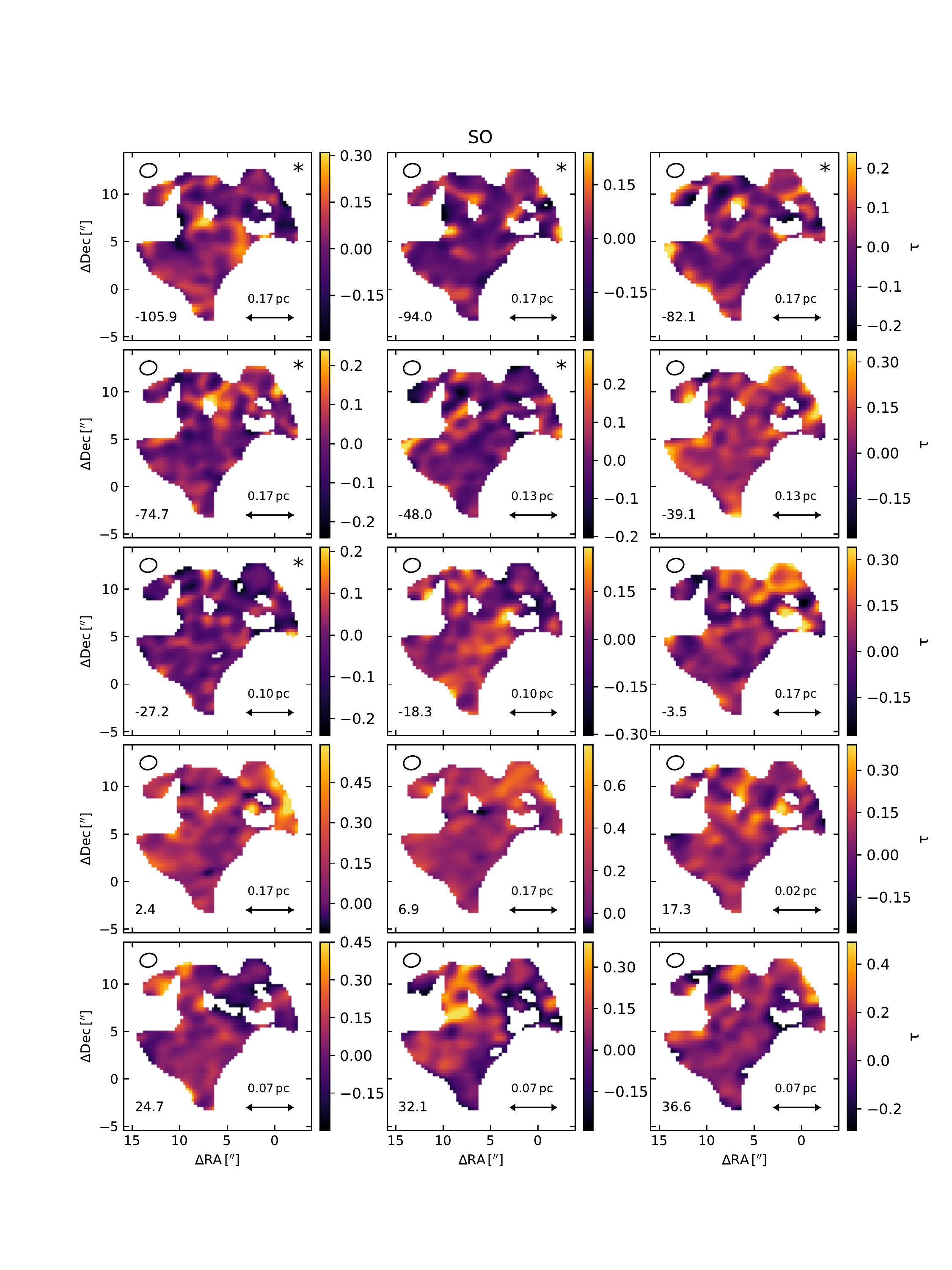}
\caption{Same as Fig.~\ref{opacity_c-c3h2}, but for SO.}
\label{opacity_so}
\end{figure*}

\begin{figure*}
\centering
\includegraphics[width=17cm, trim = 1.8cm 2.4cm 0.4cm 3.1cm, clip=True]{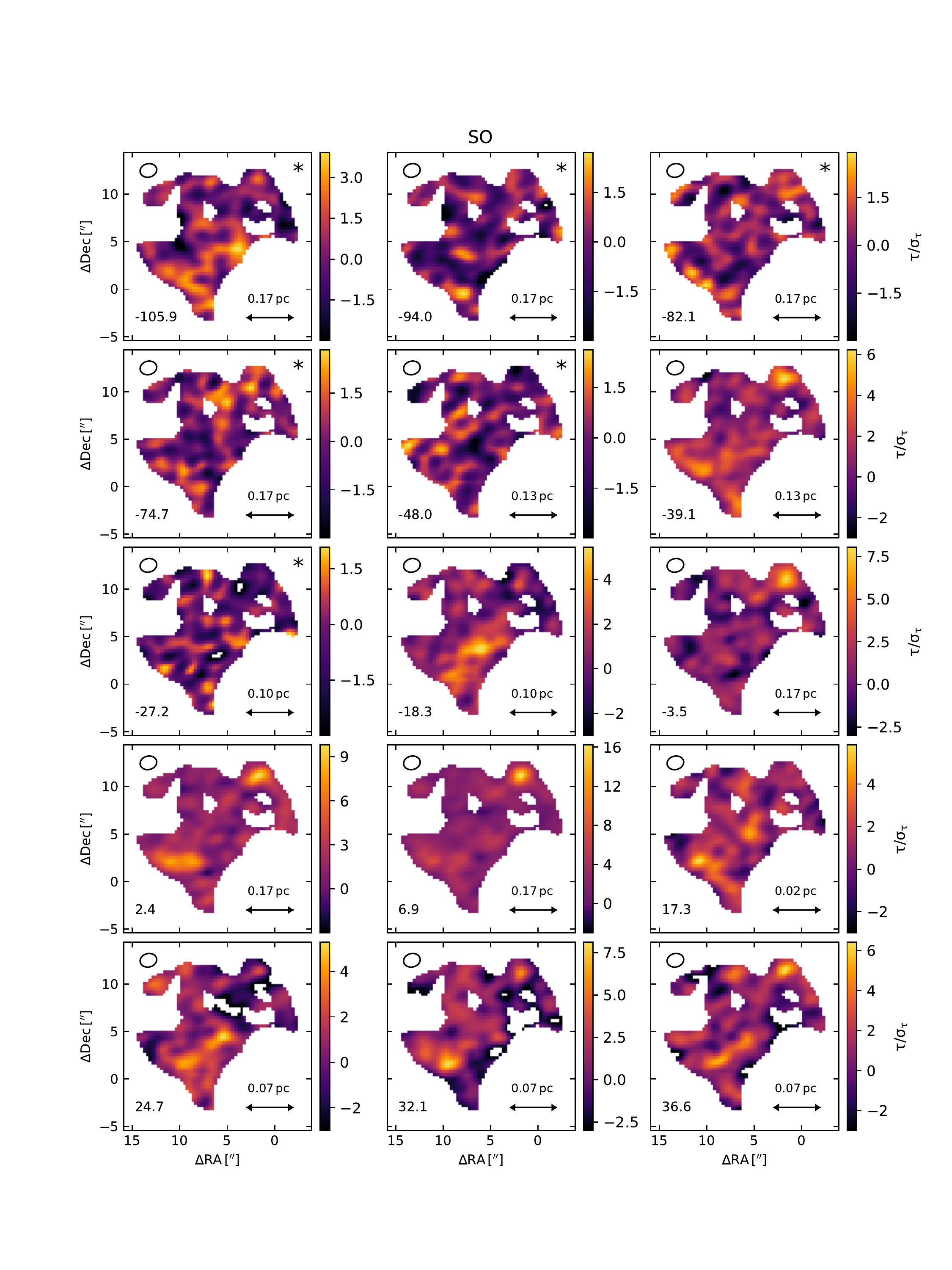}
\caption{Same as Fig.~\ref{snr_opacity_c-c3h2}, but for SO.}
\label{snr_opacity_so}
\end{figure*}

\begin{figure*}
\centering
\includegraphics[width=17cm, trim = 1.8cm 2.4cm 0.4cm 3.1cm, clip=True]{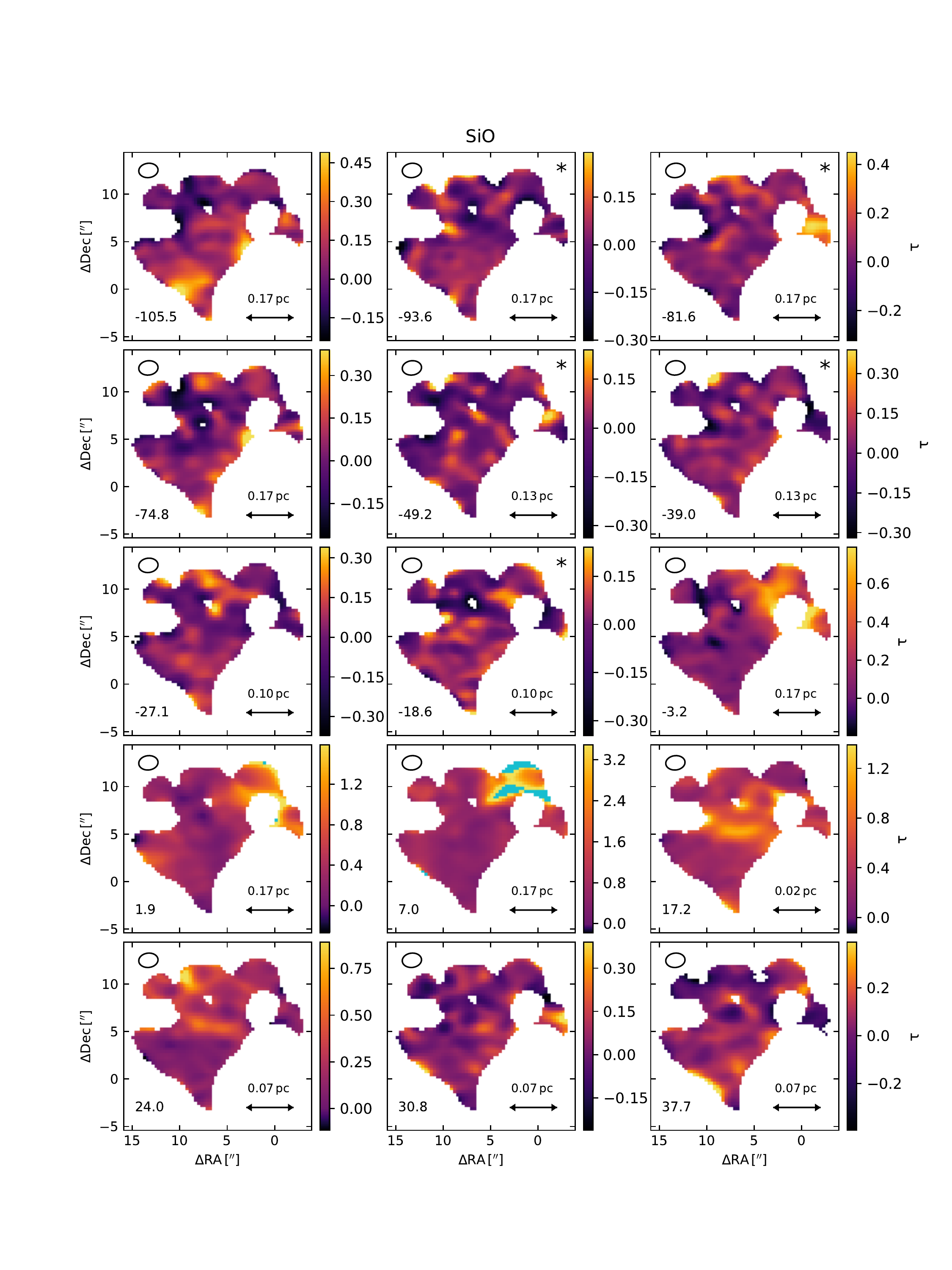}
\caption{Same as Fig.~\ref{opacity_c-c3h2}, but for SiO.}
\label{opacity_sio}
\end{figure*}

\begin{figure*}
\centering
\includegraphics[width=17cm, trim = 1.8cm 2.4cm 0.4cm 3.1cm, clip=True]{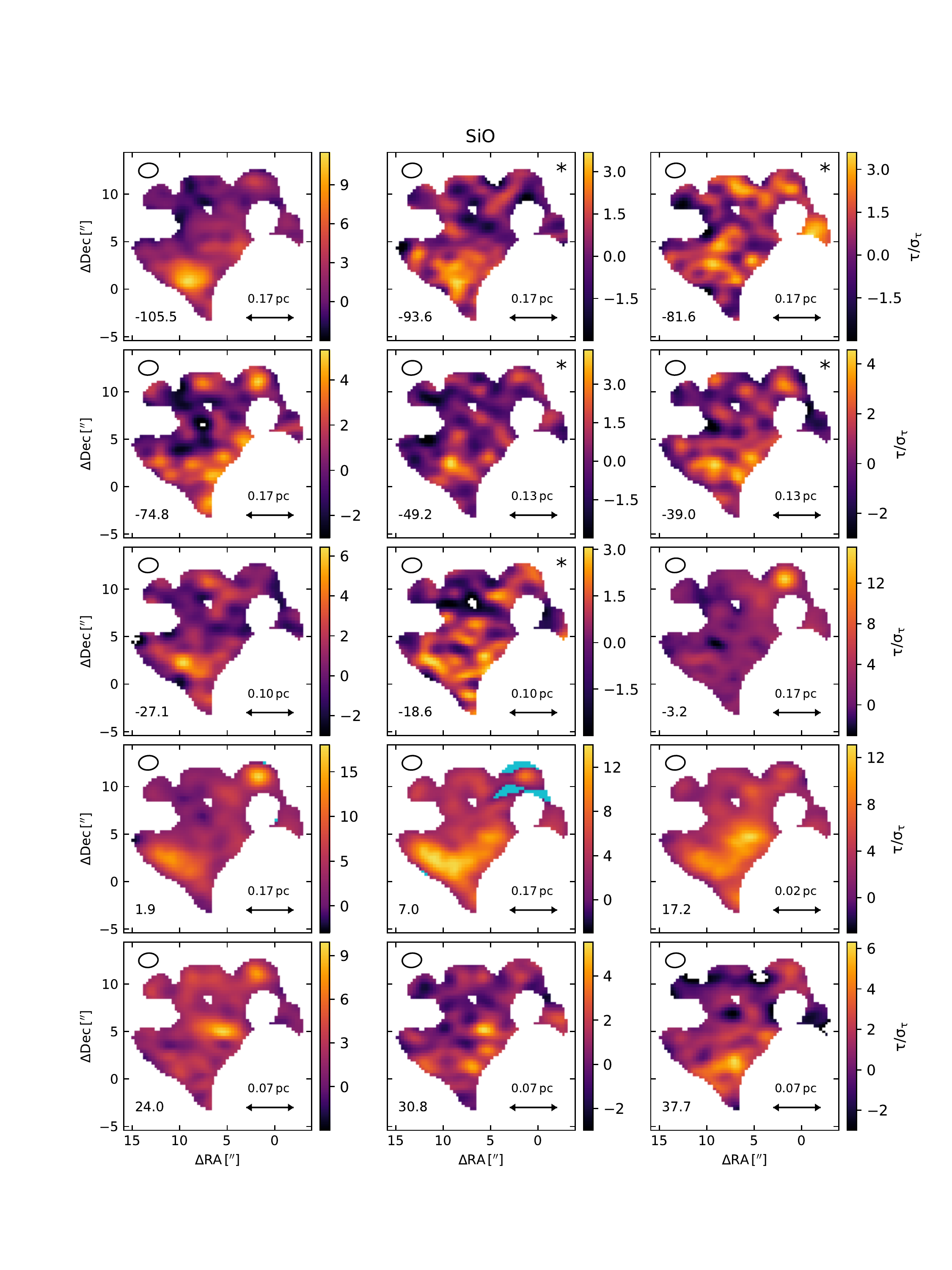}
\caption{Same as Fig.~\ref{snr_opacity_c-c3h2}, but for SiO.}
\label{snr_opacity_sio}
\end{figure*}

\begin{figure*}
\centering
\includegraphics[width=17cm, trim = 1.8cm 2.4cm 0.4cm 3.1cm, clip=True]{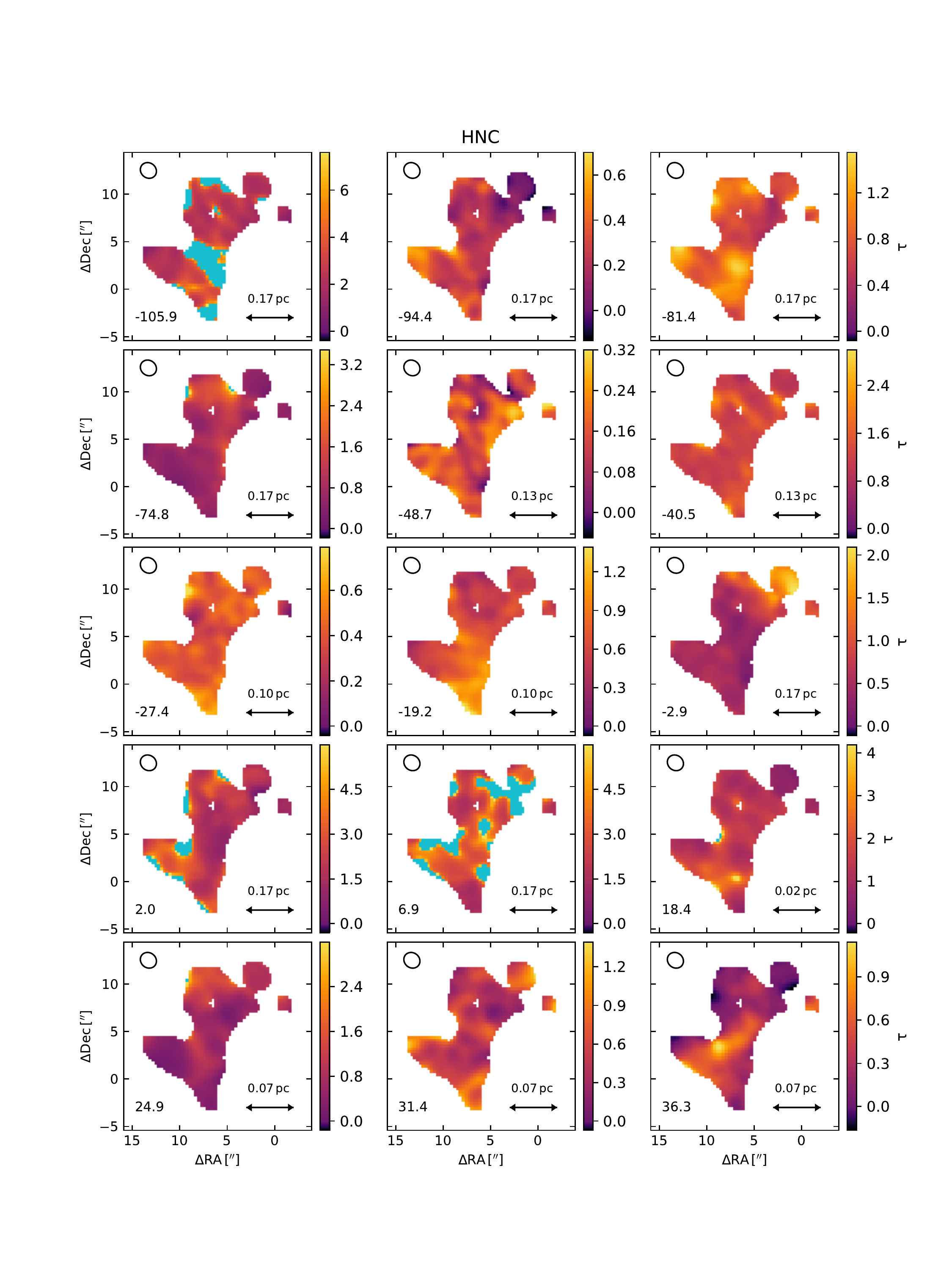}
\caption{Same as Fig.~\ref{opacity_c-c3h2}, but for HNC.}
\label{opacity_hnc}
\end{figure*}

\begin{figure*}
\centering
\includegraphics[width=17cm, trim = 1.8cm 2.4cm 0.4cm 3.1cm, clip=True]{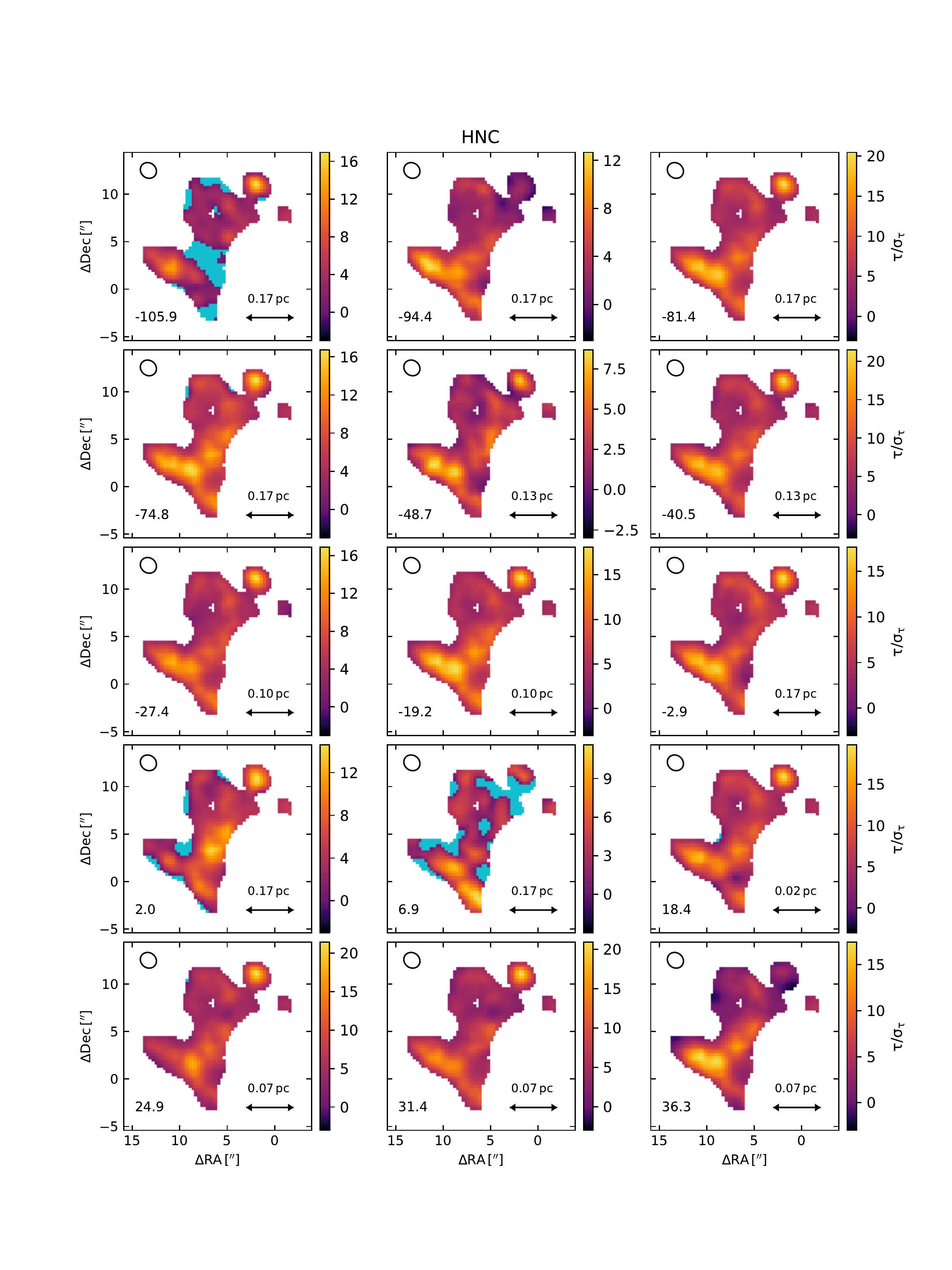}
\caption{Same as Fig.~\ref{snr_opacity_c-c3h2}, but for HNC.}
\label{snr_opacity_hnc}
\end{figure*}

\begin{figure*}
\centering
\includegraphics[width=17cm, trim = 1.8cm 2.4cm 0.4cm 3.cm, clip=True]{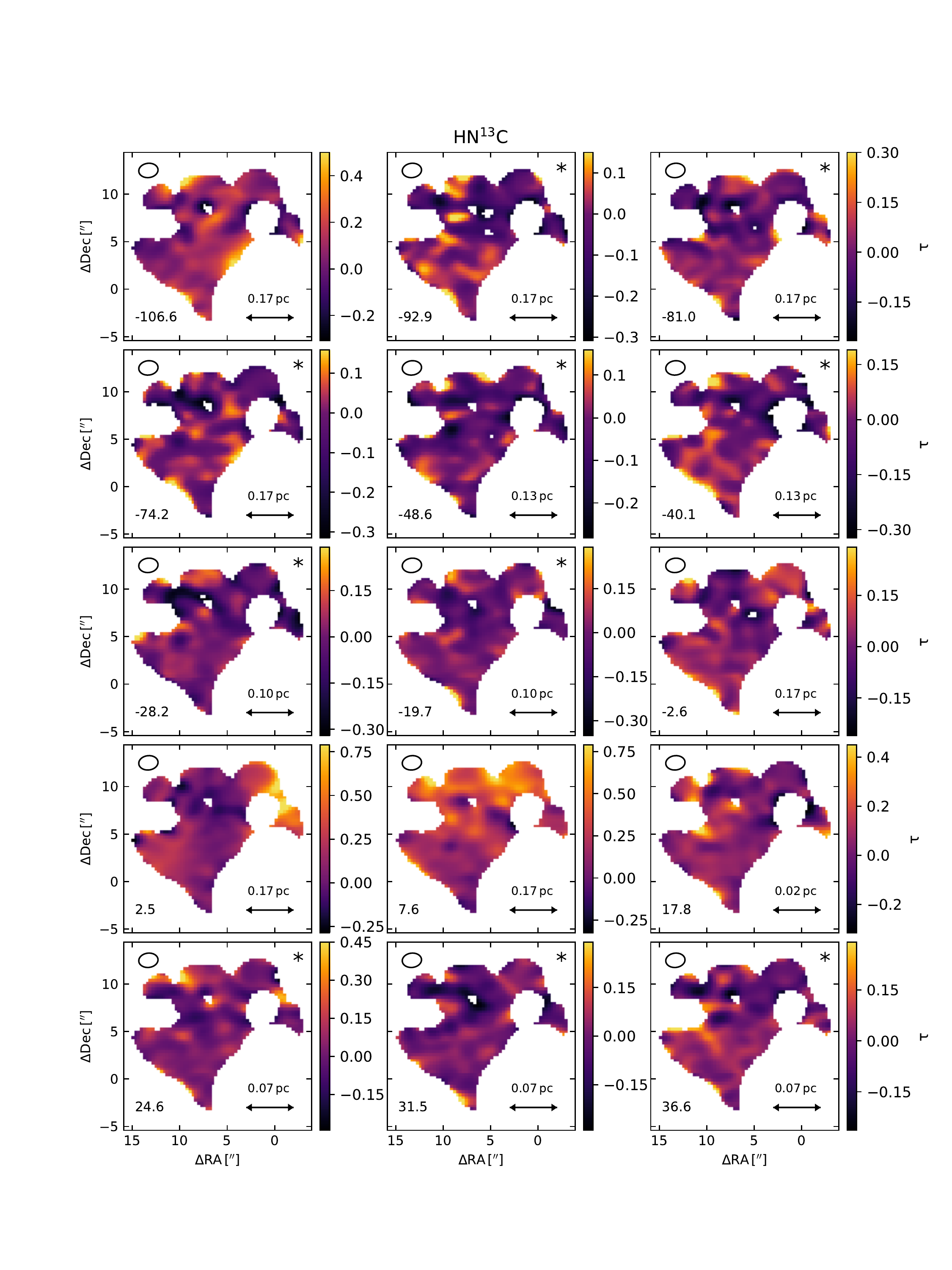}
\caption{Same as Fig.~\ref{opacity_c-c3h2}, but for HN$^{13}$C.}
\label{opacity_hn13c}
\end{figure*}

\begin{figure*}
\centering
\includegraphics[width=17cm, trim = 1.8cm 2.4cm 0.4cm 3.cm, clip=True]{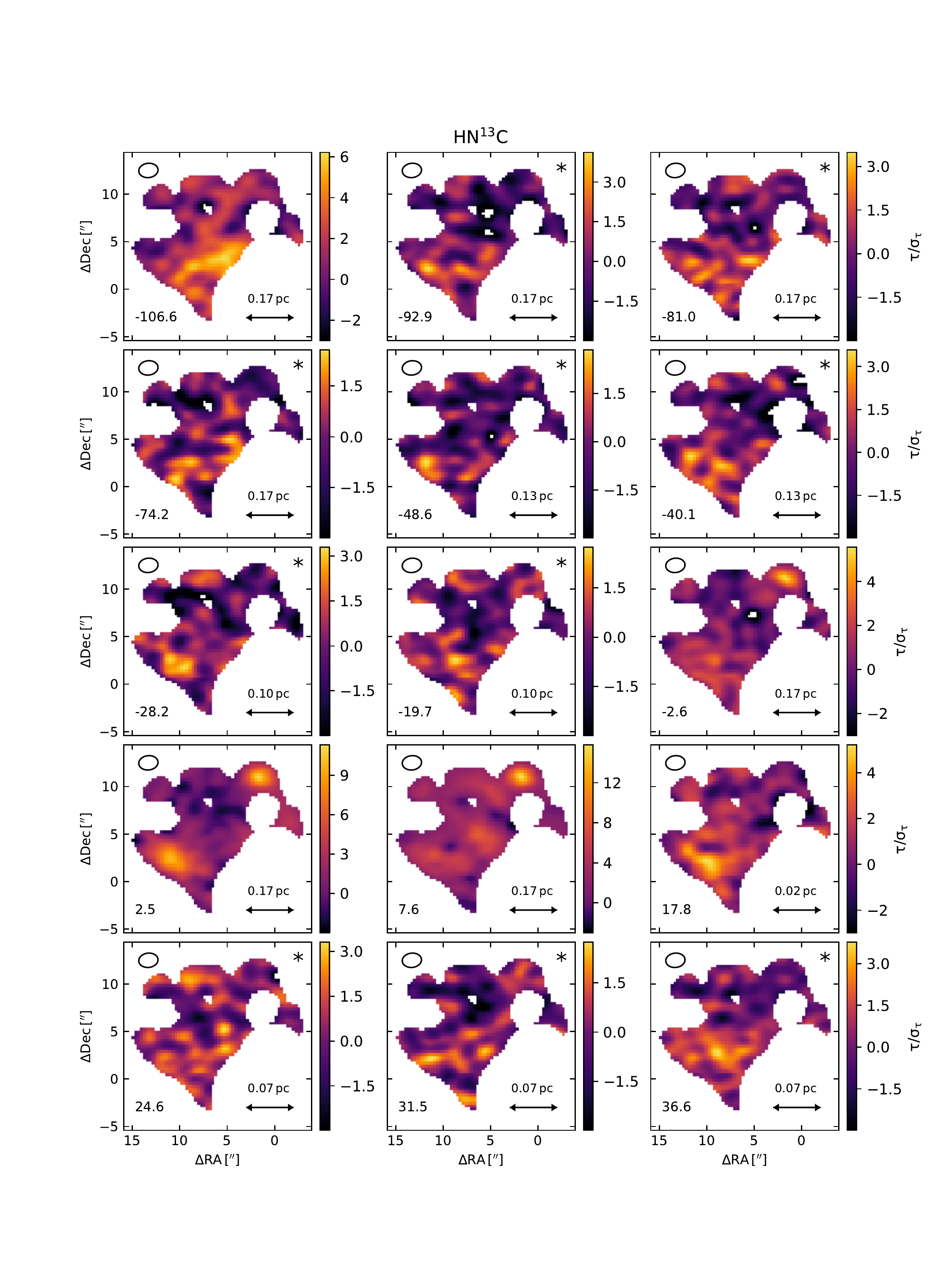}
\caption{Same as Fig.~\ref{snr_opacity_c-c3h2}, but for HN$^{13}$C.}
\label{snr_opacity_hn13c}
\end{figure*}

\begin{figure*}
\centering
\includegraphics[width=17cm, trim = 1.8cm 2.4cm 0.4cm 3.cm, clip=True]{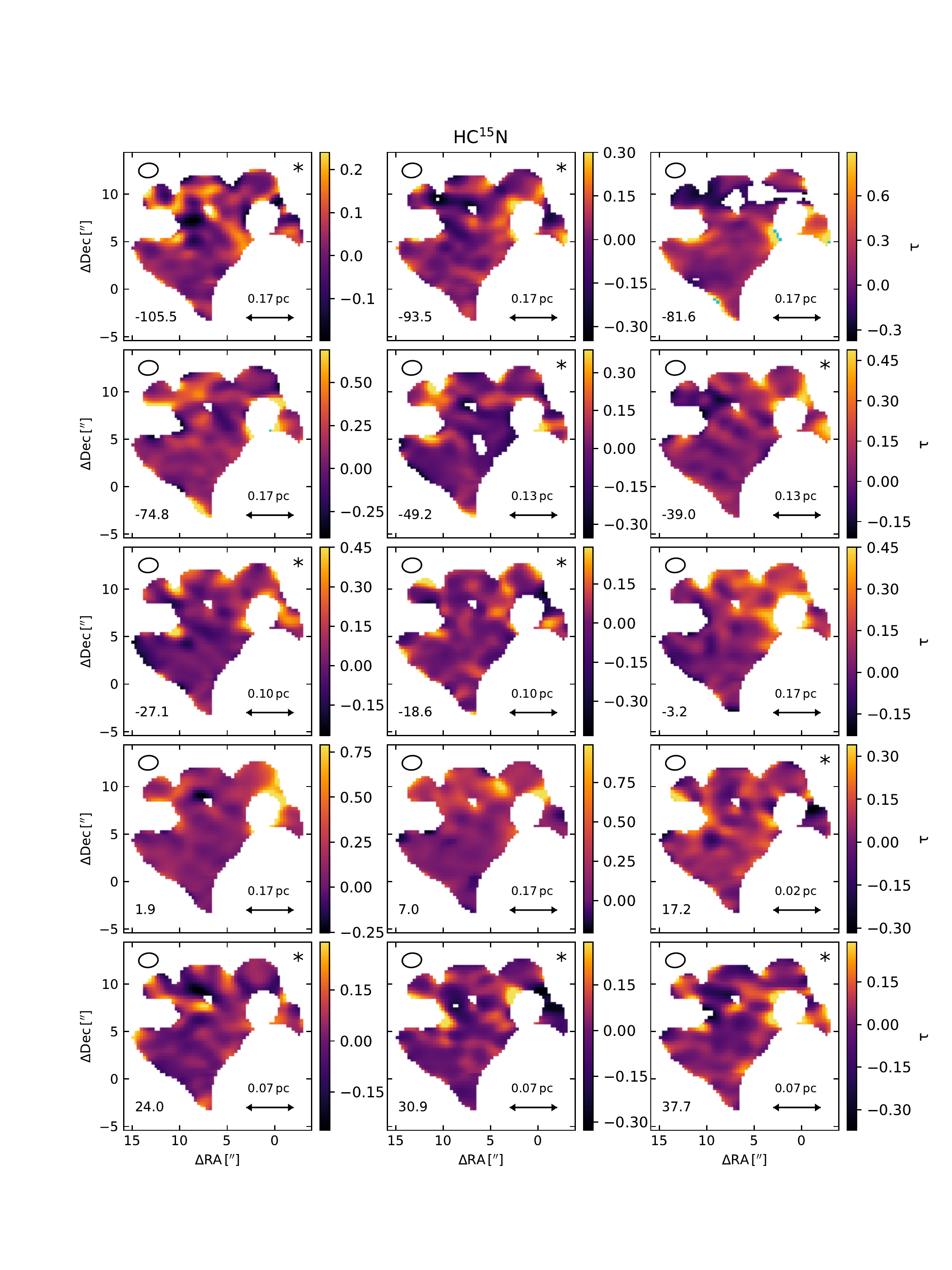}
\caption{Same as Fig.~\ref{opacity_c-c3h2}, but for HC$^{15}$N.}
\label{opacity_hc15n}
\end{figure*}

\begin{figure*}
\centering
\includegraphics[width=17cm, trim = 1.8cm 2.4cm 0.4cm 3.cm, clip=True]{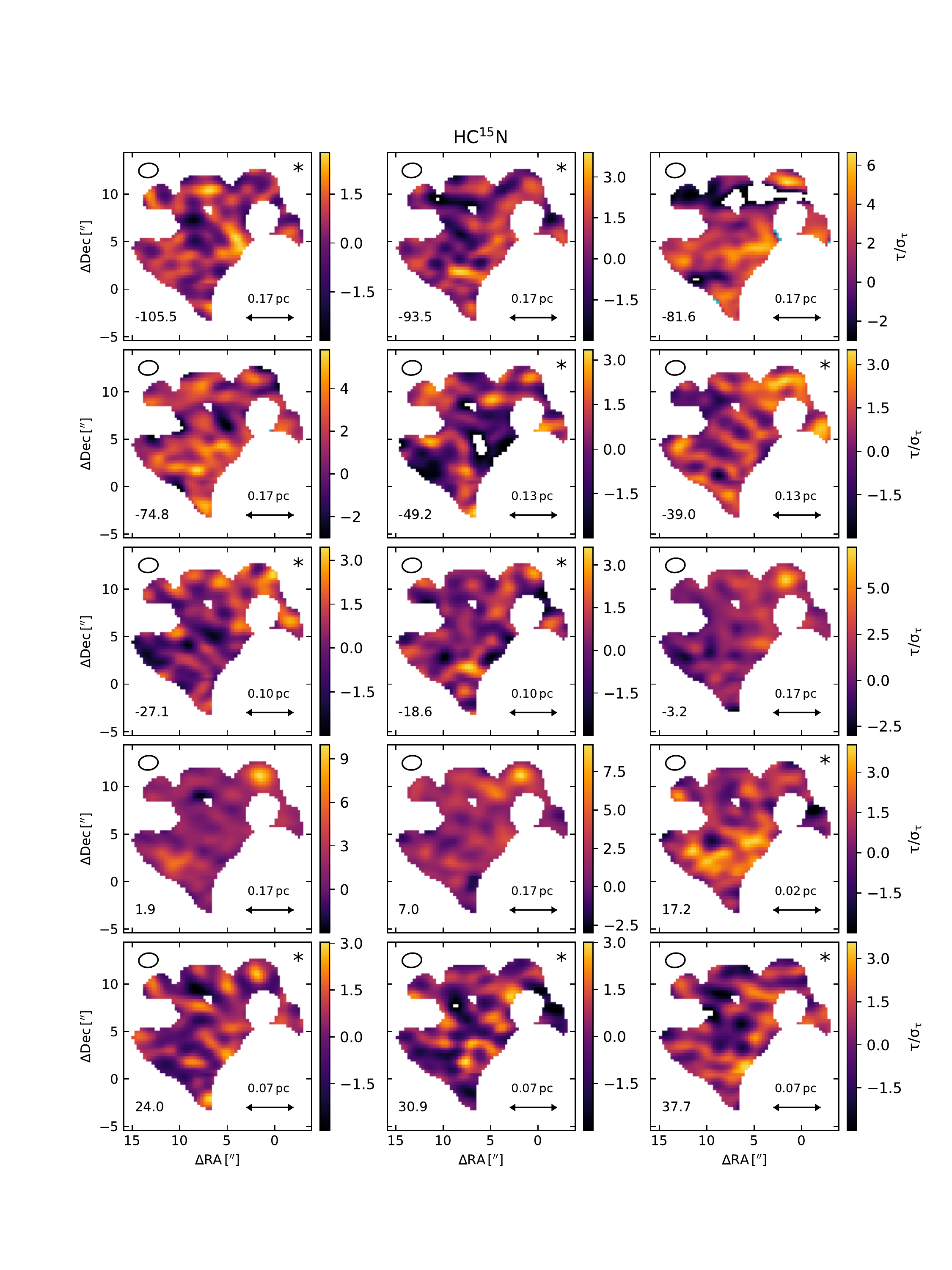}
\caption{Same as Fig.~\ref{snr_opacity_c-c3h2}, but for HC$^{15}$N.}
\label{snropacity_hc15n}
\end{figure*}

\begin{figure*}
\centering
\includegraphics[width=17cm, trim = 1.8cm 2.4cm 0.4cm 3.1cm, clip=True]{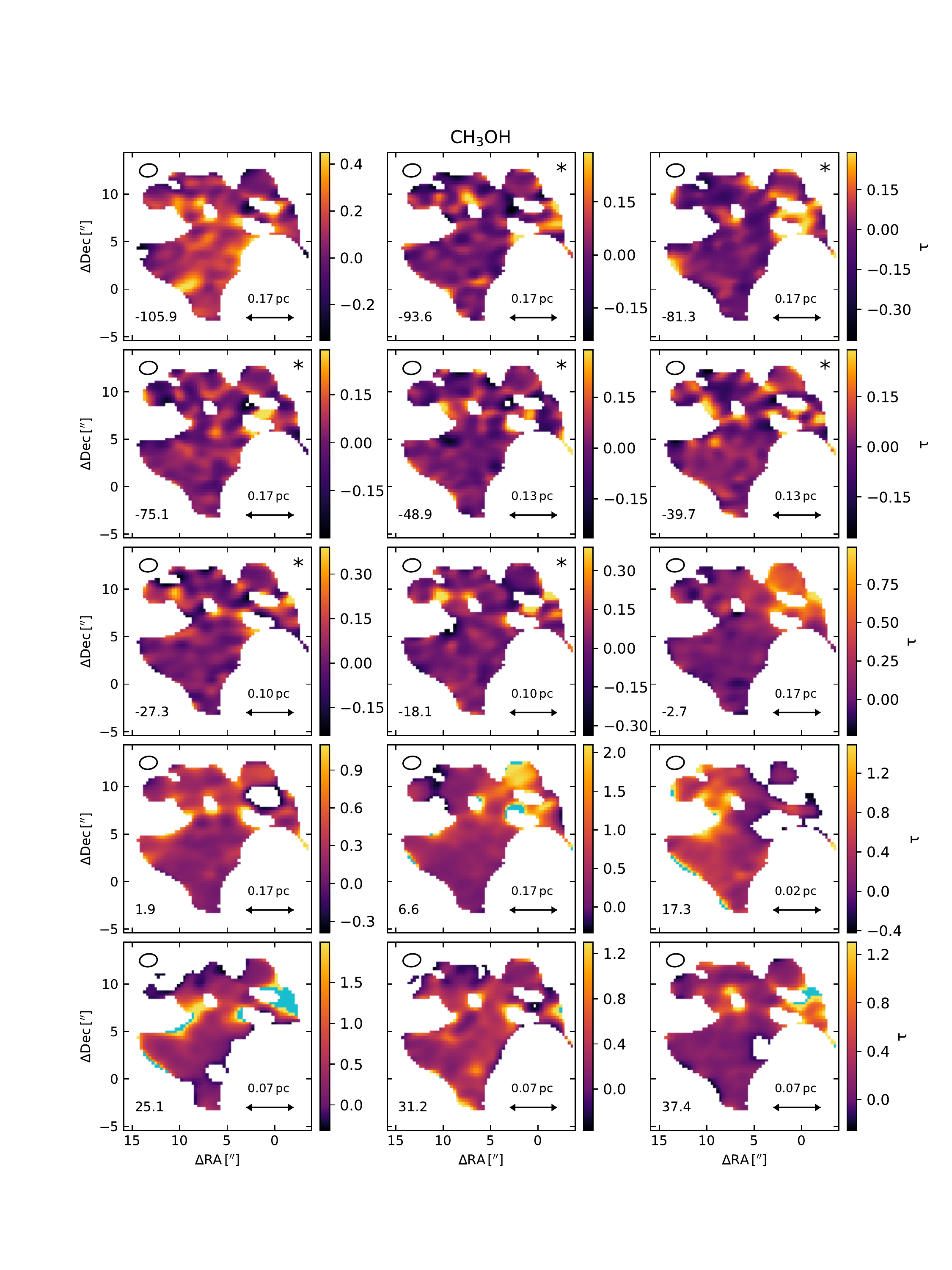}
\caption{Same as Fig.~\ref{opacity_c-c3h2}, but for CH$_3$OH.}
\label{opacity_ch3oh}
\end{figure*}

\begin{figure*}
\centering
\includegraphics[width=17cm, trim = 1.8cm 2.4cm 0.4cm 3.1cm, clip=True]{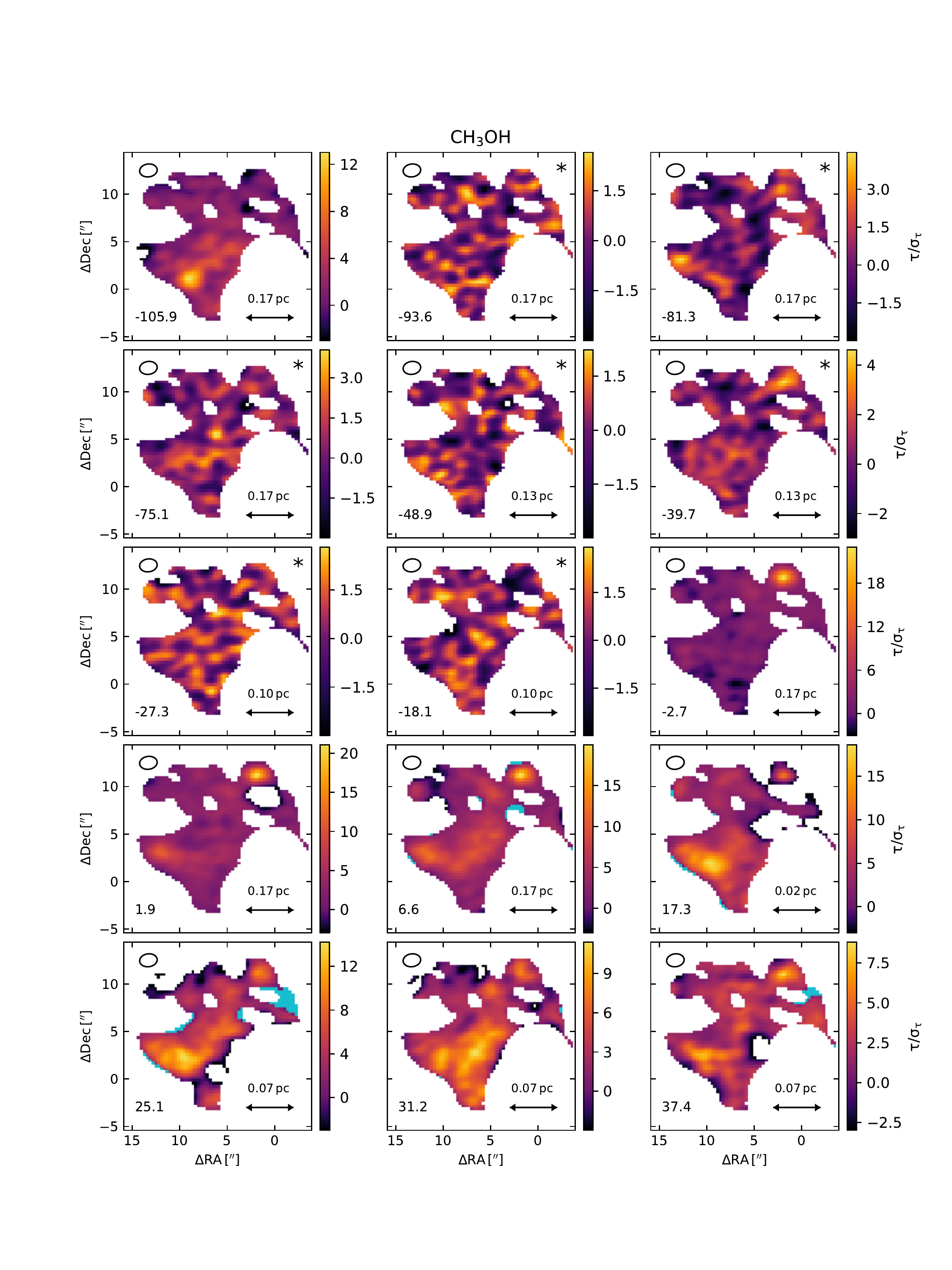}
\caption{Same as Fig.~\ref{snr_opacity_c-c3h2}, but for CH$_3$OH.}
\label{snr_opacity_ch3oh}
\end{figure*}

The opacity and signal-to-noise ratio maps of all molecules except c-C$_3$H$_2$ are shown in Figs.~\ref{opacity_h13cop}--\ref{snr_opacity_ch3oh}.










%
%
%
%
%
%
%
%
\section{Two-point auto-correlation functions}

\begin{figure*}
\centering
\includegraphics[width=17cm, trim = 3.cm 1.2cm 4.cm 2.0cm, clip=True]{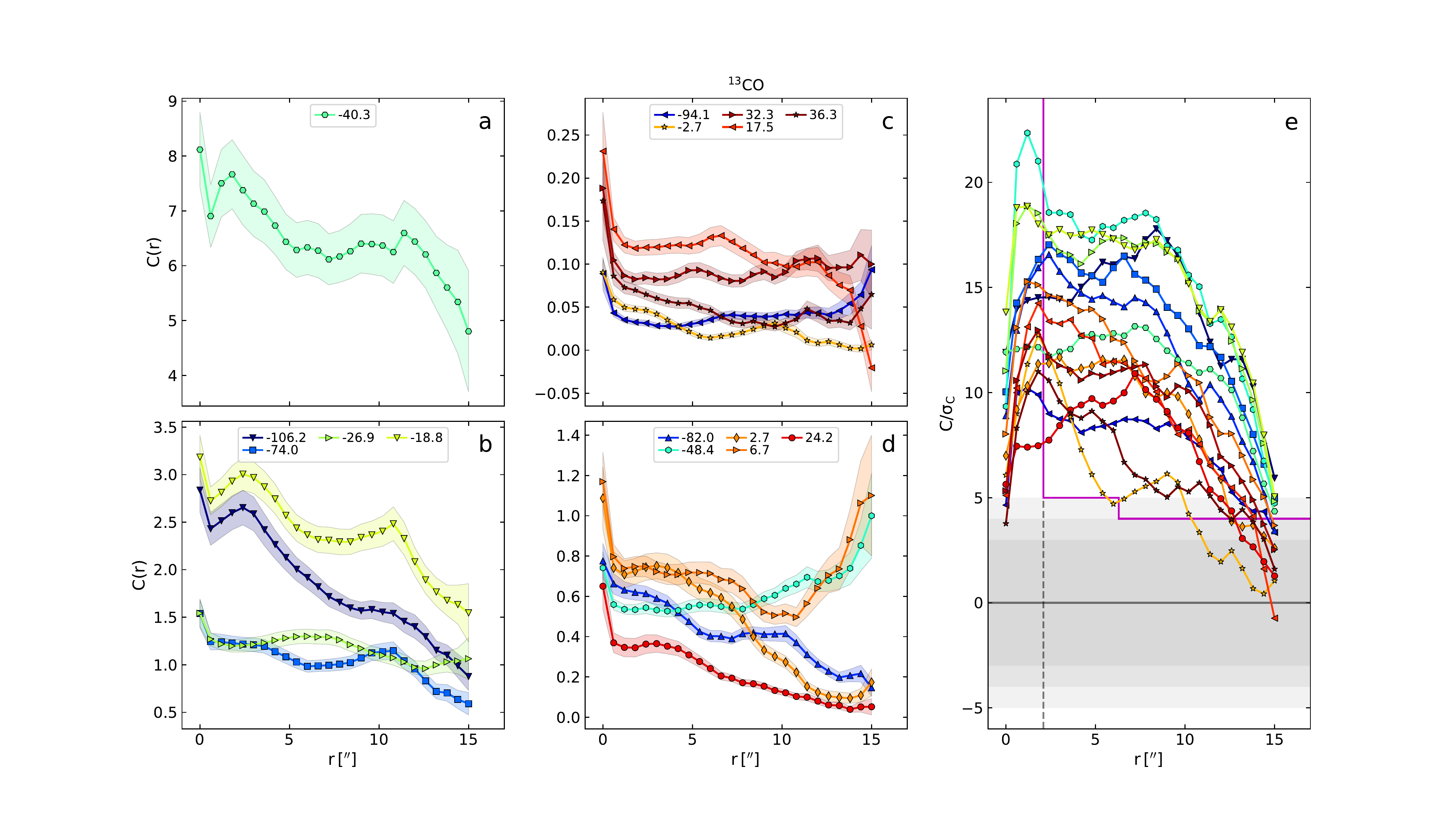}
\caption{Same as Fig.~\ref{autocorr_c-c3h2}, but for $^{13}$CO.}\label{autocorr_13co}
\end{figure*}

\begin{figure*}
\centering
\includegraphics[width=17cm, trim = 3.cm 0.5cm 3.cm 1.5cm, clip=True]{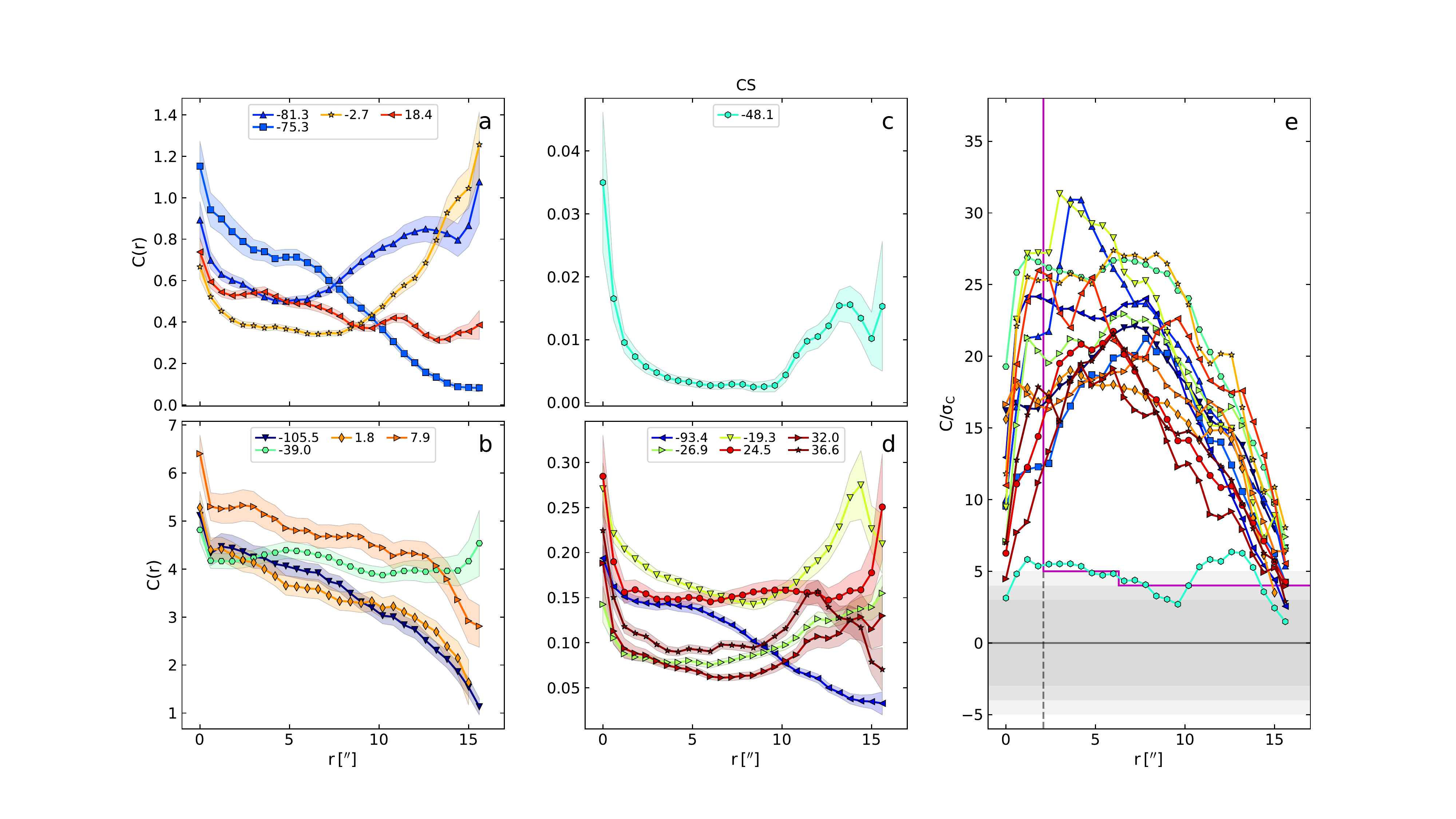}
\caption{Same as Fig.~\ref{autocorr_c-c3h2}, but for CS.} 
\label{autocorr_cs}
\end{figure*}

\begin{figure*}
\centering
\includegraphics[width=17cm, trim = 3.cm 0.5cm 3.cm 1.5cm, clip=True]{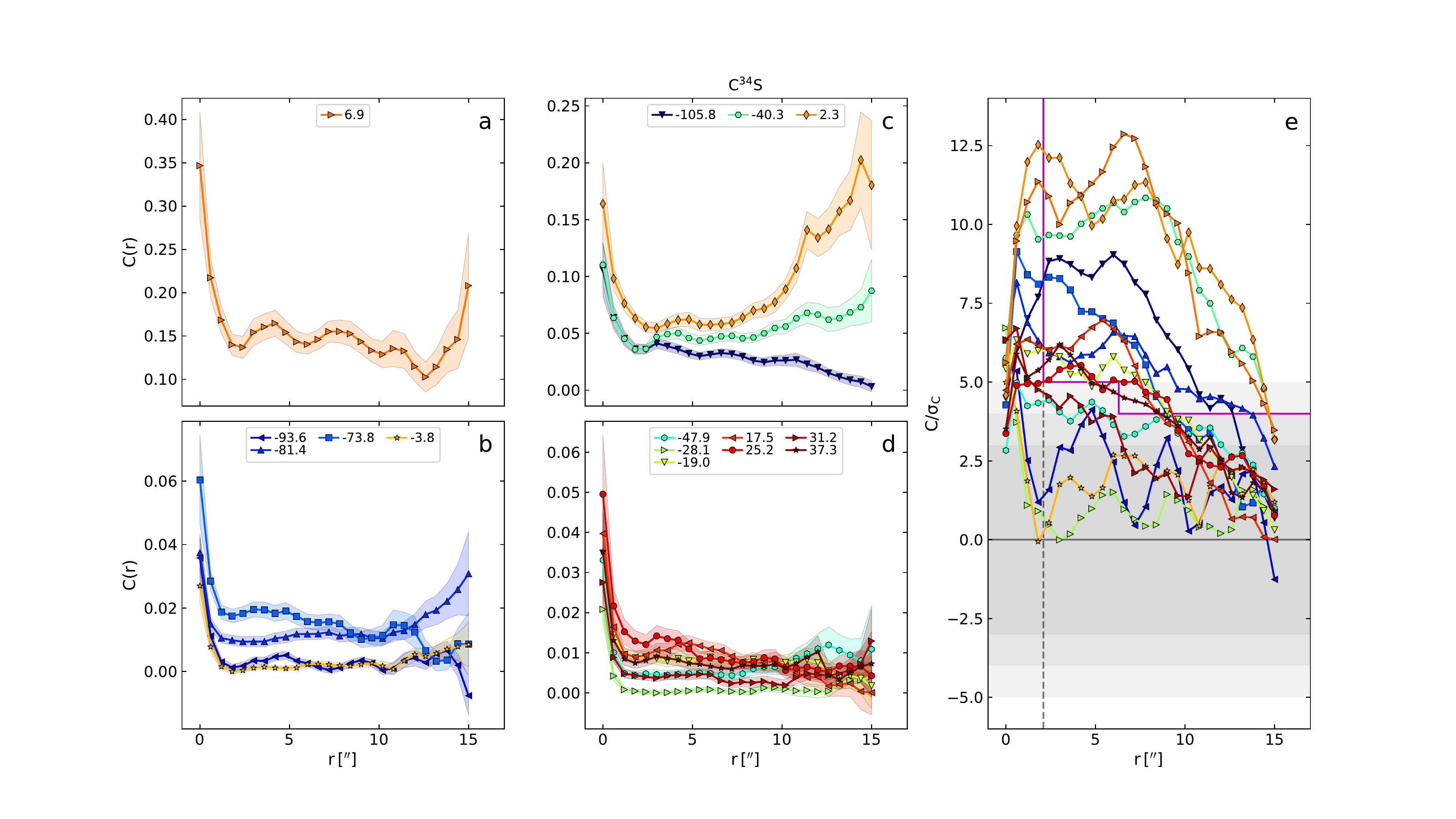}
\caption{Same as Fig.~\ref{autocorr_c-c3h2}, but for C$^{34}$S.} 
\label{autocorr_c34s}
\end{figure*}

\begin{figure*}
\centering
\includegraphics[width=17cm, trim = 3.cm 0.5cm 3.cm 1.5cm, clip=True]{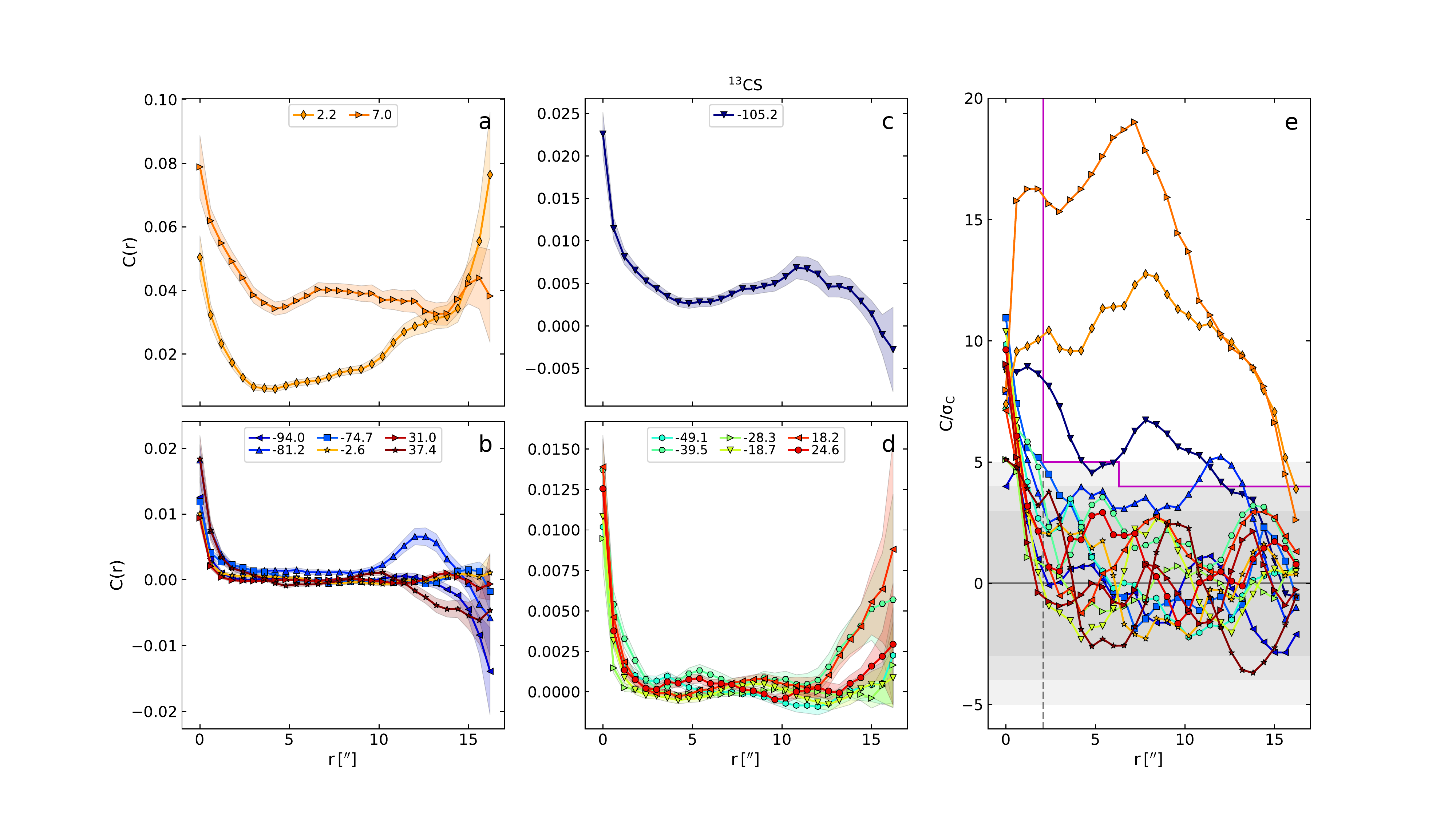}
\caption{Same as Fig.~\ref{autocorr_c-c3h2}, but for $^{13}$CS.} 
\label{autocorr_13cs}
\end{figure*}

\begin{figure*}
\centering
\includegraphics[width=17cm, trim = 3.cm 0.5cm 3.cm 1.5cm, clip=True]{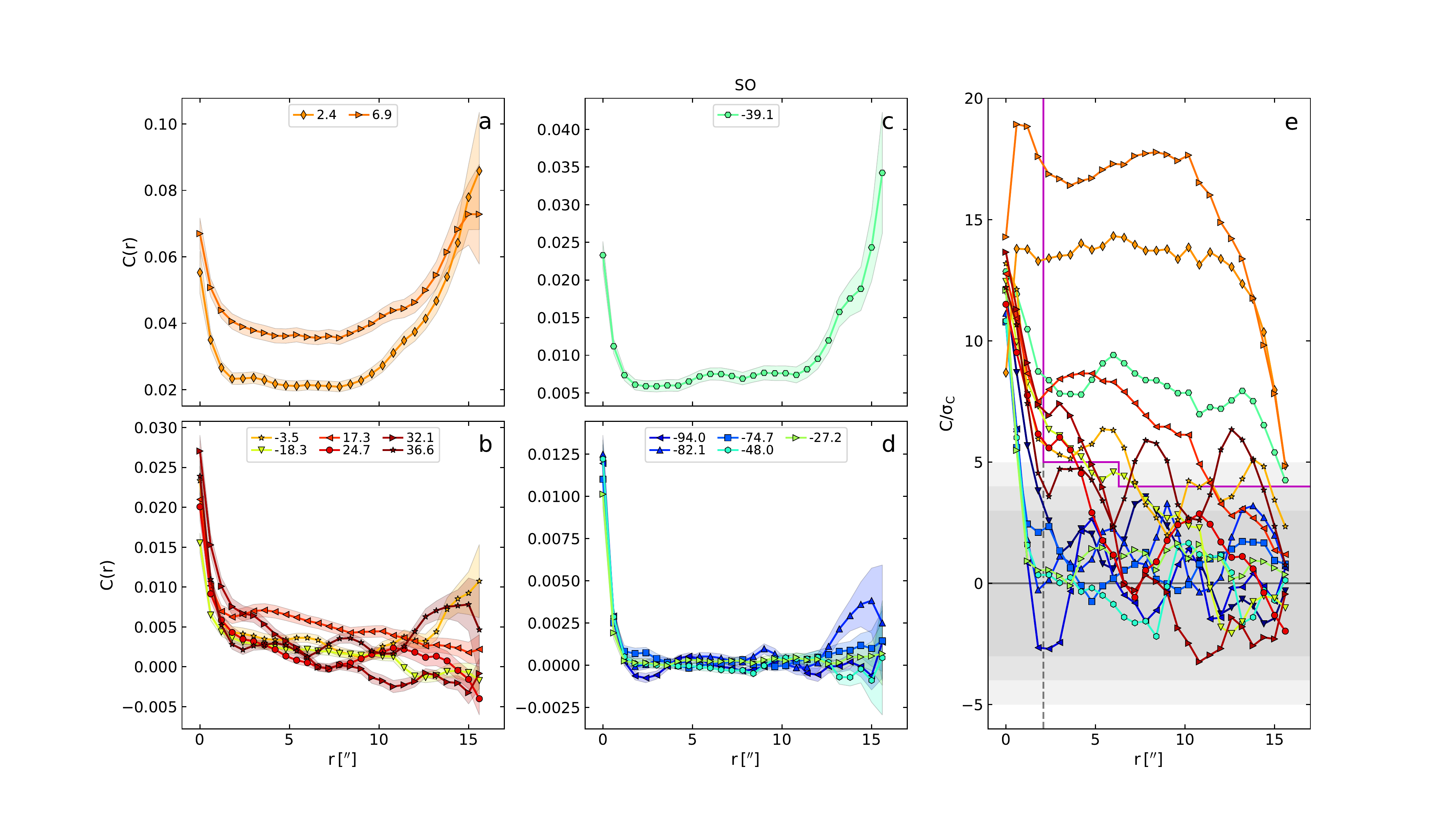}
\caption{Same as Fig.~\ref{autocorr_c-c3h2}, but for SO.} 
\label{autocorr_so}
\end{figure*}

\begin{figure*}
\centering
\includegraphics[width=17cm, trim = 3.cm 0.5cm 3.cm 1.5cm, clip=True]{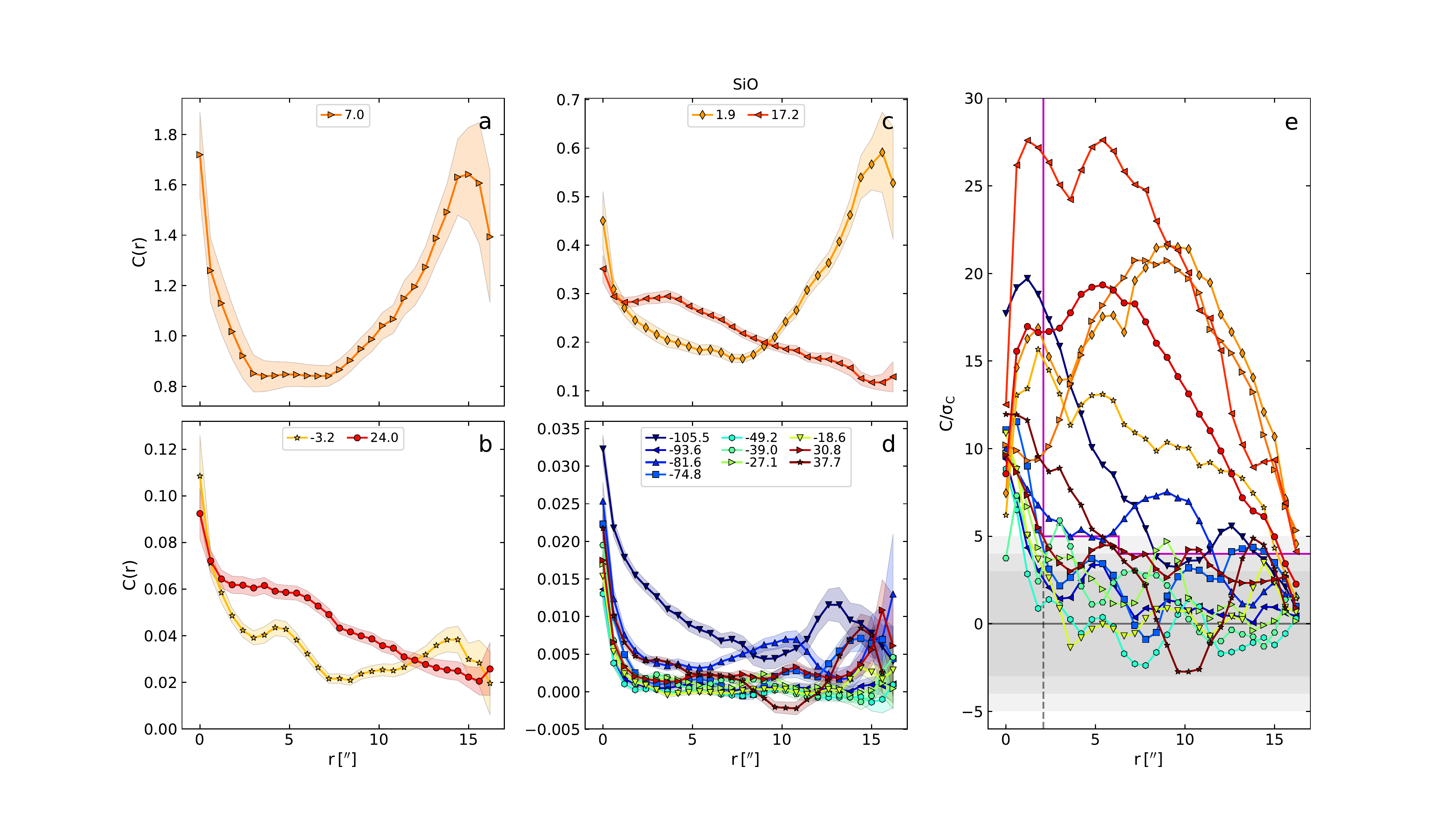}
\caption{Same as Fig.~\ref{autocorr_c-c3h2}, but for SiO.} 
\label{autocorr_sio}
\end{figure*}

\begin{figure*}
\centering
\includegraphics[width=17cm, trim = 3.cm 0.5cm 3.cm 1.5cm, clip=True]{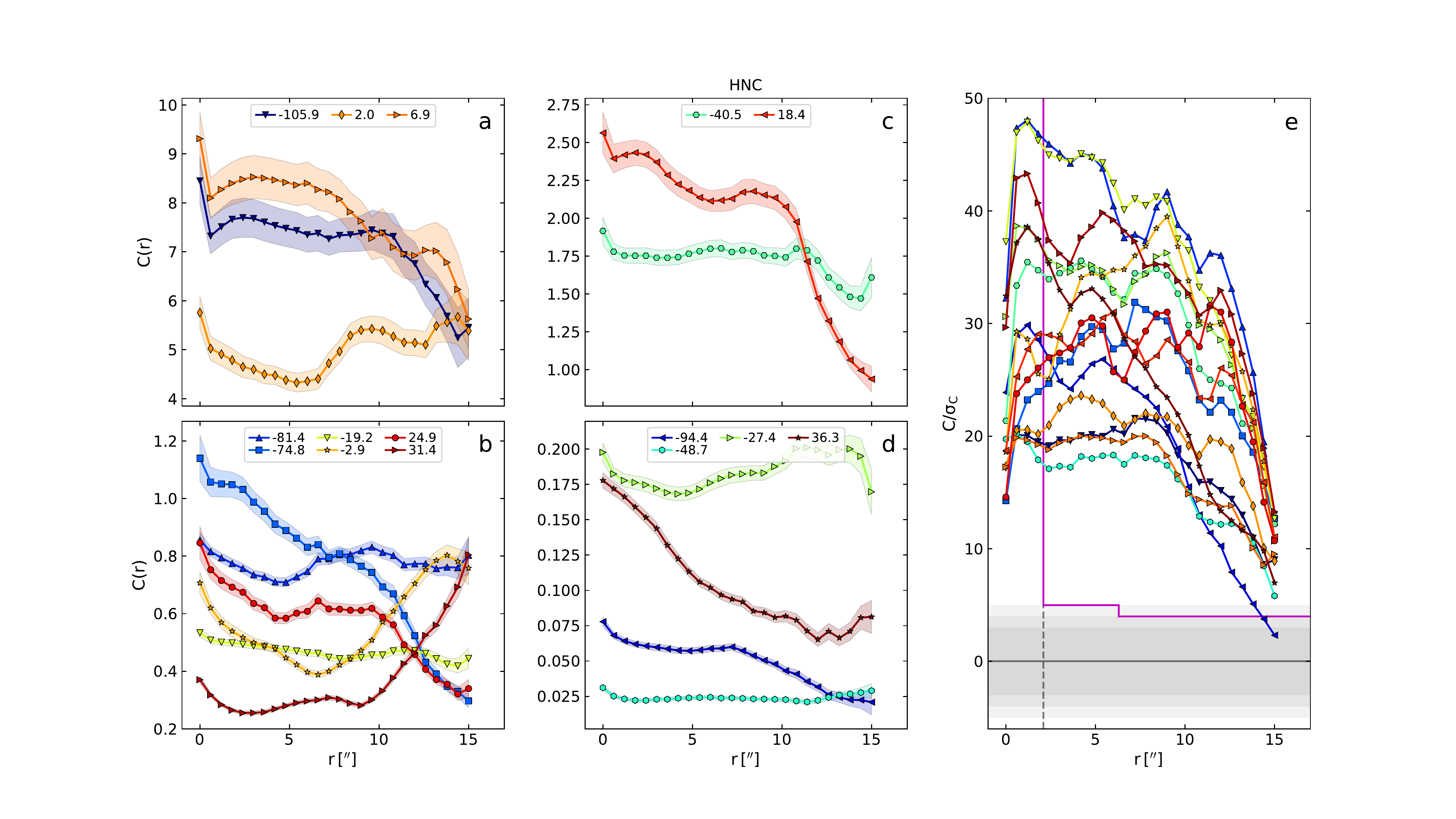}
\caption{Same as Fig.~\ref{autocorr_c-c3h2}, but for HNC.} 
\label{autocorr_hnc}
\end{figure*}

\begin{figure*}
\centering
\includegraphics[width=17cm, trim = 3.cm 0.5cm 3.cm 1.5cm, clip=True]{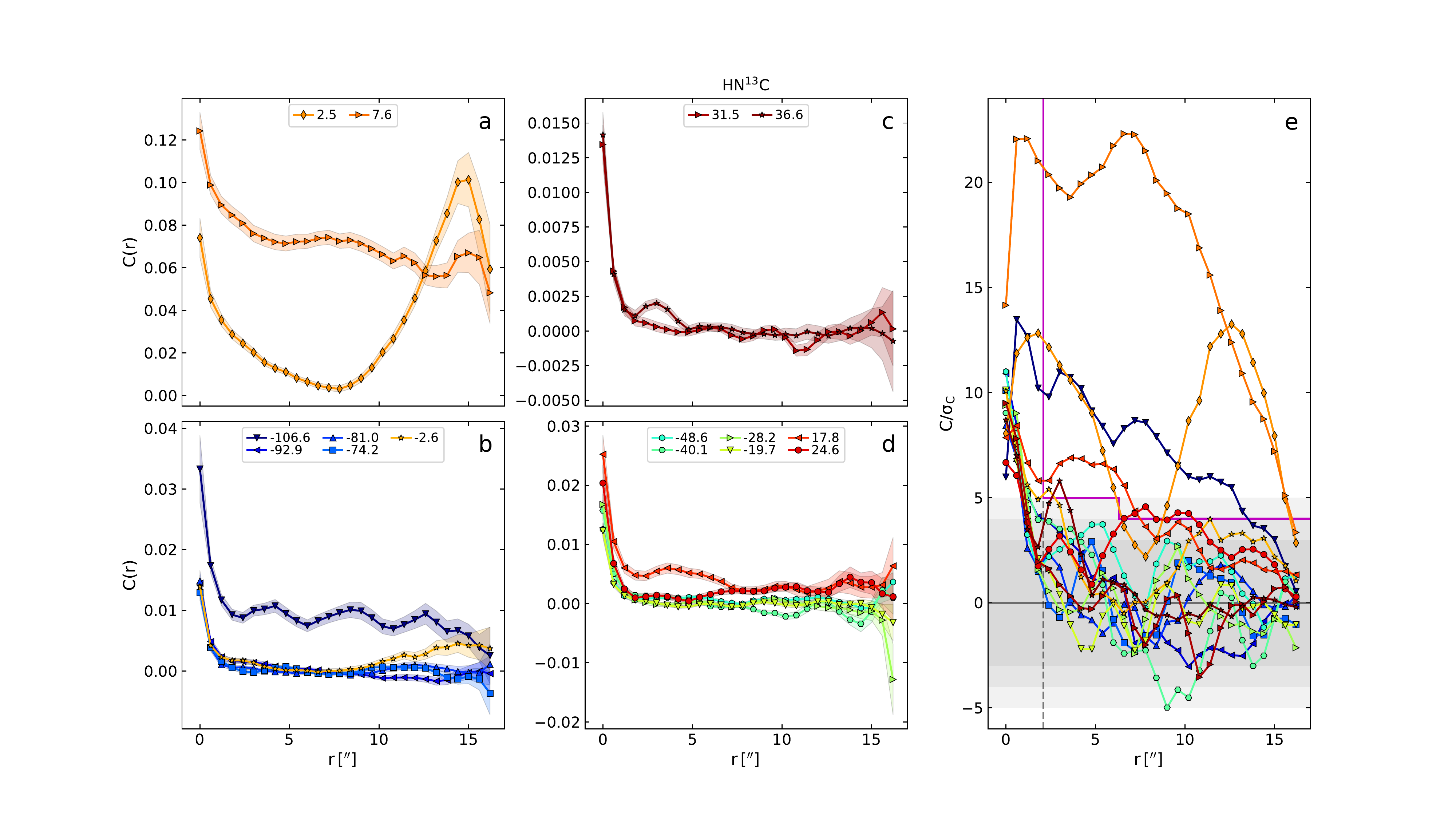}
\caption{Same as Fig.~\ref{autocorr_c-c3h2}, but for HN$^{13}$C.} 
\label{autocorr_hn13c}
\end{figure*}

\begin{figure*}
\centering
\includegraphics[width=17cm, trim = 3.cm 0.5cm 3.cm 1.5cm, clip=True]{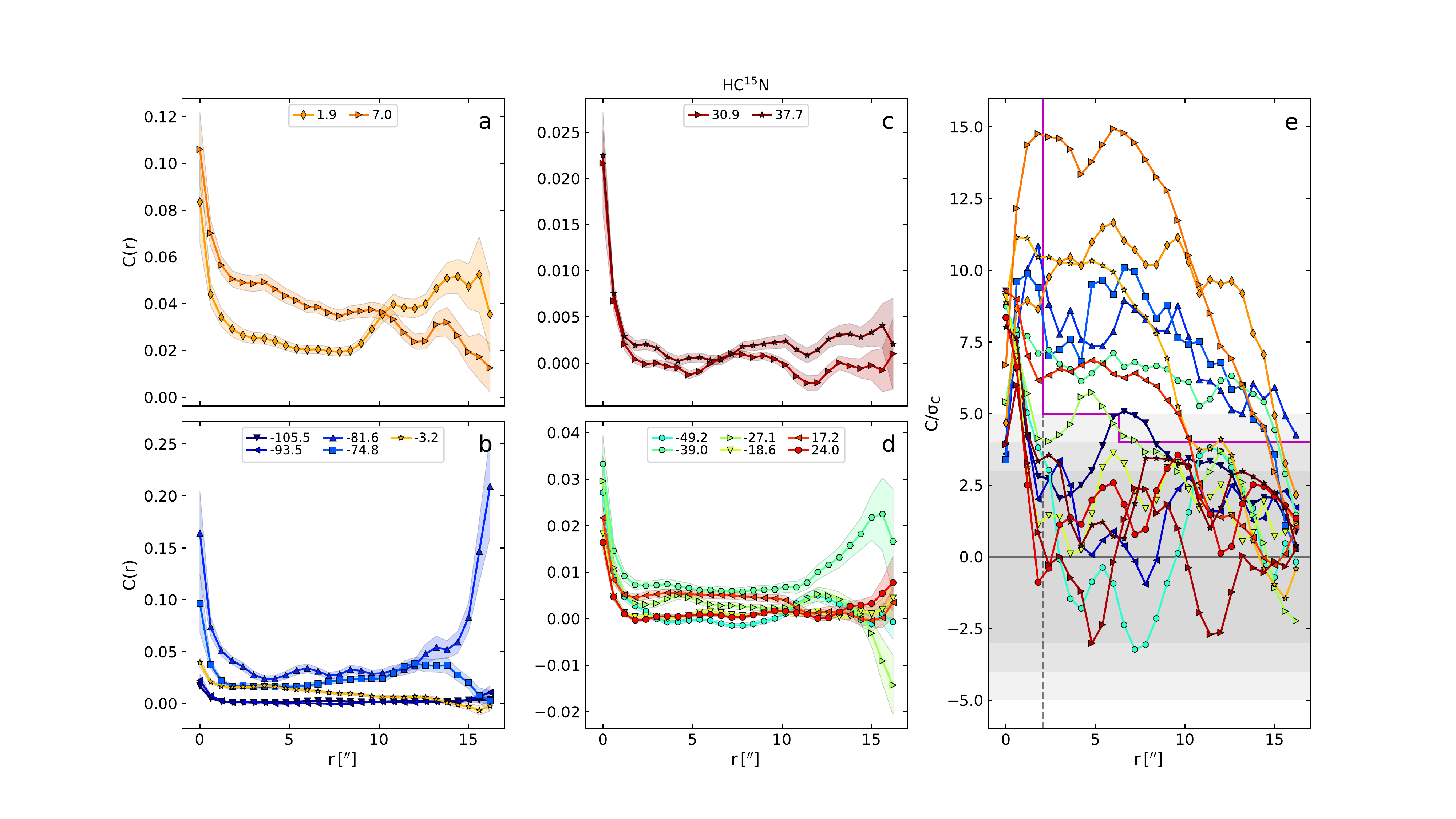}
\caption{Same as Fig.~\ref{autocorr_c-c3h2}, but for HC$^{15}$N.} 
\label{autocorr_hc15n}
\end{figure*}

\begin{figure*}
\centering
\includegraphics[width=17cm, trim = 3.cm 0.5cm 3.cm 1.5cm, clip=True]{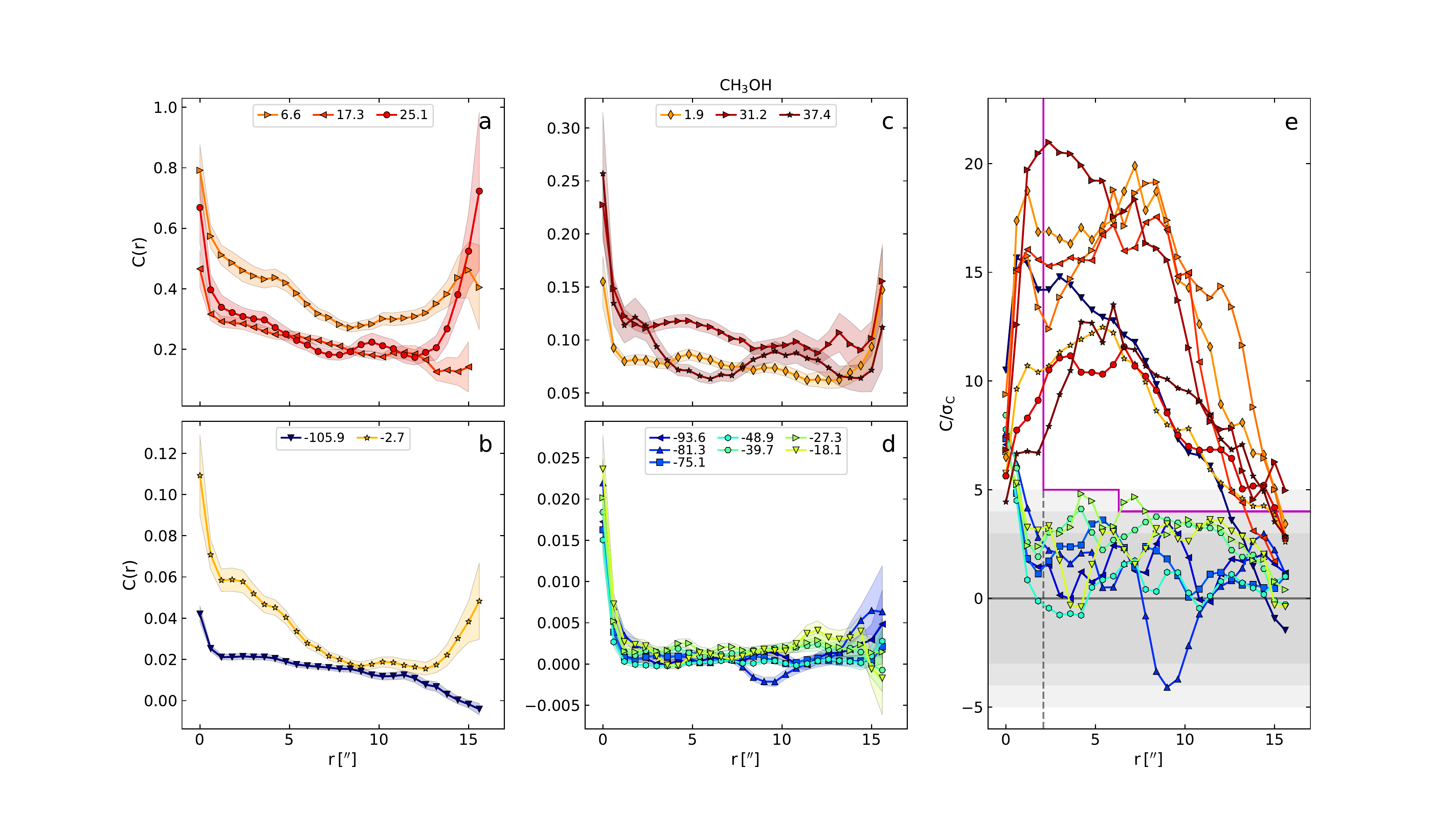}
\caption{Same as Fig.~\ref{autocorr_c-c3h2}, but for CH$_3$OH.}
\label{autocorr_ch3oh}
\end{figure*}

\begin{figure*}
\centering
\includegraphics[width=17cm, trim = 3.3cm 0.4cm 4.cm .5cm, clip=True]{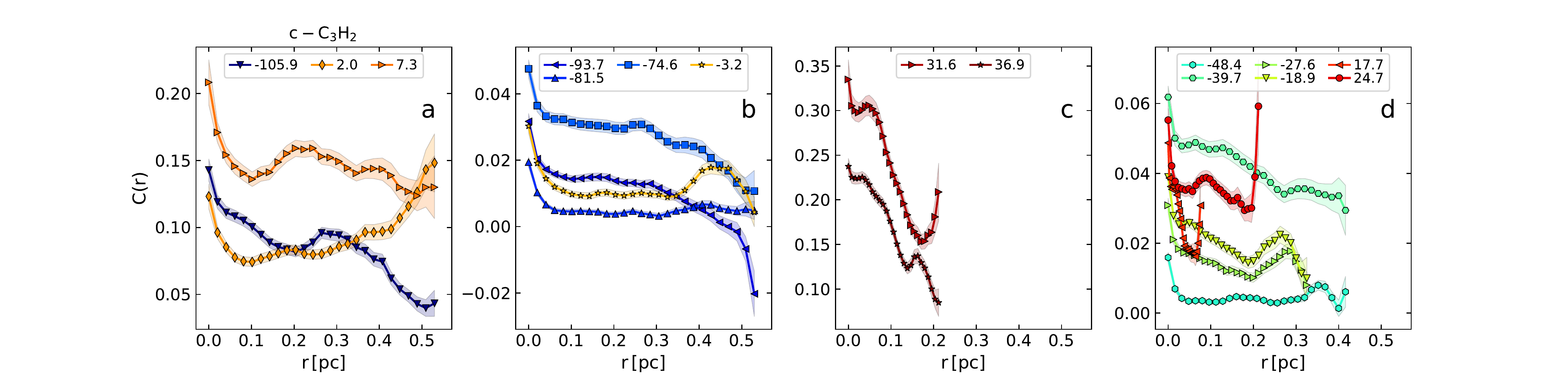}
\caption{\textbf{a--d} Two-point auto-correlation functions $C(r)$ as a function of physical distance $r$ for the velocity components traced with c-C$_3$H$_2$. The points give the mean values of the 1000 realisations and the colour-shaded regions represent the standard deviations ($1\sigma$). The centroid LSR velocities of the clouds are indicated in km~s$^{-1}$ at the top of each panel.} 
\label{autocorr_c-c3h2_small}
\end{figure*}

\begin{figure*}
\centering
\includegraphics[width=17cm, trim = 3.3cm 0.4cm 4.cm .5cm, clip=True]{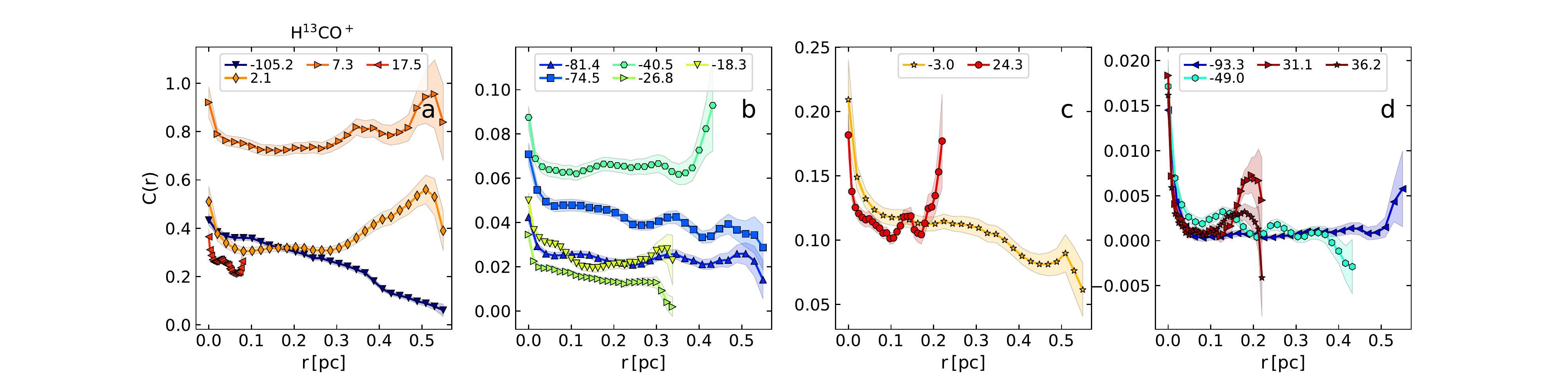}
\caption{Same as Fig.~\ref{autocorr_c-c3h2_small}, but for H$^{13}$CO$^+$.} 
\label{autocorr_h13cop_small}
\end{figure*}

\begin{figure*}
\centering
\includegraphics[width=17cm, trim = 3.3cm 0.4cm 4.cm .5cm, clip=True]{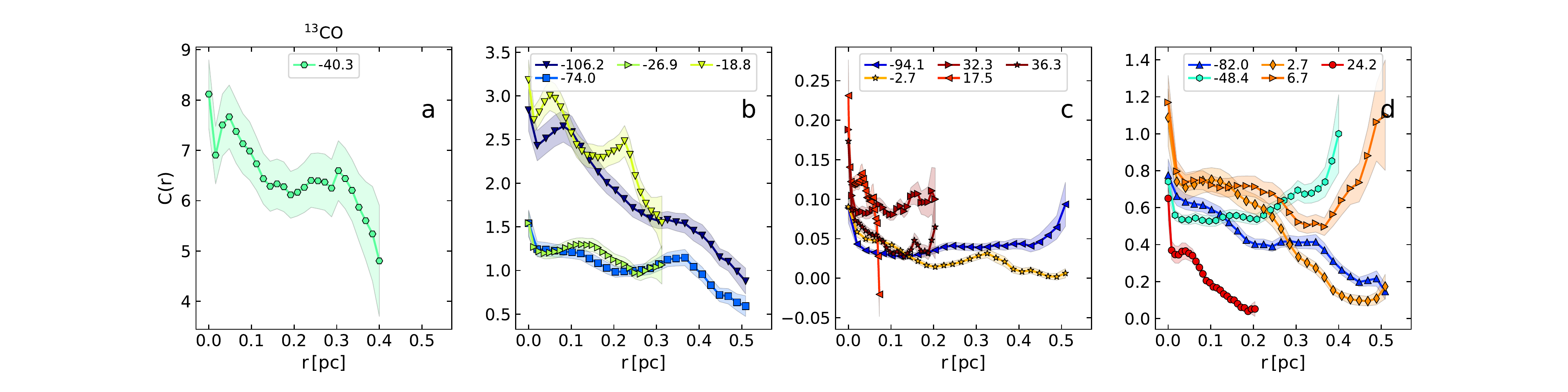}
\caption{Same as Fig.~\ref{autocorr_c-c3h2_small}, but for $^{13}$CO.} 
\label{autocorr_13co_small}
\end{figure*}

\begin{figure*}
\centering
\includegraphics[width=17cm, trim = 3.3cm 0.4cm 4.cm .5cm, clip=True]{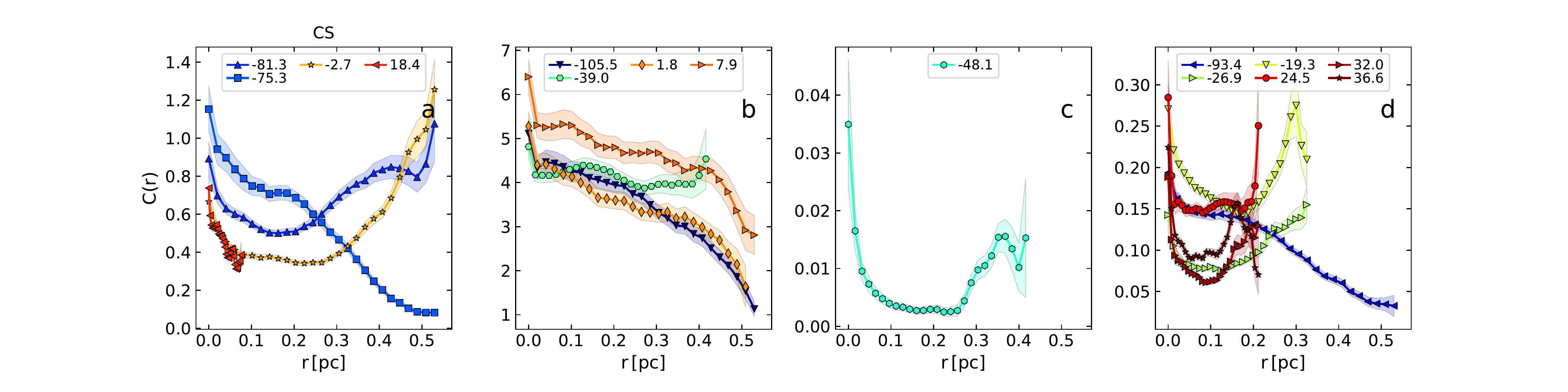}
\caption{Same as Fig.~\ref{autocorr_c-c3h2_small}, but for CS.} 
\label{autocorr_cs_small}
\end{figure*}

\begin{figure*}
\centering
\includegraphics[width=17cm, trim = 3.3cm 0.4cm 4.cm .5cm, clip=True]{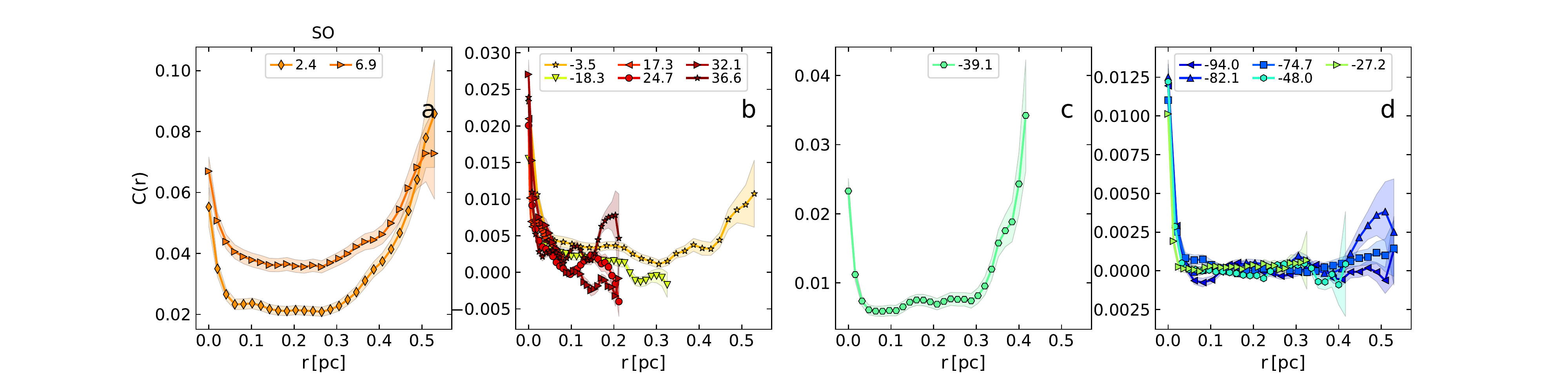}
\caption{Same as Fig.~\ref{autocorr_c-c3h2_small}, but for SO.} 
\label{autocorr_so_small}
\end{figure*}

\begin{figure*}
\centering
\includegraphics[width=17cm, trim = 3.3cm 0.4cm 4.cm .5cm, clip=True]{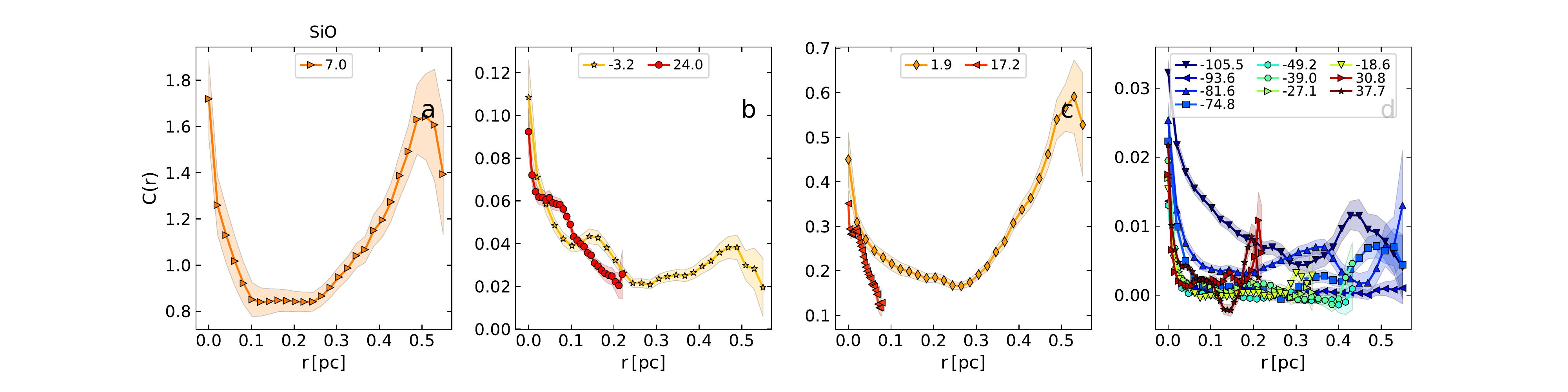}
\caption{Same as Fig.~\ref{autocorr_c-c3h2_small}, but for SiO.} 
\label{autocorr_sio_small}
\end{figure*}

\begin{figure*}
\centering
\includegraphics[width=17cm, trim = 3.3cm 0.4cm 4.cm .5cm, clip=True]{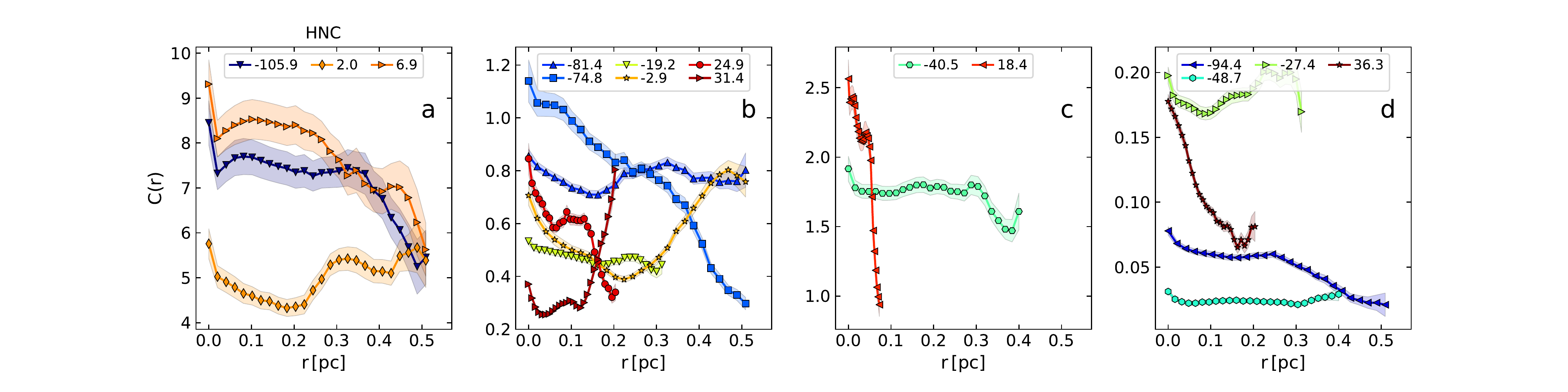}
\caption{Same as Fig.~\ref{autocorr_c-c3h2_small}, but for HNC.} 
\label{autocorr_hnc_small}
\end{figure*}

\begin{figure*}
\centering
\includegraphics[width=17cm, trim = 3.3cm 0.4cm 4.cm .5cm, clip=True]{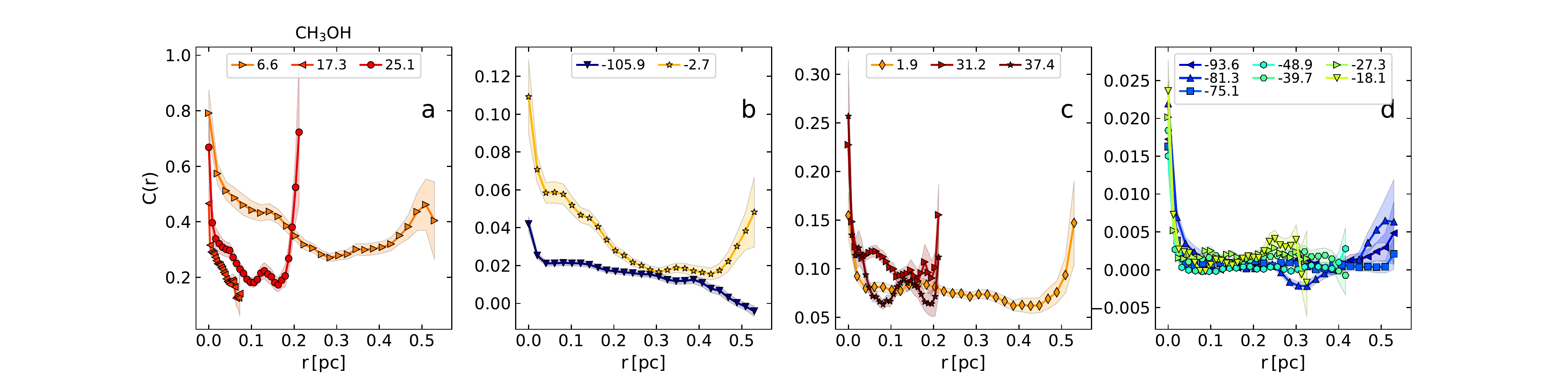}
\caption{Same as Fig.~\ref{autocorr_c-c3h2_small}, but for CH$_3$OH.} 
\label{autocorr_ch3oh_small}
\end{figure*}

The two-point auto-correlation functions as a function of angular distance are shown in Figs.~\ref{autocorr_13co}--\ref{autocorr_ch3oh} for all molecules except for c-C$_3$H$_2$ and H$^{13}$CO$^+$. The two-point auto-correlation functions as a function of physical distance are shown in Figs.~\ref{autocorr_c-c3h2_small}--\ref{autocorr_ch3oh_small} for the eight strongest molecules (c-C$_3$H$_2$, H$^{13}$CO$^+$, $^{13}$CO, CS, SO, SiO, HNC, and CH$_3$OH).

\FloatBarrier
\afterpage{\clearpage}

\FloatBarrier
 
\section{Analysis of two-point auto-correlation functions}\label{auto_appendix_mol}

We analyse here the two-point auto-correlation functions of all molecules except the ones of SiO and SO which are analysed in Sect.~\ref{autocorrel}. c-C$_3$H$_2$ and HNC are detected in all investigated velocity components and trace structures that are more extended than the investigated area ($\gtrsim 15\arcsec$) for all components but one, at $-93.7$~km~s$^{-1}$, for which the size derived from c-C$_3$H$_2$ is $\sim 11\arcsec$. This component is one of the noisiest for this molecule (with peak SNR below 7) so it is unclear whether this smaller size is significant compared to the other components, especially because the size traced with HNC ($14.4\arcsec$) is very close to the size of the investigated area. We conclude that c-C$_3$H$_2$ and HNC trace structures that are more extended than $\sim 15\arcsec$ for all clouds, which corresponds to physical sizes between 0.08 pc in the nearby Sagittarius arm and 0.5~pc in the GC. 

$^{13}$CO and CS also show structures that have sizes larger than ($\gtrsim 15\arcsec$) or similar to ($\sim 13$--$15\arcsec$) the size of the field of view. The only exceptions are the component at $-2.7$\,km\,s$^{-1}$ for $^{13}$CO and the one at $-48.1$\,km\,s$^{-1}$ for CS. In the former case, the two-point auto-correlation function drops just below the significance threshold at $6\arcsec$, but increases above this threshold at larger separations and drops below it again for sizes larger than $10\as8$ (see Fig.~\ref{autocorr_13co}). The corresponding opacity map (see Fig.~\ref{opacity_13co}) shows two compact clumps at ($6\as5$,$0.5^{\prime\prime}$) and ($9\as5$,$10^{\prime\prime}$), hence with a separation of $10\arcsec$, but one has a SNR below 3 so it may be a noise artefact and the correlation at larger pixel separation may not be real. In the case of CS, the exception concerns the noisiest component, with a peak SNR below 6. The correlation function drops below the significance threshold at $4\as8$, but increases above it again at $\sim 10\arcsec$ (see Fig.~\ref{autocorr_cs}). However, it remains close to the significance threshold so its shape is not well constrained. Therefore, overall, $^{13}$CO and CS trace structures that are more extended than or nearly as extended as the field of view, like c-C$_3$H$_2$ and HNC.

H$^{13}$CO$^+$ traces structures with similar extent as the previous tracers. Four exceptions occur, with very small sizes of 2--3$\arcsec$ or unresolved. All have low SNR in their opacity maps, with a peak SNR below 6 for the component at 31.1~km~s$^{-1}$ and below 5 for the other three components. The sensitivity of our data is therefore not sufficient to characterise the structures traced by H$^{13}$CO$^+$ for these four components.

CH$_3$OH and HC$^{15}$N behave in the same way: all components with a peak SNR higher than 5 in their opacity maps, reveal structures that are more extended than the field of view, or at least have sizes larger than 13$\arcsec$ and 11$\arcsec$, respectively.

Only six and three components traced by C$^{34}$S and $^{13}$CS, respectively, have a peak SNR in their opacity map higher than 5 (Figs.~\ref{snr_opacity_c34s} and \ref{snr_opacity_13cs}). Most of them trace extended structures of size similar to the field of view (Figs.~\ref{opacity_c34s} and \ref{opacity_13cs}). The 17.7~km~s$^{-1}$ component of C$^{34}$S and the $-105.9$~km~s$^{-1}$ component of $^{13}$CS have smaller $\Delta r_{\rm max}$ but their peak SNR is barely above 5.
The sensitivity of our data is not high enough to characterise the structures of the other velocity components on the basis of these tracers. Because C$^{34}$S and $^{13}$CS are expected to be present in the same regions as CS, that traces extended structures, they clearly show that tracers with a limited sensitivity are not appropriate to constrain the underlying cloud structure. This should be kept in mind when analysing the two-point auto-correlation functions of the other molecules that have low SNR. The case of HN$^{13}$C is very similar to C$^{34}$S and $^{13}$CS. Only five components have a peak SNR in their opacity map higher than 5 (Fig.~\ref{snr_opacity_hn13c}). Two of them trace extended structures and two of them have small or unresolved $\Delta r_{\rm max}$ but with a peak SNR barely above 5. The only exception is the component at 2.5 km~s$^{-1}$, with a correlation length of $7^{\prime\prime}$ and a peak SNR of $\sim$11 in its opacity map. For this velocity component, the main isotopologue, HNC, traces an extended structure with a correlation length larger than the field of view so the shorter correlation length derived from HN$^{13}$C results from a lack of sensitivity of our data.

%
%
%
%
%
\section{Probability distribution functions}\label{pdf_gaussians}
\begin{figure*}
\centering
\includegraphics[width=17cm, trim = 1.2cm 11.8cm 0.7cm 3.cm, clip=True]{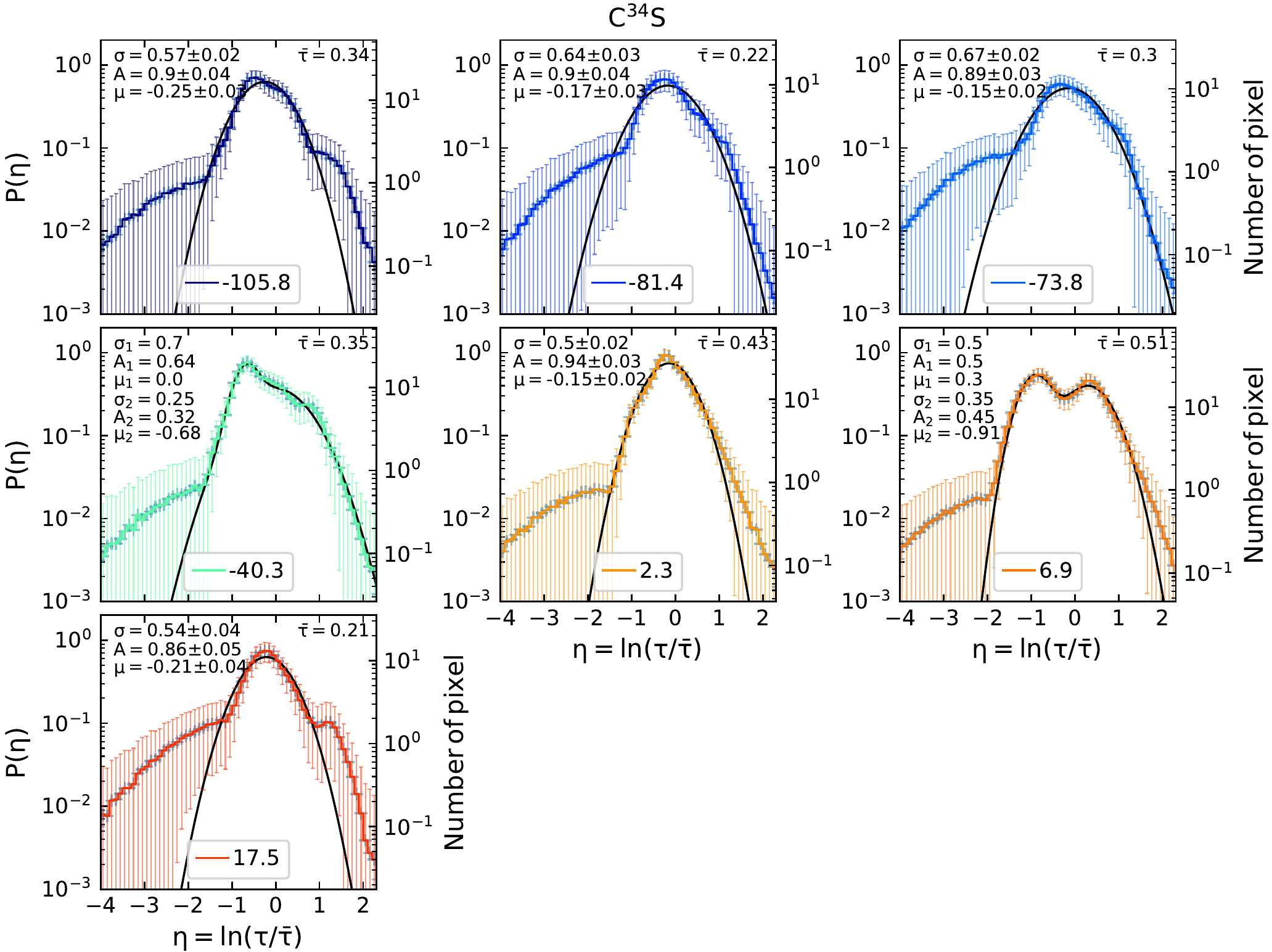}
\caption{Same as Fig.~\ref{pdf_c-c3h2}, but for C$^{34}$S. }
\label{pdf_c34s}
\end{figure*}

\begin{figure*}
\centering
\includegraphics[width=17cm, trim = 1.2cm 2.2cm 0.7cm 3.cm, clip=True]{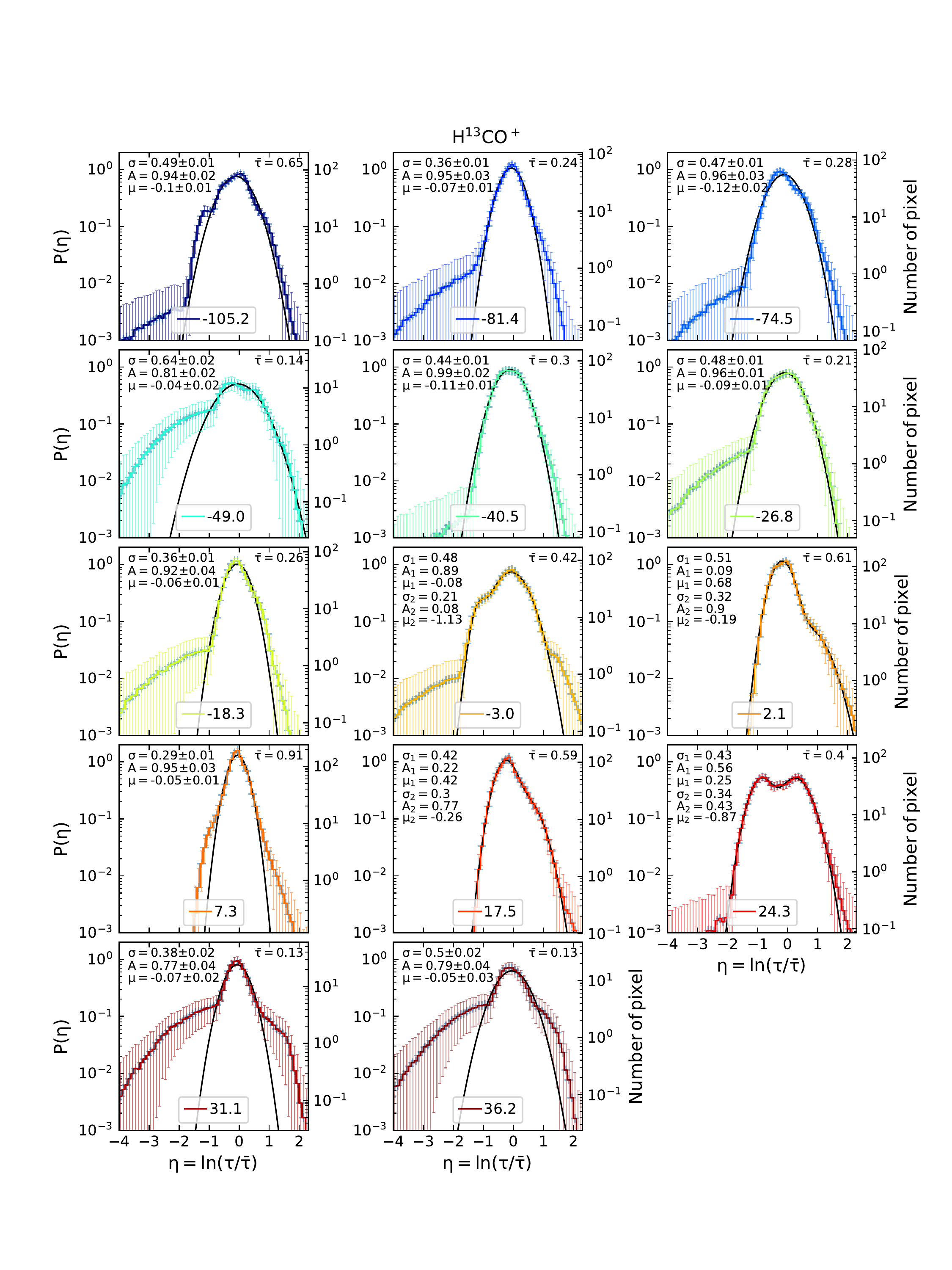}
\caption{Same as Fig.~\ref{pdf_c-c3h2}, but for H$^{13}$CO$^+$. }
\label{pdf_h13cop}
\end{figure*}

\begin{figure*}
\centering
\includegraphics[width=17cm, trim = 1.2cm 2.2cm 0.7cm 3.cm, clip=True]{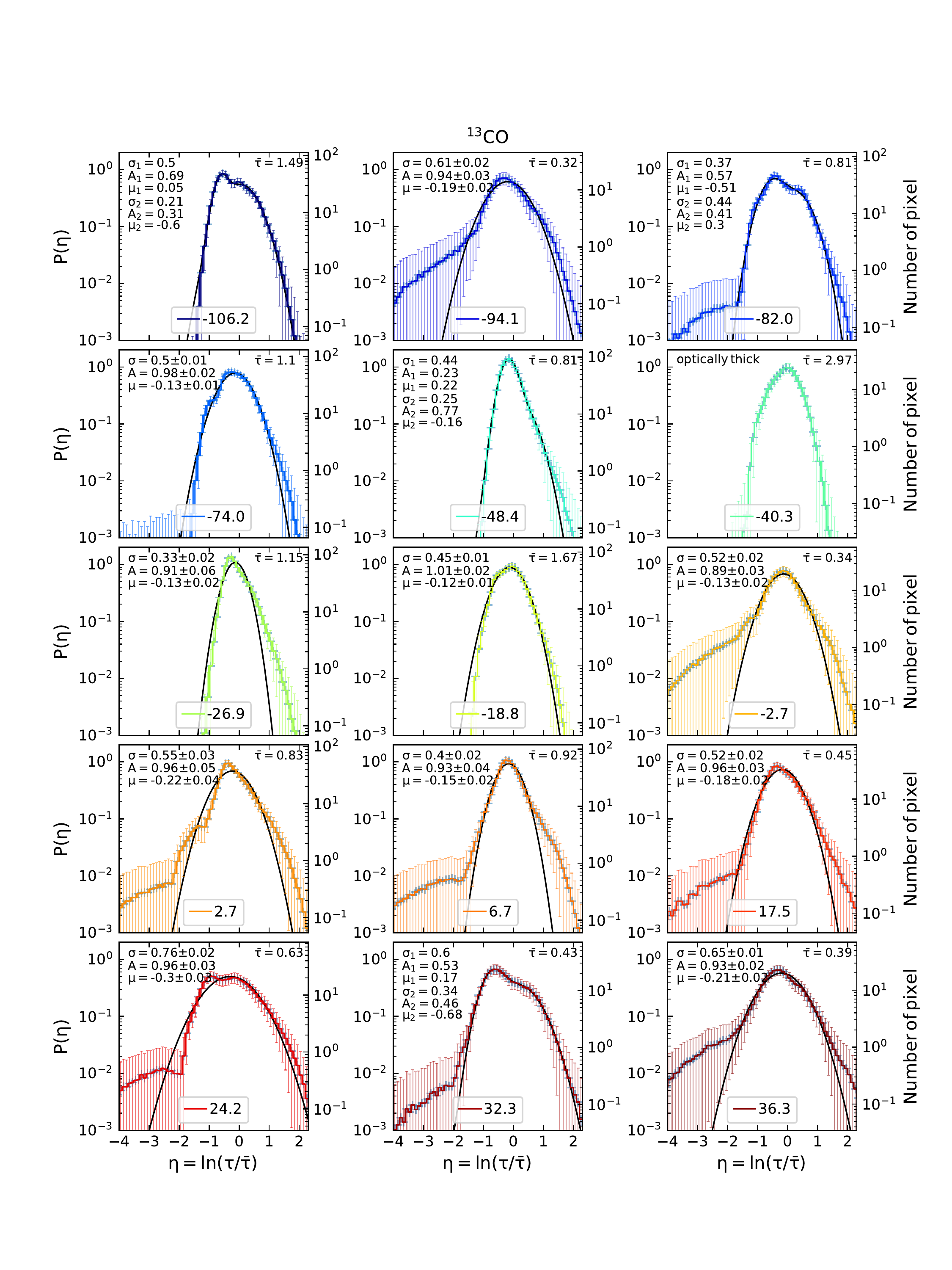}
\caption{Same as Fig.~\ref{pdf_c-c3h2}, but for $^{13}$CO.}
\label{pdf_13co}
\end{figure*}

\begin{figure*}
\centering
\includegraphics[width=17cm, trim = 1.2cm 2.2cm 0.7cm 3.cm, clip=True]{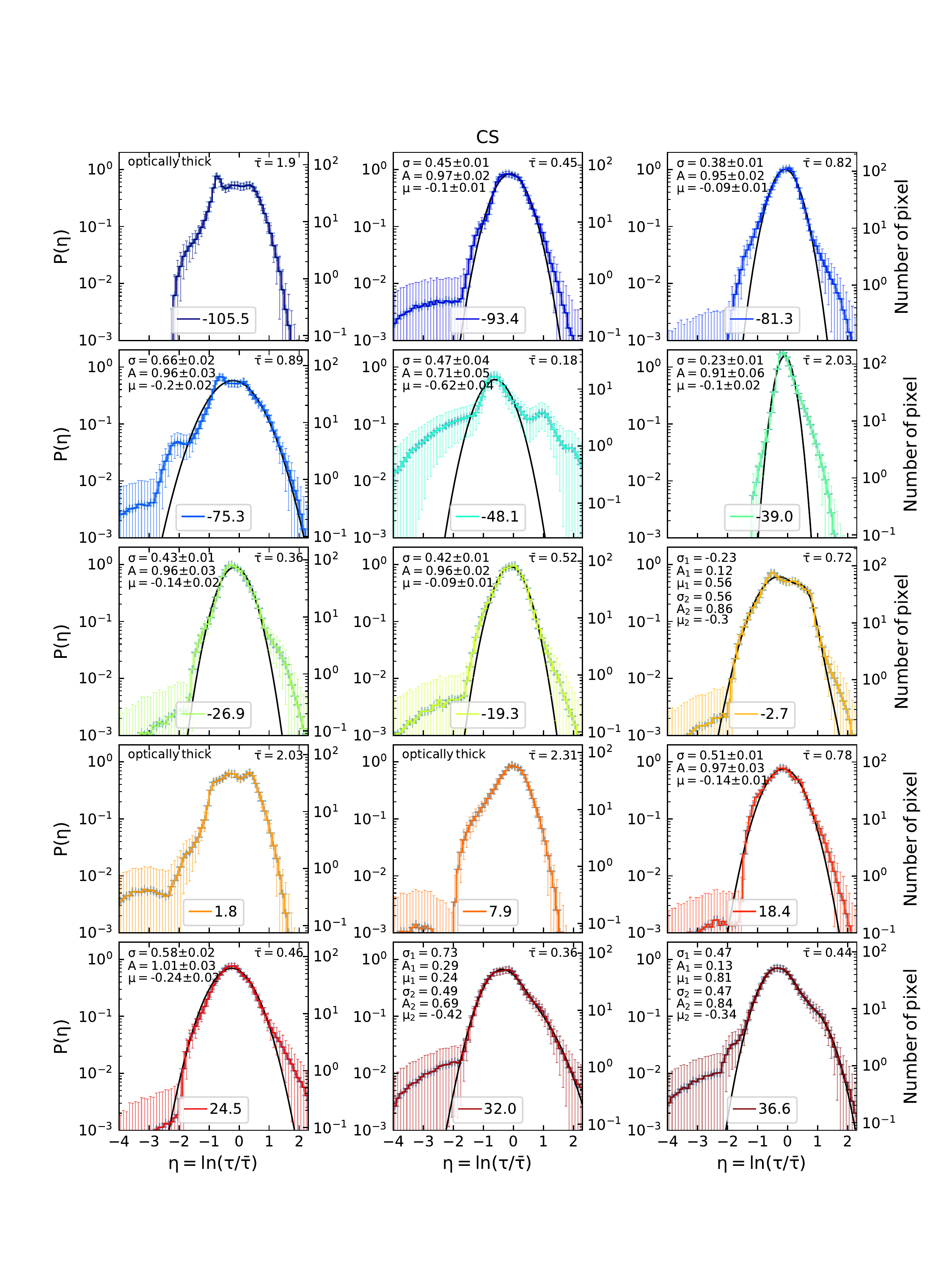}
\caption{Same as Fig.~\ref{pdf_c-c3h2}, but for CS. }
\label{pdf_cs}
\end{figure*}

\begin{figure*}
\centering
\includegraphics[width=17cm, trim = 1.2cm 2.0cm 0.7cm 3.cm, clip=True]{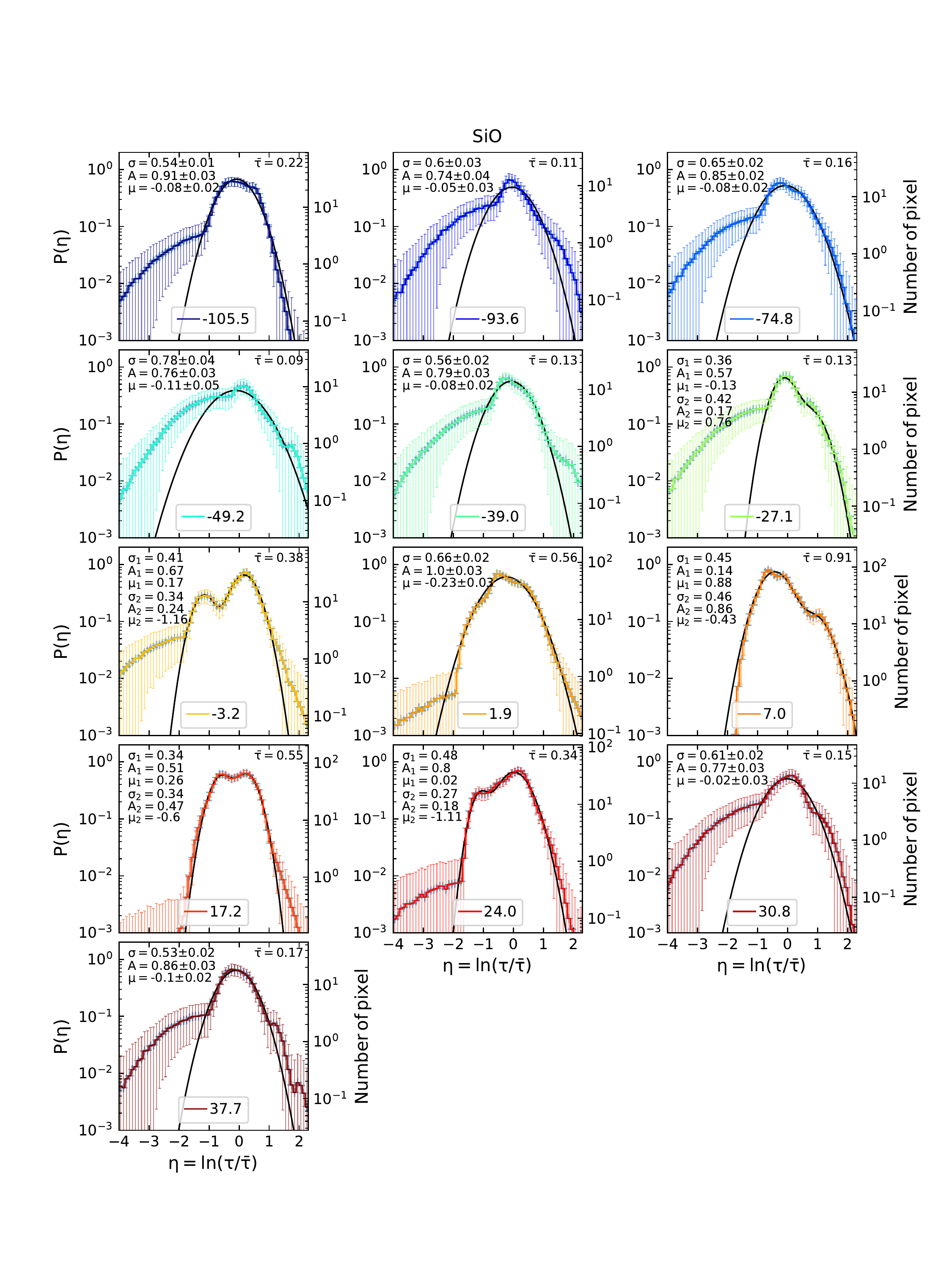}
\caption{Same as Fig.~\ref{pdf_c-c3h2}, but for SiO. }
\label{pdf_sio}
\end{figure*}

\begin{figure*}
\centering
\includegraphics[width=17cm, trim = 1.2cm 11.8cm 0.7cm 3.cm, clip=True]{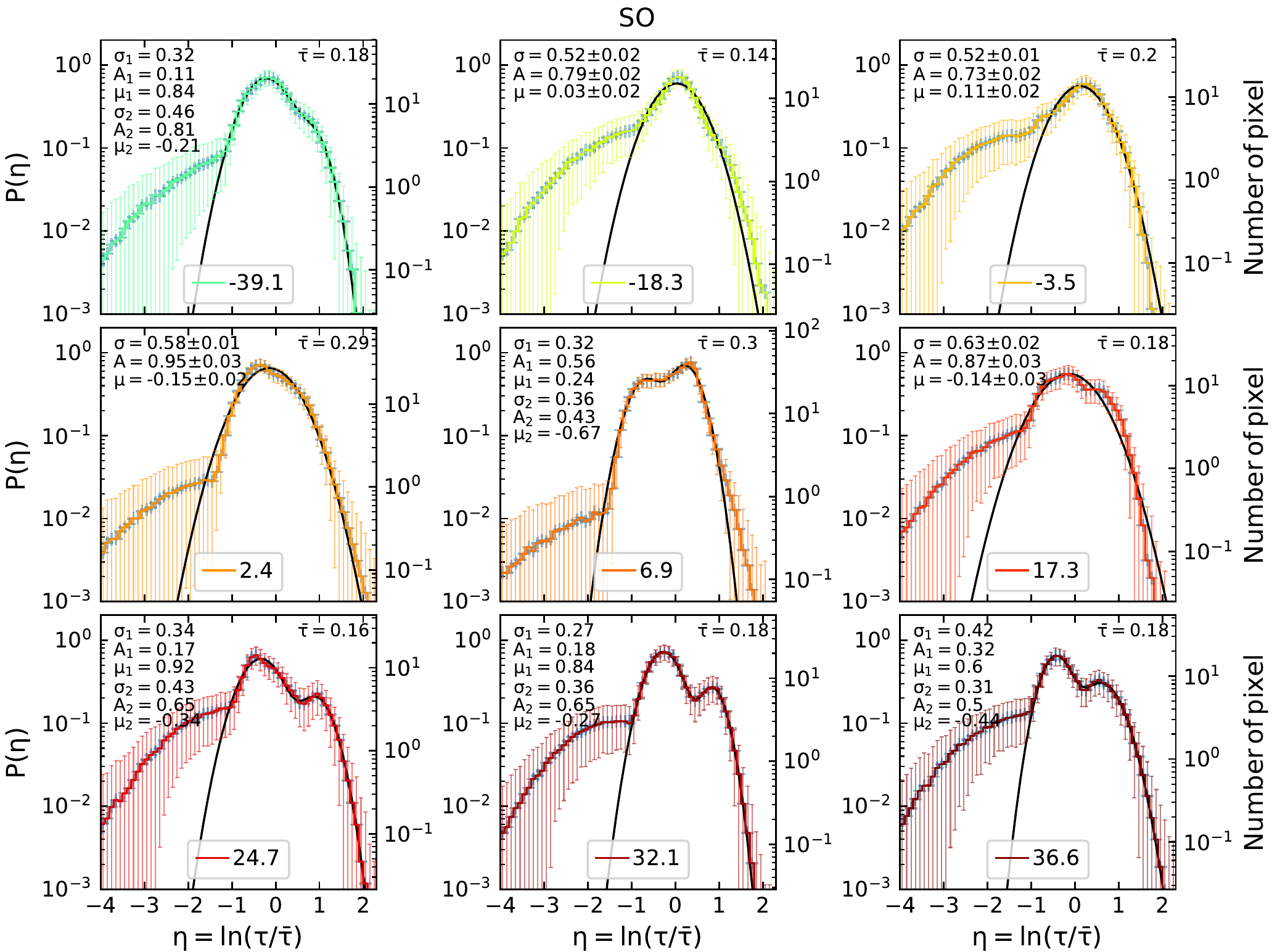}
\caption{Same as Fig.~\ref{pdf_c-c3h2}, but for SO. }
\label{pdf_so}
\end{figure*}

\begin{figure*}
\centering
\includegraphics[width=17cm, trim = 1.2cm 16.6cm 0.7cm 3.cm, clip=True]{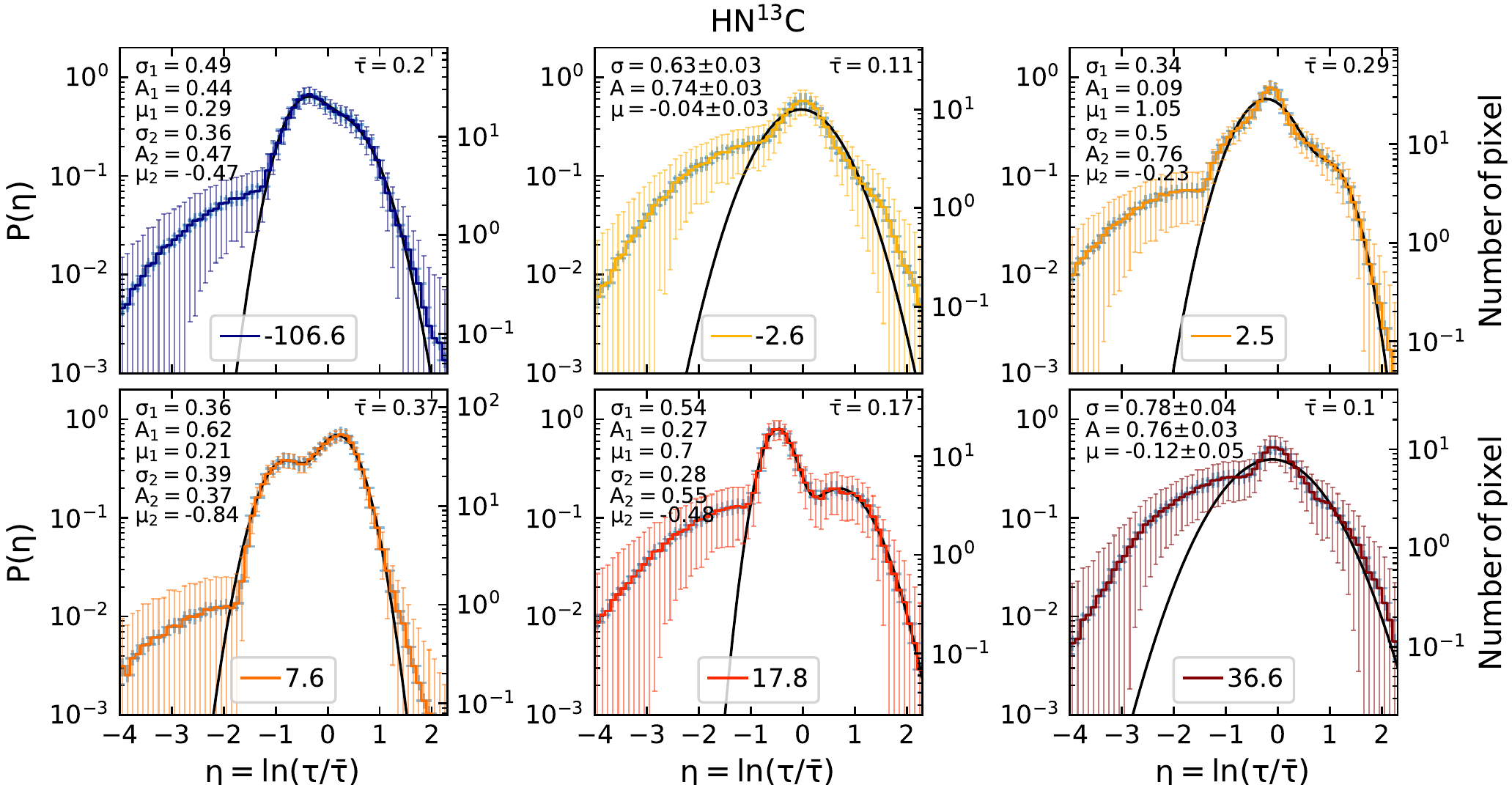}
\caption{Same as in Fig.~\ref{pdf_c-c3h2}, but for HN$^{13}$C. }
\label{pdf_hn13c}
\end{figure*}

\begin{figure*}
\centering
\includegraphics[width=17cm, trim = 1.2cm 2.2cm 0.7cm 3.cm, clip=True]{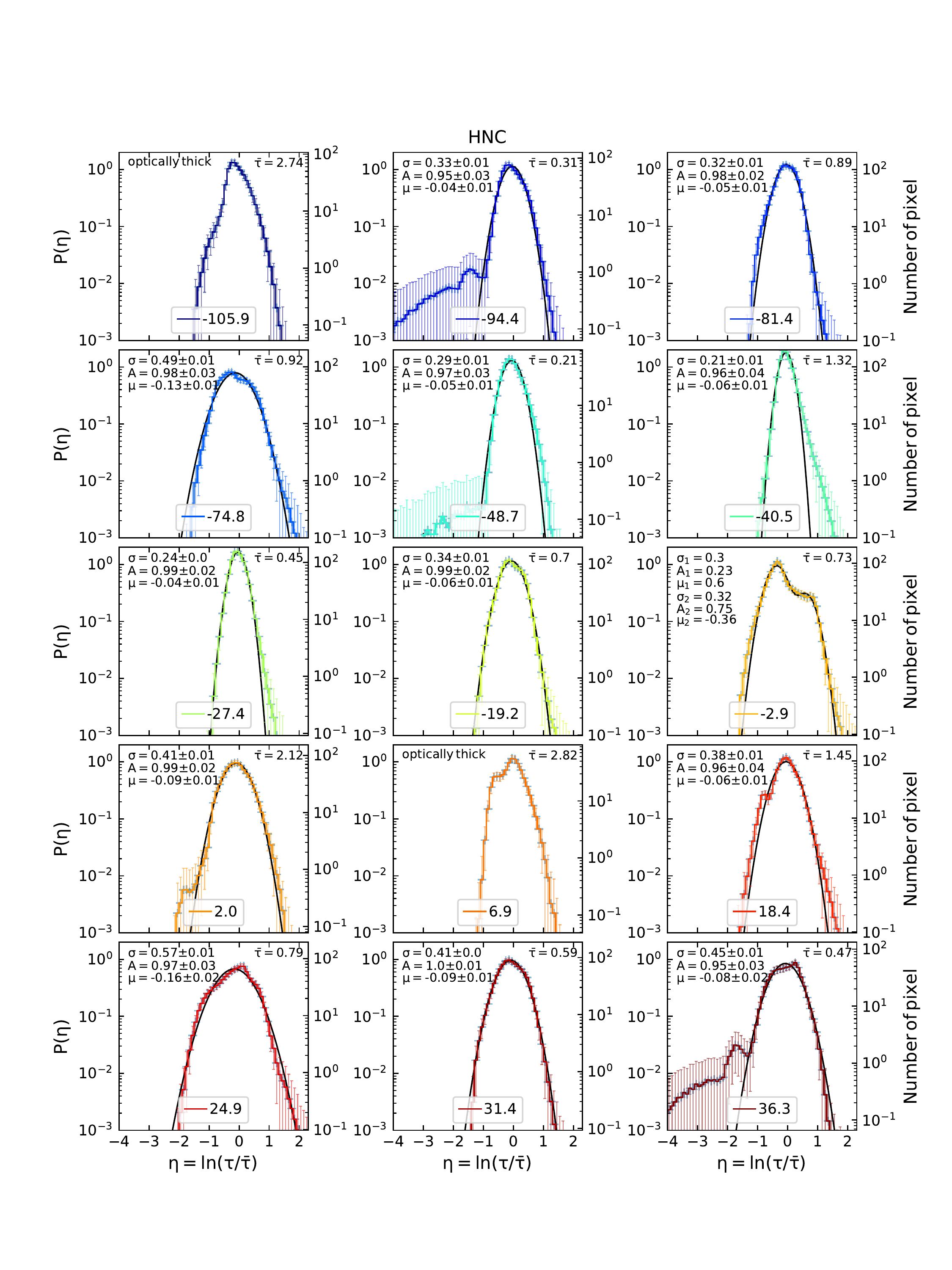}
\caption{Same as Fig.~\ref{pdf_c-c3h2}, but for HNC. }
\label{pdf_hnc}
\end{figure*}

\begin{figure*}
\centering
\includegraphics[width=17cm, trim = 1.2cm 16.5cm 0.7cm 2.0cm, clip=True]{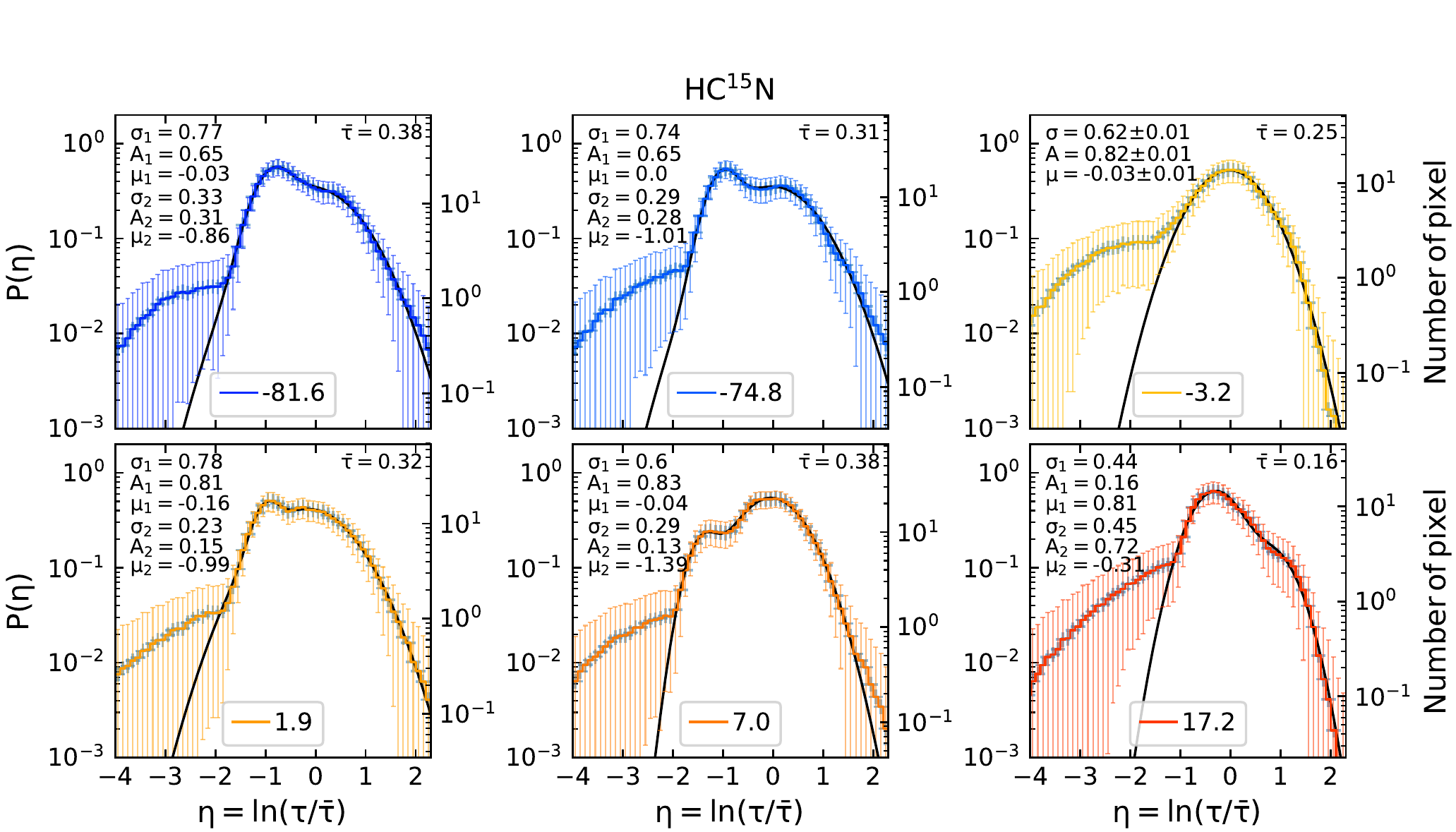}
\caption{Same as Fig.~\ref{pdf_c-c3h2}, but for HC$^{15}$N. }
\label{pdf_hc15n}
\end{figure*}

\begin{figure*}
\centering
\includegraphics[width=17cm, trim = 1.2cm 11.8cm 0.7cm 3.cm, clip=True]{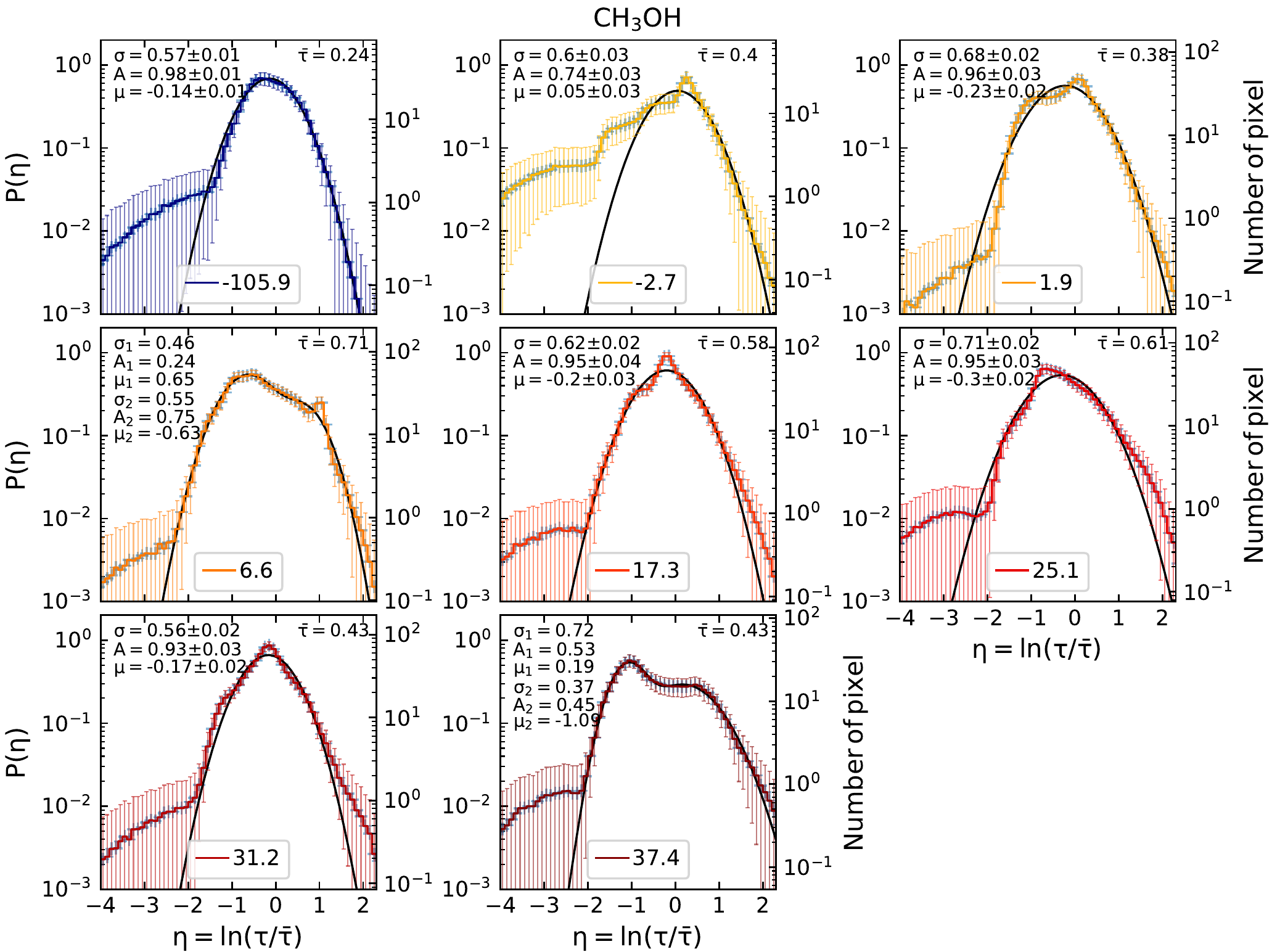}
\caption{Same as Fig.~\ref{pdf_c-c3h2}, but for CH$_3$OH. }
\label{pdf_ch3oh}
\end{figure*}

\begin{figure*}
\centering
\includegraphics[width=17cm, trim = 1.2cm 21.2cm 0.7cm 3.cm, clip=True]{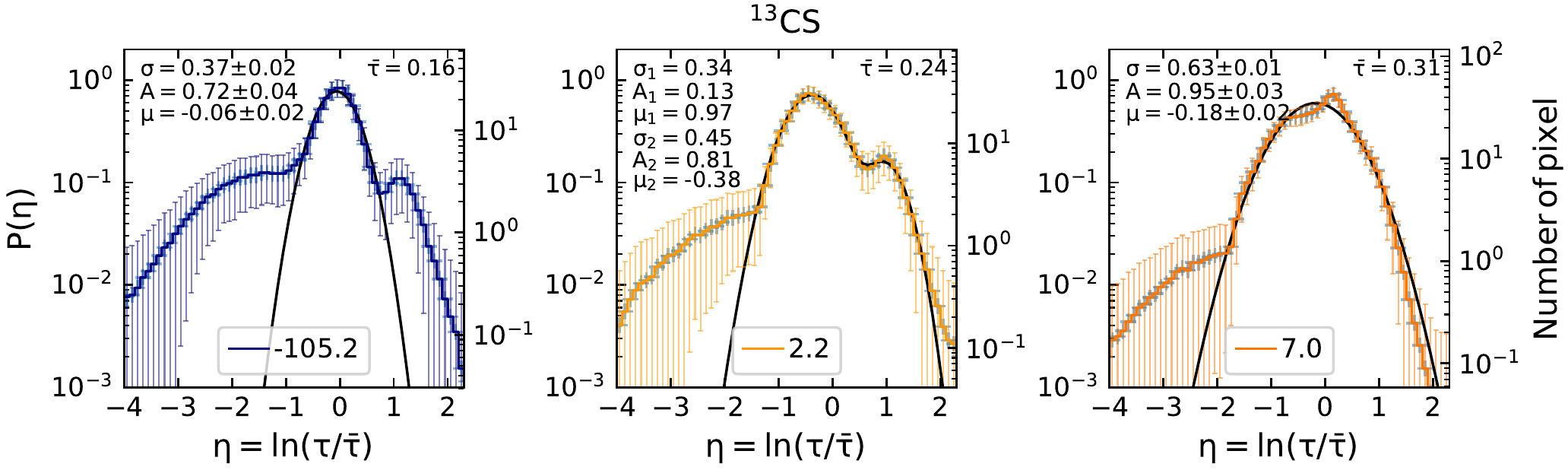}
\caption{Same as Fig.~\ref{pdf_c-c3h2}, but for $^{13}$CS. }
\label{pdf_13cs}
\end{figure*}

\begin{figure}
   \resizebox{\hsize}{!}{\includegraphics[width=0.5\textwidth,trim = 1.cm 1.0cm 2cm 2.95cm, clip=True]{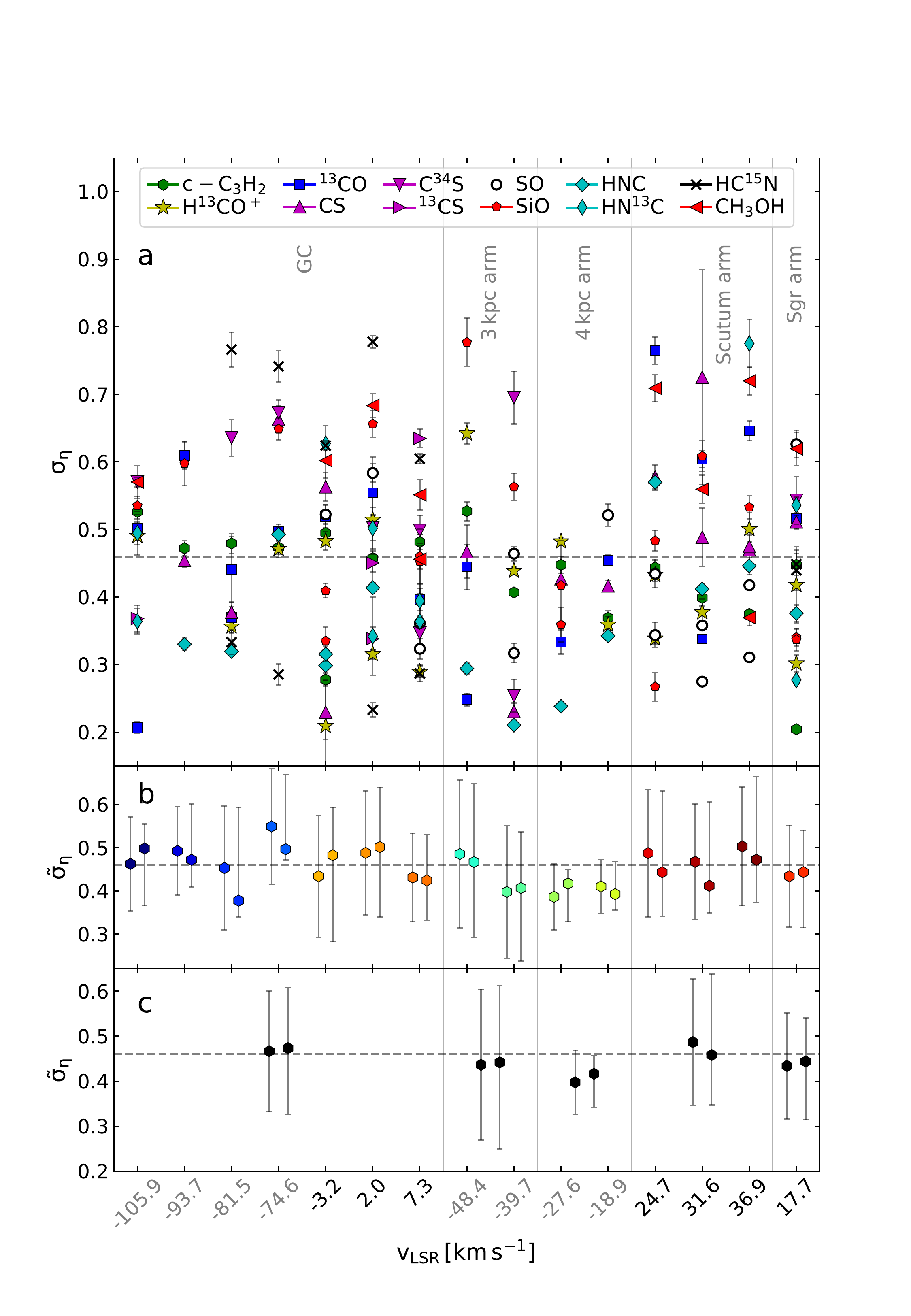}}
 \caption{\textbf{a} Widths of the Gaussians fitted to the PDFs of all molecules for 15 velocity components, roughly sorted by their distance to the Galactic centre. \textbf{b} Mean (left) and median (right) values for each velocity component. \textbf{c} Mean (left) and median (right) values for each sub-sample of clouds, from left to right: Galactic centre, 3~kpc arm, 4~kpc arm, Scutum arm, Sagittarius arm. The uncertainties represent the standard deviation for the mean and the corresponding percentiles for the median. The dashed line in panels b and c marks the mean value of all data points shown in panel a. Velocity components belonging to Category I and II are coloured in grey and black, respectively.}\label{sigma_pdf_vel_gauss}
\end{figure}

\begin{figure}
   \resizebox{\hsize}{!}{\includegraphics[width=0.5\textwidth,trim = 1.cm 0.3cm 2cm 2.2cm, clip=True]{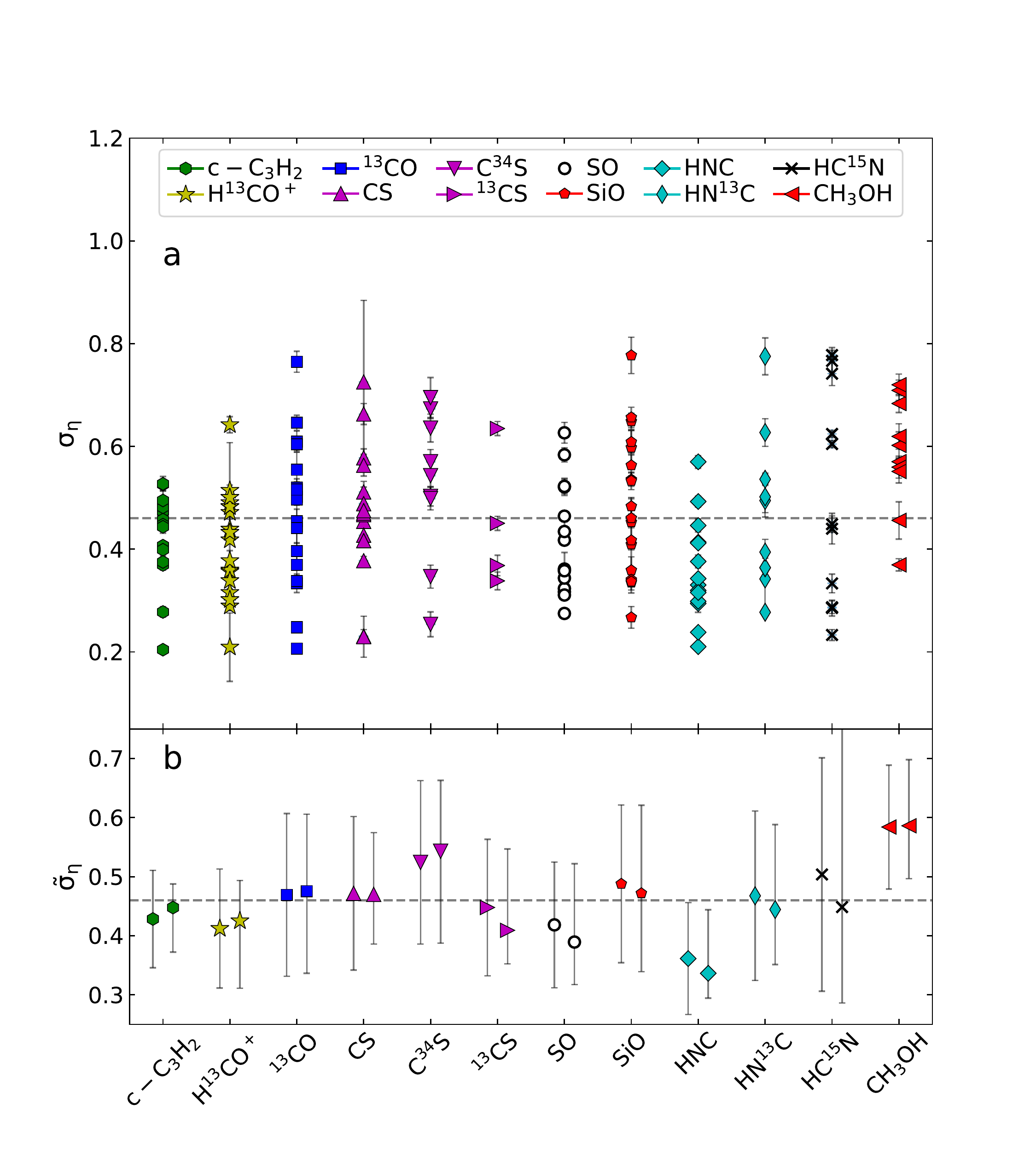}}
 \caption{\textbf{a} Widths of the Gaussians fitted to the PDFs of 15 velocity components sorted by molecule. \textbf{b} Mean (left) and median (right) values for each molecule. The uncertainties represent the standard deviation for the mean and the corresponding percentiles for the median. The dashed line marks the mean value for all components (see Fig.~\ref{sigma_pdf_vel_gauss}). }\label{sigma_pdf_mol_gauss}
\end{figure}

The PDFs of all molecules but c-C$_3$H$_2$ are shown in Figs.\ref{pdf_c34s} to \ref{pdf_13cs}. The results of the Gaussian fits to the PDFs are shown in Fig.~\ref{sigma_pdf_vel_gauss} and \ref{sigma_pdf_mol_gauss}, and listed in Tables~\ref{sigma_pdf_spiralarms_gauss} and \ref{sigma_pdf_molecules_gauss}.

\begin{table*}
\caption{Number of Gaussians fitted to the PDF.}
\label{number_gaussians}
\centering
\begin{tabular}{rcccccccccccc}       
\hline               
velocity & c-C$_3$H$_2$-o & H$^{13}$CO$^+$ & $^{13}$CO & CS & C$^{34}$S & $^{13}$CS & SO & SiO & HNC & HN$^{13}$C & HC$^{15}$N & CH$_3$OH\\

[km\,s$^{-1}$] & & & & & & & & & & & & \\
\hline\hline
  \multicolumn{13}{c}{Galactic centre}\\
$-105.9$& 1 & 1 & 2 & o.t. & 1 & 1 & n & 1 & o.t. & 2 & n & 1 \\
$-93.7$ & 1 & n & 1 & 1 & n & n & n & 1 & 1 & n & n & n \\
$-81.5$ & 1 & 1 & 2 & 1 & 1 & n & n & n & 1 & n & 2 & n \\
$-74.6$ & 1 & 1 & 1 & 1 & 1 & n & n & 1 & 1 & n & 2 & n \\
$-3.2$  & 2 & 2 & 1 & 2 & n & n & 1 & 2 & 2 & 1 & 1 & 1 \\
$2.0$   & 1 & 2 & 1 & o.t. & 1 & 2 & 1 & 1 & 1 & 2 & 2 & 1 \\
$7.3$   & 1 & 1 & 1 & o.t. & 2 & 1 & 2 & 2 & o.t. & 2 & 2 & 2 \\
 \hline
  \multicolumn{13}{c}{3\,kpc arm}\\
$-48.4$ & 1 & 1 & 2 & 1 & n & n & n & 1 & 1 & n & n & n \\
$-39.7$ & 1 & 1 & o & 1 & 2 & n & 1 & 1 & 1 & n & n & n \\
 \hline
  \multicolumn{13}{c}{4\,kpc arm}\\
$-27.6$ & 1 & 1 & 1 & 1 & n & n & n & 2 & 1 & n & n & n \\
$-18.9$ & 1 & 1 & 1 & 1 & n & n & 1 & n & 1 & n & n & n \\
 \hline
  \multicolumn{13}{c}{Scutum arm}\\
$24.7$  & 1 & 2 & 1 & 1 & n & n & 2 & 2 & 1 & n & n & 2 \\
$31.6$  & 1 & 1 & 2 & 2 & n & n & 2 & 1 & 1 & n & n & 1 \\
$36.9$  & 1 & 1 & 1 & 2 & n & n & 2 & 1 & 1 & 1 & n & 2 \\
 \hline
  \multicolumn{13}{c}{Sagittarius arm}\\
$17.7$  & 2 & 2 & 1 & 1 & 1 & n & 1 & 2 & 1 & 2 & 2 & 1 \\
\hline                      
\end{tabular}
\tablefoot{The components for which the SNR is too low are marked with n and the ones for which more than $10\%$ are optically thick are labelled with o.t..}
\end{table*}

\begin{table}
\caption{Mean ($\bar{\sigma}$) and median ($\tilde{\sigma}$) widths of the Gaussians fitted to PDFs for each velocity component.}
\label{sigma_pdf_spiralarms_gauss}
\centering
\begin{tabular}{r c c c c}       
\hline               
\multicolumn{1}{c}{$\varv_\mathrm{LSR}$\tablefootmark{a}} & $\bar{\sigma}$ & $\tilde{\sigma}$ & $\bar{\sigma}$\tablefootmark{b} & $\tilde{\sigma}$\tablefootmark{b}\\

 [km\,s$^{-1}$] & & & & \\
\hline\hline   
\multicolumn{5}{c}{Galactic centre}\\
$-105.9$ & $0.46\pm 0.11$ & $0.50\substack{+0.06 \\ -0.13}$ & \multirow{7}{*}{$0.47\pm0.13$} & \multirow{7}{*}{$0.47\substack{+0.13 \\ -0.15}$}\\
 $-93.7$ & $0.49\pm 0.10$ & $0.47\substack{+0.13 \\ -0.06}$ & &\\
 $-81.5$ & $0.45\pm 0.14$ & $0.38\substack{+0.22 \\ -0.04}$ & &\\
 $-74.6$ & $0.55\pm 0.13$ & $0.50\substack{+0.17 \\ -0.02}$ & &\\
 $-3.2$ & $0.43\pm 0.14$ & $0.48\substack{+0.11 \\ -0.20}$ & &\\
 $2.0$ & $0.49\pm 0.14$ & $0.50\substack{+0.14 \\ -0.16}$ & &\\
 $7.3$ & $0.43\pm 0.10$ & $0.42\substack{+0.11 \\ -0.09}$ & &\\
 \hline
 \multicolumn{5}{c}{3\,kpc arm}\\
$-48.4$ & $0.49\pm 0.17$ & $0.47\substack{+0.18 \\ -0.18}$ & \multirow{2}{*}{$0.44\pm0.17$} & \multirow{2}{*}{$0.44\substack{+0.17 \\ -0.19}$}\\
$-39.7$ & $0.40\pm 0.15$ & $0.41\substack{+0.13 \\ -0.17}$ & &\\
 \hline
 \multicolumn{5}{c}{4\,kpc arm}\\
 $-27.6$ & $0.39\pm 0.08$ & $0.42\substack{+0.03 \\ -0.09}$ & \multirow{2}{*}{$0.40\pm0.07$} & \multirow{2}{*}{$0.42\substack{+0.04 \\ -0.07}$}\\
$-18.9$ & $0.41\pm 0.06$ & $0.39\substack{+0.08 \\ -0.04}$ & &\\
 \hline
  \multicolumn{5}{c}{Scutum arm}\\
$24.7$ & $0.49\pm 0.15$ & $0.44\substack{+0.19 \\ -0.10}$ & \multirow{3}{*}{$0.49\pm0.14$} & \multirow{3}{*}{$0.46\substack{+0.18 \\ -0.11}$}\\
 $31.6$ & $0.47\pm 0.13$ & $0.41\substack{+0.19 \\ -0.06}$ & &\\
 $36.9$ & $0.50\pm 0.14$ & $0.47\substack{+0.19 \\ -0.10}$ & &\\
 \hline
  \multicolumn{5}{c}{Sagittarius arm}\\
 \hline
$17.7$ & $0.43\pm 0.12$ & $0.44\substack{+0.10 \\ -0.13}$ & \multirow{1}{*}{$0.43\pm0.12$} & \multirow{1}{*}{$0.44\substack{+0.10 \\ -0.13}$}\\
\end{tabular}
\tablefoot{\tablefoottext{a}{Channel velocities of c-C$_3$H$_2$.} \tablefoottext{b}{Mean and median values for each sub-sample of clouds.}}
\end{table}

\begin{table}
\caption{Mean ($\bar{\sigma}$) and median ($\tilde{\sigma}$) dispersion of the PDFs of the molecules.}
\label{sigma_pdf_molecules_gauss}
\centering
\begin{tabular}{l c c}       
\hline               
molecule & $\bar{\sigma}$ & $\tilde{\sigma}$ \\
\hline\hline   
c-C$_3$H$_2$-o & $0.43\pm 0.08$ & $0.45\substack{+0.04 \\ -0.08}$\\
H$^{13}$CO$^+$ & $0.41\pm 0.10$ & $0.43\substack{+0.07 \\ -0.11}$\\         
$^{13}$CO & $0.47\pm 0.14$ & $0.48\substack{+0.13 \\ -0.14}$\\
CS & $0.47\pm 0.13$ & $0.47\substack{+0.10 \\ -0.08}$\\
C$^{34}$S & $0.52\pm 0.14$ & $0.54\substack{+0.12 \\ -0.16}$\\
$^{13}$CS & $0.45\pm 0.12$ & $0.41\substack{+0.14 \\ -0.06}$\\
SO & $0.42\pm 0.11$ & $0.39\substack{+0.13 \\ -0.07}$\\
SiO & $0.49\pm 0.13$ & $0.47\substack{+0.15 \\ -0.13}$\\
HNC & $0.36\pm 0.09$ & $0.34\substack{+0.11 \\ -0.04}$\\
HN$^{13}$C & $0.47\pm 0.14$ & $0.44\substack{+0.14 \\ -0.09}$\\
HC$^{15}$N & $0.50\pm 0.20$ & $0.45\substack{+0.30 \\ -0.16}$\\
CH$_3$OH & $0.58\pm 0.10$ & $0.59\substack{+0.11 \\ -0.09}$\\
\hline                      
\end{tabular}
\end{table}


\section{Maps and coefficients of PCA}
\begin{figure*}
\centering
\includegraphics[width=17cm, trim = 4.cm 21.5cm 4.cm 3.2cm, clip=True]{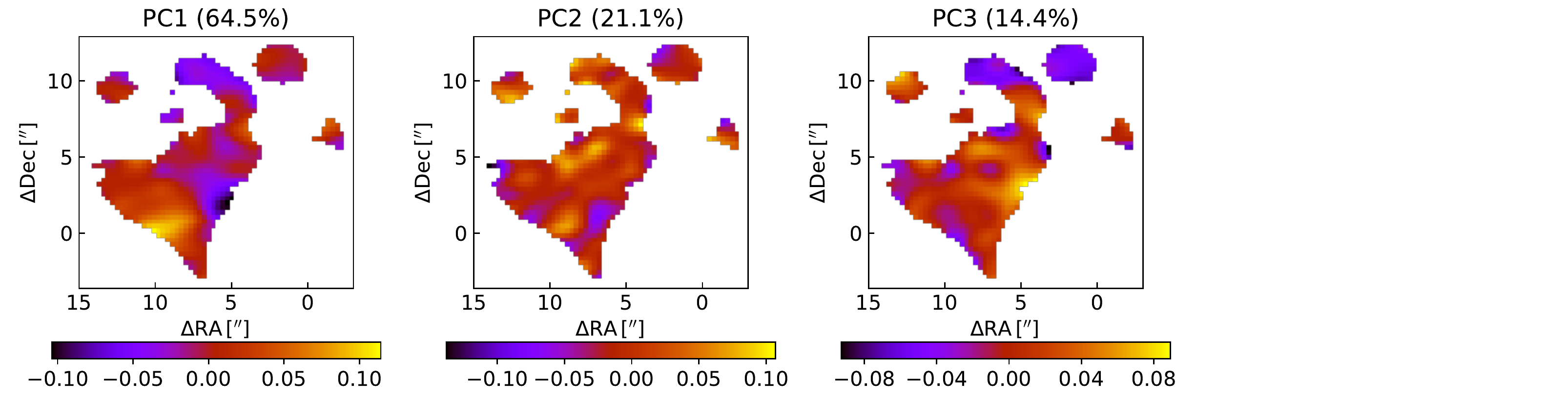}
\caption{Same as Fig.~\ref{pca_components_channel12} $\varv_\mathrm{LSR}=$-105.9\,km\,s$^{-1}$. } 
\label{pca_components_channel0}
\end{figure*}

\begin{figure*}
\centering
\includegraphics[width=17cm, trim = 4.cm 18.9cm 4.cm 3.cm, clip=True]{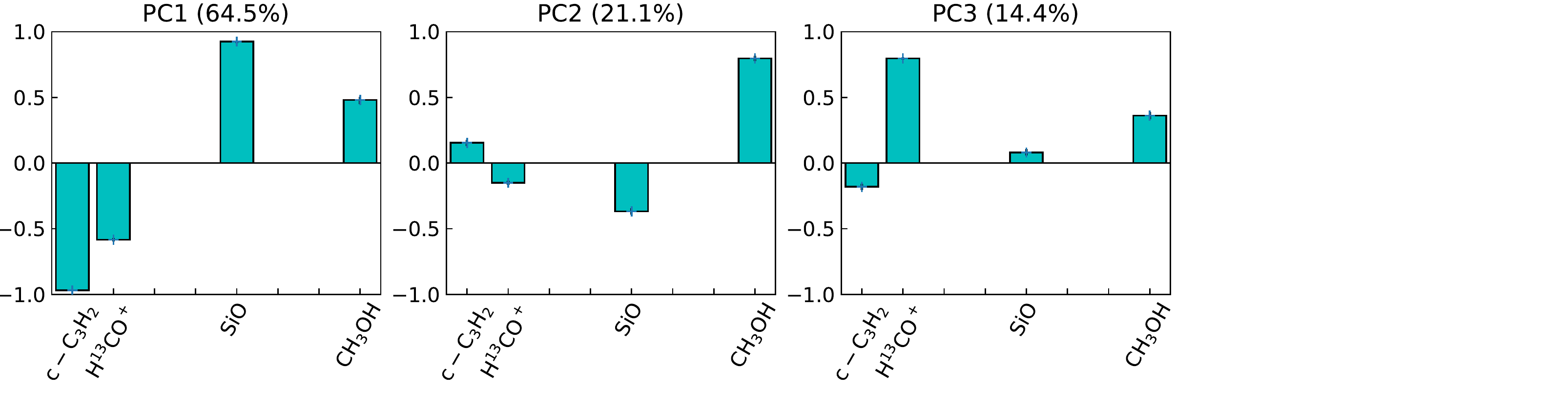}
\caption{Same as Fig.~\ref{pca_contr_channel12} but for $\varv_\mathrm{LSR}=$-105.9\,km\,s$^{-1}$. } 
\label{pca_contr_channel0}
\end{figure*}

\begin{figure*}
\centering
\includegraphics[width=17cm, trim = 4.cm 12.5cm 4.cm 3.2cm, clip=True]{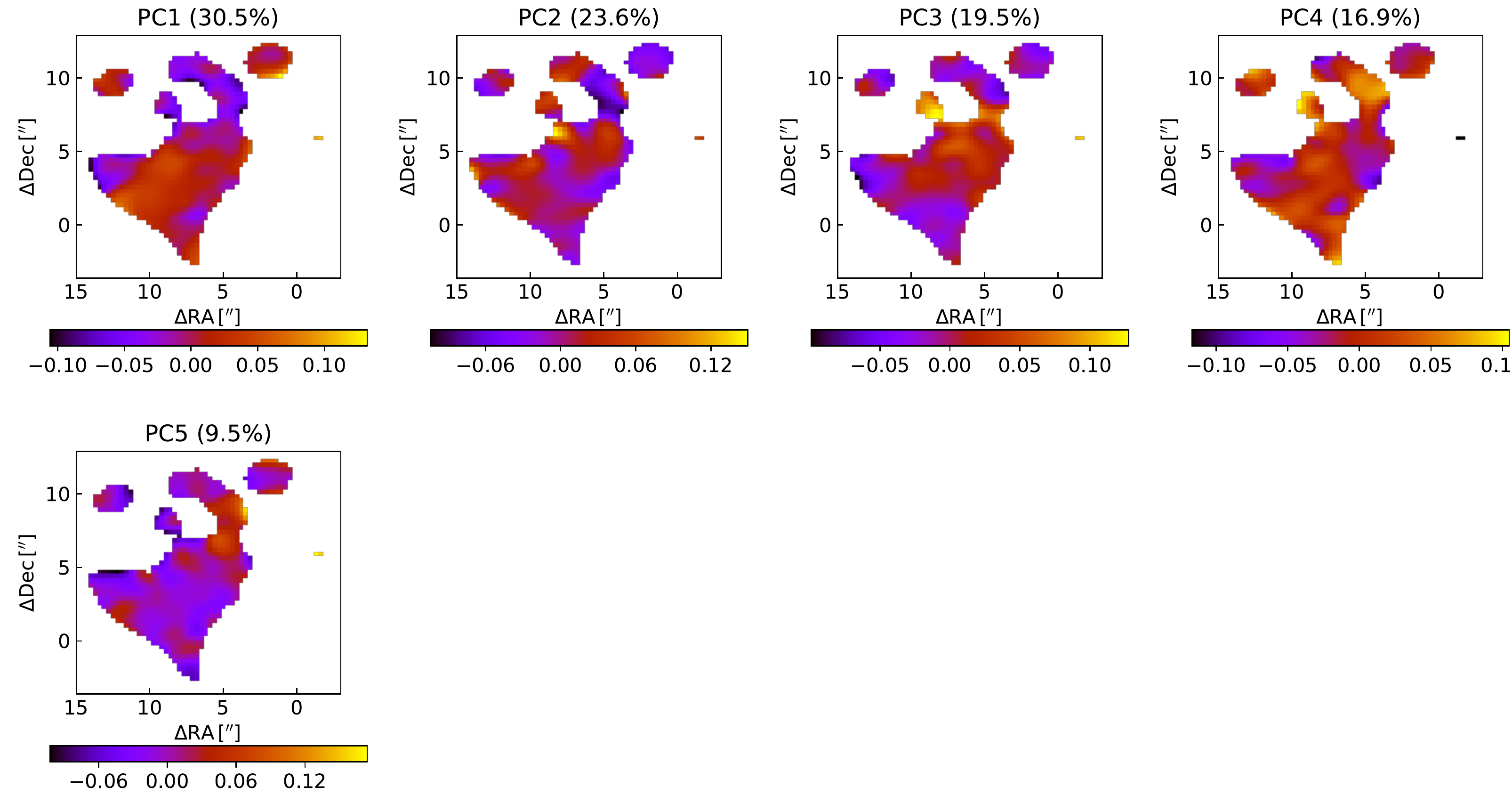}
\caption{Same as Fig.~\ref{pca_components_channel12} $\varv_\mathrm{LSR}=$2.0\,km\,s$^{-1}$. } 
\label{pca_components_channel9}
\end{figure*}

\begin{figure*}
\centering
\includegraphics[width=17cm, trim = 4.cm 9.8cm 4.cm 3.cm, clip=True]{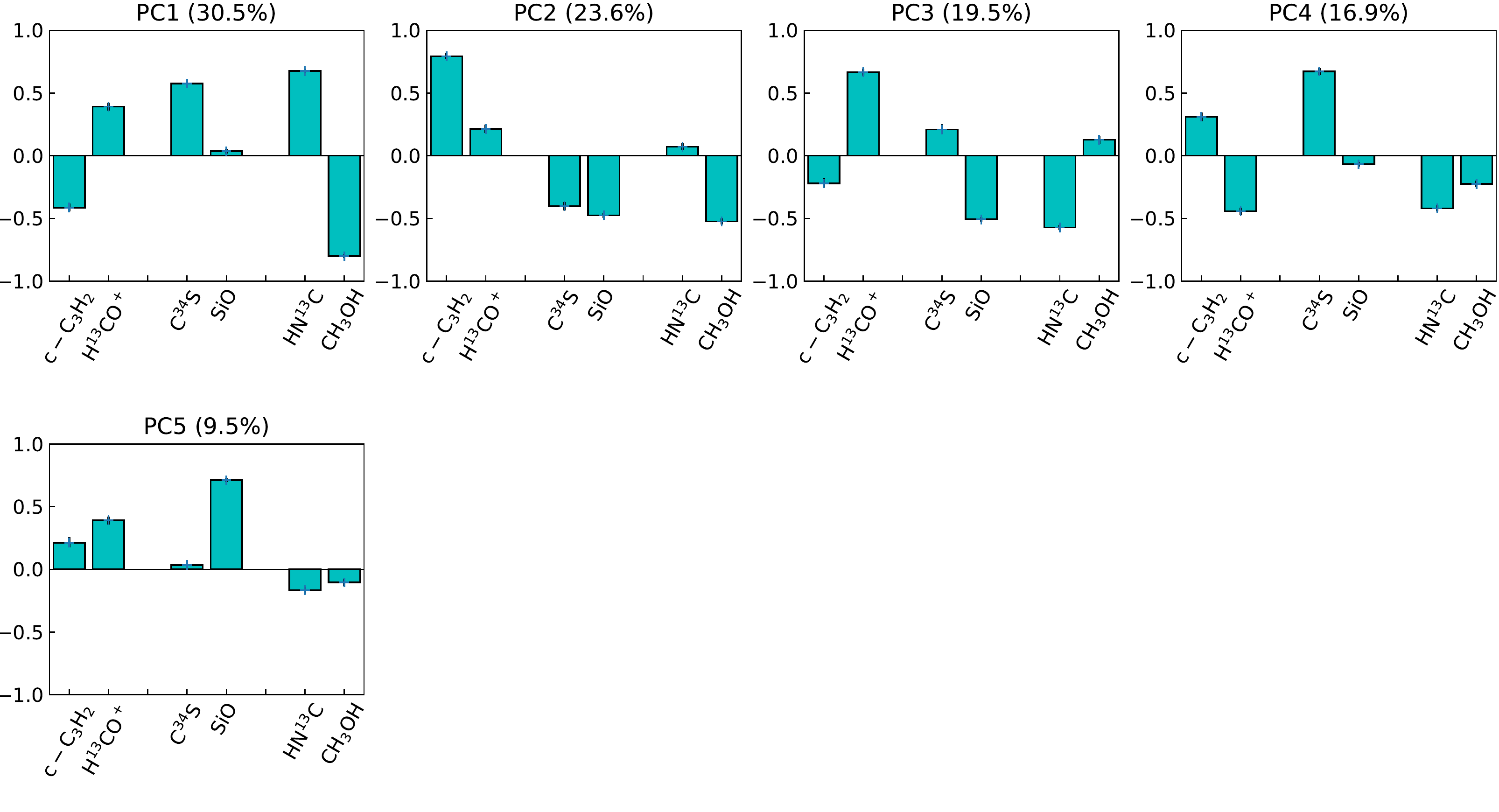}
\caption{Same as Fig.~\ref{pca_contr_channel12} but for $\varv_\mathrm{LSR}=$2.0\,km\,s$^{-1}$. } 
\label{pca_contr_channel9}
\end{figure*}

\begin{figure*}
\centering
\includegraphics[width=17cm, trim = 4.cm 21.5cm 4.cm 3.2cm, clip=True]{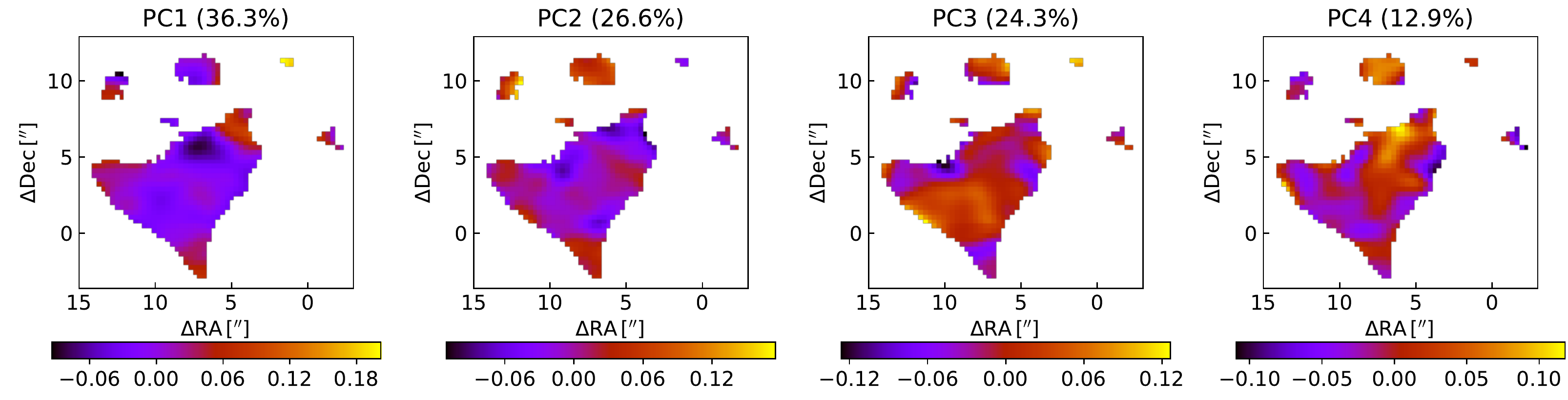}
\caption{Same as Fig.~\ref{pca_components_channel12} $\varv_\mathrm{LSR}=$7.3\,km\,s$^{-1}$. } 
\label{pca_components_channel10}
\end{figure*}

\begin{figure*}
\centering
\includegraphics[width=17cm, trim = 4.cm 18.9cm 4.cm 3.cm, clip=True]{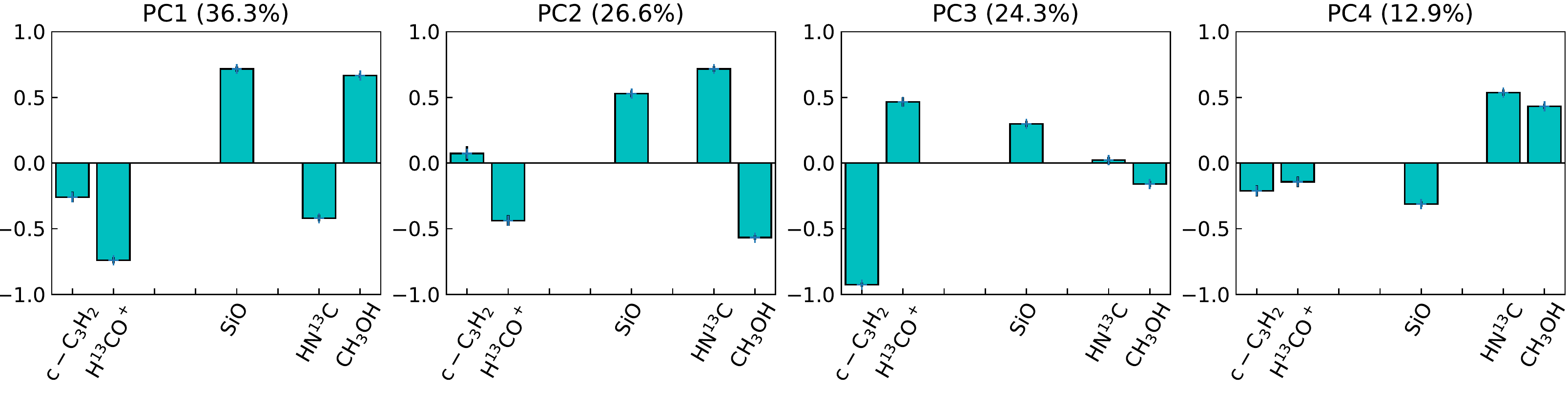}
\caption{Same as Fig.~\ref{pca_contr_channel12} but for $\varv_\mathrm{LSR}=$7.3\,km\,s$^{-1}$. } 
\label{pca_contr_channel10}
\end{figure*}

\begin{figure*}
\centering
\includegraphics[width=17cm, trim = 4.cm 21.5cm 4.cm 3.2cm, clip=True]{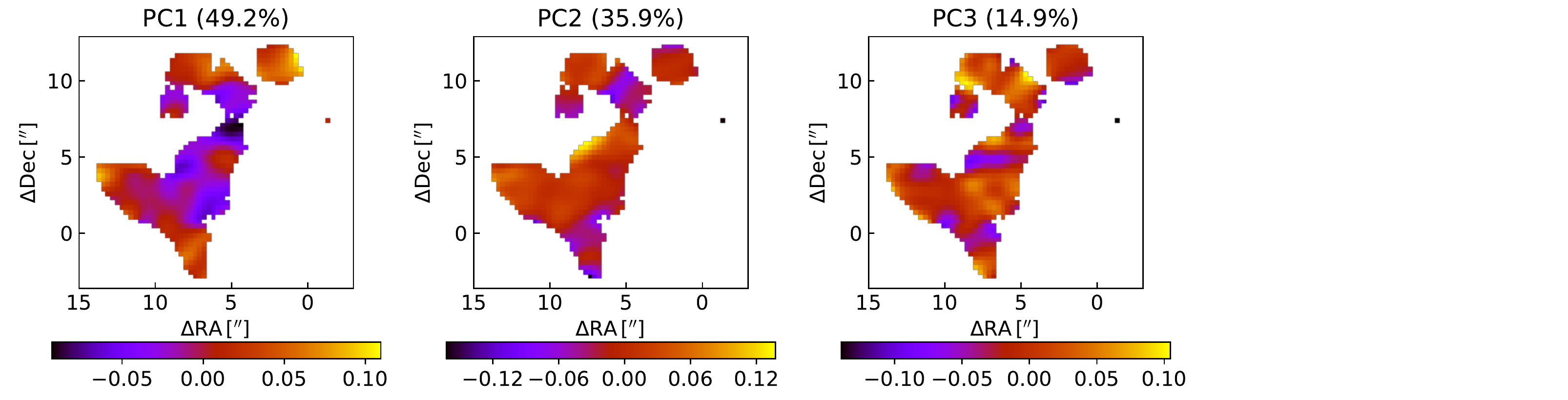}
\caption{Same as Fig.~\ref{pca_components_channel12} $\varv_\mathrm{LSR}=$31.6\,km\,s$^{-1}$. } 
\label{pca_components_channel13}
\end{figure*}

\begin{figure*}
\centering
\includegraphics[width=17cm, trim = 4.cm 18.9cm 4.cm 3.cm, clip=True]{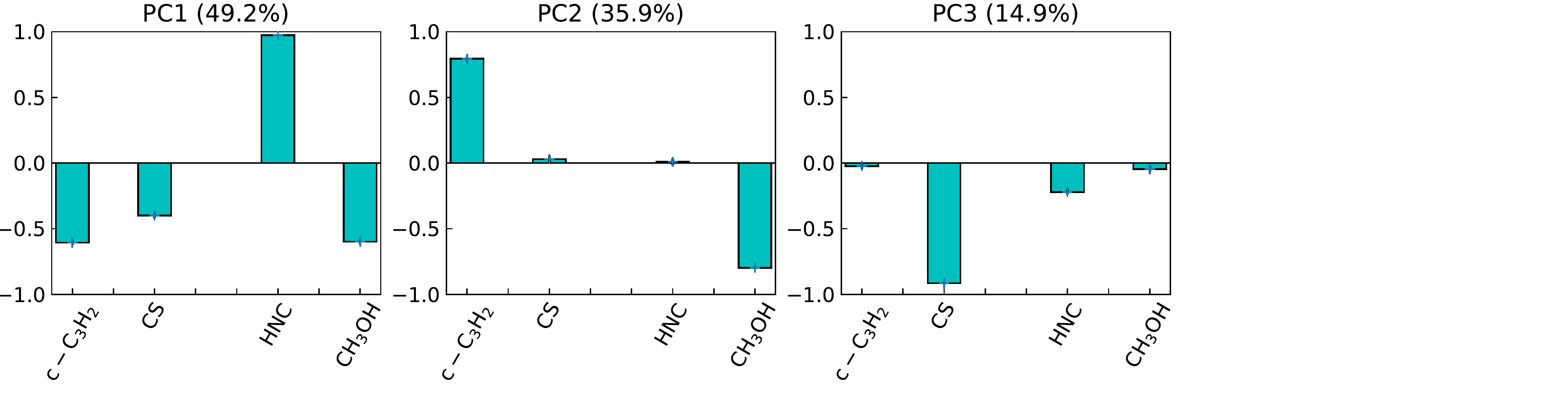}
\caption{Same as Fig.~\ref{pca_contr_channel12} but for $\varv_\mathrm{LSR}=$31.6\,km\,s$^{-1}$. } 
\label{pca_contr_channel13}
\end{figure*}

\begin{figure*}
\centering
\includegraphics[width=17cm, trim = 4.cm 12.5cm 4.cm 3.2cm, clip=True]{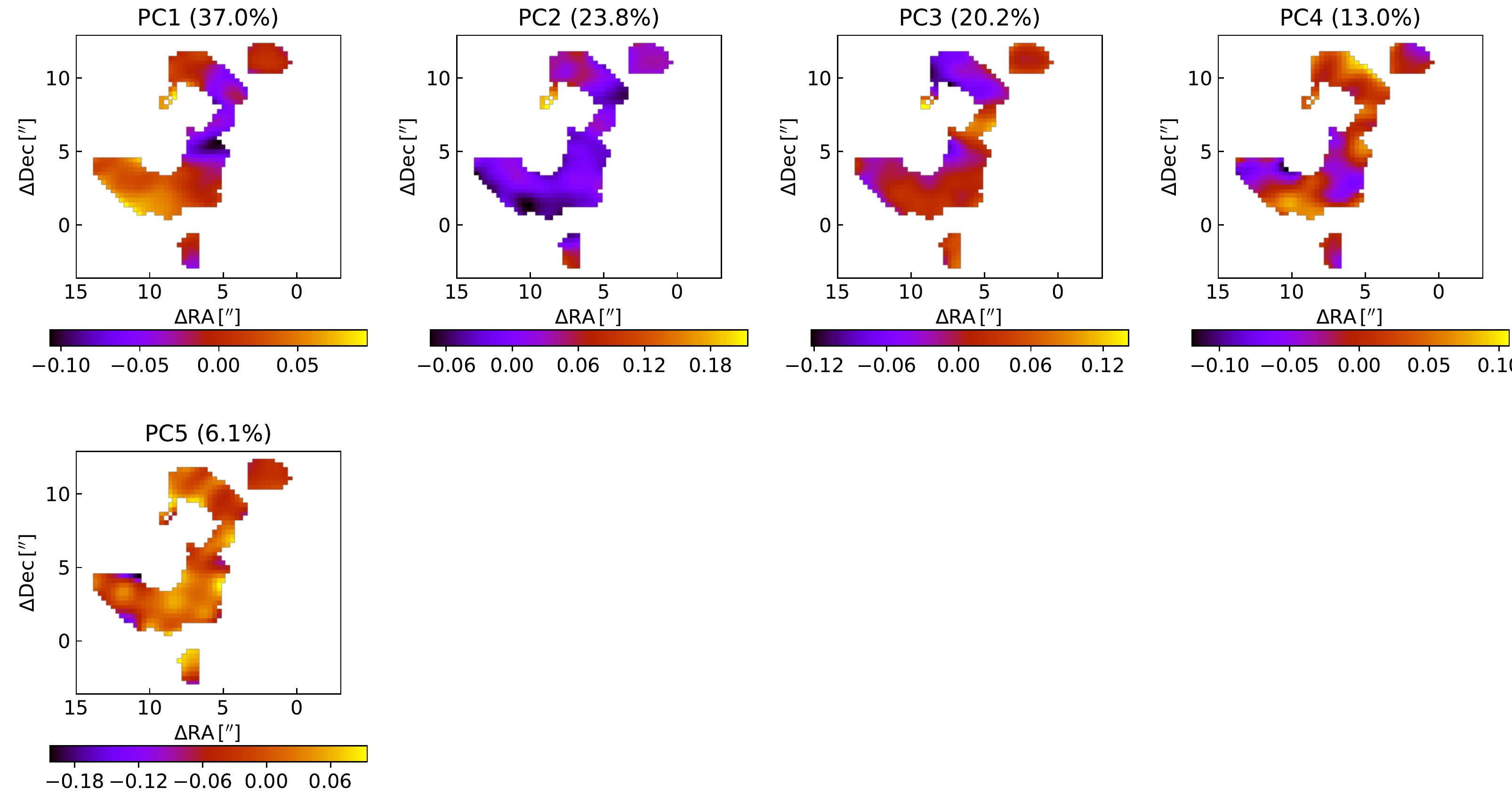}
\caption{Same as Fig.~\ref{pca_components_channel12} but for $\varv_\mathrm{LSR}=$17.7\,km\,s$^{-1}$. } 
\label{pca_components_channel11}
\end{figure*}

\begin{figure*}
\centering
\includegraphics[width=17cm, trim = 4.cm 9.8cm 4.cm 2.9cm, clip=True]{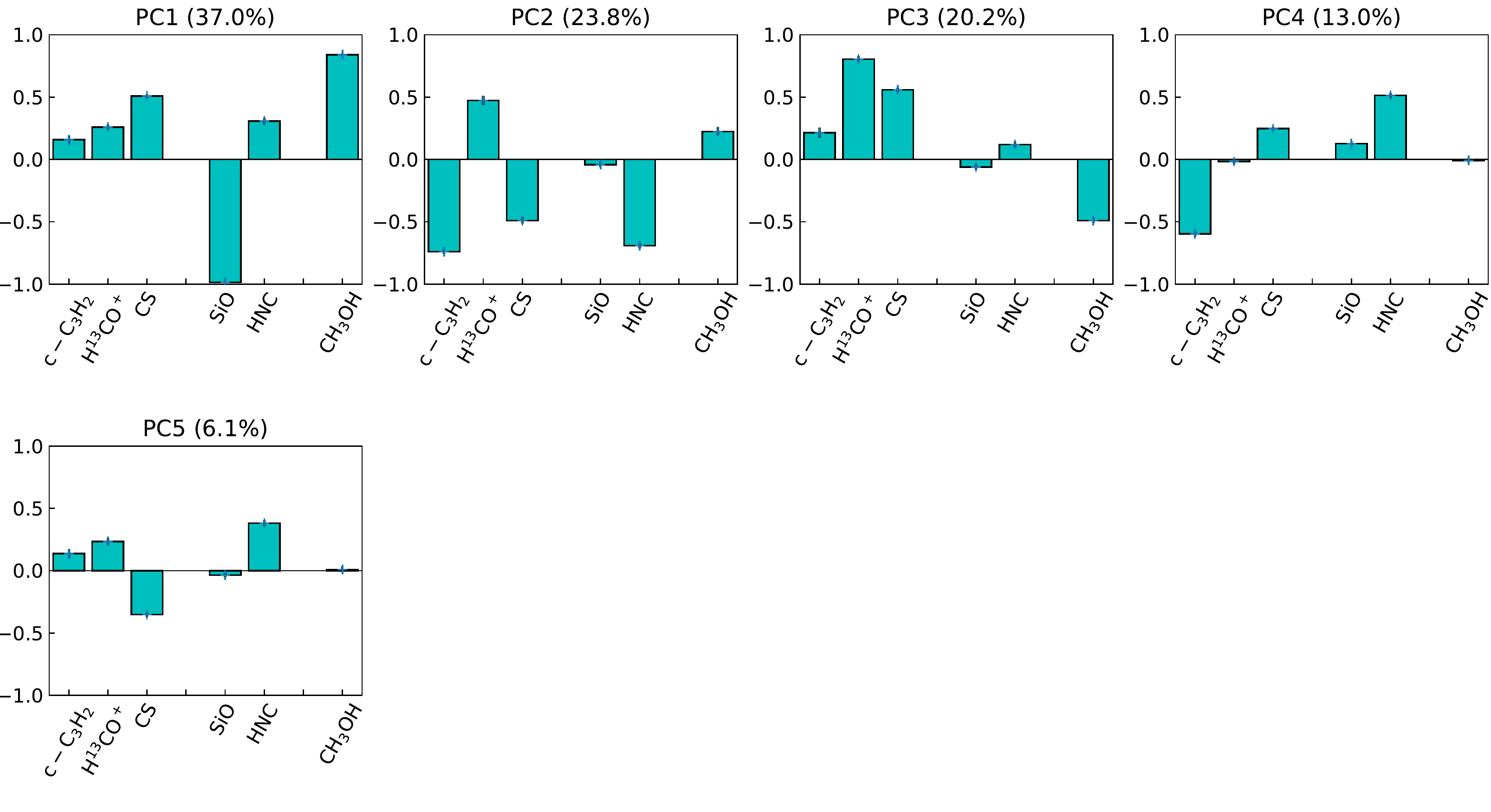}
\caption{Same as Fig.~\ref{pca_contr_channel12} but for $\varv_\mathrm{LSR}=$17.7\,km\,s$^{-1}$. } 
\label{pca_contr_channel11}
\end{figure*}

The principal components and principal component coefficients for the investigated velocities except for $\varv_\mathrm{LSR}=24.7$\,km\,s$^{-1}$ are shown in Figs.~\ref{pca_components_channel0}--\ref{pca_contr_channel11}.

\section{Mopra maps and spectra}
We show here channel maps of four velocity components seen in $^{13}$CO 1--0 emission with Mopra, and the averaged Mopra emission spectra of HNC 1--0 and CS 2--1 towards Sgr B2.

\begin{figure*}
\centering
\includegraphics[width=17cm, trim = 1.5cm 11.cm 4.cm 3.cm, clip=True]{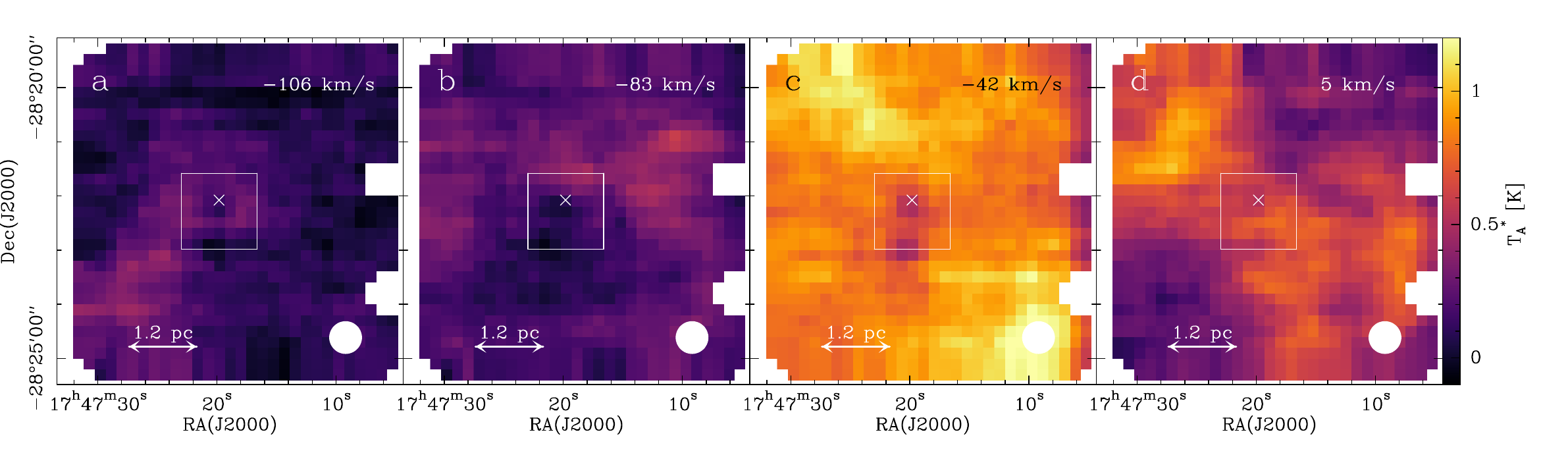}
\caption{Same as Fig.~\ref{hcop_hnc_cs_mopra_map}, but for $^{13}$CO 1--0 at 11\,km\,s$^{-1}$ (panel \textbf{a}), $-41$\,km\,s$^{-1}$ (\textbf{b}), $-83$\,km\,s$^{-1}$ (\textbf{c}), and $-106$\,km\,s$^{-1}$ (\textbf{d}). }
\label{13co_mopra_map}
\end{figure*}

\begin{figure}[t]
   \resizebox{\hsize}{!}{\includegraphics[width=0.5\textwidth,trim = 1.0cm 3.cm 5.cm 2.5cm, clip=True]{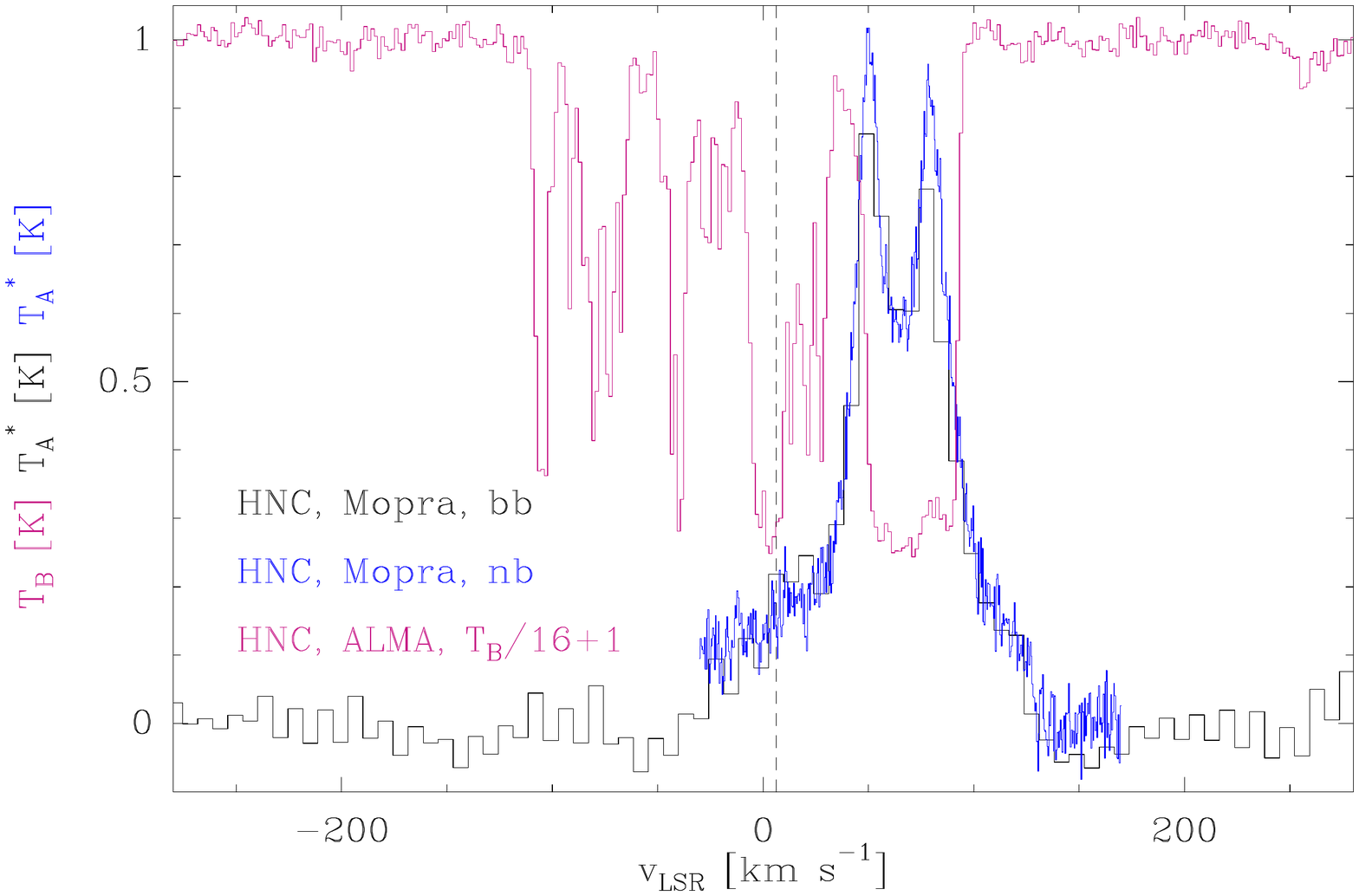}}
 \caption{Same as Fig.~\ref{spectrum_mopra_hcop}, but for HNC 1--0.}\label{spectrum_mopra_hnc}
\end{figure}

\begin{figure}[t]
   \resizebox{\hsize}{!}{\includegraphics[width=0.5\textwidth,trim = 1.0cm 3.cm 5.cm 2.5cm, clip=True]{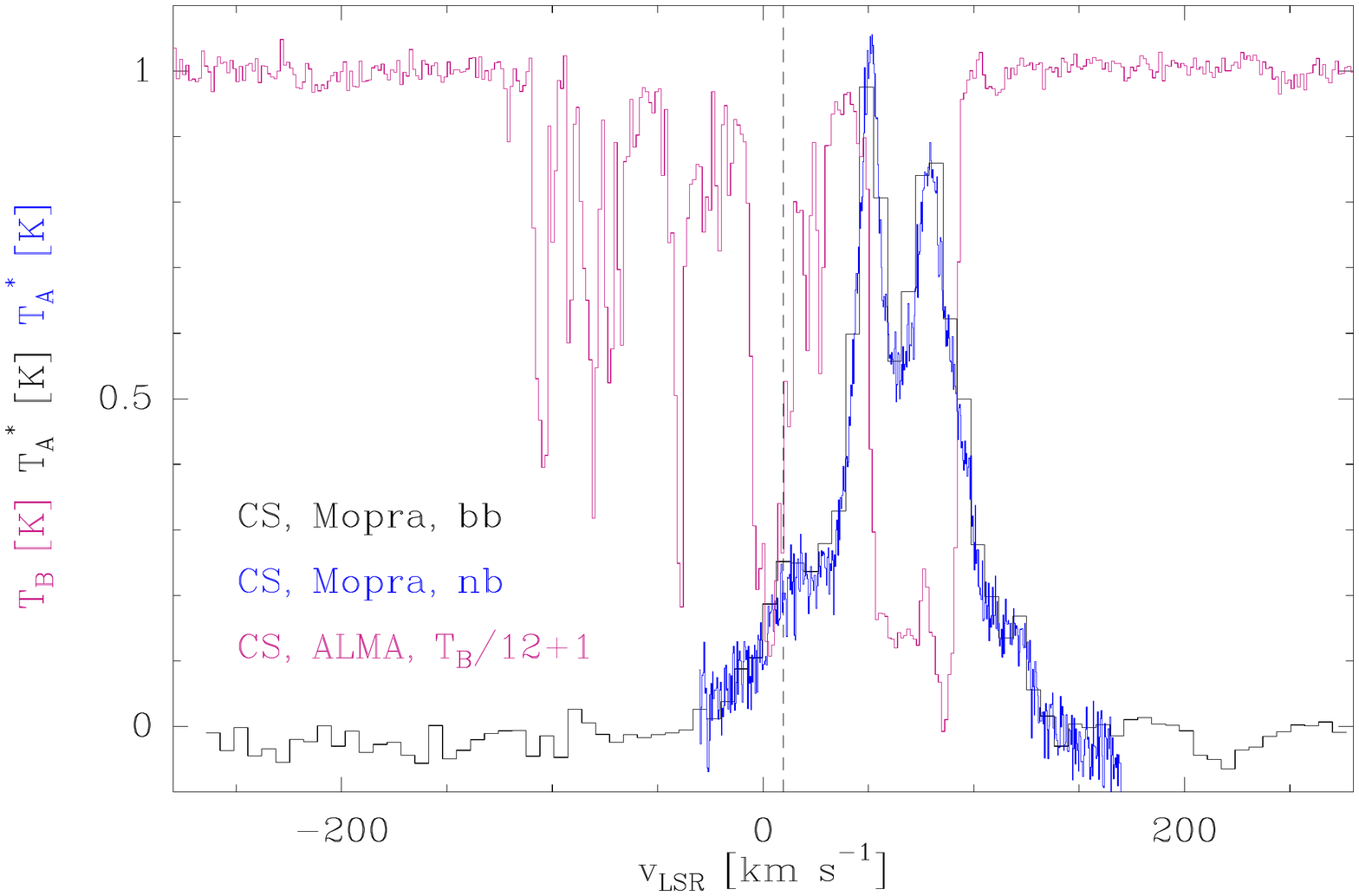}}
 \caption{Same as Fig.~\ref{spectrum_mopra_hcop}, but for CS 2--1.}\label{spectrum_mopra_cs}
\end{figure}

\section{RADEX models}
\begin{figure*}
\centering
\includegraphics[width=17cm, trim = 2.9cm 0.7cm 1.5cm 1.2cm, clip=True]{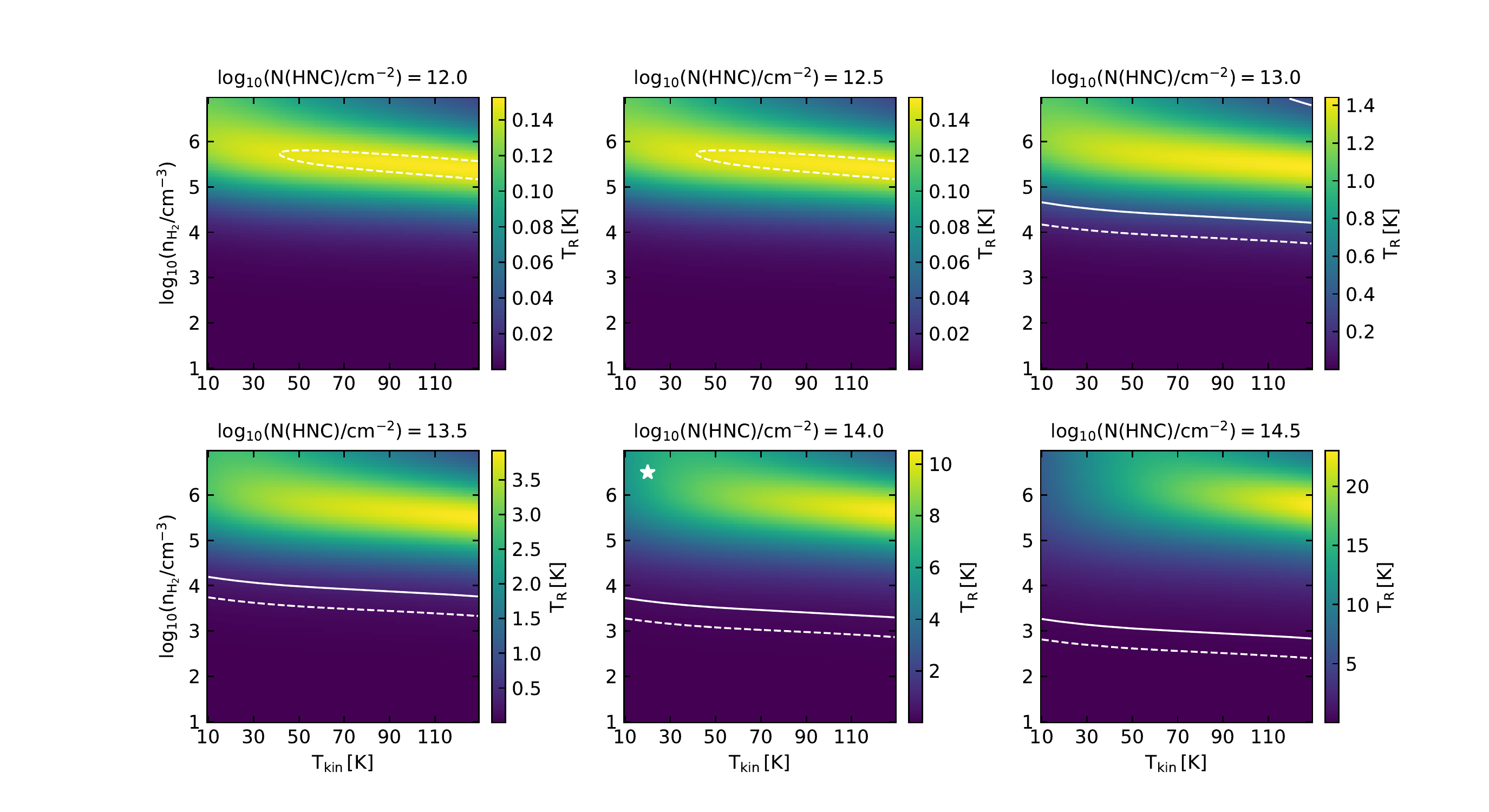}
\caption{Same as Fig.~\ref{radex_hcop}, but for HNC 1--0.} 
\label{radex_hnc}
\end{figure*}

\begin{figure*}
\centering
\includegraphics[width=17cm, trim = 2.9cm 0.7cm 1.5cm 1.2cm, clip=True]{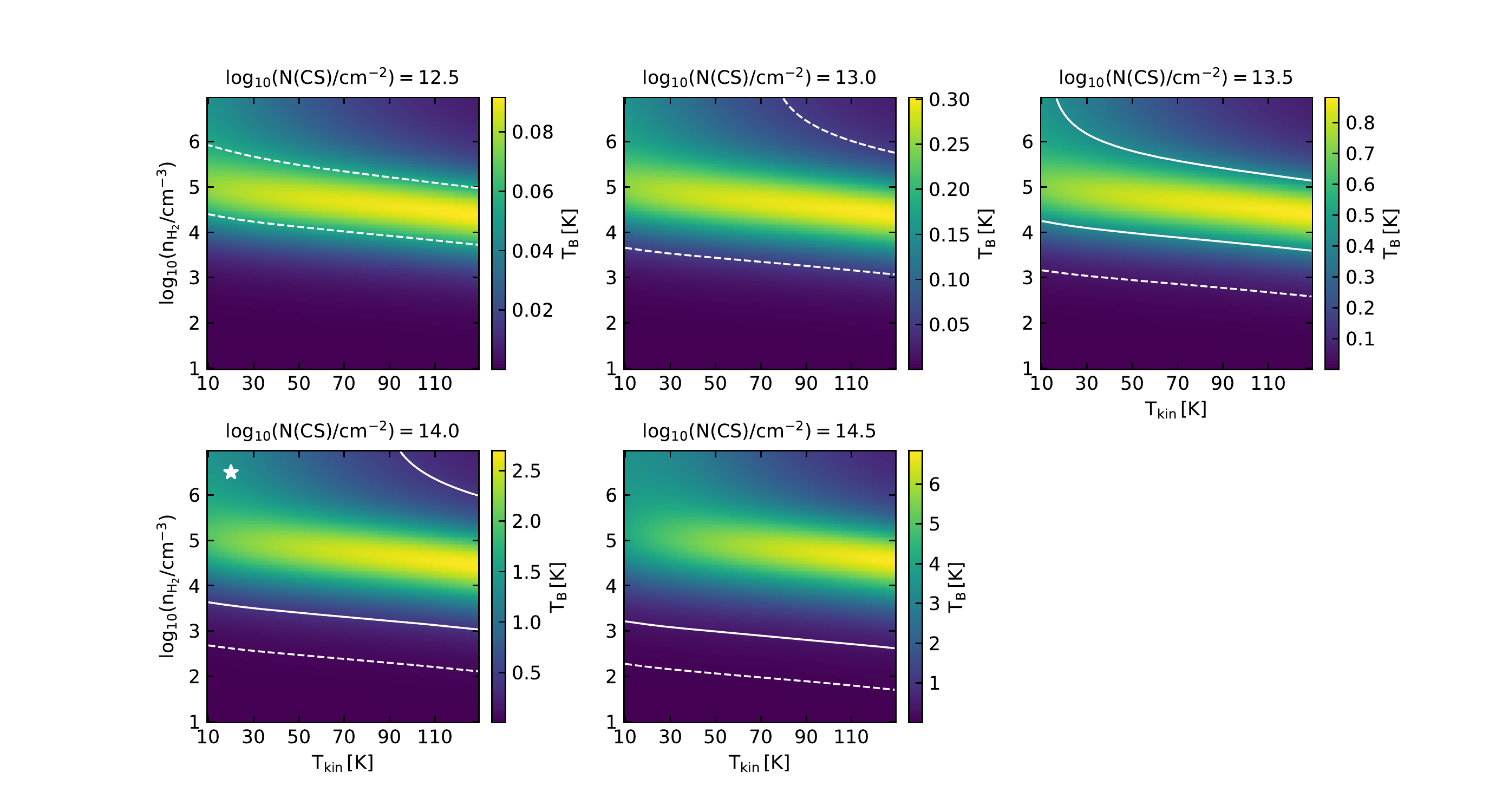}
\caption{Same as Fig.~\ref{radex_hcop}, but for CS 2--1.} 
\label{radex_cs}
\end{figure*}

\begin{figure*}
\centering
\includegraphics[width=17cm, trim = 2.9cm 0.7cm 1.5cm 1.2cm, clip=True]{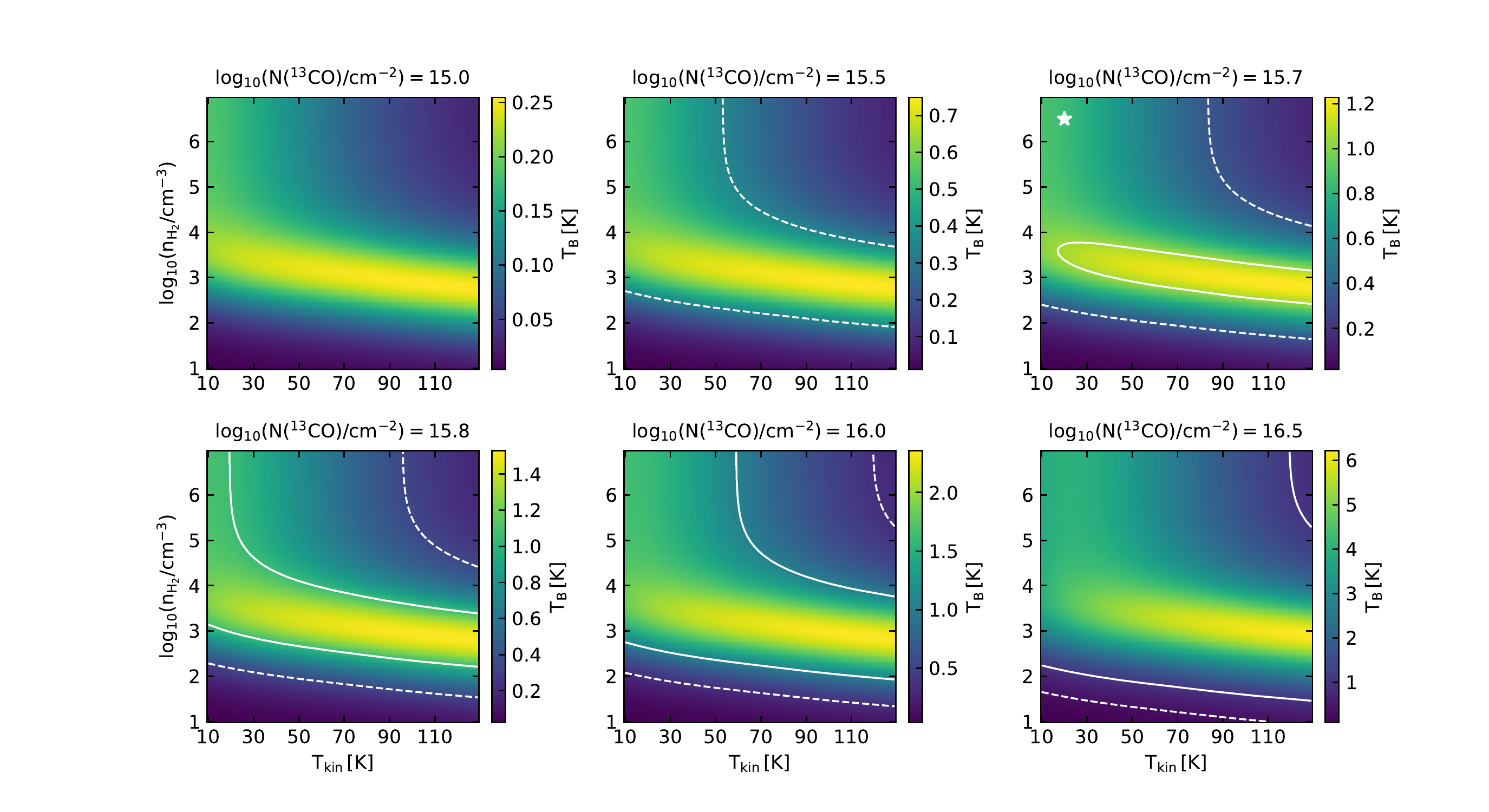}
\caption{Same as Fig.~\ref{radex_hcop}, but for $^{13}$CO 1--0. The dashed lines indicate the main beam temperature measured for $\varv_\mathrm{LSR}=-83$\,km\,s$^{-1}$.} 
\label{radex_13co}
\end{figure*}

Results of our RADEX calculations for HNC 1--0, CS 2--1, and $^{13}$CO 1--0 are shown in Figs~\ref{radex_hnc}--\ref{radex_13co}.

\section{Nature of l.o.s. clouds}
\begin{figure*}
\centering
\includegraphics[width=17cm, trim = 0.cm 0.cm 5.cm 1.8cm, clip=True]{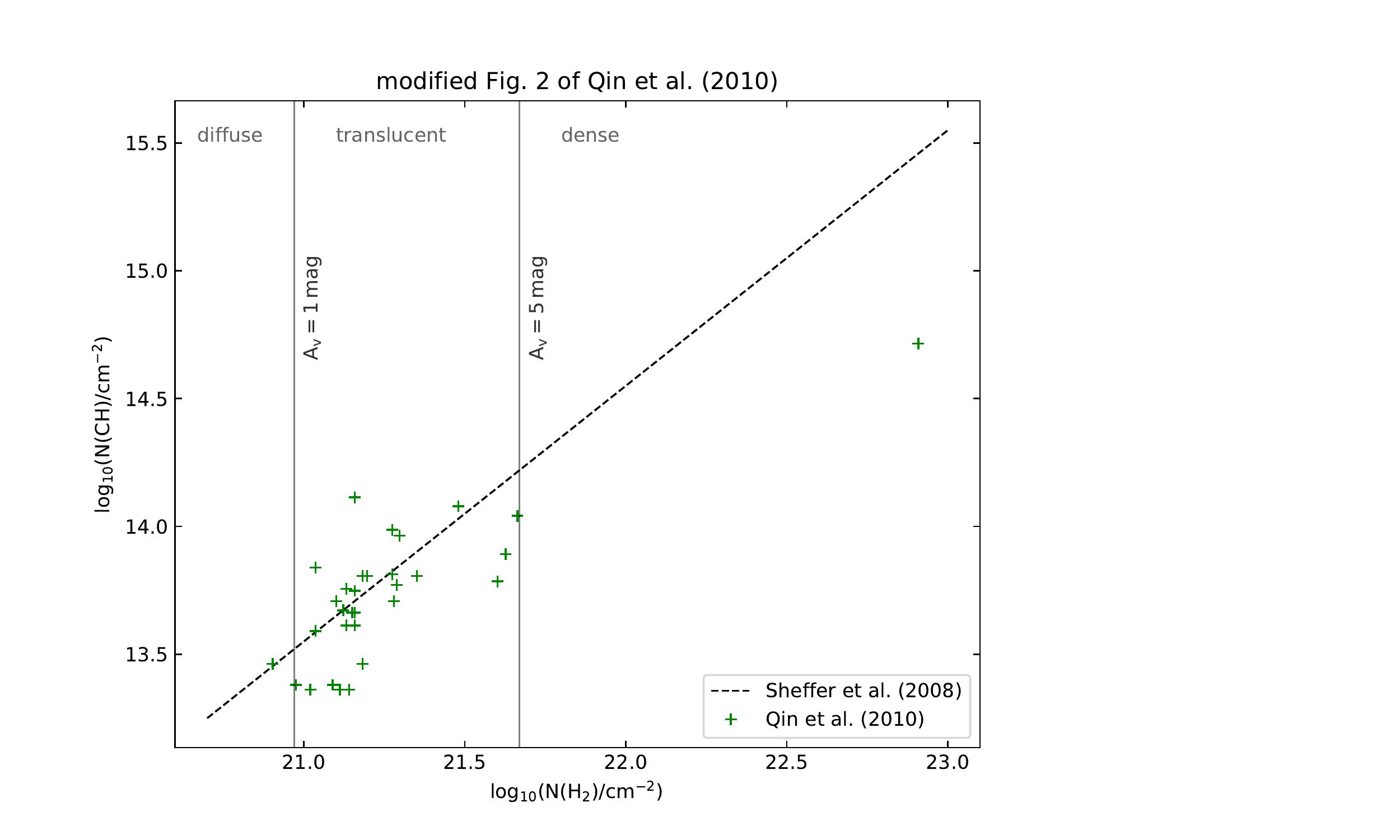}
\caption{Based on Fig.~2 {of} \citet{qin2010}. We rescaled the H$_2$ values, which they derived from HCO$^+$ assuming a uniform abundance, using the HCO$^+$ abundance profile discussed in Sect.~\ref{subsect_discuss_types}. The CH column densities in the translucent regime now follow the same correlation with H$_2$ as derived for diffuse clouds by \citet{sheffer2008}.} 
\label{ch_h2}
\end{figure*}

\begin{figure*}
\centering
\includegraphics[width=17cm, trim = 1.6cm 0.6cm 1.7cm 1.7cm, clip=True]{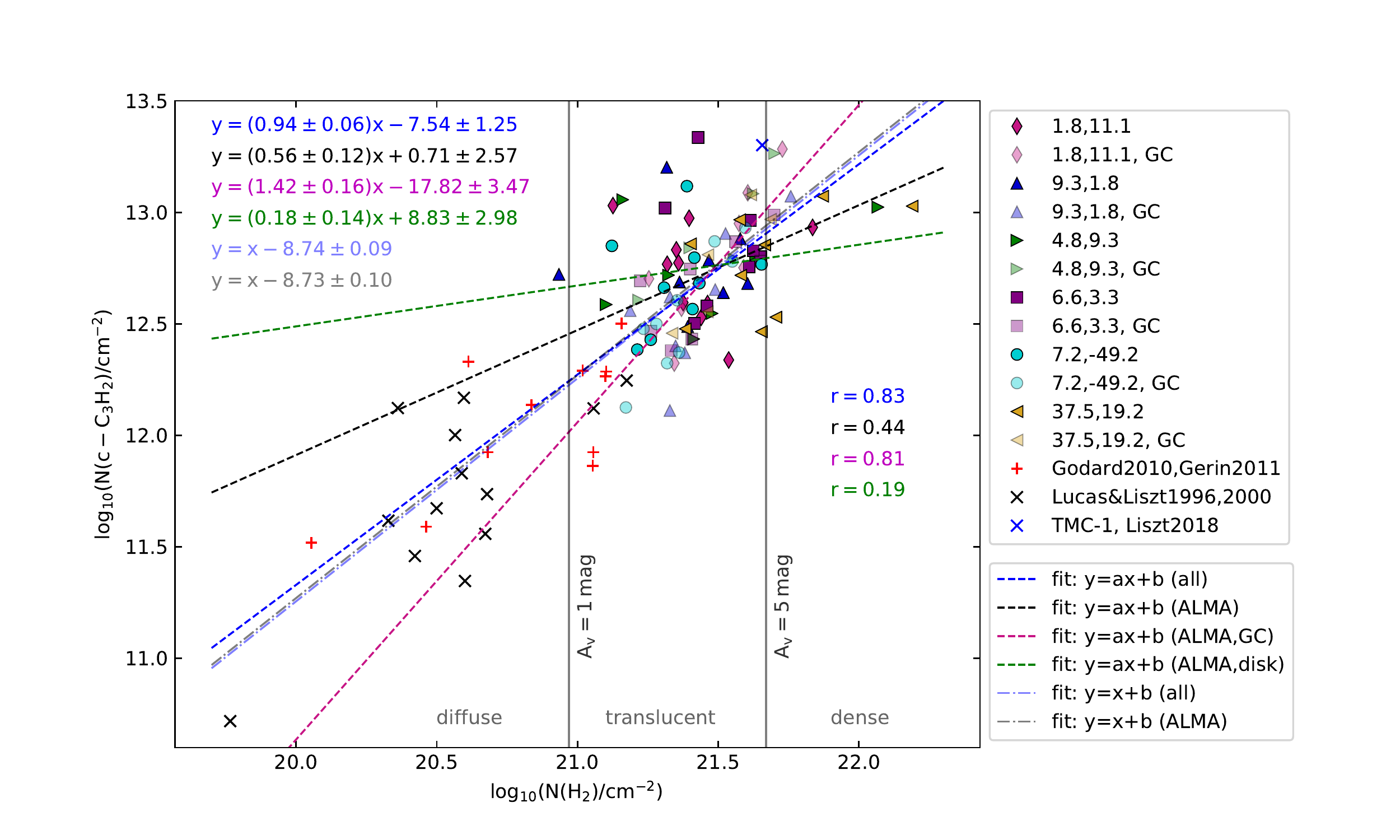}
\caption{Same as Fig.~\ref{cch_hcop}, but for c-C$_3$H$_2$. The blue cross represents the translucent cloud TMC-1 \citep{liszt2018}. } 
\label{c3h2_hcop}
\end{figure*}

\begin{figure*}
\centering
\includegraphics[width=17cm, trim = 1.6cm 0.6cm 1.7cm 1.7cm, clip=True]{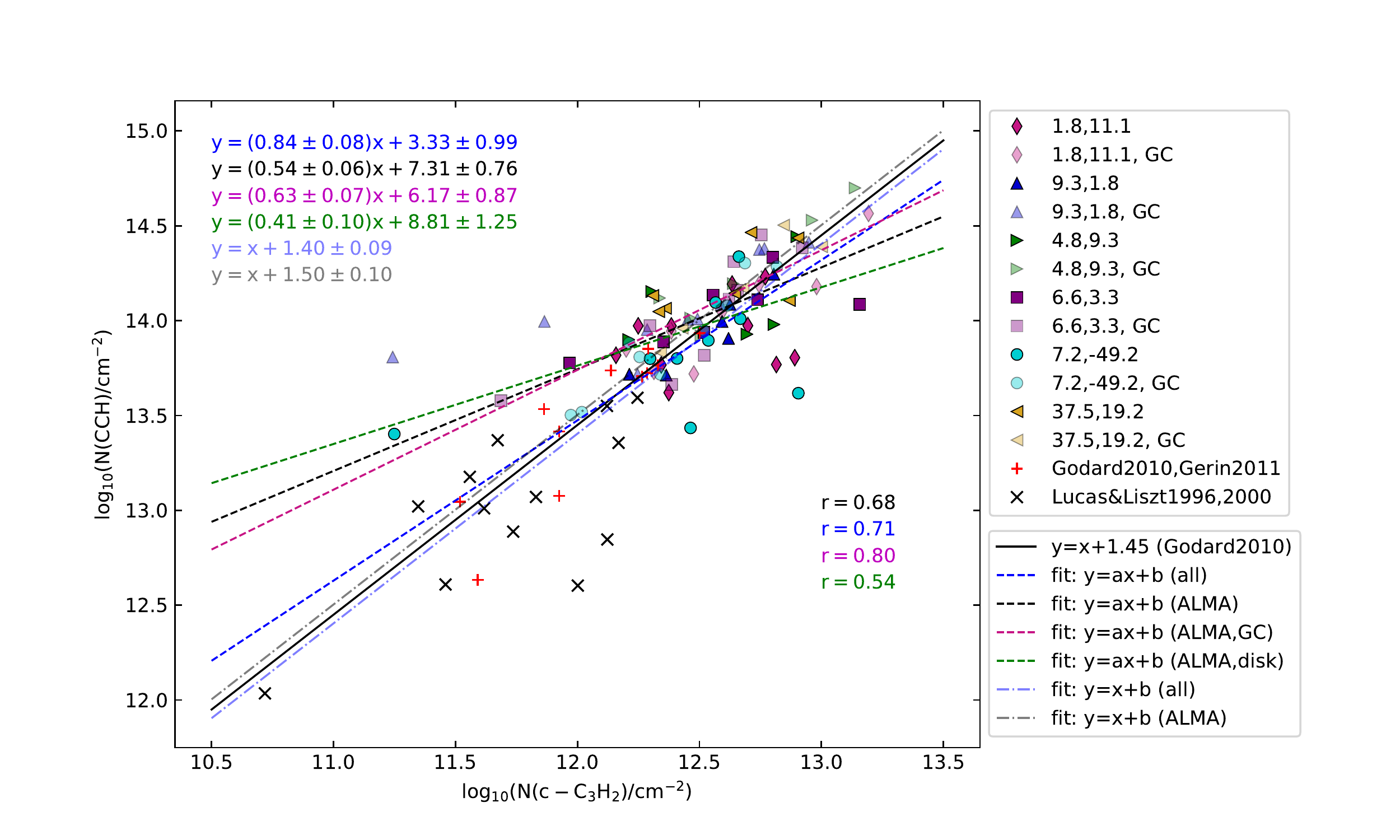}
\caption{Same as Fig.~\ref{cch_hcop}, but CCH plotted against c-C$_3$H$_2$. The black line represents a linear fit using only data of diffuse molecular clouds \citep{godard2010}.}
\label{c3h2_cch}
\end{figure*}

A modified version of Fig.~2 of \citet{qin2010} is plotted in Fig.~\ref{ch_h2} after correcting the H$_2$ column densities for the non-uniform abundance profile of HCO$^+$ discussed in Sect.~\ref{subsect_discuss_types}. The column densities of c-C$_3$H$_2$ and H$_2$ calculated from HCO$^+$ are plotted against each other in Fig.~\ref{c3h2_hcop}. The column densities of CCH and c-C$_3$H$_2$ are plotted against each other in Fig.~\ref{c3h2_cch}.

\end{appendix}
\end{document}